# Formalizing Patient-Set Semantics for OMOP CDM-Based Clinical Trial Feasibility Assessment: A Neuro-Symbolic Methodology for Traceable Site-Feasibility Patient Counts

Sohyeon Jeon

sohyeonclaire@gmail.com

https://orcid.org/0009-0002-3357-7060

**Background.** Clinical trial site feasibility assessment often requires estimating how many potentially eligible patients are available at a candidate institution. This available patient count is not a simple database quantity to be retrieved. It becomes computable only after a protocol-derived eligibility criterion has been specified as an executable patient-set meaning within a particular data environment. Natural-language criteria are clinically interpretable but under-specified for OMOP CDM execution, requiring choices about standard concepts, temporal anchors, observation windows, proxy variables, exclusion scope, local data availability, patient-set operations, and SQL. The central informatics problem is therefore not only query generation, but prior specification of the clinical quantity to be counted as a reviewable and computable patient-set object.

**Objective.** This study reframes the site-level count of potentially eligible patients not as a raw database output, nor as the discovery of a hidden true cohort, but as the cardinality of a declared executable patient-set semantics. The central question is how protocol-derived feasibility criteria can be specified as executable patient-set definitions over OMOP CDM, and how preservation of their meaning can be evaluated across counts, person sets, traceability, and feasibility-relevant site decisions.

**Methods.** This study proposes NS-CoT, a neuro-symbolic framework for specifying and executing patient-set definitions over OMOP CDM. The framework separates language-model-based interpretation from deterministic symbolic computation, so that eligibility criteria are represented as typed cohort semantics before being grounded, composed, and compiled into SQL. The resulting artifact links each count to its concept sets, temporal and exclusion logic, patient-set operations, SQL, execution trace, and intermediate cohort construction. Evaluation focused on whether the specified semantics were preserved in counts, person sets, failure localization, and feasibility-oriented site decisions.

**Results.** Explicit patient-set semantics enabled cohort computation to be evaluated beyond SQL executability. NS-CoT showed higher fidelity in temporal-inclusion and anchor-bound exclusion queries, and comparable fidelity in template-derived OMOP queries. Person-set verification showed that exact count agreement corresponded to agreement in the underlying patient sets. In a multi-schema feasibility simulation, intermediate artifacts supported failure localization, targeted repair, and improved preservation of counts, patient sets, and site rankings. Protocol-derived case studies showed that feasibility counts were interpretable only when computability boundaries, proxy choices, local executability, and exclusion semantics were explicit.

**Conclusion.** This study reframes clinical trial site feasibility assessment as an informatics problem of specifying, executing, and evaluating patient-set definitions over OMOP CDM. Site-level counts are defined not as raw database outputs or endpoints of SQL generation, but as the cardinality of declared executable patient-set semantics. By separating the input criterion, semantic representation, OMOP concept grounding, executable cohort logic, compiled SQL, executed patient set, final count, and trace metadata, the framework makes the meaning and execution path behind each count reviewable. More generally, the study shows that once clinical meaning is specified as an explicit computable object, its patient set, count, execution trace, mismatch, semantic fidelity, and decision impact can be measured and compared. This provides a reusable methodological basis for evaluating computable clinical objects beyond trial feasibility.

# Table of Contents

# List of Figures

# List of Tables

# Chapter 1. Introduction

## 1.1. Site-Level Patient-Count Estimation in Clinical Trial Site Feasibility Assessment

Clinical trial site feasibility assessment is commonly conducted using structured questionnaires sent by sponsors or contract research organizations to candidate sites[1,2]. These questionnaires evaluate multiple dimensions of trial readiness, including investigator experience, study-team capacity, institutional review board and contract timelines, facilities, equipment, laboratory capability, budgetary issues, competing trials, and prior trial performance[2,3]. Within this broader process, one component depends particularly on institutional clinical data: estimating how many patients at a candidate site may satisfy the protocol-derived inclusion and exclusion constraints that are computable from available structured data[1,4].

This study focuses on the data-dependent patient-count component of site feasibility assessment. Specifically, it addresses an upstream computation: how patient-count questions derived from clinical trial inclusion and exclusion criteria can be transformed into executable and reviewable patient-set logic over standardized institutional research data. The goal is not to automate the entire site feasibility questionnaire, evaluate institutional readiness as a whole, conduct direct patient recruitment, or screen individual patients for enrollment. The inputs considered in this study are feasibility-oriented patient-count questions, not unrestricted eligibility criteria intended for full clinical adjudication.

In practice, site-feasibility patient-count estimates are often derived from investigator experience, local knowledge, historical recruitment performance, or informal assessment of the clinical population[1,3]. Such estimates are operationally convenient, but they may be unreliable when the patient-count question involves multiple diagnoses, drug exposures, laboratory thresholds, temporal windows, and exclusion conditions[4,5]. EHR/EMR systems contain the source clinical records that could support more evidence-based estimates, and querying the institutional EHR is the most direct way to obtain them[6]. However, direct EHR use for feasibility assessment is constrained by privacy and governance requirements, legal and operational barriers, and substantial heterogeneity across local information systems[7–9]. These constraints increasingly motivate the use of standardized, governed research-data infrastructure rather than ad hoc EHR queries: source records are harmonized to a common data model and pseudonymized or de-identified for research use, and, in the distributed-research-network paradigm, analytic code is executed locally so that only aggregate results, not patient-level data, leave each institution[10–13].

This study therefore uses the OMOP Common Data Model (OMOP CDM) as a standardized data model and vocabulary layer within governed institutional research-data repositories[14,15]. OMOP CDM provides a common schema and standard vocabularies for representing clinical records across institutions, but its availability does not by itself make site-level patient-count estimation straightforward[15–17]. The inclusion and exclusion constraints underlying site feasibility patient-count questions are usually expressed as natural-language sentences or bullet-point criteria[18]. Although these statements may be clinically understandable at the protocol-design or site-assessment stage, they cannot be directly executed against structured clinical data[4]. To identify the corresponding patient set in an institutional OMOP CDM, these patient-count questions must first be expressed as computable cohort logic[17].

In this study, a patient count is treated as the cardinality produced when an explicitly specified cohort definition is executed over an institutional OMOP CDM. The focus is therefore not the count alone, but the methodology by which site feasibility patient-count questions based on clinical trial inclusion and exclusion criteria are transformed into patient-set logic executable over OMOP CDM. From the perspective of clinical trial design and feasibility assessment, the core problem is not simply SQL generation[19,20]. It is the semantic operationalization of computable protocol-derived constraints into executable cohort logic over structured clinical data[10,16].

In this sense, making a site-level patient count available is itself a semantic and computational act[21,22]. A feasibility questionnaire may ask for a single number, but the number is interpretable only after the clinical criterion has been transformed into a declared patient-set object: which clinical events count, which standard concepts represent them, which date serves as the anchor, which observation window is used, which exclusions are scoped to which events, which proxies are allowed, and which local data elements are available for execution[23–27]. Without this prior commitment, count accuracy is not stably defined, because different plausible operationalizations of the same protocol sentence may denote different patient sets and therefore different *available N* values[21]. This study therefore treats feasibility counting as the construction, execution, and evaluation of computable patient-set semantics, rather than as the retrieval of a pre-existing database number.

It is useful to distinguish this site-level patient-count estimation from statistical sample size calculation. Sample size calculation determines the number of participants a study requires to achieve adequate statistical power for a target effect size, and operates on design parameters rather than institutional data[4]. Site-level patient-count estimation, in contrast, asks how many potentially eligible persons are available within a specific institution's recorded clinical data[1]. Its input is therefore not a pre-specified statistical target but a free-text eligibility description, which must be interpreted and operationalized as executable patient-set logic over that institution's recorded clinical history. This is precisely why natural-language criteria are central to the task: the difficulty lies not in arithmetic over a known target, but in committing an ambiguous clinical description to a reviewable, executable interpretation before any count can be produced.

This transformation becomes necessary because site feasibility assessment often requires patient-count questions in which clinical trial inclusion and exclusion criteria function as computable constraints[1,5]. In governed research-planning contexts, investigators and data teams may need to derive corresponding counts, and in some cases intermediate patient sets, from institutional data[3,4]. In standardized clinical data environments such as OMOP CDM, such retrieval ultimately requires SQL or SQL-compilable cohort definitions[15–17]. However, constructing this logic remains a substantial technical barrier for clinicians and clinical researchers[18,19]. It requires familiarity with OMOP table structures, standard vocabularies, concept hierarchies, domain-specific date fields, temporal windows, and exclusion semantics[14–17]. For this reason, one common line of prior work has approached the task as direct or LLM-assisted text-to-SQL translation from eligibility text[19,20].

## 1.2. Why Direct Text-to-SQL Is an Incomplete Framing

The transformation from site feasibility patient-count questions to executable cohort logic is sometimes framed as a text-to-SQL problem: given a site feasibility patient-count question, generate a database query that returns the corresponding patient set or count[19,20,28]. This framing is useful because it emphasizes executability[19,23]. However, it is incomplete for site-level patient-count estimation over OMOP CDM[4]. In this setting, the methodological challenge lies not only in generating syntactically valid SQL, but also in operationalizing eligibility constraints, grounding that interpretation in standard vocabularies, and composing the resulting constraints as patient-set logic before SQL execution[29,30].

A simple example illustrates the issue. Consider a patient-count question constrained by the criterion “patients with diabetes on insulin therapy.” At the level of natural language, this statement may appear straightforward. In an OMOP CDM environment, however, it can correspond to several distinct executable interpretations[23,31]. It may refer to patients with diabetes who are currently receiving insulin therapy, patients with diabetes who have any prior history of insulin exposure, or patients whose insulin exposure occurred on or after diabetes diagnosis[30]. Each interpretation can be expressed in SQL, but each requires a different index event, drug-exposure window, temporal anchor, and patient-set operation[29]. In a site feasibility setting, these are not merely alternative technical implementations[17]. They may produce different estimates of the site-level count of potentially protocol-eligible patients[23].

The same issue becomes more pronounced when patient-count questions contain multiple eligibility constraints involving clinical events, exclusions, measurement thresholds, or temporal dependencies[4,30]. A criterion such as "patients with diabetes who are either on metformin without insulin therapy or on insulin with HbA1c > 9%" requires patient-set intersection, union, exclusion, and measurement-threshold logic[29]. A criterion such as "started an ACE inhibitor within three months after hypertension diagnosis" requires an anchor diagnosis event, a drug exposure event, and a bounded temporal relation between them[30]. These examples show that site-level patient-count estimation is not simply a matter of translating surface text into SQL syntax[28]. It requires reconstructing the clinical meaning units that must be preserved before the criterion can be compiled into executable cohort logic[23,29].

For this reason, query execution or count agreement alone may not be sufficient to evaluate whether a cohort-definition system has preserved the intended criterion[32]. A generated SQL query may execute successfully while using an overly broad concept, omitting descendant concepts, assigning a criterion to the wrong OMOP domain, applying an unintended time window, or implementing an exclusion as unscoped negation rather than anchor-bound absence[14,20,31,33]. Such errors may still yield a numeric count, but the count may reflect a cohort interpretation different from the one intended in the feasibility question[23].

The concern, therefore, is not that SQL is unnecessary. SQL remains the execution form through which patient sets and counts are obtained from OMOP CDM[14,16]. Rather, the concern is that direct or LLM-centric query-generation approaches can leave distinct informatics decisions—clinical concept selection, standard-vocabulary grounding, concept hierarchy expansion, OMOP domain-table routing, temporal anchoring, exclusion semantics, patient-set algebra, and SQL composition—implicit within the query-generation process[28,31]. When these decisions are not represented explicitly, it becomes difficult to inspect how the patient count was produced, reproduce the same interpretation later, or localize errors when the generated count is incorrect.

This study therefore treats SQL generation as the final compilation step, not as the primary methodological object[16]. The primary object is the explicit operationalization of cohort semantics before execution: the declared patient-set meaning to which generated SQL should be accountable. In benchmark settings, predefined OMOP-executable reference logic is used to evaluate functional fidelity. In real feasibility settings, however, the same principle means that clinical, site-specific, and data-availability choices must be externalized as a reviewable semantic commitment rather than hidden inside SQL. In this framing, site-level patient-count estimation over OMOP CDM is a clinical research informatics problem of semantic operationalization, with SQL serving as the executable target language[14,23,34,35].

## 1.3. The Computability Gap Between Feasibility Patient-Count Questions and OMOP CDM-Executable Cohort Logic

The preceding sections framed the problem as site-level patient-count estimation rather than direct patient recruitment or simple SQL generation. The remaining question is why this transformation remains difficult even when standardized clinical data are available. The difficulty arises from a computability gap between the way site feasibility patient-count questions are expressed in natural language and the way patient-level evidence must be represented and executed in OMOP CDM[19,29,36].

This gap is not merely theoretical. Prior studies have repeatedly shown that clinical trial eligibility criteria are only partially computable in structured data environments[36–39]. In an ASCO 2021 study of 2,954 oncology trials, computable attributes could be extracted from eligibility criteria, but a substantial portion required customized concepts rather than standard terminologies[39]. CriteriaMapper similarly reported that only a subset of unique eligibility-criteria attributes could be normalized[23]. Evaluations of OHDSI ATLAS have also shown that only part of trial inclusion and exclusion criteria can be fully implemented as structured cohort definitions[33]. Taken together, these findings suggest that the difficulty is not limited to SQL authoring or automated query generation[19]. It reflects linguistic, semantic, terminological, temporal, and data-model barriers that arise when human-readable trial criteria must be reconstructed as executable cohort logic[29,36].

This gap is especially consequential in site feasibility assessment[11,37]. In this setting, the practical question is not only whether a criterion can be translated into some executable query, but whether the resulting count corresponds to a clinically defensible operationalization of the site feasibility patient-count question[19,33]. A candidate site may appear feasible or infeasible depending on how expressions such as “current treatment,” “newly diagnosed,” “within the past six months,” “without prior exposure,” or “potentially protocol-eligible patients” are operationalized[30]. These expressions require choices about standard concepts, descendant expansion, index events, observation windows, event dates, temporal scopes, and exclusion logic[29,30,36]. Different choices can produce different site-level patient-count estimates[33,40].

The problem addressed here is therefore not patient recruitment itself. A recruitment system asks how eligible patients can be contacted, screened, and enrolled[4,18]. The problem addressed in this study is upstream: how a natural-language site feasibility patient-count question can be transformed into executable and reviewable cohort logic over an institutional OMOP CDM[19,29]. The count and, where governance permits, the intermediate patient set produced by such logic may later inform recruitment planning, but patient contact, enrollment workflow, and clinical adjudication of individual eligibility remain outside the direct scope of this work[11,18].

A standardized data model reduces part of the difficulty, but it does not remove the computability gap[29,33,36]. OMOP CDM provides structural interoperability by representing heterogeneous clinical records through a common table structure and standard vocabulary framework[15,29]. However, a standardized schema does not automatically determine how a natural-language criterion should be interpreted[19,36]. Expressions such as “diabetes,” “insulin therapy,” “severe renal impairment,” “after diagnosis,” and “no prior statin use” must be represented as OMOP concept sets, domain-specific event tables, date columns, anchor events, temporal windows, value constraints, and patient-set exclusions[29,30,36]. Thus, semantic interoperability in this setting extends beyond mapping terms to standard vocabularies[23,29]. It concerns whether the clinical meaning relevant to the feasibility question can be preserved as computable patient-selection logic[19,29]. This framing builds on prior work that treated clinical trial eligibility criteria as objects requiring formal knowledge representation, including expression languages, encoded clinical concepts, and patient data models, but focuses specifically on the denotation and decision-level use of OMOP-executable patient-set definitions[41].

This study uses the term semantic fidelity in a restricted computational sense. It does not refer to chart-reviewed clinical eligibility, clinical phenotype validity, or prospective recruitment performance. In benchmark settings, it refers to the extent to which the computationally relevant meaning embedded in a site feasibility patient-count question is preserved with respect to an executable reference definition. In real-trial and local deployment settings, where a single clinical gold standard may not be self-evident, the same concept refers to whether the declared executable interpretation and its assumptions are made explicit and preserved through grounding, patient-set logic, SQL compilation, and execution. This includes, at minimum, clinical concept identity, standard-vocabulary grounding, descendant concept expansion, OMOP domain-table routing, event-time anchoring, temporal-window interpretation, inclusion and exclusion roles, value and unit constraints where applicable, and patient-set composition[29,30,36].

A feasibility count is therefore not considered meaningful merely because a SQL query returns a number[33]. It is meaningful only to the extent that the cohort logic producing the number can be inspected, reconstructed, and related back to the intended executable interpretation[19,33,36]. Operationally, this means that inclusion conditions are treated as cohort-retaining requirements, whereas exclusion conditions are treated as exclusion-triggering constraints; the formal representation is introduced in Chapter 3.

In this sense, the central informatics problem is not whether standardized clinical data exist, nor whether SQL can be generated[19,28]. It is whether the feasibility question can be made computable under explicitly stated assumptions and with sufficient functional fidelity to the specified executable cohort semantics to justify the resulting count as a feasibility input[11,16,20,21]. This computability gap motivates the methodological thesis of this study: site feasibility patient-count questions with inclusion and exclusion constraints should be transformed through explicit intermediate representations, OMOP vocabulary grounding, event-time relations, exclusion semantics, patient-set algebra, and deterministic SQL compilation rather than leaving these decisions implicit inside a single or predominantly generative query-generation step.

## 1.4. Research Aim and Methodological Thesis

This study formalizes the executable patient-set meaning of feasibility-oriented clinical trial criteria over OMOP CDM. The objective is not merely to generate SQL from eligibility text, but to define the patient-set semantics that SQL is expected to implement before the resulting count is interpreted as feasibility evidence. NS-CoT is proposed as a neuro-symbolic reference implementation of this formalization.

In this formulation, the executable interpretation is not treated as a hidden clinical truth discovered by the system. It is a declared semantic commitment that can be inspected, revised, and agreed upon for a given feasibility context.

The methodological thesis is as follows. For the subset of site feasibility patient-count questions whose constraints can be represented using structured fields in OMOP CDM, those constraints can be operationally formalized over the event-based representation of OMOP CDM through standard-vocabulary grounding, concept hierarchy traversal, temporal logic, exclusion semantics, and patient-set algebra[14,15,29,30]. This formalization can then be implemented through a neuro-symbolic architecture that separates the natural-language interpretation capacity of large language models from the execution semantics of a deterministic symbolic compiler[20,42,43].

To implement this thesis, the transformation from natural-language site feasibility patient-count questions to executable cohort-computation logic is decomposed into a sequence of medical informatics decisions: clinical concept selection, standard-vocabulary grounding and expansion, OMOP domain-table routing, temporal anchoring, inclusion and exclusion semantics, value and unit constraint handling where applicable, patient-set algebra, and SQL composition[19,29,30,36]. LLM-centric query-generation approaches often leave many of these decisions implicit within prompt-mediated or predominantly generative SQL construction[20,28]. In contrast, NS-CoT restricts language-model use to semantic interpretation, criterion decomposition, reasoning-guided planning, typed intermediate-representation extraction, and structured validation of candidate concepts. OMOP vocabulary traversal, local CDM-aware executable concept filtering, temporal and exclusion reasoning, patient-set algebra, and SQL compilation are then handled by deterministic symbolic modules.

This architectural design is not intended to exclude language models from cohort-definition workflows. Rather, it uses them for the tasks to which they are well suited: interpreting flexible natural-language criteria and producing structured semantic representations[20,28,36]. The execution semantics of cohort computation are then fixed in explicit, reviewable, and reproducible structures grounded in OMOP standard vocabularies, event-time relations, and patient-level set operations[14,15,29,30].

## 1.5. Contributions and Scope

The contributions of this study can be summarized in four parts. First, at the problem-framing level, this study reframes site feasibility patient-count questions over OMOP CDM as a problem of executable patient-set semantics rather than as investigator estimation, direct raw EMR screening, or simple text-to-SQL generation. This framing does not assume that each free-text feasibility criterion has a single self-evident gold-standard patient set. Rather, it treats the executable interpretation of a criterion as a declared semantic commitment that must be made explicit before a site-level count can be interpreted as feasibility evidence. In this sense, the contribution is not the discovery of a hidden true cohort, but the externalization of the concept, temporal, exclusion, proxy, observation-window, and patient-set operation choices that determine which patients are counted.

Second, at the formalization level, this study models the eligibility constraints embedded in feasibility-oriented patient-count questions as executable cohort semantics over the longitudinal event representation of OMOP CDM. In this formalization, clinical mentions must be grounded to OMOP concept sets, events must be associated with domain-specific tables and timestamps, temporal expressions must be represented as anchor-bound event-time relations, exclusions must be interpreted as patient-level absence conditions within a defined scope, and the final cohort must be constructed through patient-set operations[29,30,33,36]. When a criterion cannot be represented directly in structured OMOP CDM fields, the framework treats proxy selection as an explicit semantic decision rather than as an invisible implementation detail. SQL is therefore treated as the final execution target, not as the primary semantic representation.

Third, at the methodological and implementation level, this study implements NS-CoT as a neuro-symbolic reference architecture. Natural-language patient-count questions are first structured into typed intermediate representations by an LLM-mediated layer. These representations are then normalized and passed to symbolic modules that perform OMOP vocabulary grounding, concept hierarchy traversal, local CDM-aware executable concept filtering, temporal and exclusion reasoning, patient-set algebra, and deterministic CTE-chain SQL compilation. Within the supported scope, the resulting system produces not only a patient count, but also the concept sets, temporal specifications, exclusion logic, proxy decisions where applicable, generated SQL, and trace artifacts needed to inspect and reconstruct how the count was produced.

Fourth, at the empirical level, this study evaluates whether the proposed architecture preserves declared executable patient-set semantics across complementary evaluation settings. CQ-Bench evaluates semantic components deliberately emphasized by this study, including temporal anchoring, drug exposure, anchor-bound exclusion, and patient-set operations. FOEM evaluates count and patient-set reproduction in an external template-derived OMOP query setting. COVIC extends the evaluation to a multi-schema feasibility simulation, asking whether system-generated site-level patient counts preserve benchmark reference-defined counts, patient sets, site-feasibility labels, and site rankings across candidate OMOP schemas. Protocol-derived clinical trial case studies further examine how trace artifacts expose computability boundaries, proxy-anchor limitations, local executability, and executable-empty concept-set behavior in realistic protocol-derived scenarios. Together, these evaluations assess functional fidelity to declared executable semantics and feasibility-oriented decision information, rather than chart-reviewed clinical phenotype validity or prospective recruitment performance.

This study is methodologically connected to prior first-author work on chain-of-thought-based prompt engineering for medical multiple-choice question answering[44]. That study examined how structured reasoning prompts affect answer selection and reasoning traces in medical question answering. The present study does not extend the same task. Rather, it transfers the methodological concern with reasoning structure to a different output setting: executable OMOP-based cohort computation. In NS-CoT, chain-of-thought is not used as a free-form rationale for a final answer. Instead, reasoning-guided system prompts structure LLM-mediated interpretation into machine-facing strategy fields, atomic facts, temporal and exclusion cues, and typed execution steps that can be consumed by deterministic symbolic modules.

The study introduces a computable evaluation paradigm for this setting. Instead of scoring query correctness alone, it measures functional fidelity in layers: count fidelity; person-set fidelity, since identical counts may denote different patient sets; and, in a multi-site setting, the preservation of the reference-defined site-feasibility decision, measured by AUROC/AUPRC, and of the site ranking, measured by rank correlation and top-site agreement. Recasting feasibility-count reproduction as a measurable decision- and ranking-preservation problem situates the evaluation outside both standard natural-language-to-SQL evaluation, such as execution match and semantic similarity, and phenotype-algorithm validation, such as predictive value against a clinical reference. It is this measurement construct, turning "does this cohort logic preserve the decision it informs?" into a quantitatively evaluable question, that constitutes the study's most transferable contribution, independent of the specific system used to produce the cohort logic.

The scope of this study is intentionally bounded. NS-CoT is not designed to automate the entire site feasibility questionnaire, replace investigator judgment, conduct patient contact or recruitment, perform clinical adjudication of individual eligibility, or demonstrate regulatory compliance[1,4,5,18]. It also does not claim to interpret all possible clinical trial eligibility criteria or all possible site feasibility questions. The present work is limited to selected OMOP-expressible patient-count questions with structured inclusion and exclusion constraints that can be represented using structured CDM fields, standard vocabularies, event dates, value constraints where available, and patient-level set operations[14,15,29,30]. Criteria requiring unstructured notes, imaging interpretation, subjective clinical judgment, unavailable local data elements, or complex multi-step clinical reasoning beyond the supported representation remain outside the direct scope of this implementation[18,45].

The claim is therefore not that NS-CoT is universally more accurate than existing systems across all query distributions or that it fully automates clinical trial eligibility assessment. The central claim is more specific: feasibility-oriented patient-count questions can be transformed into declared, executable, and reviewable patient-set semantics over OMOP CDM. This makes the resulting site-level count an inspectable feasibility input, not an enrollment forecast, a clinical eligibility determination, or a claim to a single hidden gold-standard cohort.

# Chapter 2. Background and Related Work

## 2.1. Clinical Trial Site Feasibility Assessment and the Need for Evidence-Based Patient-Count Estimation

Clinical trial site feasibility assessment is a process for determining, before trial initiation, whether a candidate institution can realistically conduct a planned clinical trial[46]. Sponsors and contract research organizations (CROs) commonly send site feasibility questionnaires or site assessment questionnaires to candidate sites to evaluate trial readiness[1,2,46]. These questionnaires typically assess multiple dimensions of site capability, including investigator experience, study-team composition, institutional review board and contract timelines, facilities and equipment, laboratory capacity, budgetary issues, competing trials, timeline feasibility, and the potential to recruit the target study population[2,3,46,47]. Site feasibility is therefore not a single technical task, but a practical process for assessing the operational and research capacity of a candidate institution[46,47].

The scope of the present study is limited to one data-dependent component of this broader process: site-level estimation of potentially protocol-eligible patients. During the feasibility stage, sponsors and investigators often need to estimate how many patients at a candidate institution may satisfy major protocol-derived inclusion and exclusion constraints, at least to the extent that those constraints are computable from available institutional data[1,4,46]. Actual feasibility questionnaires often ask sites to estimate the number of patients with a given disease, the number of active patients or visits within a specified period, the number of patients satisfying particular diagnostic or laboratory criteria, the number of patients expected to be enrolled within a future recruitment period, and the likely sources of participant identification[2,46,48]. These questions differ in meaning. A count of patients seen in routine care, a count of patients satisfying computable protocol-derived constraints, an expected enrollment estimate, and an actual enrolled count are not interchangeable. This study focuses on the computable patient-count component, not on the full recruitment funnel.

A central problem is that feasibility-stage patient-count estimates do not always predict actual recruitment performance. In the CapaCiTY programme experience, sites selected through feasibility questionnaires frequently failed to meet recruitment expectations during the subsequent trial period[5]. Some sites overestimated their recruitment potential, whereas others recruited no participants. Such findings illustrate that inaccurate feasibility estimates can lead to recruitment delays, underperforming or non-recruiting sites, and inefficient use of time and research resources[5,46]. For the present study, the relevant question is not whether a site will ultimately enroll a given number of participants, but whether the number of patients satisfying a specified computable operationalization of protocol-derived criteria can be estimated in a credible, evidence-based, and reproducible way.

Efforts have also been made to standardize site selection and feasibility assessment[2,46–48]. The standardized automated site assessment survey instrument (SASI), for example, was developed for multicenter trial site selection and collected information such as site commitment time, study-team experience, feasibility of meeting tight timelines, local medical community equipoise, and institutional research service contacts[48]. Such approaches show that site feasibility assessment should not depend solely on individual investigator impressions or informal local judgment[46–48]. Standardized survey instruments can improve the collection of site-readiness and operational information[2,46–48]. However, even when operational readiness is assessed more systematically, the question of how to compute the number of patients satisfying inclusion and exclusion constraints from institutional data remains a distinct informatics problem.

EHR/EMR systems are important potential data sources for making patient-count estimates more evidence-based[1,4]. These systems contain source clinical records generated during routine care, including diagnoses, prescriptions, laboratory tests, procedures, and visits. An interview study of pharmaceutical companies and CROs in the Nordic countries reported that sponsors often relied on historical trial performance databases, investigator relationships, local intelligence, and investigator estimates for site identification and recruitment projections, whereas actual use of EHR data remained limited[1]. Many sites provided patient-count estimates based on investigator experience, but some sites reviewed EHR data to support their estimates, and such EHR-supported evidence was perceived by sponsors as faster and more reliable.

A direct precedent for this problem is the EHR4CR protocol feasibility system evaluated by Soto-Rey et al[40]. In that multi-country study, free-text clinical trial eligibility criteria were simplified by feasibility experts into non-ambiguous feasibility criteria and translated into EHR-executable patient-count queries. The study compared automated EHR4CR counts with conventional clinician questionnaire estimates and an alloyed gold standard derived from manual chart review. It showed that automated patient-count generation could substantially reduce the time needed to obtain feasibility counts, but that count accuracy varied across sites, trials, and data elements. Mismatches were attributed to issues such as insufficiently structured source data, local terminology, missing laboratory or test information, and eligibility-relevant information available only in free text. This finding is highly relevant to the present study: automated patient counting is feasible, but the clinical and data-model assumptions behind the count must be made explicit if the count is to be interpreted as feasibility evidence.

However, direct use of raw EHR/EMR data for feasibility assessment is constrained by important practical, legal, and organizational barriers[1,4]. These systems are source clinical systems designed primarily for care delivery and hospital operations, and they contain highly sensitive personal health information. Direct access by sponsors or CROs is generally not feasible, and even within institutions, data access is governed by security policies, institutional review and data-access procedures, privacy regulations, research-use approvals, and the operational burden placed on hospital information systems[1,4]. In addition, EHR systems differ across institutions in data structures, local codes, documentation practices, clinical workflows, and extractable fields. These differences make it difficult to execute the same site feasibility patient-count question with the same meaning across sites using raw EHR data alone.

For these reasons, this study does not assume direct access to raw EHR/EMR systems. Instead, it focuses on an institutional OMOP CDM-based research database in which source clinical records are extracted, standardized, quality controlled, and managed under pseudonymization or de-identification policies for research use. OMOP CDM transforms heterogeneous source EHR data into a common table structure and standard vocabulary framework, allowing clinical events such as diagnoses, drug exposures, laboratory measurements, procedures, and visits to be represented in a comparable and reproducible form[14–16,49]. Importantly, OMOP CDM itself is not a privacy-preserving mechanism by definition. Rather, within institutional research-data governance, an OMOP CDM instance can serve as a standardized research database operated under appropriate pseudonymization or de-identification controls. In this sense, OMOP CDM does not eliminate all privacy, governance, or institutional heterogeneity issues, but it provides a practical environment in which cohort definitions can be executed in a standardized way and their definitions and outputs can be reviewed and reproduced[14–16,49].

This direction is also consistent with recent discussions on platform trials and federated research networks[10,50]. Projects such as EU-PEARL have emphasized the need to use EHR data to improve feasibility assessment, site selection, and recruitment estimates, and have discussed federated or standardized-data approaches for representing feasibility questions as executable cohort queries. These efforts show that the relevant challenge in clinical trial feasibility is not merely the existence of patient data, but whether the same clinical criteria can be represented and executed in standardized data environments in a way that produces reproducible site-level patient-count estimates[10,50].

Nevertheless, the availability of OMOP CDM does not by itself solve the site-level patient-count computation problem. The constraints underlying these patient-count questions are usually provided by sponsors or protocols as natural-language inclusion and exclusion criteria[18–20]. They are not simple disease-name searches, but patient-count questions with inclusion and exclusion constraints involving drug exposure, laboratory thresholds, temporal windows, medical history, current treatment status, and exclusion conditions[19,20,36]. To become executable over OMOP CDM, such natural-language criteria must be transformed into standard concepts, domain-specific event tables, event dates, value and unit constraints where available, temporal windows, exclusion scopes, patient-set operations, and executable SQL[19,20,29,30].

Accordingly, the present study does not aim to automate the entire site feasibility assessment process, perform final adjudication of individual enrollment eligibility, or implement a patient recruitment workflow[18,51,52]. Its focus is the methodology for reproducibly estimating a site-level count of potentially protocol-eligible patients over OMOP CDM, limited to protocol-derived constraints that can be represented using available structured data. Specifically, this study addresses how such patient-count questions can be transformed into OMOP standard concepts, event-time relations, exclusion semantics, value constraints, patient-set operations, and executable SQL, so that the resulting feasibility count and its cohort-defining assumptions can be inspected.

## 2.2. Standardized RWD Infrastructure and OMOP CDM as a Cohort Execution Environment

Before the widespread use of OMOP CDM in real-world data analytics, cohort discovery platforms such as i2b2 had already demonstrated the practical importance of enabling clinical researchers to query institutional clinical data for patient identification and translational research[53]. Extensions such as SHRINE further showed how i2b2-based institutional repositories could be connected through federated query networks to support aggregate patient-count queries across sites[54]. These systems established cohort discovery as a core biomedical informatics problem: clinical researchers need usable, governed, and reproducible mechanisms for identifying patient populations from complex clinical data environments[53,54]. However, they do not resolve the specific transformation problem addressed in this study: converting natural-language trial eligibility criteria into executable, standard-vocabulary-grounded OMOP CDM cohort logic with explicit temporal, exclusion, and patient-set semantics.

More broadly, large-scale clinical research data infrastructures have developed around common data models and distributed analytic networks. PCORnet and Sentinel illustrate non-OMOP common-data-model approaches for multi-institutional clinical research and regulatory evidence generation, whereas OHDSI, EHDEN, and DARWIN EU illustrate the expanding use of OMOP CDM or OMOP-compatible distributed real-world data infrastructures for standardized real-world data analytics[55–60]. These infrastructures show that clinical data can move beyond local institutional storage and become reusable and comparable research resources through common structures, standard vocabularies, explicit parameters, and reproducible analytic programs[55–60]. The present study builds on this broader infrastructure context but restricts its execution target to OMOP CDM, because the methodological problem studied here is the semantic compilation of eligibility-constrained patient-count questions into OMOP-executable cohort logic.

OMOP CDM represents diverse clinical events, including diagnoses, procedures, drug exposures, measurements, observations, visits, and devices, through a common table structure and standardized vocabulary framework[15,16,57]. By harmonizing heterogeneous data models, vocabularies, and coding schemes, OMOP CDM improves the feasibility of reproducible and comparable multi-institutional research. OMOP-based networks support semantic harmonization by mapping institution-specific native terminologies into a shared standard vocabulary framework. This harmonization provides the foundation for large-scale real-world data analysis and cohort definition.

The FDA Sentinel System provides a useful contrast[58,61]. Sentinel does not directly use OMOP CDM, but it is a major regulatory infrastructure built around a common data model-based distributed data network, parameterizable programs, publicly available analytic tools, and documentation designed to support transparency, reproducibility, and replicability. Sentinel's parameterizable programs separate analytic logic from input parameters, allowing the same analysis to be executed consistently across multiple data partners. This principle is closely related to the motivation of the present work. In clinical research and regulatory settings, the important issue is not simply whether a query can be executed once to obtain a patient count. It is whether the criteria, parameters, data structure, and resulting output can be documented and re-executed.

OHDSI ATLAS Cohort Definition JSON illustrates a representative structured artifact for representing computable cohort definitions over OMOP CDM after eligibility criteria have been interpreted[62,63]. ATLAS JSON includes concept sets, inclusion criteria, exclusion criteria, cohort entry and exit events, observation windows, occurrence counts, and related metadata. Concept sets contain arrays of concept IDs together with metadata such as includeDescendants and includeMapped[63]. Inclusion and exclusion criteria are structured through event-based temporal semantics and Boolean logic[62,63]. Unlike a simple SQL string, an ATLAS cohort definition represents a cohort as a structured object that combines standard vocabularies, temporal conditions, and patient-set operations.

## 2.3. Natural-Language Eligibility Criteria in Trial Protocols and Registries

The mismatch between structured cohort artifacts and natural-language eligibility criteria is already visible in clinical trial registration and protocol documentation. In clinical trial protocols and regulatory or registry documentation, eligibility criteria are generally written as natural-language bullet points under inclusion and exclusion criteria[64,65]. The SPIRIT 2025 guideline identifies participant eligibility criteria as an essential protocol item, but it does not require these criteria to be represented in a specific structured data format[65]. Oncology-focused efforts, including FDA eligibility guidance and LUNGevity-led eligibility standardization work, aim to improve consistency in eligibility-criteria wording and reduce unnecessary restrictiveness[66–68]. They do not, however, convert eligibility criteria themselves into executable cohort logic.

ClinicalTrials.gov illustrates the same issue. Under Section 801 of the FDA Amendments Act, trial registration requires structured data elements[69–71]. Yet these structured elements primarily concern protocol and registry metadata such as trial title, study description, intervention, outcome measures, recruitment status, phase, and eligibility module fields[70,71]. Eligibility criteria remain a free-text or narrative data element entered as inclusion and exclusion criteria[70]. The WHO Trial Registration Data Set similarly includes key inclusion and exclusion criteria as a registration item, rather than as executable cohort logic[72]. As a result, clinical trial information may be registered and publicly accessible, but the inclusion and exclusion criteria required to identify participants in real databases remain natural-language rules that must be interpreted before they can be executed.

## 2.4. From Free-Text Eligibility Criteria to Computable RWD Cohort Logic

The existence of standardized data models and distributed analytic infrastructure does not automatically make natural-language eligibility criteria computable. In clinical trial protocols and ClinicalTrials.gov, eligibility criteria exist as natural-language text[64,65,70]. In RWD infrastructures and cohort-definition environments such as OHDSI ATLAS, DARWIN EU, and Sentinel, analytic logic must instead be represented through explicit cohort definitions, parameters, standard concepts where applicable, temporal windows, inclusion and exclusion rules, and reproducible analytic specifications[58,60,62,63]. On one side of this gap are free-text criteria that are clinically interpretable. On the other side are structured cohort definitions that are executable and reproducible. The mismatch between these two representation systems constitutes the computability gap between natural-language trial criteria and structured RWD analytics.

A foundational review by Weng et al. framed this problem as one of formal knowledge representation for clinical trial eligibility criteria[41]. They identified five dimensions of computable eligibility-criteria representation—intended use, criteria classification, expression language, concept encoding, and patient data modeling—and emphasized expression language, concept codification, and patient data modeling as essential constructs for formal representation. The present study builds on this framing but shifts the unit of analysis from formal eligibility-criteria representation in general to OMOP CDM-executable patient-set semantics for site-feasibility counting: once an eligibility-derived criterion is made computable, the remaining question is what patient set it denotes over a local OMOP CDM instance and whether the cardinality of that set can serve as traceable feasibility evidence.

This gap becomes particularly visible in clinical trial feasibility assessment. The practical question in feasibility assessment is how many patients in a given institutional database satisfy a specified operationalization of eligibility criteria[1,4,46]. Such estimates may inform site selection, recruitment planning, protocol refinement, and assessment of trial feasibility[1,46,48]. Yet eligibility criteria are written for human clinical interpretation, whereas institutional clinical data are stored as coded event-based structures such as diagnoses, drug exposures, measurements, procedures, and visits[14–16]. Feasibility assessment therefore requires more than database lookup. It requires an informatics process that transforms natural-language criteria into computable cohort definitions capable of producing a count and, where governance permits, an intermediate patient set for review[19,20,29,30,36].

This distinction between full eligibility criteria and feasibility-oriented computable criteria has practical precedent. In EHR4CR, feasibility experts did not necessarily implement every original protocol criterion. Instead, they extracted the subset of eligibility criteria considered necessary and computable for protocol feasibility, such as criteria involving diagnoses, medications, laboratory tests, Boolean logic, temporal conditions, and inclusion or exclusion roles[40]. Some original criteria were removed or simplified because they were not essential for feasibility counting or could not be reliably represented in structured EHR data. This supports the framing used in the present study: the target is not full clinical eligibility adjudication, but the construction of a computable feasibility query whose assumptions, omissions, and approximations are explicit.

The key point is that a computable cohort definition should not be reduced to the SQL string itself. In trial feasibility, the central output is a site-level count derived from a specified patient set satisfying an explicitly operationalized eligibility criterion. SQL is the execution form used to derive that patient set and count[19,20]. The essential question is therefore not "Can natural language be converted into SQL?" but "Which standard vocabulary, which OMOP domain, which temporal anchor, which inclusion or exclusion rule, and which patient-set operation should represent this criterion?"[29,30,33,36]. In this sense, eligibility criteria translation should be framed not as a direct text-to-SQL problem, but as a standard-vocabulary-grounded cohort-computability problem.

## 2.5. Empirical Evidence of Limited Eligibility-Criteria Computability

The computability gap in eligibility criteria is not merely a theoretical concern. It has been repeatedly observed in empirical studies. In an ASCO 2021 report of 2,954 oncology trials, 256 computable attributes were extracted from trial eligibility criteria. However, only 132 of these attributes could be encoded using standard terminologies, while 124 required customized concepts[39]. This finding shows that even when computable attributes can be identified from trial criteria, a substantial portion cannot be directly represented using existing standard terminologies.

The CriteriaMapper study reported a similar limitation. Among 640 unique eligibility-criteria attributes, only 367, or 57.3%, could be normalized. Of these, 174 could be encoded using standard terminologies, whereas 193 required in-house reference tables[23]. This suggests that a large portion of eligibility criteria cannot be fully represented using standard vocabularies alone and may depend on institution-specific reference tables or local rules. Such local normalization can support implementation within a specific setting, but it limits portability and creates tension with the goal of semantic interoperability in standardized RWD networks.

Evidence from an ATLAS implementation study points to the same issue at the cohort-definition level. Blasini et al. evaluated the implementation of 223 eligibility criteria from 30 clinical trials in ATLAS and reported that only 51% could be fully implemented, while only 10% of the 30 studies could be fully represented[33]. Criteria that were difficult to implement included those requiring subjective clinical judgment, unstructured pathology or imaging findings, complex temporal logic, and temporally constrained negation. This indicates that even when a structured cohort-definition tool such as ATLAS is available, translating natural-language eligibility criteria into that format remains an unresolved informatics problem.

Taken together, these findings show that the computability gap in eligibility criteria is not simply a matter of automating SQL authoring. It reflects a combination of linguistic, semantic, structural, and practical barriers that arise when natural-language clinical criteria must be reconstructed as executable patient-selection logic over a standardized data model[19,23,33,36,39]. Clinical trial criteria are written as free text, whereas RWD analytic infrastructures require explicit standard concepts, event anchors, temporal windows, Boolean logic, and patient-set operations. This mismatch is the core informatics problem addressed in the present work.

## 2.6. Semantic Annotation Resources for Eligibility Criteria

A related line of work has focused on making clinical trial eligibility criteria amenable to semantic annotation and machine learning. These studies address an important upstream problem: before eligibility criteria can be converted into database queries or cohort definitions, the clinical entities, attributes, relations, temporal expressions, negation cues, comparison operators, and logical structures embedded in free text must be identified[36,73–75]. In this sense, semantic annotation resources provide the training and evaluation substrate for natural-language understanding of eligibility criteria.

Early OMOP-oriented information extraction work includes EliIE[36], an open-source eligibility-criteria information extraction system designed to parse and formalize free-text clinical research eligibility criteria according to OMOP CDM version 5.0. EliIE processes eligibility criteria through clinical entity and attribute recognition, negation detection, relation extraction, concept normalization, and output structuring. Its annotated corpus was developed from Alzheimer's disease trials and included entities, attributes, and relations aligned with OMOP-relevant categories such as conditions, observations, drugs or substances, and procedures or devices. EliIE is important because it explicitly recognized that eligibility criteria contain more than isolated clinical concepts: temporal constraints, measurements, modifiers, negation, and relations between entities and attributes must also be extracted. However, EliIE also illustrates the limitations of treating eligibility formalization primarily as an information-extraction problem. Its query representation was simplified, and more complex interactions among multiple clinical entities were not fully represented as executable cohort logic. Thus, EliIE moved eligibility criteria closer to OMOP-compatible structure, but it did not fully address patient-set algebra, anchor-bound temporal exclusion, deterministic SQL compilation, or execution over an institutional OMOP CDM to produce reproducible patient counts.

Large annotated corpora have further advanced this line of work. Chia is a large corpus of clinical trial eligibility criteria extracted from interventional Phase IV trials and annotated with entities and relationships[73]. Its annotation scheme represents each criterion as a directed acyclic graph that can be transformed into Boolean logic. This design is valuable because it captures the fact that eligibility criteria are not simple bags of concepts but structured semantic objects composed of entity types, relationship types, modifiers, and logical dependencies. Similarly, the Leaf Clinical Trials Corpus was developed as a human-annotated resource for query generation from clinical trial eligibility criteria[75]. It uses highly granular labels to capture biomedical phenomena relevant to downstream query generation, including temporal and contextual distinctions that cannot be captured by concept extraction alone. For example, “diabetes,” “diabetes for at least one year,” and “diabetes in the past six months” may involve the same clinical concept but require different query logic. Such resources show that eligibility-criteria automation requires more than named entity recognition; it requires preserving the semantic structure that determines how a criterion should be queried[73,75].

COVIC extends this annotation-oriented line of work into an explicitly OMOP CDM-related setting. It provides an OMOP CDM-compliant annotated corpus for COVID-19 clinical trial eligibility criteria and uses a hierarchical annotation schema spanning study cohort, eligibility criterion, named entity, and standard concept levels[74]. This makes COVIC particularly relevant to the present study because it links free-text eligibility criteria to OMOP-oriented semantic categories and standard concepts. Nevertheless, COVIC remains primarily an annotated corpus and semantic normalization resource. Its purpose is to support the representation and extraction of eligibility semantics, not to construct local executable concept sets, traverse the OMOP concept hierarchy for cohort execution, evaluate local CDM data availability, compile temporal and exclusion-aware patient-set logic, or execute the resulting cohort definition against an institutional OMOP CDM.

These studies also clarify the role of named entity recognition in the present work. NS-CoT does not introduce a separate BIO-tagging NER module and does not evaluate its contribution through token-level NER F1. Instead, NER-like extraction is embedded in the LLM-mediated typed decomposition layer. Clinical mentions are extracted together with execution-relevant fields such as OMOP domain, role, polarity, temporal cue, and value or unit constraint. This design reflects the objective of the study. The goal is not to optimize entity-span detection as an independent NLP task, but to determine whether the extracted semantic units can be grounded to OMOP standard concepts, composed through temporal and exclusion-aware patient-set logic, compiled into SQL, and executed to reproduce a predefined executable reference cohort count.

The distinction between annotation-level semantics and executable cohort computation is therefore central. If an annotation system identifies “bleeding” as a condition, “anticoagulants” as a drug class, or “current use” as a temporal modifier, this does not yet determine how the criterion should be executed over OMOP CDM. Additional decisions are required: which OMOP standard concept identifiers should represent the clinical mention; whether descendant concepts should be included; which OMOP domain and table should be queried; whether the relevant concepts are actually present in the local CDM instance; how a temporal modifier such as “current” should be translated into an event window; whether an exclusion should be implemented as simple negation or as anchor-bound absence; and how the final patient set should be assembled through intersection, union, or set difference. These are not merely annotation decisions. They are execution-level semantic decisions that determine the resulting patient set and patient count.

In relation to this literature, the contribution of NS-CoT is not to replace semantic annotation corpora, but to extend beyond the layer they primarily address. Annotation resources such as EliIE, Chia, Leaf, and COVIC help define what semantic information should be extracted from eligibility text[36,73–75]. NS-CoT addresses the subsequent informatics problem: how eligibility-derived semantic structure can be transformed into executable OMOP CDM cohort logic. In NS-CoT, natural-language criteria are first decomposed into typed intermediate representations, then grounded to OMOP concepts, interpreted over the event-time representation of OMOP CDM, composed through patient-set algebra, compiled into deterministic CTE-chain SQL, and executed to produce patient counts and trace artifacts. Thus, whereas semantic annotation resources make eligibility criteria more machine-interpretable, NS-CoT aims to make eligibility-criteria-derived cohort definitions database-executable, inspectable, and reproducible.

## 2.7. Natural-Language Interfaces and LLM-Based Text-to-Query Approaches

Early efforts to address this gap include Criteria2Query, which proposed a natural-language interface for transforming eligibility criteria into clinical database queries[19]. Criteria2Query is important because it clearly identified cohort-definition construction as a bottleneck requiring expertise in clinical terminology and clinical data models. However, its evaluation reported an entity normalization accuracy of 0.514 and an attribute normalization accuracy of 0.793, indicating that grounding clinical entities extracted from natural language to OMOP standard concepts remained a major challenge. Criteria2Query demonstrated the feasibility of converting natural-language criteria into queries, but it did not fully resolve the gap in semantic normalization and concept grounding.

Subsequent human-machine collaboration approaches sought to improve the practical usability of query construction by allowing users to review and correct criteria selection, parsing, and concept mapping[76]. These approaches show that clinical judgment and automation are both needed when eligibility criteria are translated into computable logic. At the same time, they also reveal that the transformation from natural-language criteria to structured cohort logic still requires expert review and manual correction. Human-in-the-loop systems can improve adoption and usability, but they do not by themselves provide a formal solution to the gap between natural-language criteria and executable cohort definitions.

More recently, generative LLMs have been incorporated into this line of work. Criteria2Query 3.0 showed that GPT-4-based concept extraction, SQL generation, and reasoning can be applied to clinical trial eligibility query generation[20]. This work is important because it demonstrates that LLMs can assist with interpreting eligibility criteria and generating SQL. However, Criteria2Query 3.0 also reported logic-related SQL-generation errors involving relation, temporality, quantity, and negation. This suggests that even when LLM-based text-to-SQL systems improve SQL generation, the stable preservation of temporal anchoring, absence semantics, Boolean structure, and numeric constraints remains an open challenge. This distinction is important even in light of rapid progress in general text-to-SQL benchmarks.

For example, the BIRD Single-Model Leaderboard reports Gemini-SQL2 at 80.04% test execution accuracy, with human performance listed at 92.96%, and Google Cloud emphasizes that BIRD evaluates execution-verified accuracy rather than merely plausible SQL[77,78]. However, such progress addresses whether SQL can be generated and executed against a database; it does not eliminate the prior clinical informatics question of what patient-set semantics the SQL is intended to implement. Recent LLM-based studies have directly examined the automation of clinical trial eligibility criteria conversion into OMOP CDM queries, showing both the feasibility of LLM-assisted query generation and the persistence of challenges in concept grounding, schema alignment, temporal logic, and executable cohort fidelity[79]. This reinforces the distinction between SQL executability and cohort-semantic fidelity.

Related evidence from medical question answering also suggests that the structure of LLM prompting can materially affect medical-language reasoning. This prior work is relevant not because the task is identical, but because it motivates structured reasoning for medical-language interpretation while also showing why free-form reasoning traces must be converted into typed, executable representations for cohort computation[44]. However, medical question answering and cohort-definition generation require different output semantics. A medical question-answering system may produce a textual answer, whereas an OMOP cohort-definition system must produce executable and inspectable patient-set logic. The relevance of this prior work to the present study is therefore methodological rather than task-identical: it motivates the use of structured reasoning for medical-language interpretation, while also highlighting the need to move beyond free-form reasoning traces toward typed, executable, and reproducible intermediate representations.

Beyond direct end-to-end text-to-SQL, recent studies have also explored approaches that first decompose or structure eligibility criteria before generating queries. For example, Chen et al. used LLMs to decompose eligibility criteria into smaller semantic units or study traits and to structure components such as entities, modifiers, constraints, and negation[80]. Such decomposition-and-parsing workflows show that LLMs can be useful for extracting semantic units from natural-language criteria. However, the subsequent question remains: how should these decomposed units be reliably compiled into OMOP standard concepts, event tables, temporal anchors, exclusion semantics, patient-set operations, and executable cohort logic?

EC2Seq2Sql is another recent approach closely related to the motivation of the present work. It proposes a two-stage method in which free-text clinical trial criteria are first converted into structured pattern sequences and then passed to an LLM agent for SQL generation[28]. This recognizes the limitation of single-step text-to-SQL and introduces an intermediate structure before query generation. In contrast, NS-CoT fixes OMOP vocabulary traversal, temporal and exclusion semantics, patient-set algebra, and SQL compilation in deterministic symbolic modules. The recent literature therefore points toward the need for intermediate representations, but the central claim of the present work is that such intermediate structures must be tied to the execution semantics of OMOP CDM and compiled into reproducible cohort logic.

LeafAI represents one of the most closely related systems to the present work. It extends the Leaf cohort discovery ecosystem, a model-agnostic cohort discovery platform for translational biomedical research, by generating cohort discovery queries from free-text clinical trial eligibility criteria[75,81,82]. It extends the Leaf cohort discovery ecosystem by generating cohort discovery queries from free-text clinical trial eligibility criteria[82]. Its architecture combines BERT-based named entity recognition and relation extraction, T5-based logical-form transformation, recursive parsing of logical forms into an abstract syntax tree, UMLS-based concept normalization, knowledge-base reasoning through linked biomedical ontologies, semantic metadata mapping between UMLS concepts and database schema elements, and query compilation. LeafAI is therefore not a simple direct text-to-SQL system. It explicitly introduces intermediate logical forms and schema-aware mapping before query generation, and it was evaluated against a human database programmer using institutional clinical-trial data at the authors' institution.

LeafAI is the closest methodological comparator to NS-CoT because it moves beyond direct SQL generation by introducing logical forms, concept normalization, knowledge-base reasoning, semantic metadata mapping, and query compilation[82]. However, the two systems make different methodological commitments. LeafAI pursues data model-agnostic cohort discovery query generation, whereas NS-CoT focuses on OMOP CDM-specific feasibility-oriented cohort computation. By restricting the execution target to OMOP CDM, NS-CoT explicitly represents standard-concept grounding, concept hierarchy traversal, local CDM executability, event-time anchoring, temporal exclusion semantics, patient-set algebra, deterministic SQL compilation, and traceable patient-count outputs.

These studies show that natural-language interfaces and intermediate-representation-based systems can lower technical barriers for cohort query construction[19,20,28,76,79,80,82]. However, they also differ in how explicitly they represent the execution semantics needed for reproducible cohort computation. As a result, even when generated SQL is syntactically executable, it may be difficult to determine which interpretation of the criterion was used, how that interpretation was grounded in standard vocabulary, which temporal logic was applied, and where an error occurred when the output is wrong.

The preceding systems motivate intermediate representations for query generation; a related line of clinical neuro-symbolic work motivates a stricter division between language-model extraction and symbolic verification. A recent clinical neuro-symbolic AI study provides a useful methodological precedent for this design choice, although its task differs from the present study. Prenosil et al. combined GPT-4 with a Prolog-based expert system to extract structured information from free-text PET/CT reports. In that system, the language model extracted candidate facts from narrative reports, whereas the symbolic expert system verified those facts against medical rules and produced deterministic, auditable labels[83]. The study showed that such a hybrid architecture could outperform GPT-4 alone and provide traceable reasoning for clinical information extraction. The present work adopts a related division of labor, but the symbolic layer is used not to verify report-level labels, but to ground protocol-derived criteria, define patient-set operations, and compile OMOP-executable cohort logic. This supports the broader premise that LLM-mediated interpretation can be made more reliable when its outputs are constrained, checked, and operationalized by an explicit symbolic layer.

Recent work on neural-symbolic biomedical concept mapping further supports this architectural direction. Zhang et al. proposed Medical Concept Mapping (MCM), a neural-symbolic agent workflow for biomedical concept mapping[84]. MCM uses language-model components to decide whether an ambiguous biomedical mention should be expanded, generate an explicit expansion, and validate semantic equivalence, while symbolic retrieval and linking modules map the original or expanded mention to standardized terminology identifiers. This design shows that even at the mention-to-concept level, LLM outputs benefit from being constrained and checked by symbolic components. However, MCM operates at the concept-mapping level and is evaluated primarily through concept-linking metrics such as Recall@K. It does not address criterion-level cohort computation, temporal anchoring, exclusion scope, local CDM executability, patient-set algebra, SQL compilation, or site-level count estimation. In this sense, MCM is complementary to the present work: it strengthens the case for neural-symbolic concept grounding, whereas NS-CoT formalizes the downstream transformation from grounded clinical criteria to executable OMOP patient-set logic.

The remaining gap beyond mention-level mapping becomes clearer when eligibility criteria contain semantic ambiguity. For instance, “patients with diabetes on insulin therapy” may refer to current insulin therapy, any prior insulin exposure, or insulin exposure anchored to diabetes diagnosis. Such distinctions require different index events, exposure windows, temporal anchors, and patient-set operations in OMOP CDM. This illustrates why eligibility criteria should not be treated as SQL-generation problems alone.

## 2.8. Criteria Normalization, Concept Knowledge Bases, and Patient-Trial Matching

CriteriaMapper represents another approach to trial cohort identification through eligibility-criteria normalization and matching against patient clinical features[23]. It converts natural-language criteria into normalized attributes and matches them to patient characteristics, thereby demonstrating the potential of automated trial cohort identification. However, as noted above, only 57.3% of unique eligibility-criteria attributes could be normalized, and many of those required in-house reference tables. This indicates that criteria normalization can be useful for patient-trial matching, but it remains distinct from the generation of portable, standard-vocabulary-grounded executable cohort definitions.

TrialGPT and other agent-based patient-trial matching frameworks likewise show that LLMs can be useful for interpreting eligibility criteria and patient information[85]. However, these systems primarily focus on determining whether a specific patient is eligible for a given trial, or on ranking trials for a patient. Patient-trial matching is therefore closer to the question, "Does this patient match this trial?" The problem addressed here is different: "How many patients in an institutional OMOP CDM satisfy this specified computable operationalization of protocol-derived criteria?" These matching-oriented systems demonstrate the promise of LLMs for clinical trial matching, but their primary objective is not the generation of executable, OMOP standard-vocabulary-grounded cohort definitions for site-level patient-count computation.

Other work, such as the Clinical Trial Knowledge Base, has demonstrated the potential to link eligibility criteria to OMOP standard concepts[30]. Studies have extracted OMOP CDM standard concepts from large collections of clinical trials and used them to support EHR phenotype knowledge engineering. These efforts are valuable for organizing clinical concepts from eligibility criteria into reusable knowledge resources. However, they do not provide a complete methodology for transforming those concepts into executable cohort definitions with temporal logic, patient-set algebra, SQL, or ATLAS-compatible structured artifacts over an actual OMOP CDM database. A gap therefore remains between concept-level knowledge extraction and executable cohort-logic construction.

Separate from patient-trial matching, some early studies attempted to represent eligibility criteria themselves in OMOP CDM-compatible structures. Si and Weng[29], for example, proposed a method for representing free-text eligibility criteria from Alzheimer's disease clinical trials in a structured relational database using medical terminology and OMOP CDM version 5. This work is important because it showed that clinical terms in eligibility criteria can be linked to a standardized data model. However, such approaches are closer to criteria-level knowledge representation or relational encoding. They do not fully cover the transformation from a natural-language criterion into temporal cohort logic, exclusion semantics, patient-set algebra, SQL compilation, and executable patient counts over OMOP CDM.

Thus, concept-level normalization and executable cohort-logic construction are related but not identical problems. The focus here is the latter: transforming standard-vocabulary-grounded clinical criteria into executable patient-set logic over the OMOP event-time representation.

## 2.9. Structured Cohort Definitions and Formal Clinical Logic

## Representation

Alongside natural-language query systems, tools for managing cohort definitions as structured artifacts have also developed. ATLAS provides a widely used interface for defining cohorts over OMOP CDM data[62]. Yet previous evaluations have shown that only a subset of clinical trial inclusion and exclusion criteria can be fully implemented in ATLAS[33]. This suggests that the challenge is not simply the lack of a cohort-definition interface. Rather, eligibility criteria must still be interpreted as standard concepts, temporal constraints, Boolean logic, event anchors, and patient-set operations before they can become computable cohort definitions.

A closely related system within the OMOP-compatible cohort-query space is the Advanced Cohort Engine (ACE) developed by Callahan et al[86]. ACE combines a temporal query language with an in-memory patient-object datastore to support fast, scalable, time-aware search over longitudinal patient records. It can ingest OMOP CDM data, supports Boolean and temporal operations, and allows knowledge-graph-based query expansion. Importantly, ACE also describes clinical-trial recruitment as a use case in which inclusion and exclusion criteria can be represented as query variables and executed to estimate an eligible patient pool or monitor newly matching records. ACE therefore shows that temporal cohort logic can be treated as a first-class computational object rather than reduced to ordinary SQL execution. However, ACE assumes that users have already expressed the intended cohort logic in its temporal query language. The present study addresses a complementary upstream problem: how natural-language site-feasibility criteria should be interpreted, grounded, and fixed as OMOP-executable patient-set semantics before SQL is generated and before a patient count is treated as feasibility evidence.

Formal representation of cohort definitions and clinical logic has also been actively studied outside the OMOP ecosystem. Clinical Quality Language (CQL) and the Expression Logical Model (ELM) are representative standards for expressing clinical quality measures and cohort logic in declarative, machine-executable forms[87]. They aim to improve traceability between clinical intent and execution logic. Brandt et al.'s PhEMA Workbench combines FHIR and CQL to support the authoring, execution, and validation of standards-compliant phenotype definitions[88]. Li et al. proposed an approach that combines openEHR and CQL to represent clinical quality measures in a configurable form[89]. Berens et al. also proposed a framework for formalizing cohort definitions using FHIR resources, showing the need to explicitly represent cohort-modeling elements such as patient demographics, clinical terminology, outcomes, temporal expressions, and logical combinations[90]. Together, these studies support the idea that computable cohort definitions should be structured representations linking clinical meaning and execution semantics, not merely SQL strings.

However, most of these studies focus on formalizing clinical logic or executing quality measures within standards such as FHIR, CQL, and openEHR[87–90]. The focus of the present work is different: transforming eligibility-constrained cohort-count questions for site feasibility assessment into executable cohort logic for OMOP CDM-based RWD analytics. The approach developed here adopts the broader lesson from CQL/FHIR-oriented work—that formal computable representations are necessary—but adapts that lesson to the OMOP CDM setting through standard vocabulary, concept hierarchy traversal, domain-specific event tables, temporal anchors, exclusion logic, patient-set algebra, and SQL compilation.

DARWIN EU is another important example of how distributed RWD networks seek to achieve reproducibility and standardization in cohort definition and evidence generation[60]. In DARWIN EU, study definitions, phenotype algorithms, and analytic specifications are managed within a distributed real-world data network rather than as isolated local queries. However, DARWIN EU is fundamentally a distributed RWD evidence-generation infrastructure and phenotype curation workflow. In such workflows, the transformation from natural-language eligibility criteria into executable cohort logic generally remains a manual or expert-curated step. Thus, DARWIN EU illustrates the importance of reusable phenotype definitions and distributed execution, but it is not itself a method for automatically transforming natural-language eligibility criteria into OMOP-executable cohort logic.

The FDA Sentinel System carries a similar implication[58,61]. Sentinel has developed a regulatory-oriented analytic infrastructure through a common data model-based distributed data network, modular programs, explicit parameters, standardized cohort-identification algorithms, validation studies, and extensive documentation. Its importance does not lie in automatic conversion of natural-language criteria into queries. Rather, Sentinel shows that once clinical criteria are to be analyzed in a regulatory setting, they must be translated into standardized queries, explicit parameters, and reproducible analytic programs. In this sense, Sentinel is not a solution to the computability gap itself; it illustrates the kind of transparent analytic infrastructure that becomes necessary after that gap has been bridged.

Studies on the variability of EHR-based phenotype definitions further reinforce this point. Brandt et al. analyzed rule-based phenotype definitions and found that many definitions include narrative descriptions or flow diagrams but do not provide sufficiently executable artifacts[91]. This suggests that reusable phenotype or cohort definitions require more than natural-language descriptions. They must include logical criteria, standard terminologies, locally adaptable components, and executable representations. Work by Chondrogiannis et al. and Fernández-Breis et al. similarly highlights the need for semantic representations, standard data models, ontologies, and semantic web technologies in eligibility criteria interpretation and patient cohort identification[92,93].

This line of work aligns with the premise of the present study: eligibility criteria translation is not merely a problem of natural-language understanding or SQL generation. It is a knowledge-engineering problem in which clinical semantics must be operationalized as computable representations before they can be executed against a standardized data model.

## 2.10. Vocabulary Grounding, Local Executability, and Mapping Transparency

Research on OMOP CDM transformation and data harmonization raises a closely related issue: mapping transparency. Several studies have pointed out that researchers often lack sufficiently detailed guidance for harmonizing data into OMOP CDM[94]. In response, stepwise processes have been proposed that include dataset specification, data profiling, vocabulary identification, coverage analysis, semantic mapping, structural mapping, ETL, qualitative data quality assessment, and quantitative data quality assessment. ETL framework studies similarly emphasize the transparency of the mapping process and the reuse of transformation logic across institutions, often proposing metadata-driven approaches that can be applied across heterogeneous source datasets[18,94,95]. The use of structured mapping logic in OMOP CDM transformation reflects the same principle: reliable clinical data analysis requires not only a final SQL query or final result, but also reusable documentation of the intermediate steps through which data harmonization and semantic mapping were performed[16,94,95].

Work from the OHDSI NLP community also points in this direction. When textual data such as clinical notes are transformed into OMOP CDM tables, the task is not simply information extraction. It is also necessary to document how a textual expression is mapped to a standard concept and how that concept is placed into a CDM table structure[45]. The need to document each step in the process of extracting data from clinical notes into OMOP CDM applies equally to the transformation of natural-language eligibility criteria into computable cohort logic. To use clinical text within a standardized data model, the process must make visible how natural-language interpretation, standard-vocabulary mapping, temporal semantics, exclusion logic, and SQL composition were performed.

More broadly, semantic ambiguity is not eliminated by the adoption of standardized clinical data models or exchange resources. Recent work by Ceusters and Rabenberg on semantic difficulties in FHIR Condition resources illustrates that even standardized representations of diagnoses, problems, and clinical states may retain ambiguity when status, verification, and clinical assertion are combined[96]. This supports the premise of the present study that OMOP CDM standardization provides an execution environment, but does not by itself determine the executable patient-set meaning of a natural-language criterion.

From this perspective, OMOP vocabulary grounding is not a simple terminology lookup. It is a core component of computable cohort definition. Expressions such as “diabetes,” “insulin therapy,” “metastatic disease,” and “statin use” must be connected not only to OMOP standard concepts, but also to descendant expansion, domain routing, and, for executable cohort computation, local data availability[14,31,97]. Vocabulary grounding is therefore a central step in preserving functional fidelity to predefined executable cohort semantics when natural-language criteria are transformed into executable patient-set logic.

## 2.11. Reproducibility, Provenance, and Post-Hoc Verification in Cohort Computation

Recent work in RWE methodology has increasingly emphasized the importance of linking study protocols, data extraction, analysis code, and decision provenance. RWE-BRIDGE, for example, proposed an approach for organizing study specifications into machine-readable structures and connecting them to analytic pipelines in order to bridge protocol content and code programming[98]. OpenSAFELY is another representative platform in which data extraction and codelist definitions are standardized, with reproducibility built into the analytic workflow as a design principle[99]. The Staging and Clean Room framework aims to improve transparency and reduce bias by pre-specifying and documenting major decision points in real-world comparative effectiveness studies[100], while the SPIFD2 framework supports the structured assessment of target trial specification, data fitness, design choices, and potential bias for regulatory-grade RWE generation[101].

These studies do not directly address the conversion of natural-language eligibility criteria into OMOP cohort queries. Nevertheless, they make clear that in clinical research and RWE settings, the final result alone is not sufficient. The protocol logic, data definitions, code, parameters, and decision trail that produced the result must be reconstructible[98–101]. The design adopted in this study— preserving typed intermediate representations, concept sets, temporal operators, generated SQL, vocabulary version, and compiler version—follows the same broader direction of reproducibility by design and computational provenance.

Recent debates around AI deployment in health care also show why evidence, transparency, and scope must be kept aligned. In discussing Doctronic's authorization in a regulatory sandbox, Celi et al. cautioned that performance claims can become misleading when evidence generated in one clinical context is communicated as if it supports a different deployment context[102]. A similar concern applies to cohort-definition systems: a count-generation system should clearly distinguish executable-reference fidelity, clinical phenotype validity, recruitment utility, and regulatory compliance. The present study therefore treats NS-CoT's evaluation as evidence of functional fidelity to predefined OMOP-executable cohort logic, not as evidence of autonomous clinical recruitment safety.

Regulatory environments further reinforce the importance of traceability and reproducibility, although the present study does not claim regulatory compliance. Clinical trial data-integrity discussions by FDA, MHRA, and related regulatory authors emphasize the importance of reliable, reconstructable, and well-documented clinical research data[103–105]. FDA 21 CFR Part 11 and related clinical research documentation studies also highlight the role of secure, computer-generated, time-stamped audit trails in regulated electronic-record environments[106,107]. These principles provide design motivation for the present work: a feasibility count should not remain an isolated number or a one-time SQL output, but should be linked to the cohort-defining assumptions, concept sets, temporal windows, exclusion logic, generated query, and execution metadata that produced it. The trace artifacts generated in this study should therefore be understood as computational provenance artifacts for research review and debugging, not as formal Part 11 audit trails or regulatory essential records.

From the perspective of IRB review and research ethics, the issue is not merely technical. SPIRIT and SPIRIT-Path similarly emphasize that trial protocols should document key trial procedures and, where applicable, the personnel, processes, and standards for specimen- and pathology-related components of the trial[64,65,108]. These documentation and verification concerns reflect the long-standing importance of participant protection, accountability, and post hoc reviewability in clinical research governance. Cohort definitions in clinical trials should therefore not remain only as natural-language descriptions or opaque SQL outputs. They should be treated as informatics objects that make it possible to reconstruct how the study population was operationally defined and what data resources and assumptions supported that definition.

In this context, the limitation of LLM-based medical AI extends beyond accuracy. It is also a problem of reproducibility and decision provenance. LLMs may be updated, prompts may differ across users and time points, and OMOP vocabulary releases may change concept hierarchies. If only the final patient count is stored, without preserving the intermediate representation, concept sets, SQL, vocabulary version, and compiler version, it may be difficult to reconstruct the same cohort later from the same eligibility criterion. The present study does not empirically demonstrate FDA- or MFDS-level regulatory compliance. The relevance to regulatory settings should therefore be understood as design motivation rather than as a claim of compliance. The goal is to build cohort-computation infrastructure that is compatible with the need for later reconstruction, explanation, and review[98–101,103–105].

## 2.12. Summary of the Remaining Informatics Gap

Taken together, the prior work reviewed in this chapter shows that the computability gap between natural-language eligibility criteria and RWD analytic infrastructure persists at multiple levels. Criteria2Query and related systems demonstrated that natural-language eligibility criteria can be converted into clinical database queries, but they also revealed persistent challenges in concept normalization, relation extraction, temporality, quantity, and negation[19,76,82]. Criteria2Query 3.0 further showed the promise of LLM-based SQL generation, while its reported relation, temporality, quantity, and negation errors illustrate the limitations of LLM-centric query generation when temporal, quantitative, and negation logic must be preserved[20].

In a closely related setting, Lee et al. developed an automated system that converts free-text ClinicalTrials.gov eligibility criteria into OMOP CDM-compatible SQL using GPT-4, validated it against institutional OMOP CDM data, and systematically evaluated hallucinations across eight LLMs[79]. They reported hallucination rates of 21–50% and introduced a severity-ordered hallucination taxonomy spanning nonexistent concept identifiers, valid concepts assigned to incorrect OMOP domains, and natural-language text emitted into concept fields. These findings are directly relevant to the present work: they empirically show that when concept selection, domain routing, and SQL composition are bundled into a predominantly generative query-construction process, the resulting errors can be difficult to localize and review—a failure mode that motivates the NS-CoT design by separating LLM-mediated typed decomposition from deterministic symbolic compilation.

LLM-based decomposition and agentic SQL-generation studies suggest that natural-language criteria can be broken down into more structured intermediate forms[28,80]. However, they do not fully resolve the next step: how such intermediate structures should be deterministically compiled into OMOP CDM execution semantics, including standard-vocabulary grounding, event-time representation, exclusion semantics, and patient-set algebra. Work based on CQL, FHIR, openEHR, and semantic web technologies demonstrates the importance of computable clinical logic and formal representation, but most of these studies focus on representing phenotypes or quality measures within specific standard frameworks rather than building an end-to-end pipeline from natural-language trial criteria to executable OMOP cohort definitions[87–90,92,93].

Human-machine workflows improve parsing and concept mapping through user intervention, but they do not remove the need for expert review and manual correction[76]. CriteriaMapper showed the practical reach of eligibility criteria normalization, while also demonstrating that many normalized attributes depend on in-house reference tables[23]. ATLAS and DARWIN EU provide structured formats for cohort definitions and phenotype workflows, but the transformation from natural-language criteria into those structures generally remains an expert-curated step[33,60,62]. Sentinel demonstrates the value of explicit parameters and modular analytic programs, but it does not provide a methodology for automatically converting natural-language protocol criteria into computable cohort logic[58,61].

RWE methods such as RWE-BRIDGE, OpenSAFELY, Staging and Clean Room, and SPIFD2 emphasize protocol-to-code linkage, reproducibility, computational provenance, transparency, and documented decision processes[98–101,109]. These frameworks strengthen the motivation for the present study, but their focus is generally on how to manage already specified protocol logic, data definitions, analytic workflows, code, and parameters in a transparent and reproducible way. They do not directly address the semantic operationalization of natural-language eligibility criteria themselves.

The remaining gap is therefore clear. Existing research has helped structure eligibility criteria, normalize selected concepts, assist SQL generation, support patient-trial matching, and manage manually authored cohort definitions. Yet few approaches have formalized the full semantic structure required to transform natural-language trial criteria into executable OMOP CDM cohort logic: standard-vocabulary grounding, concept hierarchy traversal, OMOP domain-table routing, temporal anchoring, absence and exclusion semantics, Boolean composition, patient-set algebra, and SQL composition. The computability gap is not merely a text-to-SQL performance problem. It is a clinical research informatics problem concerning what must be preserved when clinical language becomes patient-selection logic over a standardized data model.

This gap also motivates a neuro-symbolic framing. Neuro-symbolic AI has been proposed as a way to combine the strengths of neural methods, which are effective for learning from flexible and noisy inputs, with symbolic methods, which support explicit knowledge representation, reasoning, and inspectability[42,43,110,111]. In this view, the relevant contrast is not between using or not using neural models, but between leaving semantic decisions inside an opaque learned mapping and separating language interpretation from explicit symbolic computation. This distinction is central to the present study: NS-CoT uses the LLM for natural-language interpretation and typed semantic decomposition, while OMOP vocabulary traversal, temporal predicates, exclusion semantics, patient-set algebra, and SQL compilation are handled by deterministic symbolic modules.

LeafAI is the closest methodological comparator among the reviewed systems because it combines eligibility-text interpretation, logical forms, concept normalization, reasoning, schema mapping, and query generation. NS-CoT differs by fixing OMOP CDM as the execution target and by making local executability, event-time predicates, exclusion semantics, patient-set algebra, deterministic SQL compilation, and traceable count outputs explicit.

The key issue is not whether SQL can be generated. It is what clinical-semantic decisions are made before SQL is generated, and whether those decisions are preserved through standard vocabularies, temporal logic, exclusion logic, and patient-set operations. “Diagnosed within the past six months” requires an index date and temporal window. “No prior systemic therapy” requires absence semantics and an observation-period definition. “ECOG performance status 0–1” requires determining whether the relevant structured field exists or whether proxy variables or NLP extraction would be needed. “Metastatic disease” requires appropriate concept hierarchy traversal and descendant concept inclusion. These are not problems of SQL syntax. They are problems of semantic formalization required for computable cohort definition.

The literature reviewed in this chapter has addressed parts of the overall problem: natural-language interpretation, concept mapping, cohort representation, SQL execution, patient-trial matching, formal clinical logic, and distributed reproducibility. What remains insufficiently addressed is the problem of formalizing the semantic structure of clinical trial criteria as patient-set algebra, temporal logic, concept hierarchy traversal, local executability checks, and exclusion semantics, and then deterministically compiling that structure into OMOP CDM-based execution logic. The unresolved gap is therefore not simply the absence of a better SQL generator. It is the absence of an explicit layer that fixes the executable patient-set meaning of a natural-language clinical criterion before SQL generation. This study addresses that gap by formalizing the intermediate semantic commitments—concept grounding, local executability, temporal anchoring, exclusion scope, and patient-set operations—that determine which patient set a criterion denotes over OMOP CDM.

Taken together, the related work defines several adjacent but distinct layers. Semantic annotation resources make eligibility criteria more machine-interpretable by identifying entities, attributes, temporal expressions, negation, and logical structure. Natural-language query systems and cohort builders help translate interpreted criteria into executable database queries or structured cohort artifacts. Phenotype validation frameworks, such as PheValuator and other structured-data phenotype evaluation approaches, evaluate whether an already specified algorithm identifies a clinically defined condition or event with acceptable operating characteristics[112,113]. These layers are important, but they do not fully address the problem studied in this study.

The present work focuses on the semantic operationalization layer that precedes both SQL execution and clinical phenotype validation. In site feasibility assessment, a free-text protocol criterion does not always correspond to a single self-evident patient set before interpretive choices are made. Different reviewers may reasonably choose different concept sets, temporal anchors, proxy variables, exclusion scopes, observation windows, or patient-set operations, and different OMOP implementations may make different parts of the criterion computable. The problem is therefore not only whether an executable query can be produced, nor whether an already specified phenotype algorithm is clinically valid. The prior question is how the criterion is committed to a declared executable patient-set semantics over a local OMOP CDM instance.

This distinction defines the position of NS-CoT. The framework does not replace semantic annotation corpora, OMOP cohort construction tools, or phenotype validation methods. Instead, it externalizes the semantic commitments that connect free-text feasibility criteria to executable cohort logic: concept grounding, local executability, temporal anchoring, proxy decisions, exclusion scope, value constraints, and patient-set algebra. Once these commitments are explicit, generated SQL, patient sets, counts, and site-feasibility decisions can be inspected, reproduced, compared, and revised. The contribution is therefore a medical informatics methodology for making the interpretive step between protocol text and OMOP-executable cohort logic explicit, rather than leaving it implicit inside a text-to-SQL or cohort-generation process.

# Chapter 3. Formalization and System Architecture

As discussed in Chapter 2, the central gap between protocol-derived feasibility questions and OMOP CDM-based cohort computation is not simply a lack of SQL-generation capability. Prior work has shown that natural-language eligibility criteria can be transformed into database queries or cohort-discovery logic, but it has also revealed persistent challenges in concept normalization, temporal interpretation, negation, and execution semantics[19,20,23,28,33,36,82]. The more fundamental problem is that the clinical meaning embedded in site feasibility patient-count questions with protocol-derived inclusion and exclusion constraints must be explicitly operationalized as those criteria are transformed into executable patient-set logic over institutional OMOP CDM data. This transformation requires clinical concept selection, standard-vocabulary grounding, OMOP domain-table routing, event-time anchoring, temporal-window interpretation, exclusion semantics, value constraints, Boolean composition, and patient-set operations. The task addressed in this chapter is therefore not direct text-to-SQL generation, but the operational formalization and deterministic compilation of feasibility-oriented, eligibility-criteria-derived cohort definitions into executable OMOP CDM cohort logic.

This chapter presents the formalization and system architecture of NS-CoT (Figure 1, Figure 2). The framework does not use the LLM as an end-to-end SQL generator. Instead, the LLM is used in the front-end interpretation layer: it interprets site feasibility patient-count questions with protocol-derived inclusion and exclusion constraints, decomposes them into typed intermediate representations, and, where applicable, supports semantic compatibility checks for candidate standard concepts. After the typed IR boundary, the remaining stages are handled by a deterministic symbolic layer. These stages include OMOP vocabulary traversal, descendant concept expansion, local CDM-aware executable concept filtering where applicable, temporal predicate construction, exclusion semantics, patient-set algebra, and CTE-chain SQL compilation. This architectural separation follows the broader neuro-symbolic motivation of combining flexible neural interpretation with explicit symbolic representation, reasoning, and inspectability[42,43,110,111].

The formalization in this chapter should be understood as an operational formalization with typed representations and compilation semantics. It is not a proof-theoretic formal semantics of all clinical trial eligibility criteria, nor a claim of soundness or completeness with respect to unrestricted clinical eligibility semantics. Rather, it is a methodological abstraction of the transformation carried out by the NS-CoT pipeline. A natural-language cohort criterion is first converted into a typed intermediate representation, in which clinical mentions, OMOP domains, roles, polarities, value constraints, and temporal dependencies are made explicit. The typed representation is then grounded to OMOP-compatible concept sets and execution metadata. Grounded criteria are interpreted as symbolic patient-set operations over the OMOP event-time representation, and the resulting cohort logic is finally serialized as deterministic SQL for execution. SQL is therefore not the methodological starting point of the pipeline. It is the final execution form of typed, grounded, and symbolically composed cohort semantics.

The purpose of this formalization is to define a reviewable computational object before SQL is generated. In NS-CoT, this object is the declared OMOP-executable cohort logic $h_q$ for a given input criterion $q$. Once $h_q$ has been specified through typed facts, grounded concept sets, temporal predicates, polarity-aware patient-set operations, and compilation assumptions, the resulting patient count can be inspected not only as a scalar output but also as a cohort-definition artifact. This artifact makes explicit the concept sets, local executability decisions, temporal anchors, boundary policies, exclusion scopes, patient-set operations, generated SQL, and trace metadata that produced the count.

The remainder of the chapter is organized as follows. Section 3.1 describes the design objective and architectural commitment of NS-CoT. Section 3.2 formally defines the problem of transforming site feasibility patient-count questions with protocol-derived inclusion and exclusion constraints into executable OMOP CDM cohort logic. Section 3.3 abstracts structured OMOP clinical records as a typed longitudinal event-time representation. Section 3.4 defines the typed intermediate representation extracted from natural-language cohort criteria. Section 3.5 describes standard-vocabulary grounding, descendant expansion, drug-mention subtype routing, and local executable concept filtering. Section 3.6 defines single-anchor temporal predicates, inclusion and exclusion polarity, anchor-bound temporal exclusion, and patient-set algebra. Section 3.7 explains how grounded IR is compiled into deterministic CTE-chain SQL. Section 3.8 describes the traceability and provenance artifacts generated during this process. Finally, Sections 3.9 and 3.10 describe the reference implementation, LLM backend, methodological assumptions, and scope boundaries. Figure 2 summarizes the operational formalization pipeline that connects the input criterion, intermediate representations, cohort logic, SQL compilation, execution, and traceable output artifacts.

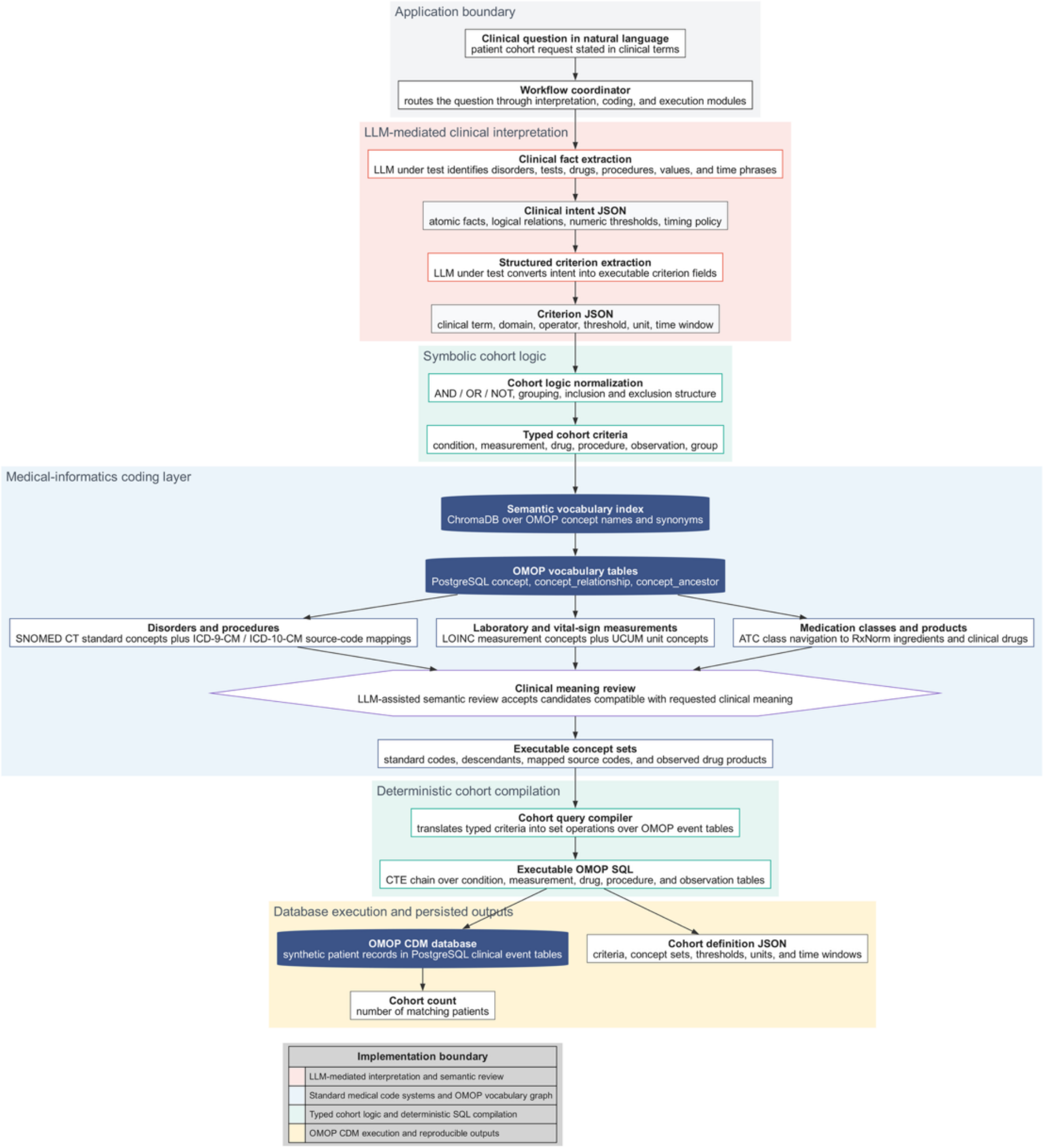

**Figure 1. NS-CoT architecture overview.** The framework separates LLM-mediated semantic interpretation from deterministic symbolic cohort computation. Natural-language criteria are decomposed into typed representations, while OMOP vocabulary grounding, temporal and exclusion reasoning, patient-set algebra, and SQL compilation are handled by the symbolic layer.

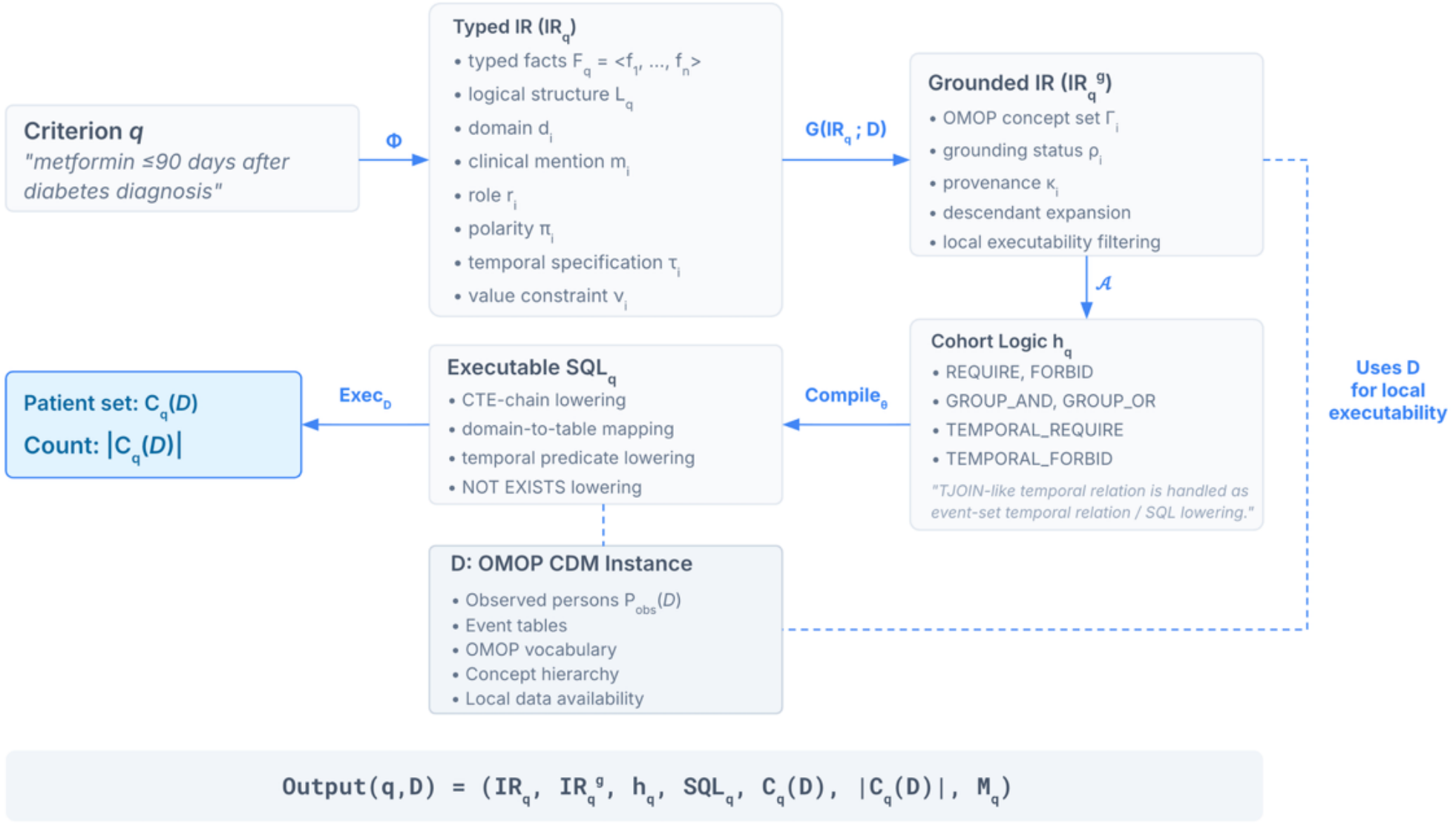

**Figure 2. NS-CoT operational formalization pipeline.** A natural-language cohort criterion q is transformed into a typed intermediate representation $\boldsymbol{IR_q}$, grounded to OMOP-executable concepts as $\boldsymbol{IR_q^g}$, assembled as patient-set algebraic cohort logic $\boldsymbol{h_q}$, deterministically compiled into $\boldsymbol{SQL_q}$, and executed over an OMOP CDM instance D to produce the patient set $\boldsymbol{C_q(D)}$ and count $|\boldsymbol{C_q(D)}|$. The final output is a traceable cohort-definition artifact rather than only a scalar count.

## 3.1. Design Objective and Architectural Commitment

NS-CoT is neuro-symbolic in the architectural sense: neural language interpretation and deterministic symbolic cohort computation are coupled through a typed intermediate representation. The design objective of NS-CoT is to transform site feasibility patient-count questions with protocol-derived inclusion and exclusion constraints into executable OMOP CDM cohort logic, while preventing clinically important semantic decisions from being hidden inside SQL. In a direct text-to-SQL approach, clinical concept selection, standard-vocabulary grounding, OMOP domain-table routing, temporal-window specification, exclusion handling, patient-set operations, and SQL composition may be bundled into a single predominantly generative query-construction process. Even when the resulting SQL is syntactically executable, it can be difficult to determine which standard concepts were selected, whether descendant concepts were included, which event date was used as the anchor, whether an exclusion was implemented as simple negation or as anchor-bound absence, and how patient-set operations were ordered.

To address this problem, NS-CoT adopts a neuro-symbolic architectural separation. The LLM-mediated layer interprets natural-language cohort criteria and produces typed semantic structure, whereas the symbolic layer performs OMOP vocabulary traversal, temporal and exclusion reasoning, patient-set algebra, and deterministic SQL compilation. The two components are coupled through the typed intermediate representation rather than fused into a single model. This design follows the broader neuro-symbolic motivation that neural and symbolic systems provide complementary capabilities: neural models are useful for interpreting flexible natural-language inputs, whereas symbolic representations make domain knowledge, logical constraints, temporal rules, and execution procedures explicit and inspectable.

Thus, NS-CoT is not a fully integrated neural reasoning engine in which symbolic inference occurs inside the neural network. It is neuro-symbolic in the architectural sense: neural language interpretation is separated from, and constrained by, explicit symbolic cohort execution semantics. NS-CoT applies this principle to OMOP CDM-based feasibility cohort computation. The neural component interprets the protocol-derived cohort criterion, but the symbolic component determines how the interpreted criterion is grounded, constrained, combined, and executed.

The LLM-mediated layer decomposes a natural-language cohort criterion into typed facts and extracts fields such as domain, mention, role, polarity, grounding candidate, temporal specification, and value constraint where applicable. This layer is necessary because protocol-derived criteria are linguistically diverse, clinically elliptical, and often written in non-standard forms. However, the LLM does not directly generate the final SQL and is not treated as the authority over cohort execution semantics. Once the typed IR has been produced, the symbolic layer takes over. It uses the OMOP vocabulary graph to resolve standard and descendant concepts, constructs locally executable concept sets when applicable, interprets temporal relations and exclusion semantics as typed operators, assembles cohort logic through patient-set algebra, and compiles the result into executable SQL.

The “CoT” component of NS-CoT reflects this methodological lineage from structured reasoning in medical-language tasks. In contrast to free-form chain-of-thought prompting evaluated in prior medical question-answering work, NS-CoT treats reasoning as typed semantic decomposition rather than as an unconstrained textual rationale[44]. The output of the LLM-mediated layer is therefore not a final answer or a final SQL query, but a constrained intermediate representation consumed by downstream symbolic modules.

The architecture rests on three core commitments. First, a feasibility-oriented natural-language cohort criterion must be transformed into a typed semantic representation before it is compiled into SQL. The typed representation is not a free-form explanation. It is a constrained, machine-facing representation consumed by downstream symbolic code. Whenever possible, domain, clinical mention, polarity, temporal relation, anchor policy, value constraint, and logical structure are represented as explicit fields.

Second, standard-vocabulary validity must be distinguished from local CDM executability. A concept may be clinically valid and represented as an OMOP standard concept, but that does not mean that the concept or its descendants are actually observed in a particular institutional OMOP CDM instance. This distinction is especially important for drug-class criteria. A pharmacologic class expanded through the vocabulary graph may yield a set of candidate drug concepts, but only a subset of those concepts may be present in the local drug_exposure table. NS-CoT therefore separates vocabulary-level semantic grounding from local data-level executability wherever this distinction is relevant.

Third, temporal inclusion and temporal exclusion are defined as problems of event-time relations and patient-set algebra, not as problems of SQL syntax. Expressions such as "drug exposure after diagnosis," "no prior exposure," "within 90 days after anchor," and "more than 180 days after diagnosis" are not generated directly as SQL fragments. They are first represented as typed temporal predicates between an anchor event and an outcome event. The deterministic compiler later lowers these predicates into SQL date predicates, correlated EXISTS, or correlated NOT EXISTS structures.

In this framework, a cohort is defined as a person-level set. Each typed fact extracted from the natural-language criterion is interpreted as a patient-level predicate over the OMOP event-time representation. The final cohort is constructed through patient-set operations such as REQUIRE, FORBID, GROUP_AND, and GROUP_OR, which are defined formally in Section 3.6. Exclusion criteria are not treated as surface-level negation. Anchor-bound temporal exclusion is defined as the absence of a forbidden event satisfying a specified temporal relation to an anchor event. At SQL compilation time, this is lowered into a correlated NOT EXISTS pattern rather than a simple global negation or unscoped set subtraction. This design preserves exclusion semantics explicitly and makes it possible to localize errors to concept grounding, temporal interpretation, exclusion handling, or SQL composition.

The aim of NS-CoT is not to build a general-purpose recruitment system capable of automatically interpreting every clinical trial eligibility criterion. The aim is more specific: to provide a methodology for transforming selected OMOP-expressible, selected OMOP-expressible, feasibility-oriented cohort criteria into typed IR, grounding them in standard vocabulary and local CDM evidence, representing single-anchor temporal predicates and patient-set operations explicitly, and compiling the resulting cohort logic deterministically. This design addresses the computability gap between protocol-derived feasibility questions and executable OMOP cohort definitions and provides the methodological foundation for the functional-fidelity and count-reproduction evaluation presented in later chapters.

## 3.2. Formal Problem Definition

The problem addressed in this study is the transformation of a site feasibility patient-count question with protocol-derived inclusion and exclusion constraints into executable OMOP CDM patient-set logic. The input is a protocol-derived cohort criterion written in natural language. The output is not merely a SQL string. It includes the person-level patient set satisfying the specified executable interpretation of the criterion, the cardinality of that set, the executable SQL used to compute it, and traceable artifacts that make the transformation path inspectable. The task is therefore not to map a sentence directly into SQL syntax, but to operationalize the clinical semantics contained in the criterion as patient-set logic over the structured event-time representation of an institutional OMOP CDM.

Let $q$ denote a natural-language site feasibility patient-count question with protocol-derived inclusion and exclusion constraints and let $D$ denote an OMOP CDM database instance. The database instance $D$ contains a set of persons, observation periods, structured clinical event tables, and OMOP vocabulary tables including concept hierarchy relations. Let $P_D$ denote the set of all persons in $D$, and let $P_{obs}(D) \subseteq P_D$ denote the set of persons with valid observation periods. The objective is to construct an OMOP-executable cohort logic $h_q$ such that, after deterministic compilation and execution over $D,$ it yields a patient set $C_q(D)$, where:

$$C_q(D) \subseteq P_{obs}(D)$$

The corresponding feasibility-oriented patient count is the cardinality of this patient set:

$$count(q, D) = | C_q(D) |$$

In this study, $h_q$ is not treated as a free-form SQL string. Rather, it is constructed through a sequence of typed and grounded transformations. A natural-language cohort criterion $q$ is first mapped to a typed intermediate representation, then grounded to OMOP-compatible concept sets, interpreted as event-time predicates, composed through patient-set algebra, and finally lowered to executable SQL. The transformation pipeline is defined as:

$$q \xrightarrow{\Phi} IR_q \xrightarrow{G} IR_q^g \xrightarrow{\mathcal{A}} h_q \xrightarrow{Compile} SQL_q \xrightarrow{Exec_D} C_q(D),\ count(q,D) = \mid C_q(D) \mid$$

Here, $\Phi$ denotes the natural-language-to-typed-IR transformation. It corresponds to the LLM-mediated interpretation stage, in which the criterion is decomposed into typed facts and execution-relevant semantic fields. $G$ denotes standard-vocabulary grounding, descendant expansion, and local CDM-aware executable concept filtering. $IR_q^g$ denotes the grounded intermediate representation, which includes typed fields, resolved OMOP concept sets, domain information, temporal specifications, and execution-relevant metadata. $\mathcal{A}$ denotes the patient-set algebraic construction that interprets grounded criteria as operations over subsets of $P_{obs}(D)$. $Compile$denotes deterministic SQL generation from the grounded cohort representation, and $Exec_D$ denotes execution over the OMOP CDM instance $D$. Because $q$ may admit multiple clinically plausible interpretations, $C_q(D)$ and count$(q,D)$ are defined with respect to the declared executable cohort logic $h_q$. Equivalently, the executable result of a criterion can be summarized as:

$$IR_q^g = G(IR_q; D)$$

$$h_q = \mathcal{A}(IR_q^g)$$

$$SQL_q = Compile_\theta(h_q)$$

$$C_q(D) = Exec_D(SQL_q)$$

$$count(q,D;h_q) = \mid C_q(D) \mid$$

The role of the LLM in this formulation is restricted primarily to the natural-language interpretation stage and, where applicable, structured semantic validation of candidate concepts. The LLM-mediated layer decomposes the input criterion into typed facts, identifies clinical mentions, assigns tentative domains and roles, extracts temporal expressions, captures inclusion or exclusion polarity, and proposes structured semantic information. However, the subsequent stages are not left to unconstrained language generation. Vocabulary traversal, descendant expansion, local executable concept filtering where applicable, temporal predicate construction, exclusion semantics, patient-set algebra, and SQL composition are deterministic with respect to the grounded IR, the OMOP CDM instance, the vocabulary snapshot, the database dialect, the configuration, and the compiler version. This separation is central to the methodology of NS-CoT.

The semantic target of this transformation is not unrestricted clinical truth, final enrollment eligibility, or a chart-reviewed gold-standard phenotype. Rather, the target is OMOP-executable cohort semantics under a specified interpretation of the criterion and under the representational capacity of the available OMOP CDM data. In other words, the question is whether a feasibility-oriented natural-language cohort criterion has been transformed into the intended executable patient-set logic over OMOP CDM. Semantic fidelity is therefore defined as the preservation of computationally relevant clinical meaning during the transformation from natural language to executable cohort logic. In this study, semantic fidelity includes at least the following components: clinical concept identity, standard-vocabulary grounding, descendant concept expansion, OMOP domain-table routing, event-time anchoring, temporal-window interpretation, inclusion or exclusion polarity, value or quantity constraints when supported, local CDM executability where applicable, and patient-set composition. In this chapter, semantic fidelity is therefore defined relative to a specified executable cohort interpretation, not relative to all possible clinical interpretations of the natural-language criterion.

This distinction is important because two SQL queries may both execute successfully while representing different clinical meanings. For example, a criterion such as “drug exposure after diagnosis” may be operationalized using the first diagnosis date, any diagnosis date, an index date, or a diagnosis date constrained by observation period. Similarly, “no prior drug exposure” may mean absence before cohort entry, absence before diagnosis, or absence within a specified look-back window. These interpretations can lead to different patient sets and therefore different feasibility-level patient counts even when all resulting SQL queries are syntactically valid. Therefore, the correctness target of NS-CoT is not SQL validity alone, but the preservation of explicitly represented cohort semantics prior to SQL compilation.

Given a grounded representation $IR_q^g$, patient-set algebra first constructs an OMOP-executable cohort logic $h_q$. Deterministic SQL compilation is then defined as a function from this cohort logic to executable SQL:

$$\mathrm{A}(IR_q^g) = h_q$$

$$Compile: h_q \rightarrow SQL_q$$

For a fixed $h_q$ and compilation environment $\theta$, NS-CoT produces the same SQL structure, parameter binding, and execution semantics:

$$Compile_{\theta}(h_q) = SQL_q$$

where $\theta$ denotes the fixed compilation environment, including compiler code, configuration, SQL dialect, vocabulary snapshot, and database schema assumptions. This property does not imply that the original natural-language interpretation is always clinically correct or that the resulting cohort is a clinically adjudicated phenotype. Rather, it ensures that once the interpretation has been represented in the grounded IR, the subsequent execution semantics are reproducible, inspectable, and localizable to specific transformation stages.

Within the scope of this study, the formal problem is restricted to a clinically important subset of OMOP-expressible, eligibility-criteria-derived cohort definitions. The primary supported domains are condition occurrence and drug exposure, with additional support for selected measurement, procedure, demographic, and observation-based criteria when representable in structured OMOP fields. The supported temporal semantics are primarily single-anchor predicates in which an outcome event is evaluated relative to one anchor event, such as drug exposure on or after diagnosis, drug exposure within a bounded interval after diagnosis, or absence of drug exposure after an anchor. The current framework does not claim unrestricted clinical trial eligibility interpretation, unstructured-note-dependent criteria, subjective clinical judgment, final enrollment screening, or multi-anchor sequential temporal chains in which an intermediate outcome event becomes the anchor for a subsequent temporal predicate. This boundary is methodological rather than incidental: the objective is to formalize and implement an executable-semantics-preserving compilation framework for selected OMOP-expressible cohort criteria used in site feasibility patient-count estimation, not to provide a general-purpose clinical trial recruitment system. Table 1 summarizes the symbols used in the formal problem definition.

| Symbol | Meaning |
| --- | --- |
| $q$ | Feasibility-oriented natural-language cohort criterion with protocol-derived inclusion and exclusion constraints |
| $D$ | Institutional OMOP CDM database instance used as the execution environment |
| $P_D$ | Set of all persons represented in database instance $D$ |
| $P_{obs}(D)$ | Set of persons in $D$ with valid observation periods |
| $IR_q$ | Typed intermediate representation extracted from criterion $q$ |
| $IR_q^g$ | Grounded intermediate representation of $q$, including OMOP concept sets, domain information, temporal specifications, value constraints where applicable, and execution metadata |
| $h_q$ | OMOP-executable cohort logic, represented through patient-set algebra, constructed from grounded criterion $IR_q^g$ before SQL compilation |
| $\Phi$ | LLM-mediated natural-language-to-typed-IR transformation |
| $G$ | Standard-vocabulary grounding, descendant expansion, and local CDM-aware executable concept filtering |
| $\mathcal{A}$ | Patient-set algebraic construction from grounded cohort semantics |
| $C_q(D)$ | Resulting patient set for criterion $q$ executed over database instance $D$ |
| $Compile_\theta$ | Deterministic compilation from OMOP-executable cohort logic $h_q$ to executable SQL |

| Symbol | Meaning |
| --- | --- |
| $SQL_q$ | Executable SQL query generated by compiling cohort logic $h_q$ under compilation environment $\theta$ |
| $Exec_D$ | SQL execution over database instance $D$ |
| $\theta$ | Fixed compilation environment: compiler version, configuration, SQL dialect, OMOP vocabulary snapshot, and schema assumptions |
| $M_q$ | Metadata and trace artifact associated with criterion $q$, including decomposition, grounding provenance, temporal predicates, SQL generation, execution status, and count output. |

**Table 1. Symbols used in the formal problem definition for OMOP-CDM-based feasibility cohort computation.**

## 3.3. OMOP CDM as a Typed Longitudinal Event–Time Representation

The formal problem defined in Section 3.2 requires a data model in which site feasibility patient-count questions can be interpreted as computable patient-level predicates over institutional OMOP CDM data. In this study, OMOP CDM is not treated merely as a collection of relational tables. Instead, it is abstracted as a typed longitudinal event-time representation over persons, concepts, and event dates. This abstraction is necessary because site feasibility patient-count questions with protocol-derived inclusion and exclusion constraints are commonly expressed in terms of clinical events and their temporal relations: a patient is diagnosed with a disease, receives a drug, undergoes a procedure, has a measurement value, or has no prohibited exposure before or after an anchor event. Therefore, the relevant computational unit for cohort construction is not a table row in isolation, but a typed clinical event associated with a person, an OMOP concept, and an event time.

Let $D$ denote an OMOP CDM database instance, as defined in Section 3.2. Let $P_D$ denote the set of all persons in $D$, and let $P_{obs}(D) \subseteq P_D$ denote the set of persons with valid observation periods. In this study, the initial universe for cohort construction is $P_{obs}(D)$, rather than all persons in the PERSON table, because eligibility-criteria-derived cohort definitions are evaluated over observed clinical history. A person without a valid observation period cannot be interpreted in the same way as a person whose clinical events are observable within the CDM instance. Thus, cohort construction begins from the set of observed persons:

$$C_0 = P_{obs}(D)$$

Let $\mathcal{D}_{evt}$ denote the set of OMOP event domains considered by the cohort construction framework. These domains include, in principle, Condition, Drug,

Measurement, Procedure, Observation, Visit, Device, and Death. In the empirical scope of this study, however, the primary event domains form a core subset of $\mathcal{D}_{evt}$:

$$\mathcal{D}_{core} = \{Condition, Drug\} \subset \mathcal{D}_{evt}$$

This distinction is important. The event-time abstraction can be applied broadly across structured OMOP event tables, but the central methodological and empirical claims of this study are focused on condition-based anchor or inclusion events and drug exposure inclusion or exclusion. Measurement, Procedure, Observation, Visit, Device, and Death domains are included in the formal abstraction for completeness and extensibility, but they are not the primary basis of the empirical evaluation unless explicitly stated.

For each supported event domain $d \in \mathcal{D}_{evt}$, let $\mathcal{V}_d$ denote the set of OMOP concepts eligible to represent events in domain $d$ under the vocabulary and domain assumptions of this study. A structured OMOP record is abstracted as a typed event-time relation:

$$E_d(p, c, t)$$

where $p \in P_{obs}(D)$ is a person, $c \in \mathcal{V}_d$ is an OMOP concept identifier, and $t \in \mathcal{T}$ is the event time. Here, $\mathcal{T}$ denotes the date or timestamp domain represented in the OMOP CDM instance, such as condition start dates, drug exposure start dates, measurement dates, procedure dates, visit start dates, or other domain-specific event-time fields. The expression $E_d(p, c, t)$ means that person $p$ has an observed clinical event in domain $d$, represented by concept $c$, at time $t$. This representation provides a common interface for evaluating eligibility criteria across different OMOP event tables while preserving the domain-specific table structure through explicit mappings.

For condition events, the event-time relation is defined over the CONDITION_OCCURRENCE table:

$$E_{Condition}(p, c, t)$$

where:

$p =$ condition_occurrence.person_id
$c =$ condition_occurrence.condition_concept_id
$t =$ condition_occurrence.condition_start_date

Thus, a natural-language expression such as “patients with diabetes” is not interpreted as a keyword search over a diagnosis label. It is interpreted as the existence of at least one condition event whose concept belongs to the grounded diabetes concept set.

For drug exposure events, the event-time relation is defined over the DRUG_EXPOSURE table:

$$E_{Drug}(p, c, t)$$

where:

$$p = \text{drug_exposure.person_id}$$
$$c = \text{drug_exposure.drug_concept_id}$$
$$t = \text{drug_exposure.drug_exposure_start_date}$$

Drug exposure duration may additionally be represented by drug_exposure_end_date when a criterion requires an exposure interval. However, in the single-anchor temporal predicates evaluated in this study, the primary event time for drug criteria is the exposure start date. This choice reflects the supported criterion types, such as drug exposure after diagnosis, drug exposure within a bounded interval after an anchor, or absence of exposure after an anchor. Table 2 summarizes the event-time mappings used in the formalization.

| Clinical event domain | OMOP table | Concept field | Event-time field |
|---|---|---|---|
| Condition | condition_occurrence | condition_concept_id | condition_start_date |
| Drug | drug_exposure | drug_concept_id | drug_exposure_start_date |
| Measurement | measurement | measurement_concept_id | measurement_date |
| Procedure | procedure_occurrence | procedure_concept_id | procedure_date |
| Observation | observation | observation_concept_id | observation_date |
| Visit | visit_occurrence | visit_concept_id | visit_start_date |
| Device | device_exposure | device_concept_id | device_exposure_start_date |
| Death | death | cause_concept_id, where applicable | death_date |

**Table 2. OMOP event-time mappings for structured clinical domains.**

*Note. Person field (person_id). For Death, cause_concept_id is relevant only when a cause-of-death concept is used; death occurrence itself is primarily represented by person_id and death_date.*

Although Table 2 includes multiple structured OMOP domains, the core implementation and evaluation in this work are centered on Condition and Drug events. This boundary should not be interpreted as a limitation of the event-time abstraction itself. Rather, it reflects the methodological focus of the study: preserving eligibility semantics for condition-based anchors, drug exposure inclusion, and drug exposure exclusion.

Value-bearing events require an extension of the basic event-time tuple. Measurement criteria, for example, cannot be represented solely by the existence of a measurement concept. A criterion such as "HbA1c < 7%" requires the measurement concept, a numeric value, a comparison operator, and a unit. For such cases, the event-time relation can be generalized as:

$$E_d(p, c, t, v, u)$$

where $v$ denotes the recorded value and $u$ denotes the corresponding unit. Under this generalized representation, a measurement criterion is evaluated by combining concept membership with value and unit constraints where applicable. For example, an HbA1c criterion requires both a grounded HbA1c concept set and a predicate over the measured value. This is an intentionally simplified abstraction; concrete implementation may use *value_as_number, value_as_concept_id, unit_concept_id, or unit_source_value* depending on the domain and criterion. In the present study, this generalized representation is treated as an extensible component of the framework, whereas the primary evaluated domains remain Condition and Drug. The comparison operator is not treated as part of the event record itself; it belongs to the criterion-level value predicate applied to the recorded value and unit.

The event-time abstraction is closely tied to standard-vocabulary grounding. A mention extracted from a feasibility-oriented natural-language cohort criterion does not directly define an executable event predicate. It must first be grounded to a domain-specific concept set. Let $m$ denote a clinical mention extracted from an eligibility criterion, and let $\Gamma_d(m) \subseteq \mathcal{V}_d$ denote the grounded concept set associated with mention $m$ in domain $d$. In this section, $\Gamma_d(m)$ is used as a preliminary shorthand for the concept set associated with mention $m$. Section 3.5 refines how this concept set is obtained by distinguishing vocabulary-level grounding $G_d^{vocab}(m)$from local CDM-aware executable grounding $G_d^{exec}(m; D)$.

The corresponding event set for a person $p$ can be defined as:

$$\mathcal{E}_d(m, p) = \{(c, t) \mid E_d(p, c, t) \wedge c \in \Gamma_d(m)\}$$

For example, if $m =$"diabetes" and $d =$ Condition, then:

$$\mathcal{E}_{\text{Condition}}(\text{diabetes}, p) = \{(c, t) \mid E_{\text{Condition}}(p, c, t) \wedge c \in \Gamma_{\text{Condition}}(\text{diabetes})\}$$

Similarly, if $m =$"insulin" and $d =$ Drug, then:

$$\mathcal{E}_{\text{Drug}}(\text{insulin}, p) = \left\{(c, t) \mid E_{\text{Drug}}(p, c, t) \wedge c \in \Gamma_{\text{Drug}}(\text{insulin})\right\}$$

This formulation separates clinical language interpretation from event retrieval. The mention "diabetes" is not assumed to correspond to a single database code, and the mention "insulin" is not assumed to correspond to a single drug concept. Instead, each mention is grounded to a concept set, potentially expanded through OMOP vocabulary hierarchy, and then evaluated over the corresponding event-time relation.

Temporal cohort criteria are defined over these event sets. Let $A(p)$ denote the set of anchor event times for person $p$. For example, if diabetes diagnosis is used as an anchor, then $A_{\text{diabetes}}(p)$ is the set of event times associated with grounded diabetes condition events:

$$A_{\text{diabetes}}(p) = \{t \mid \exists c \in \Gamma_{\text{Condition}}(\text{diabetes}) : E_{\text{Condition}}(p, c, t)\}$$

In the current framework, the representative anchor policy is the first-anchor policy. If $A(p) \neq \emptyset$, the first anchor time is defined as:

$$a(p) = \min A(p)$$

If $A(p) = \emptyset$, then $a(p)$ is undefined, and the person cannot satisfy an anchor-based criterion that depends on that anchor. For example, if diabetes diagnosis is used as the anchor, then the first diabetes diagnosis date is:

$$a_{\text{diabetes}}(p) = \min A_{\text{diabetes}}(p) \text{ provided that } A_{\text{diabetes}}(p) \neq \emptyset.$$

This anchor definition allows temporal criteria to be expressed as predicates over event times rather than as ad hoc SQL date fragments. A criterion such as "metformin within 90 days after diabetes diagnosis" can be represented as the existence of a drug exposure event whose event time falls within a bounded interval after $a_{\text{diabetes}}(p)$. Conversely, a criterion such as "no statin exposure after diagnosis" can be represented as the absence of a drug exposure event satisfying a temporal predicate relative to the same anchor.

More generally, let $\Theta(t_e, t_a)$ denote a temporal predicate between a target event time $t_e$ and an anchor time $t_a$. Examples include:

$$\Theta_{\text{after}}(t_e, t_a) \equiv t_e \geq t_a$$

$$\Theta_{\text{within } N \text{ days after}}(t_e, t_a) \equiv t_a \leq t_e \leq t_a + N$$

$$\Theta_{\text{more than } N \text{ days after}}(t_e, t_a) \equiv t_e > t_a + N$$

These predicates provide the formal bridge between natural-language temporal expressions and executable date constraints. Importantly, they are defined before SQL compilation. The compiler later lowers these predicates into SQL conditions using the appropriate domain-specific event-time columns.

This event-time representation is the reason NS-CoT treats temporal and exclusion semantics as first-class components of cohort logic. A temporal expression is not merely a string to be inserted into SQL. It is a predicate over event times. Likewise, an exclusion criterion is not merely surface-level negation. It is a patient-level condition involving the absence of a target event within a specified temporal scope. For example, “no target drug exposure after diagnosis” does not mean that the patient has no drug exposure record anywhere in the database. It means that, relative to a specified anchor event, there does not exist a target drug exposure event satisfying the corresponding temporal predicate. This distinction is essential for preserving eligibility semantics during cohort construction.

The abstraction also clarifies the methodological boundary of the current framework. NS-CoT supports single-anchor temporal predicates, in which an outcome event is evaluated relative to one anchor event. Examples include drug exposure after diagnosis, drug exposure within a bounded interval after diagnosis, and absence of drug exposure after diagnosis. However, the current representation does not claim full support for general multi-anchor sequential temporal chains in which an intermediate outcome event must be preserved as a new anchor for a later predicate. For example, a criterion such as “patients diagnosed with diabetes, started metformin within 90 days, and achieved HbA1c < 7% within 180 days after metformin start” requires the metformin start event to become a second anchor for the HbA1c predicate. Such chained anchor propagation requires additional multi-anchor temporal machinery, such as produced anchors or chained temporal joins, beyond the single-anchor representation emphasized in this research.

In summary, OMOP CDM is formalized in this study as a typed longitudinal event-time representation in which structured clinical records are interpreted as domain-specific events $E_d(p, c, t)$over persons, concepts, and event dates. This representation supports the central design of NS-CoT: natural-language criteria are first mapped to typed clinical facts, grounded to OMOP-compatible concept sets, interpreted as temporal event predicates, and then composed as patient-level cohort logic. The next section defines the typed intermediate representation used to bridge natural-language eligibility criteria and these OMOP event-time predicates.

## 3.4. Typed Intermediate Representation for Cohort Criteria

The event-time representation defined in Section 3.3 provides the data model on which cohort criteria can be evaluated. However, natural-language cohort criteria cannot be executed directly over this representation. A criterion such as "patients with diabetes who started metformin within 90 days after diagnosis" contains multiple computational decisions: the disease mention must be distinguished from the drug mention, the diagnosis event must be recognized as the temporal anchor, the drug exposure must be interpreted as the target event, and the phrase "within 90 days after diagnosis" must be represented as an explicit temporal relation. These decisions must be made before vocabulary grounding, patient-set algebra, and SQL compilation can occur.

In NS-CoT, this intermediate layer is represented by a typed intermediate representation, denoted $IR_q$. The typed IR is the interface between natural-language interpretation and symbolic cohort construction. It is not a free-form explanation of the criterion, nor is it a SQL query. Rather, it is a constrained machine-facing representation that encodes the computationally relevant structure of a feasibility-oriented cohort criterion in typed fields that can be validated, grounded, and passed to downstream symbolic cohort construction.

Given a feasibility-oriented natural-language cohort criterion $q$, the LLM-mediated interpretation layer produces a typed representation:

$$\Phi(q) = IR_q$$

where $IR_q$ consists of a collection of typed facts and their logical or temporal dependencies. For notational clarity, the typed IR can be written as:

$$IR_q = (F_q, L_q)$$

where $F_q = \langle f_1, f_2, \dots, f_n \rangle$ is an ordered collection of typed facts, and $L_q$ represents the logical structure among these facts, such as conjunction, disjunction, grouping, and temporal dependency. The ordering of facts preserves the decomposition structure produced from the original criterion, but execution semantics are not determined by textual order alone. Instead, each fact explicitly specifies its domain, role, polarity, grounding specification, value constraint when applicable, and temporal relation. In this formulation, $L_q$ captures the global logical and dependency structure among facts, whereas $\tau_i$ stores the temporal specification attached to an individual fact.

Each typed fact $f_i$ is defined as:

$$f_i = (id_i, d_i, m_i, r_i, \pi_i, \gamma_i, \tau_i, \nu_i)$$

where $id_i$ is the fact identifier, $d_i$ is the OMOP-compatible domain, $m_i$ is the natural-language clinical mention, $r_i$ is the functional role of the fact, $\pi_i$ is its polarity, $\gamma_i$ is the grounding specification, $\tau_i$ is the temporal specification, and $\nu_i$ is an optional value constraint. The major fields are summarized in Table 3.

| Field | Symbol | Type or allowed values | Meaning |
|---|---|---|---|
| Fact identifier | $id_i$ | String | Unique identifier of the fact within the typed IR |
| Domain | $d_i$ | Condition, Drug, Measurement, Procedure, Observation, Demographic, etc. | Execution domain or predicate class to which the mention is expected to map |
| Mention | $m_i$ | Natural-language span | Clinical expression extracted from the eligibility-derived cohort criterion |
| Role | $r_i$ | anchor, criterion, outcome, modifier | Functional role of the fact in cohort construction |
| Polarity | $\pi_i$ | require, forbid | Whether the fact contributes an inclusion requirement or exclusion condition |
| Grounding specification | $\gamma_i$ | Unresolved slot, candidate concepts, class key, or concept-set specification | Information required for OMOP vocabulary grounding |
| Temporal specification | $\tau_i$ | TemporalSpec or null | Temporal relation or time window associated with the fact, including anchor-relative when applicable |
| Value constraint | $\nu_i$ | ValueSpec or null | Numeric, categorical, or unit-bearing constraint when applicable |

**Table 3. Typed fact schema.**

The domain field $d_i$ determines which event-time relation will later be used to evaluate the fact. For example, a diagnosis mention is represented as a Condition-domain fact and is later evaluated against $E_{\text{Condition}}(p, c, t)$, whereas a medication mention is represented as a Drug-domain fact and is later evaluated against $E_{\text{Drug}}(p, c, t)$. In the empirical scope of this research, the central supported domains are Condition and Drug. Other domains, including Measurement, Procedure, Observation, and Demographic, can be represented in the typed schema, but they are not the primary basis of the empirical claims unless explicitly stated. Demographic criteria differ from event-domain criteria in that they are evaluated as person-level predicates rather than as event-time predicates. Thus, in Table 3, the domain field denotes an execution domain or predicate class, not only an OMOP event domain in $\mathcal{D}_{evt}$.

The mention field $m_i$ stores the natural-language span from which the fact was derived. For instance, in the criterion "patients with diabetes who started metformin within 90 days after diagnosis," the expressions "diabetes" and "metformin" are represented as separate mentions. The mention itself is not yet an OMOP concept. It is a linguistic unit that will later be grounded to a domain-specific concept set in Section 3.5. This distinction is important because a clinical mention may correspond to a single concept, a set of descendant concepts, a pharmacologic class, or no locally executable concept at all.

The role field $r_i$ specifies how a fact participates in cohort construction. An anchor fact defines the event relative to which temporal predicates are evaluated. An outcome fact defines the target event whose existence or absence is tested relative to an anchor. A criterion fact contributes a non-temporal inclusion or exclusion predicate. A modifier fact may provide additional constraints, such as value, frequency, or demographic restrictions. For example, in "metformin within 90 days after diabetes diagnosis," the diabetes condition fact serves as the anchor, while the metformin drug exposure fact serves as the outcome event evaluated relative to that anchor.

The polarity field $\pi_i$distinguishes inclusion requirements from exclusion conditions. A fact with $\pi_i =$ requirecontributes a positive predicate: the corresponding event or person-level condition must be satisfied for a person to remain in the cohort. A fact with $\pi_i = forbid$ contributes an exclusion predicate: the corresponding event must not exist under the specified temporal or logical condition. This explicit polarity field is necessary because exclusion is not treated as a surface-level negation in the natural-language text. It is later interpreted as a patient-set operation and, when temporally anchored, as an anchor-bound absence predicate.

The grounding specification $\gamma_i$ stores the information needed to resolve a natural-language mention to OMOP-compatible concepts. At the typed IR stage, $\gamma_i$ may be empty, partially specified, or contain candidate information. For drug criteria, it may include whether the mention is interpreted as an ingredient, product, brand, or pharmacologic class. For drug-class mentions, it may include a class key or vocabulary path that guides the grounding process. However, $\gamma_i$ should not be interpreted as the final executable concept set. The final grounded concept set is determined only after the grounding stage described in Section 3.5. Once grounding is complete, the typed IR becomes a grounded IR, denoted $IR_q^g$. Thus, $\gamma_i$ denotes a grounding request or specification in the typed IR, whereas $\Gamma_i$ denotes the grounded concept set produced in the grounded IR.

The optional value constraint $v_i$ is used for criteria that involve measurement values, thresholds, categories, or units. For example, “HbA1c < 7%” should not be represented as a single concept-search phrase. Instead, the core measurement mention “HbA1c” is represented separately from the comparator “<”, the threshold “7”, and the unit “%”. This separation prevents numeric thresholds from being absorbed into concept retrieval and allows value-bearing predicates to be compiled as explicit conditions over structured measurement fields. In the present study, value constraints are included in the typed representation for completeness, but the central evaluated criteria focus on condition and drug event predicates.

Temporal dependency is represented by the temporal specification $\tau_i$. If a fact has no temporal dependency, $\tau_i = \emptyset$. If a fact is evaluated relative to an anchor, $\tau_i$ is represented as a TemporalSpec:

$$\tau_i = (\text{anchor_ref}_i, \text{relation}_i, \text{lower}_i, \text{upper}_i, \text{boundary}_i, \text{anchor_policy}_i)$$

Here, $\text{anchor_ref}_i$ identifies the anchor fact, $\text{relation}_i$ specifies the temporal relation type, $\text{lower}_i$ and $\text{upper}_i$ define the lower and upper day offsets relative to the anchor, $\text{boundary}_i$defines boundary inclusivity or exclusivity, and $\text{anchor_policy}_i$ specifies the anchor policy, such as first-anchor selection. Table 4 summarizes the TemporalSpec schema used to represent anchor-relative temporal constraints.

| Field | Symbol | Meaning |
|---|---|---|
| Anchor reference | $anchor_ref_i$ | Identifier of the anchor fact relative to which the current fact is evaluated |
| Temporal relation | $relation_i$ | Type of temporal relation, such as on_or_after, bounded_after, lower_only_after, or same_day |
| Lower offset | $l_i$ | Lower-bound day offset relative to the anchor |
| Upper offset | $u_i$ | Upper-bound day offset relative to the anchor; may be unbounded |
| Boundary specification | $b_i$ | Inclusive or exclusive boundary policy |
| Anchor policy | $policy_i$ | Anchor selection policy, such as first_anchor |

**Table 4. TemporalSpec schema.**

The stable temporal relation types used in the current framework are:

$$relation_i \in \{\text{on_or_after}, strict_after, \text{bounded_after}, \text{lower_only_after}, \text{same_day}\}$$

The relation on_or_after represents events occurring on or after the anchor date.
The relation bounded_after represents events occurring within a lower and upper day offset after the anchor. The relation lower_only_after represents events occurring more than a specified number of days after the anchor.The mathematical semantics of these temporal operators are defined in Section 3.6. In this section, they are introduced only as typed labels in the intermediate representation.

TemporalSpec is intentionally explicit because natural-language temporal expressions often under-specify boundary conditions. For example, “after diagnosis” may be interpreted as strictly after the diagnosis date or as on-or-after the diagnosis date, depending on the operational convention. Similarly, “within 3 months” must be normalized to a fixed day offset or another predefined temporal unit before deterministic compilation. In NS-CoT, these choices are not left implicit in a generated SQL string. They are represented as temporal relation, boundary, offset, and anchor-policy fields in the typed IR.

An example typed IR for the criterion “patients with diabetes who started metformin within 90 days after diagnosis” is shown below.

$$IR_q = (F_q, L_q)$$

where:

$$F_q = \langle f_1, f_2 \rangle$$

$$f_1 = (F1,\text{Condition},\text{“diabetes”},\text{anchor},\text{require}, \emptyset, \emptyset, \emptyset)$$

$$f_2 = (F2,\text{Drug},\text{“metformin”},\text{outcome},\text{require}, \emptyset, \tau_2, \emptyset)$$

and:

$$\tau_2 = (F1,\text{bounded_after}, 0,90,\text{inclusive},\text{first_anchor})$$

In JSON-like notation, the same representation can be written as:

```
{
"criterion_text":
  "patients with diabetes who started metformin within 90 days after diagnosis",
"facts": [
  {
    "id": "F1",
    "domain": "Condition",
    "mention": "diabetes",
    "role": "anchor",
    "polarity": "require",
    "grounding": null,
    "time": null,
    "value": null
  },
```

```
    {
      "id": "F2",
      "domain": "Drug",
      "mention": "metformin",
      "role": "outcome",
      "polarity": "require",
      "grounding": null,
      "time": {
        "anchor_ref": "F1",
        "relation": "bounded_after",
        "lower": 0,
        "upper": 90,
        "boundary": "inclusive",
        "policy": "first_anchor"
      },
      "value": null
    }
  ],
  "logic": {
    "operator": "AND",
    "members": ["F1", "F2"]
  }
}
```

This example illustrates the central function of the typed IR. In this JSON-like representation, AND and OR are serialization labels for logical composition; in the patient-set algebra defined in Section 3.6, they correspond to GROUP_AND and GROUP_OR. The natural-language criterion is decomposed into a condition anchor and a drug outcome. The temporal relation is represented explicitly as a bounded interval relative to the anchor, rather than being embedded directly into SQL. The actual OMOP concepts for “diabetes” and “metformin” are not finalized in this section; they are resolved in the subsequent grounding stage.

Exclusion criteria are represented using the same fact schema with $\pi_i =$ forbid. For example, the criterion “patients with diabetes who had no insulin exposure after diagnosis” can be represented as an anchor condition fact and a forbidden drug exposure fact:

```
{
  "criterion_text":
    "patients with diabetes who had no insulin exposure after diagnosis",
  "facts": [
    {
      "id": "F1",
      "domain": "Condition",
```

```
      "mention": "diabetes",
      "role": "anchor",
      "polarity": "require",
      "grounding": null,
      "time": null,
      "value": null
    },
    {
      "id": "F2",
      "domain": "Drug",
      "mention": "insulin",
      "role": "outcome",
      "polarity": "forbid",
      "grounding": null,
      "time": {
        "anchor_ref": "F1",
        "relation": "on_or_after",
        "lower": 0,
        "upper": null,
        "boundary": "inclusive",
        "policy": "first_anchor"
      },
      "value": null
    }
  ],
  "logic": {
    "operator": "AND",
    "members": ["F1", "F2"]
  }
}
```

This representation does not yet determine how the exclusion will be expressed in SQL. It only states that insulin exposure is a forbidden Drug-domain outcome relative to the diabetes anchor. In Section 3.6, this structure is interpreted as an anchor-bound temporal exclusion predicate over patient sets. In Section 3.7, it is compiled into a deterministic SQL pattern, typically using a correlated NOT EXISTS structure rather than a simple NOT IN condition.

Logical composition is also represented explicitly. A conjunction of criteria is represented as an AND structure over facts or groups. A disjunction is represented as a group with an OR operator, rather than as multiple independent top-level criteria. This distinction is necessary because sequential criteria normally narrow the current cohort, whereas an OR group represents alternative ways of satisfying the same logical condition. For example, “patients with diabetes or hyperlipidemia” should be represented as a single OR group containing two Condition-domain facts, rather than as two sequential condition filters. This prevents disjunctive criteria from being incorrectly compiled as intersections.

Before the typed IR is passed to the grounding and compilation stages, it is subject to schema-level validation. At minimum, each executable fact must have a unique identifier, a supported domain, a non-empty mention, a valid role, and a valid polarity. If $\tau_i$ is not null, its $anchor_ref_i$ must refer to an existing anchor fact, and its temporal relation, boundary, offset, and anchor policy must belong to the supported set of temporal operators and policies. A temporal outcome fact without a valid anchor reference is not compiled as a valid single-anchor predicate. Similarly, an exclusion fact must have an explicit polarity field rather than relying only on the surface occurrence of words such as "no" or "without."

The typed output is also normalized before symbolic cohort construction. This logical plan normalization step operates on typed fields and structural relations; it does not reinterpret the original free text or invent new clinical concepts. Its purpose is to make the typed representation structurally consistent and compiler-ready while preserving the semantic commitments already expressed in the IR. Typical normalization operations include removing non-executable facts such as section labels or metadata, normalizing disjunctive criteria into explicit OR groups, normalizing exclusion signals into $\pi_i =$ forbid, enforcing temporal bindings between anchor and outcome facts, and preventing executable clinical facts from being dropped because of inconsistent auxiliary fields. In this sense, logical plan normalization serves as a symbolic bridge between LLM-mediated decomposition and deterministic cohort construction.

The current typed IR is designed for single-anchor temporal cohort predicates. An outcome fact may refer to an anchor fact, but the representation does not generally support a chain in which an intermediate outcome event becomes a new anchor for a subsequent temporal predicate. For example, the criterion "patients diagnosed with diabetes, started metformin within 90 days, and achieved HbA1c < 7% within 180 days after metformin start" requires the metformin start event to become a second anchor for the HbA1c measurement predicate. Such multi-anchor sequential temporal chains require additional representational machinery, such as produced anchors, anchor propagation, or explicit chained temporal joins, and are outside the current scope of the system.

In summary, the typed IR provides a constrained and inspectable representation of the semantic structure extracted from site feasibility patient-count questions with protocol-derived inclusion and exclusion constraints. It separates natural-language interpretation from OMOP vocabulary grounding and SQL generation. Each cohort criterion is represented as typed facts with explicit domain, mention, role, polarity, grounding specification, value constraint when applicable, and temporal dependency. This representation is the input to the grounding process described in the next section, where natural-language mentions are resolved to OMOP-compatible concept sets and, when applicable, filtered according to local CDM executability.

## 3.5. Standard-Vocabulary Grounding and Local CDM-Aware Executable Concept-Set Construction

The typed IR defined in Section 3.4 specifies the semantic structure of a cohort criterion, but it is not yet executable over an OMOP CDM database. A typed fact such as “diabetes,” “metformin,” or “statin” identifies a clinical mention and its expected domain, role, polarity, and temporal dependency. However, the event-time relations defined in Section 3.3 are evaluated over OMOP concept identifiers, not over natural-language strings. Therefore, before a typed fact can be interpreted as a patient-level predicate, its clinical mention must be grounded to an OMOP-compatible concept set.

This section defines the grounding stage that converts typed clinical mentions into grounded facts. The purpose of grounding is not merely to find a vocabulary code that resembles the input phrase. It is to produce a concept set that can support cohort execution under the representational assumptions of OMOP CDM and the local database instance. In this method, grounding is treated as a staged process involving candidate retrieval, semantic and domain validation, vocabulary graph expansion, and, where applicable, local CDM-aware executable concept filtering.

Let $m_i$ denote the natural-language mention in a typed fact $f_i$, and let $d_i$ denote its assigned domain. For each domain $d$, let $\mathcal{V}_d$ be the set of OMOP-compatible concepts that can represent events in that domain. The following notation distinguishes vocabulary-level grounding, local executable grounding, and the final grounded concept set used for execution.

$$G_d^{vocab}(m) \subseteq \mathcal{V}_d$$

$$G_d^{exec}(m; \mathrm{D}) \subseteq G_d^{vocab}(m)$$

Here, vocabulary-level grounding returns concepts that are valid in the OMOP vocabulary, whereas local executable grounding restricts that vocabulary-grounded set to concepts that are executable in the target OMOP CDM instance when local filtering is applied. The final grounded concept set stored in a grounded fact is denoted by $\Gamma_i$. In general form, grounding can be represented as:

$$G_d^{vocab}(m) = Expand_d\left(Validate_d\left(Retrieve_d(m)\right)\right)$$
$$G_d^{exec}(m; D) = LocalFilter_d\left(G_d^{vocab}(m); D\right)$$
$$\Gamma_i = G_{d_i}^{exec}(m_i; D)$$

When local executable filtering is not applied for a domain, $G_d^{exec}(m; D)$ reduces to the vocabulary-grounded set $G_d^{vocab}(m)$. Here, $Retrieve_d(m)$ denotes candidate concept retrieval for mention $m$ within domain $d$. $Validate_d(\cdot)$ retains candidates that are semantically compatible with the intended clinical meaning and the expected OMOP domain. $Desc_d(\cdot)$ expands selected seed concepts through OMOP vocabulary hierarchy or relationship structure when descendant concepts are required for cohort execution. This formulation separates retrieval, validation, and expansion rather than allowing all grounding decisions to be hidden inside a generated SQL query.

For a typed fact:

$$f_i = (id_i, d_i, m_i, r_i, \pi_i, \gamma_i, \tau_i, \nu_i)$$

the grounding stage produces a grounded fact:

$$f_i^g = (id_i, d_i, m_i, r_i, \pi_i, \Gamma_i, \rho_i, \kappa_i, \tau_i, \nu_i)$$

where $\Gamma_i = G_{d_i}^{exec}(m_i; D)$ is the grounded concept set, $\rho_i$ records the grounding status, and $\kappa_i$ stores grounding provenance and execution-relevant metadata. The collection of grounded facts forms the grounded IR:

$$IR_q^g = G(IR_q; D)$$

The grounded IR is the representation consumed by the subsequent symbolic temporal and patient-set algebra. After this point, the compiler does not reinterpret the original natural-language mention. It operates on typed fields, grounded concept sets, temporal specifications, polarity, and value constraints.

| Field | Symbol | Meaning |
|---|---|---|
| Grounded concept set | $\Gamma_i$ | OMOP-compatible concept identifiers used for execution |
| Grounding status | $\rho_i$ | Status of grounding, such as resolved, unresolved, ambiguous, or executable-empty |
| Grounding provenance | $\kappa_i$ | Retrieval path, typed grounding specification $\gamma_i$, seed concepts, vocabulary relationships, descendant expansion, and local filtering metadata |

| Field | Symbol | Meaning |
|---|---|---|
| Original typed fields | $id_i, d_i, m_i, r_i, \pi_i, \tau_i, v_i$ | Preserved fields from the typed IR |

**Table 5. Grounded fact fields added during concept resolution.**

### 3.5.1. Vocabulary Graph and Concept Expansion.

OMOP vocabulary is used in this study not as a flat dictionary, but as a structured vocabulary graph. Concepts are represented as nodes, and vocabulary relationships are represented as edges. The relevant vocabulary tables include concept, concept_synonym, concept_relationship, and concept_ancestor. In this study, these tables are treated as the OMOP vocabulary infrastructure that supports standard concept selection, source-to-standard mapping, class-to-descendant expansion, relationship-based traversal, and provenance-aware inspection of concept names and synonym metadata. However, in the evaluated implementation, concept candidate retrieval was performed primarily through ChromaDB vector indexes over OMOP concept metadata, followed by domain validation, standard-concept validation, vocabulary-relationship expansion, descendant expansion where configured, and local CDM evidence filtering. Therefore, synonym metadata are part of the available OMOP vocabulary infrastructure, but a standalone synonym-expansion or lexical-search module is not claimed as an independently evaluated component of NS-CoT.

Concept expansion is necessary because eligibility-criteria-derived cohort definitions often refer to a clinical category rather than a single database concept. A disease mention may require inclusion of descendant condition concepts. A drug mention may refer to an ingredient, product, branded product, or pharmacologic class. If grounding were limited to a single concept identifier, the resulting cohort definition could be too narrow and fail to represent the intended clinical meaning. Conversely, if grounding expands too broadly, the cohort may include events that are not semantically appropriate. The role of grounding is therefore to construct a concept set that is semantically appropriate for the declared criterion and executable in the target database.

For a grounded fact with mention $m_i$, domain $d_i$, and grounded concept set $\Gamma_i$, the executable event set for person $p$ is:

$$\mathcal{E}_d(m,p) = \{(c,t) \mid E_d(p,c,t) \land c \in \Gamma_i\}$$

The quality of this event set depends directly on the grounding concept set $\Gamma_i$. If $G_d(m)$is too narrow, qualifying events may be missed. If it is too broad, clinically inappropriate events may be included. For this reason, grounding is a cohort-defining decision, not a preprocessing detail.

### 3.5.2. Condition-Domain Grounding.

For Condition-domain facts, grounding links a natural-language disease mention to a condition-compatible OMOP concept set. A condition mention such as “diabetes” or “hyperlipidemia” is first retrieved against condition-compatible standard concepts. Candidate concepts are then validated for semantic compatibility and expanded when descendant condition concepts are intended to be included.

For a Condition-domain mention $m$, the grounded concept set is:

$$\Gamma_{\text{Condition}}(m; D) \subseteq \mathcal{V}_{\text{Condition}}$$

The corresponding condition event set for person $p$ is:

$$\mathcal{E}_{\text{Condition}}(m, p) = \{(c, t) \mid E_{\text{Condition}}(p, c, t) \wedge c \in \Gamma_{\text{Condition}}(m; D)\}$$

If the condition fact has the role $r_i = \text{anchor}$, this event set later provides candidate anchor times. For example, if the mention is “diabetes,” the first-anchor policy defined in Section 3.3 can derive:

$$a_{\text{diabetes}}(p) = \min\{t \mid \exists c \in \Gamma_{\text{Condition}}(\text{diabetes}; D) \colon E_{\text{Condition}}(p, c, t)\}$$

provided that the set is not empty. Thus, condition grounding determines which diagnosis events can serve as anchor events, but the temporal interpretation of those events is handled by the symbolic temporal algebra in Section 3.6.

### 3.5.3. Drug Mention Subtype Routing.

Drug-domain grounding requires additional structure because natural-language drug mentions may refer to different levels of pharmacologic abstraction. A drug expression may denote an ingredient, a clinical drug formulation, a branded product, or a pharmacologic class. Treating all of these as a single undifferentiated Drug mention can lead to unstable grounding. For example, “statin” denotes a pharmacologic class, whereas “atorvastatin” denotes an ingredient-level concept, and “Lipitor” denotes a branded product.

NS-CoT therefore distinguishes drug mention types before constructing executable drug concept sets. In the typed IR, this distinction may be represented through drug-specific fields such as is_drug_class, class_key, or an equivalent subtype indicator. The purpose of this routing is to determine the appropriate vocabulary path before concept expansion and execution.

| Drug mention type | Examples | Primary grounding path |
|---|---|---|
| Ingredient | atorvastatin, lisinopril, meloxicam | RxNorm ingredient-compatible grounding |
| Clinical drug / formulation | atorvastatin 20 mg oral tablet | RxNorm clinical drug or formulation grounding |
| Brand or product | Lipitor, A-STIN | Brand/product grounding followed by standard mapping where required |
| Pharmacologic class | statin, ACE inhibitor, SSRI, NSAID | Class-constrained grounding through ATC, SNOMED, or linked vocabulary structures |

**Table 6. Drug mention subtype routing.**

This subtype-aware routing is a methodological component of the framework. The LLM-mediated layer may identify that a drug mention is a pharmacologic class, brand, formulation, or ingredient, but the LLM does not generate the executable drug concept set. Given the typed drug mention and the selected grounding path, executable concepts are derived through OMOP vocabulary structure and local CDM evidence. Figure 3 illustrates this grounding pathway using the drug-class mention “statin.” The example shows why drug-class resolution cannot be reduced to a generic vector search over drug names: the executable concept set is produced through vocabulary graph traversal and local CDM evidence filtering. This example also illustrates the distinction between vocabulary-level semantic validity and local CDM executability. A drug class may be valid in the OMOP vocabulary, but only the subset supported by observed local ‘drug_exposure’ records can be used as an executable cohort filter in the target database.

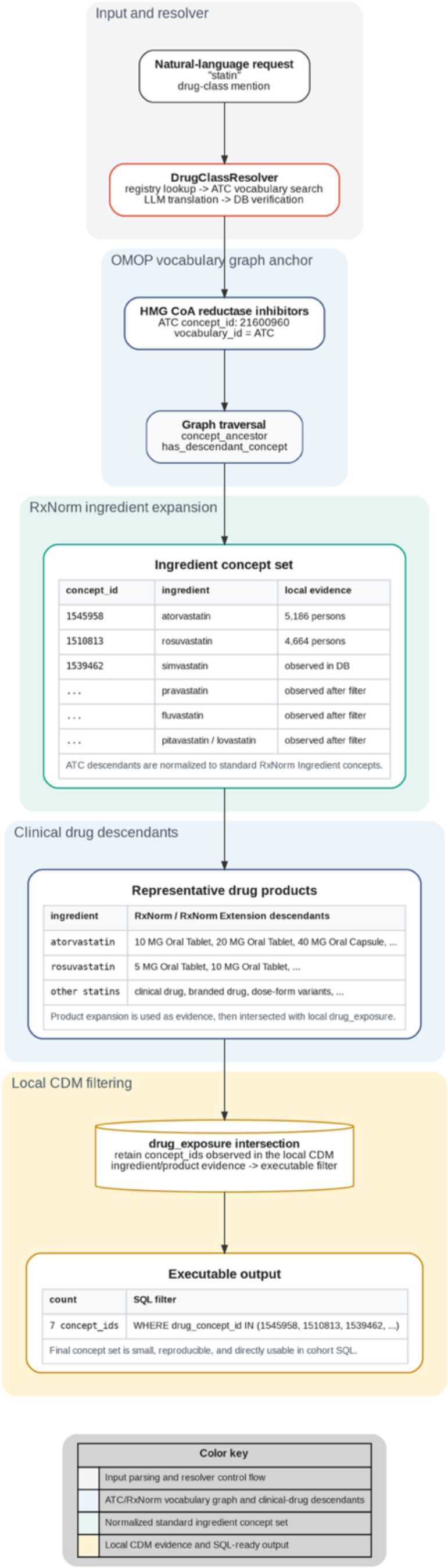

**Figure 3. Drug-class grounding and local executable concept-set construction.** The figure illustrates how NS-CoT resolves the natural-language drug-class mention "statin" into an executable OMOP concept set. The resolver first maps the mention to an ATC class concept, traverses the OMOP vocabulary graph through 'concept_ancestor', normalizes descendants to standard RxNorm ingredient concepts, expands representative clinical drug descendants, and finally intersects the resulting concept evidence with locally observed 'drug_exposure' records. The final output is a SQL-ready concept set that is both vocabulary-grounded and executable in the target CDM instance.

**3.5.4. Non-Class Drug Grounding.**

For non-class drug mentions, such as ingredients, formulations, clinical drugs, or branded products, grounding proceeds through drug-compatible vocabulary concepts. The retrieved concept may correspond to a standard concept directly, or it may require mapping to standard drug concepts through OMOP vocabulary relationships. Descendant expansion may then be applied when a broader ingredient-level or product-level concept should include more specific descendant drug concepts.

For a non-class drug mention $m$, the concept set can be written as:

$$S_{\mathrm{Drug}}^{seed}(m) = Validate_{\mathrm{Drug}}\left(Retrieve_{\mathrm{Drug}}(m)\right)$$

$$G_{\mathrm{Drug}}^{vocab}(m) = Expand_{\mathrm{Drug}}\left(S_{\mathrm{Drug}}^{seed}(m)\right)$$

The final executable concept set used in SQL is denoted $\Gamma_{\mathrm{Drug}}(m; D)$, which may be identical to $G_{\mathrm{Drug}}^{vocab}(m)$ or may be restricted through local CDM-aware filtering when such filtering is applied. When local CDM-aware filtering is applied, the executable subset is restricted to concepts that are actually observed in the target drug_exposure table. This restriction is especially important for drug classes, but the same distinction between vocabulary-level validity and local executability can also be useful for non-class drug criteria.

The resulting drug event set is evaluated over the drug_exposure table:

$$\mathcal{E}_{\mathrm{Drug}}(m, p) = \left\{(c, t) \mid E_{\mathrm{Drug}}(p, c, t) \wedge c \in \Gamma_{\mathrm{Drug}}(m; D)\right\}$$

**3.5.5. Pharmacologic Class Grounding.**

Pharmacologic class mentions require a distinct grounding path. A class mention such as “statin” should not be resolved by asking the LLM to produce a list of statin ingredients. Such a list would be difficult to verify, could omit relevant descendants, and would place cohort-defining semantics inside stochastic generation. Instead, NS-CoT treats pharmacologic classes as vocabulary-graph grounding problems.

Let $m$be a pharmacologic class mention. The class grounding process first identifies a class-level seed concept or vocabulary path, denoted $K(m)$. This may come from an explicit class key in the typed IR or from a class-constrained search over vocabulary names and synonyms. The selected class concept is then expanded through vocabulary relationships to obtain descendant or related drug concepts:

$$S_{\text{class}}(m) = Desc_{\text{Drug}}(K(m))$$

The set $S_{\text{class}}(m)$ is a vocabulary-level expansion. It represents drug concepts that are semantically reachable from the class grounding path, but it is not necessarily the set that should be inserted into the executable SQL. $Desc_{Drug}$ is used here as shorthand for configured vocabulary traversal over descendant or related drug concepts. A local OMOP CDM instance may contain only a subset of those concepts in the drug_exposure table. Therefore, the executable drug concept set is defined by intersecting the vocabulary-expanded set with locally observed drug exposure concepts:

$$O_{\text{Drug}}(D) = \{c \mid \exists p, t: E_{\text{Drug}}(p, c, t)\}$$
$$G_{\text{Drug}}^{exec}(m; D) = S_{\text{class}}(m) \cap O_{\text{Drug}}(D)$$

For the corresponding grounded fact $f_i^g$, the final grounded concept set is $\Gamma_i = G_{\text{Drug}}^{exec}(m_i; D)$. This operation separates recognition from execution. The class concept or ATC/SNOMED vocabulary node is used to recognize and expand the clinical class. The final SQL predicate, however, is evaluated against standard drug concept identifiers that are observed in the local drug_exposure table:

$$\text{drug_exposure.drug_concept_id} \in G_{\text{Drug}}^{exec}(m; D)$$

For example, the mention "statin" may correspond to a pharmacologic class such as HMG-CoA reductase inhibitors. The grounding path may identify the class through ATC or another linked vocabulary structure, expand the class through vocabulary relationships, and then restrict the resulting set to locally observed RxNorm-compatible drug concepts. The final executable concept set is not the string "statin," nor the class concept itself, but the set of locally observed standard drug concepts that instantiate the class in the target database.

This design is particularly important for cohort feasibility. Feasibility counts are computed in a specific database instance. A concept set that is valid at the vocabulary level may still be non-executable locally if no corresponding descendant concepts are observed in the target CDM. Conversely, local exposure records may appear under more specific descendant concepts rather than under the seed concept originally retrieved from the mention. Descendant expansion preserves semantic coverage; local exposure filtering preserves executability.

### 3.5.6. Grounding Status and Executable-Empty Concept Sets.

The grounding stage must represent not only successful concept resolution but also cases in which no executable concepts are found. This is especially important for drug class criteria. A pharmacologic class may be valid in the OMOP vocabulary graph, but none of its expanded descendant drug concepts may appear in the local drug_exposure table. In such cases, the executable concept set is empty:

$$\Gamma_i = G_{\mathrm{Drug}}^{exec}(m_i; D) = \emptyset$$

This state is not treated as an undefined pipeline failure. It is represented explicitly in the grounded IR as an executable-empty concept set, together with a grounding status and provenance. The patient-set consequence of an empty concept set depends on polarity and is defined formally in Section 3.6. Intuitively, a required exposure with an empty executable concept set cannot be satisfied by any patient. By contrast, a forbidden exposure with an empty executable concept set does not remove patients, because there is no locally observable event that can trigger the exclusion.

Preserving the executable-empty state is methodologically preferable to silently failing or substituting an unvalidated concept list. It allows the downstream symbolic layer to compile the criterion consistently, and it allows the analyst to inspect whether the absence of matched events is due to true cohort logic, local data absence, or vocabulary resolution failure.

**3.5.7. Grounding Provenance.**

A grounded fact contains both execution inputs and provenance. For each grounded clinical mention, NS-CoT preserves the original mention, assigned domain, selected grounding path, seed concept candidates, accepted concept set, descendant expansion results, local executable subset where applicable, and grounding status. For drug criteria, the provenance additionally records whether the mention was interpreted as an ingredient, formulation, brand, product, or pharmacologic class.

The grounding provenance is not part of the SQL predicate itself, but it is part of the cohort definition artifact. It makes the concept-set decision inspectable. If two cohort counts differ, the analyst can determine whether the difference arose from concept selection, descendant expansion, class routing, local exposure filtering, temporal interpretation, or SQL execution. This is one of the reasons NS-CoT treats grounding as a formal stage in the pipeline rather than as an implicit preprocessing step.

The output of this section is the grounded IR:

$$IR_q^g = (F_q^g, L_q)$$
$$F_q^g = \langle f_1^g, f_2^g, \dots, f_n^g \rangle$$
$$f_i^g = (id_i, d_i, m_i, r_i, \pi_i, \Gamma_i, \rho_i, \kappa_i, \tau_i, \nu_i)$$

where each grounded fact has a typed domain, polarity, role, temporal specification, optional value constraint, and execution-ready concept information. This grounded representation is the input to the symbolic temporal and exclusion algebra in the next section. There, grounded facts are no longer treated as language mentions; they become patient-set predicates over OMOP event-time relations.

## 3.6. Symbolic Temporal and Exclusion Cohort Algebra

The grounded IR produced in Section 3.5 is not directly interpreted as SQL. In NS-CoT, grounded facts are first converted into patient-level predicates over OMOP event-time relations. Cohort construction is then expressed as a restricted patient-set algebra before it is lowered into executable SQL. This section defines that symbolic layer. Collectively, applying the operations defined in this section to $IR_q^g$ yields $h_q = \mathcal{A}(IR_q^g)$, the OMOP-executable cohort logic that is passed to the deterministic SQL compiler in Section 3.7.

The term patient-set algebra is used in a restricted methodological sense. It does not claim to introduce a new general algebraic system. Rather, it refers to a set-theoretic representation of cohort construction in which cohorts are modeled as subsets of observed persons and site feasibility patient-count questions with protocol-derived inclusion and exclusion constraints are represented as operations over those subsets. This representation is useful because it makes explicit what SQL often hides: which events are required, which events are forbidden, which temporal anchor is used, and how patient sets are narrowed or combined.

Let $IR_q^g = \langle f_1^g, f_2^g, \dots, f_n^g \rangle$ denote the grounded IR for criterion $q$. Each grounded fact $f_i^g$ contains a domain $d_i$, a grounded concept set $\Gamma_i$, a role $r_i$, a polarity $\pi_i$, an optional temporal specification $\tau_i$, and an optional value constraint $\nu_i$. For an event-domain fact, the grounded concept set is evaluated over the event-time relation defined in Section 3.3.

For each grounded fact $f_i^g$, define the event-time set for person $p$as:

$$T_i(p) = \{t \mid \exists c \in \Gamma_i : E_{d_i}(p, c, t) \wedge V_i(p, c, t)\}$$

where $V_i(p, c, t)$ is the value predicate associated with $v_i$. If the fact has no value constraint, $V_i(p, c, t)$ is treated as true. For a Condition-domain fact, $T_i(p)$ is the set of dates on which person $p$has a qualifying condition occurrence. For a Drug-domain fact, $T_i(p)$ is the set of dates on which person $p$has a qualifying drug exposure. In the current implementation, drug exposure time is primarily represented by drug_exposure_start_date.

A grounded fact without temporal dependency is interpreted as a simple event-existence predicate:

$$P_i(p) \equiv T_i(p) \neq \emptyset$$

This predicate is true when person $p$has at least one qualifying event for the grounded concept set of $f_i^g$. For example, a grounded diabetes condition fact is true for person $p$ if $p$ has at least one condition occurrence whose concept belongs to the diabetes concept set. A grounded metformin drug fact is true if $p$ has at least one qualifying drug exposure record for metformin-compatible concepts.

### 3.6.1. Anchor Event Semantics.

Temporal eligibility criteria require a reference event. In NS-CoT, this reference event is represented by an anchor fact. If a grounded fact $f_a^g$ has role $r_a =$ anchor, it defines an anchor event-time set for each person:

$$A_a(p) = T_a(p)$$

The current framework uses a first-anchor policy for the single-anchor temporal predicates evaluated in this study. If $A_a(p) \neq \emptyset$, the first anchor date is defined as:

$$a(p) = \min A_a(p)$$

If $A_a(p) = \emptyset$, then $a(p)$ is undefined, and person $p$cannot satisfy a temporal criterion that depends on that anchor. Operationally, the anchor criterion is applied before dependent temporal predicates. Thus, in a criterion such as "metformin within 90 days after diabetes diagnosis," the diabetes condition event first defines the eligible anchor population, and the metformin exposure is evaluated relative to each patient's first qualifying diabetes diagnosis date.

For a specific condition mention $m_a$, the first-anchor date can be written as:

$$a_a(p) = \min\{t \mid \exists c \in \Gamma_a : E_{\text{Condition}}(p, c, t)\}$$

provided that the set is not empty. This anchor policy is not a SQL artifact; it is a cohort-definition decision represented before SQL compilation. Figure 4 visualizes how temporal inclusion and temporal exclusion are evaluated relative to an anchor event in the current single-anchor formalization.

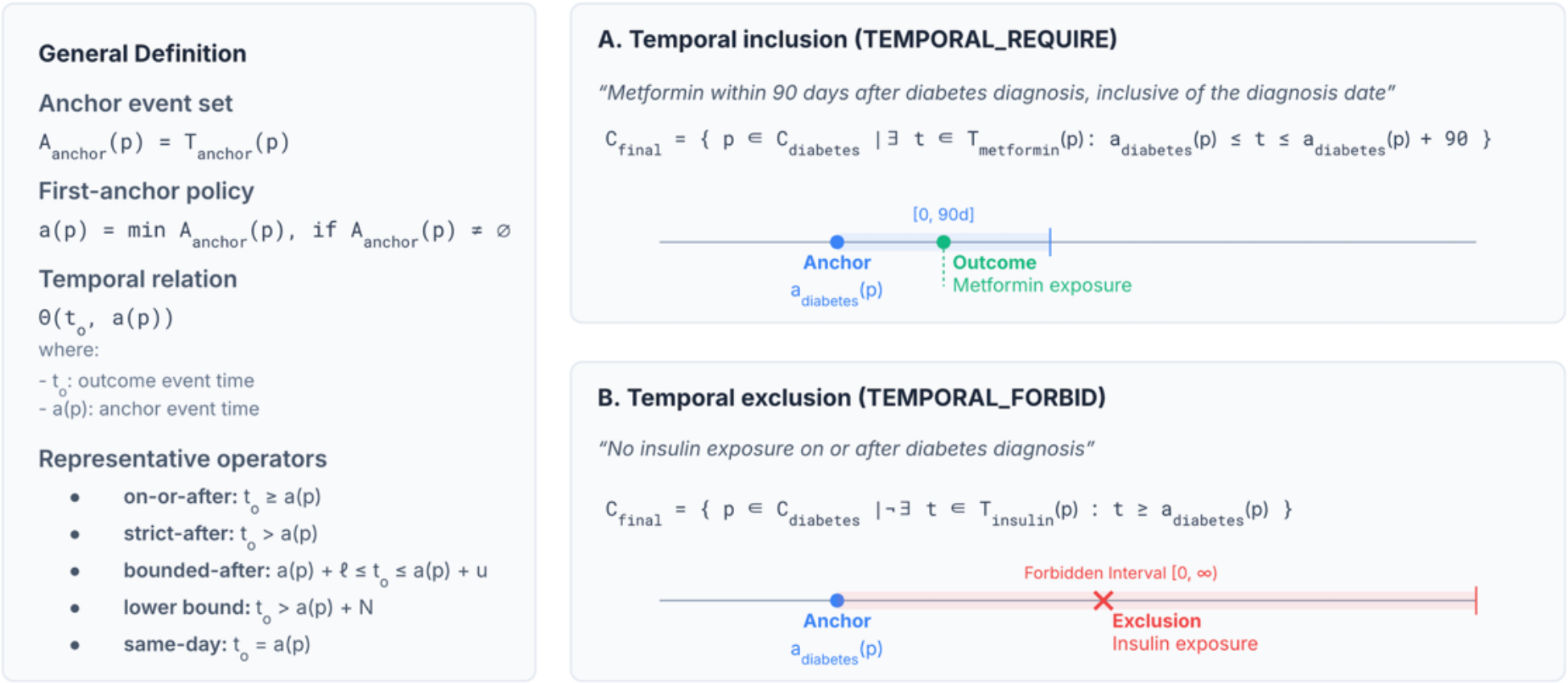

**Figure 4. Anchor-relative temporal semantics.** Temporal criteria are evaluated relative to an anchor event selected by the first-anchor policy. TEMPORAL_REQUIRE implements anchor-relative temporal inclusion, whereas TEMPORAL_FORBID implements anchor-bound temporal exclusion. The examples illustrate metformin exposure within 90 days after diabetes diagnosis and absence of insulin exposure after diabetes diagnosis.

### 3.6.2. Temporal Operators.

Temporal relations are defined over event times. Let $t_e$ denote the outcome event time and $t_a$ denote the anchor time. A temporal operator evaluates whether the outcome event satisfies the required relation to the anchor.

The implemented framework supports a small set of single-anchor temporal operators. These operators are sufficient to represent the central criteria evaluated in this work, including exposure on or after diagnosis, exposure within a bounded interval after diagnosis, exposure more than N days after diagnosis, and same-day event matching.

The on-or-after operator is defined as:

$$\Theta_{\text{on_or_after}}(t_e, t_a) \equiv t_e \geq t_a$$

This operator includes same-day events. It is useful when an operational convention treats "after diagnosis" as "on or after the diagnosis date." By contrast, a strict-after operator excludes same-day events:

$$\Theta_{\text{strict_after}}(t_e, t_a) \equiv t_e > t_a$$

A bounded-after operator is used for expressions such as "within 90 days after diagnosis." With inclusive lower and upper bounds, it is defined as:

$$\Theta_{\text{bounded_after}}^{[l,u]}(t_e, t_a) \equiv t_a + l \leq t_e \leq t_a + u$$

For example, "within 90 days after diagnosis" under an inclusive on-or-after policy can be represented as:

$$\Theta_{\text{bounded_after}}^{[0,90]}(t_e, t_a) \equiv t_a \leq t_e \leq t_a + 90$$

If the operational policy requires strict exclusion of the anchor date, the lower boundary can be made exclusive:

$$\Theta_{\text{bounded_after}}^{[0,90]}(t_e, t_a) \equiv t_a < t_e \leq t_a + 90$$

A lower-only-after operator is used for expressions such as "more than N days after diagnosis." It has a lower temporal bound but no upper bound:

$$\Theta_{\text{lower_only_after}}^{(N,\infty)}(t_e, t_a) \equiv t_e > t_a + N$$

This operator is particularly important for temporal exclusion criteria, such as "no exposure to drug more than N days after the first diagnosis of condition $m_a$."

The same-day operator is defined as:

$$\Theta_{\text{same_day}}(t_e, t_a) \equiv t_e = t_a$$

Boundary behavior is represented explicitly because natural-language expressions often under-specify inclusivity. "After diagnosis," "within 90 days," "at least 30 days after," and "more than 30 days after" are not equivalent expressions. The typed IR therefore stores temporal relation, offset, and boundary policy before SQL compilation. This prevents boundary decisions from being silently embedded inside a generated SQL fragment.

| **Temporal relation** | **Operator** | **Semantics** |
|---|---|---|
| On or after anchor | $\Theta_{\text{on_or_after}}$ | $t_e \geq t_a$ |
| Strict after anchor | $\Theta_{\text{strict_after}}$ | $t_e > t_a$ |
| Bounded after anchor | $\Theta_{\text{bounded_after}}^{[l,u]}$ | $t_a + l \leq t_e \leq t_a + u$ |

| Temporal relation | Operator | Semantics |
|---|---|---|
| Strict lower bounded after anchor | $\Theta^{(l,u]}_{\text{bounded_after}}$ | $t_a + l < t_e \le t_a + u$ |
| Lower-only after anchor | $\Theta^{(N,\infty)}_{\text{lower_only_after}}$ | $t_e > t_a + N$ |
| Same day | $\Theta_{\text{same_day}}$ | $t_e = t_a$ |

**Table 7. Temporal operators used in single-anchor cohort predicates.**

Calendar expressions are normalized before deterministic compilation. For example, a phrase such as "within 3 months" may be converted to a predefined integer day offset, such as 90 days, depending on the system convention. This normalization is part of the semantic representation, not a property of SQL syntax.

**3.6.3. Patient-Set Operations.**

The output of cohort construction is a set of persons. Let:

$$C \subseteq P_{obs}(D)$$

denote a current cohort, and let the initial cohort be:

$$C_0 = P_{obs}(D)$$

A required predicate restricts the current cohort to persons satisfying the predicate:

$$REQUIRE(C, P) = \{p \in C \mid P(p)\}$$

A forbidden predicate removes persons satisfying the predicate:

$$FORBID(C, P) = \{p \in C \mid \neg P(p)\}$$

Equivalently:

$$FORBID(C, P) = C \setminus \{p \in C \mid P(p)\}$$

Boolean composition can also be represented as patient-set operations. For predicates $P_1, \dots, P_k$ evaluated within a current cohort $C$:

$$GROUP_AND(C; P_1, \dots, P_k) = \{p \in C \mid P_1(p) \wedge \cdots \wedge P_k(p)\}$$

$$GROUP_OR(C; P_1, \dots, P_k) = \{p \in C \mid P_1(p) \vee \cdots \vee P_k(p)\}$$

Equivalently, after each predicate $P_j$ is evaluated within the same current cohort $C$ to produce $C_j = \{p \in C \mid P_j(p)\}$, the group operations can be written as set intersection or union. If $C_j = \{p \in C \mid P_j(p)\}$, then:

$$GROUP_AND(C_1, \ldots, C_k) = \bigcap_{j=1}^{k} C_j$$

$$GROUP_OR(C_1, \ldots, C_k) = \bigcup_{j=1}^{k} C_j$$

This algebra is the symbolic meaning of cohort construction in NS-CoT. SQL is generated later as a deterministic lowering of these patient-set operations. The symbolic layer therefore does not ask only whether a sentence can be translated into syntactically valid SQL. It asks which patient-level predicates should be applied, how they are anchored, whether they are required or forbidden, and how the resulting patient sets should be combined. Figure 5 summarizes the patient-set algebra used to assemble grounded cohort facts into OMOP-executable cohort logic. Event-set temporal joins, such as same-day or within-window co-occurrence, are handled as temporal-relation or SQL-lowering constructs rather than as separate core algebra operators.

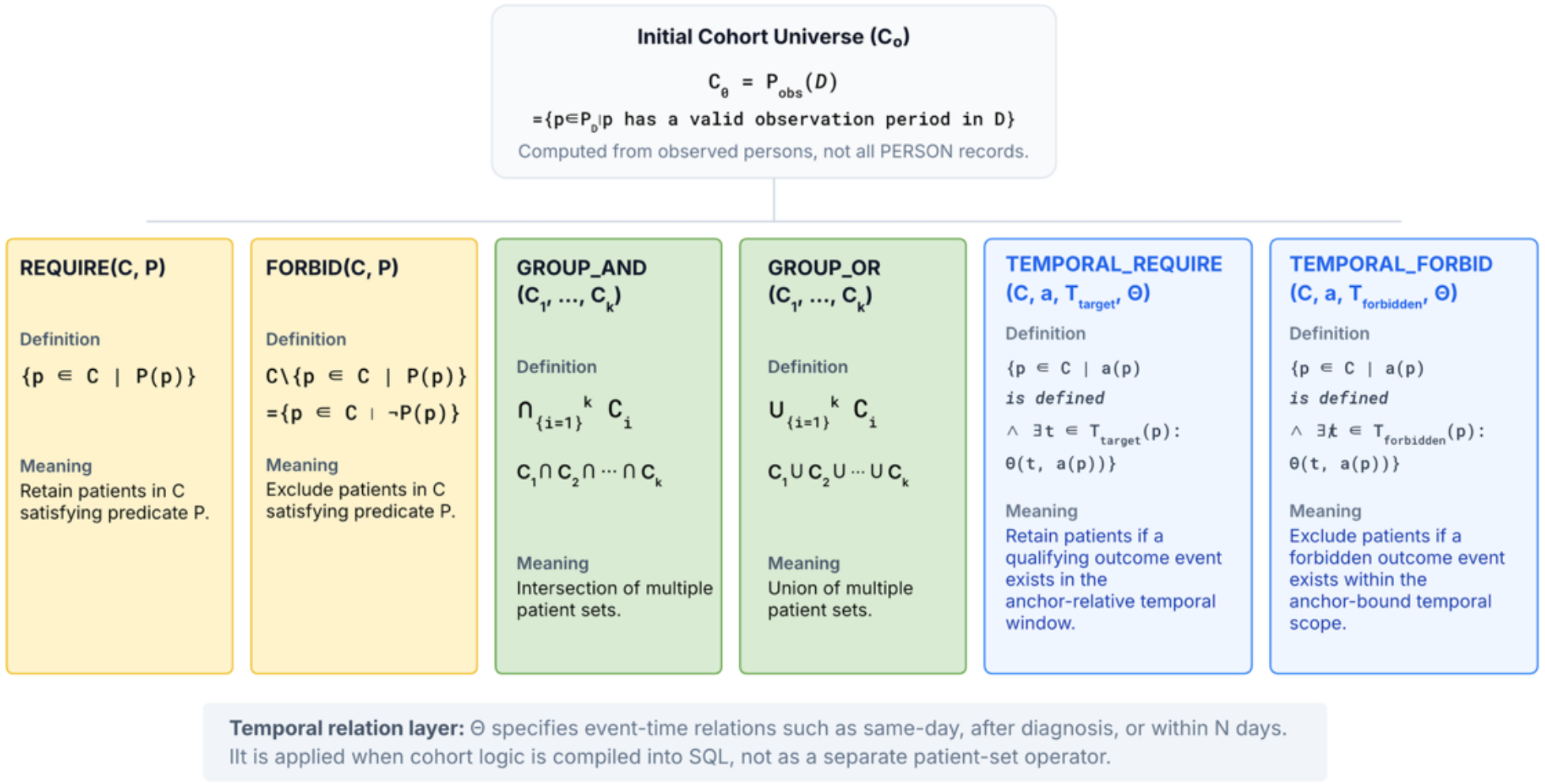

**Figure 5. Patient-set algebra for OMOP-executable cohort logic.** Starting from the observed-person universe $C_0 = Pobs(D)$, REQUIRE and FORBID implement patient-level inclusion and exclusion, GROUP_AND and GROUP_OR implement intersection and union, and TEMPORAL_REQUIRE and TEMPORAL_FORBID implement anchor-relative temporal inclusion and exclusion. Event-set temporal joins, such as same-day or within-window co-occurrence, are treated as temporal-relation or SQL-lowering constructs rather than separate core algebra operators.

#### 3.6.4. Single-Anchor Temporal Inclusion.

A grounded outcome fact with temporal dependency is interpreted as an anchor-relative event predicate. Let $f_o^g$ be an outcome fact and $f_a^g$ its anchor fact. Let $T_o(p)$ be the outcome event-time set and $a(p)$ be the first anchor date. For a temporal operator $\Theta$, the temporal existence predicate is:

$$P_{o|a}^{\Theta}(p) \equiv a(p) \text{ is defined} \wedge \exists t_e \in T_o(p): \Theta(t_e, a(p))$$

A temporal inclusion criterion is then represented as:

$$TEMPORAL_REQUIRE(C, a, o, \Theta) = \{p \in C \mid P_{o|a}^{\Theta}(p)\}$$

For example, the criterion “patients who started metformin within 90 days after diabetes diagnosis” can be expressed as follows. Let $a_{\text{diabetes}}(p)$ be the first diabetes diagnosis date and $T_{\text{metformin}}(p)$ be the set of qualifying metformin exposure dates.

The temporal predicate is:

$$P_{\text{metformin}|\text{diabetes}}^{[0,90]}(p) \equiv \exists t_e \in T_{\text{metformin}}(p): a_{\text{diabetes}}(p) \leq t_e \leq a_{\text{diabetes}}(p) + 90$$

The resulting cohort is:

$$C_{\text{metformin within 90d}} = TEMPORAL_REQUIRE(C_{\text{diabetes}}, a_{\text{diabetes}}, \text{metformin}, \Theta_{\text{bounded_after}}^{[0,90]})$$

where $C_{\text{diabetes}}$ is the cohort already restricted to persons with a qualifying diabetes anchor.

The current framework is intentionally narrower than a general temporal logic over arbitrary event chains. The supported pattern is:

anchor event → outcome event relative to that anchor

It does not generally propagate the outcome event as a new anchor for a later predicate.

**3.6.5. Temporal Exclusion and Anchor-Bound Absence.**

Temporal exclusion is one of the central semantics of NS-CoT. A simple non-temporal exclusion removes persons who satisfy a forbidden predicate:

$$FORBID(C, P_i) = \{p \in C \mid \neg P_i(p)\}$$

However, many eligibility criteria use exclusion in a temporally anchored way. For example, “no insulin exposure after diabetes diagnosis” does not mean that the patient has never had insulin exposure at any time in the database. It means that, relative to a diabetes diagnosis anchor, there is no insulin exposure satisfying the specified temporal predicate.

Let $f_o^g$ be a forbidden outcome fact and $f_a^g$ its anchor fact. The anchor-bound temporal exclusion predicate is:

$$TEMPORAL_FORBID(C, a, o, \Theta) =$$
$$\{p \in C \mid a(p) \text{ is defined} \wedge \neg\exists t_e \in T_o(p) : \Theta(t_e, a(p))\}$$

In the implemented sequential cohort construction, the anchor requirement is applied before the dependent temporal exclusion. Therefore, $C$ is typically already restricted to persons for whom $a(p)$ is defined. Under that condition, the expression can be read more simply as:

$$TEMPORAL_FORBID(C, a, o, \Theta) = \{p \in C \mid \neg\exists t_e \in T_o(p) : \Theta(t_e, a(p))\}$$

For the criterion "patients with diabetes who had no insulin exposure after diagnosis," let $C_{\text{diabetes}}$ be the cohort of persons with a qualifying diabetes anchor. Let $T_{\text{insulin}}(p)$ be the set of qualifying insulin exposure dates. Under an on-or-after policy:

$$C_{\text{final}} = \{p \in C_{\text{diabetes}} \mid \neg\exists t_e \in T_{\text{insulin}}(p) : t_e \geq a_{\text{diabetes}}(p)\}$$

This is not a global absence condition. It does not exclude a patient merely because insulin exposure exists somewhere in the patient's observable history. It excludes the patient only when the forbidden event satisfies the temporal relationship to the anchor. This is why the SQL lowering uses an anchor-relative anti-existence pattern rather than an unanchored global negation or unscoped set subtraction.

The distinction is clinically important. A patient who received insulin before the first qualifying diabetes diagnosis but not after that diagnosis may satisfy "no insulin after diagnosis" under the defined temporal semantics. A patient who received insulin after the anchor does not. Both cases require the anchor date to be preserved and used in the exclusion predicate.

**3.6.6. Empty Executable Concept-Set Semantics.**

The grounding stage described in Section 3.5 may produce an empty executable concept set. This occurs when a clinical mention is valid at the vocabulary level but has no locally observed executable descendants in the target OMOP CDM instance. Drug-class criteria are the most important example: a pharmacologic class may be recognized through the vocabulary graph, but none of its expanded drug concepts may occur in the local drug_exposure table.

If the grounded concept set for fact $f_i^g$ is empty:

$$\Gamma_i = \emptyset$$

then its event-time set is empty for every person:

$$T_i(p) = \emptyset \forall p \in P_{obs}(D)$$

The patient-set consequence depends on polarity.
For a required predicate:

$$REQUIRE(C, P_i) = \emptyset$$

because no person can have an event whose concept set is empty. The same holds for temporal inclusion:

$$TEMPORAL_REQUIRE(C, a, i, \Theta) = \emptyset$$

For a forbidden predicate:

$$FORBID(C, P_i) = C$$

because there is no executable event that can trigger the exclusion. The same logic applies to temporal exclusion:

$$TEMPORAL_FORBID(C, a, i, \Theta) = C$$

assuming $C$ has already been restricted to persons satisfying the required anchor condition. In logical terms, if $T_i(p) = \emptyset$, then the existential predicate $\exists t_e \in T_i(p): \Theta(t_e, a(p))$ is false for every $p$. Its negation is therefore true for every patient in the current cohort.

This behavior is important in local OMOP CDM settings. A valid vocabulary-level concept does not guarantee that the corresponding event exists in the target database. NS-CoT preserves the executable-empty state explicitly so that an empty local concept set is interpreted consistently rather than treated as an undefined exception or silently replaced by an unvalidated concept list.

This executable-empty state should be distinguished from an unresolved grounding state: in the former, a semantically valid concept set was found but no locally executable concept was observed; in the latter, the concept could not be resolved sufficiently for execution.

**3.6.7. Maximum Implemented Temporal-Exclusion Pattern.**

The most expressive implemented pattern in the current framework combines condition anchoring, drug grounding, local exposure filtering, lower-only temporal logic, and temporal exclusion. Its natural-language form can be written as:

patients with condition $m_a$ who had no exposure to drug mention $m_o$ more than N days after the first diagnosis of $m_a$.

Let $m_a$ be the condition anchor mention and $m_o$ be the drug outcome mention. The condition anchor event-time set is:

$$T_a(p) = \{t \mid \exists c \in \Gamma_a : E_{\text{Condition}}(p, c, t)\}$$

The first diagnosis anchor is:

$$a_{m_a}(p) = \min T_a(p)$$

provided that $T_a(p) \neq \emptyset$. The anchor cohort is:

$$C_a = \{p \in P_{obs}(D) \mid T_a(p) \neq \emptyset\}$$

The locally executable drug event-time set is:

$$T_o(p) = \{t \mid \exists c \in \Gamma_o : E_{\text{Drug}}(p, c, t)\}$$

Where $T_o$ may have been derived through ingredient, product, brand, or pharmacologic-class grounding and may have been filtered to locally observed drug concepts. The lower-only temporal predicate is:

$$\Theta_{>N}(t_e, a_{m_a}(p)) \equiv t_e > a_{m_a}(p) + N$$

The final cohort is:

$$C_{\text{final}} = \{p \in C_a \mid \neg\exists t_e \in T_o(p) : t_e > a_{m_a}(p) + N\}$$

In patient-set operation form:

$$C_0 = P_{obs}(D)$$
$$C_a = REQUIRE(C_0, P_a)$$
$$C_{\text{final}} = TEMPORAL_FORBID(C_a, a_{\text{a}}, \text{o}, \Theta_{>N})$$

This pattern is central because it requires the major components of NS-CoT to work together. The condition mention must be grounded and used as an anchor. The drug mention must be resolved through the appropriate subtype-aware grounding path. The drug concept set must be executable against the local drug_exposure table. The phrase "more than N days after diagnosis" must be represented as a lower-only temporal predicate. Finally, the exclusion must be applied as anchor-bound temporal anti-existence over the current patient set.

**3.6.8. Boundary of the Current Temporal Algebra.**

The temporal algebra defined here is intentionally scoped. It supports single-anchor predicates of the form:

anchor event → outcome event evaluated relative to that anchor

This covers criteria such as drug exposure on or after diagnosis, drug exposure within a bounded interval after diagnosis, and absence of drug exposure after or more than a specified number of days after diagnosis.

It does not provide unrestricted temporal reasoning over arbitrary clinical event chains. In particular, it does not generally support multi-anchor sequential temporal chains in which an intermediate outcome event must become a new anchor for a later predicate. Consider the criterion:

patients diagnosed with diabetes, started metformin within 90 days, and achieved HbA1c < 7% within 180 days after metformin start

The first temporal predicate evaluates metformin exposure relative to the diabetes diagnosis anchor. The second temporal predicate must then evaluate HbA1c relative to the metformin start date, not the original diabetes diagnosis date. This requires preserving an intermediate outcome event as a new anchor:

diabetes diagnosis → metformin start → HbA1c measurement

The current single-anchor representation supports the first arrow but does not generally propagate the metformin event as a new anchor for the second arrow. Supporting this pattern would require additional machinery such as produced anchors, anchor propagation, chained temporal joins, and possibly event-level rather than only patient-level intermediate state.

The algebra in this section therefore defines the execution semantics for the current study scope: grounded facts are interpreted as patient-level predicates over OMOP event-time relations; temporal relations are represented as single-anchor operators; and cohorts are constructed through REQUIRE, FORBID, GROUP_AND, GROUP_OR, TEMPORAL_REQUIRE, and TEMPORAL_FORBID operations. The next section explains how these symbolic operations are lowered into deterministic CTE-chain SQL over OMOP CDM tables.

## 3.7. Deterministic SQL Compilation and Output Artifacts

The symbolic cohort algebra defined in Section 3.6 must ultimately be executed against an OMOP CDM database. In NS-CoT, this execution is performed by lowering the OMOP-executable cohort logic $h_q$ into deterministic SQL. SQL is therefore not the starting point of interpretation. It is the final executable representation of a cohort definition that has already been typed, grounded, temporally structured, and expressed as patient-set operations.

This distinction is central to the architecture. The LLM-mediated layer does not generate SQL directly from the feasibility-oriented natural-language cohort criterion q. Instead, it produces a typed representation $IR_q$. The grounding layer resolves clinical mentions into OMOP-compatible concept sets and execution metadata, producing the grounded representation $IR_q^g$. The symbolic algebra layer then constructs the cohort logic $h_q$ from the grounded representation:

$$h_q = A\left(IR_q^g\right)$$

The SQL compiler receives this cohort algebraic representation and emits an executable CTE-chain query over OMOP CDM tables. At this stage, the compiler does not reinterpret the original natural-language text.

Formally, deterministic compilation can be written as:

$$Compile_\theta\left(h_q\right) = SQL_q$$

where $h_q$ is the OMOP-executable cohort logic constructed from the grounded intermediate representation $IR_q^g$, $SQL_q$ is the compiled SQL query, and $\theta$ denotes the fixed compilation environment. The compilation environment includes the compiler version, configuration, SQL dialect, OMOP schema assumptions, OMOP vocabulary snapshot, and database-specific execution assumptions. For a fixed $h_q$ and fixed $\theta$, the compiler produces a reproducible SQL structure and parameter binding

The compiled SQL is executed over database instance $D$:

$$Exec_D(SQL_q) = C_q(D)$$

And the feasibility-oriented patient count is obtained as:

$$count(q, D) = | C_q(D) |$$

The determinism claim applies to this lowering and execution specification. It does not imply that the original natural-language interpretation is deterministic or clinically infallible. Rather, once the criterion has been represented as grounded cohort logic $h_q$, the subsequent SQL compilation step is reproducible and inspectable.

### 3.7.1. Compiler Input and CTE-Chain Construction.

Formally, the compiler input is $h_q$. In the reference implementation, $h_q$ is represented as an ordered sequence of executable steps:

$$h_q \equiv S_q = \langle s_1, s_2, \dots, s_k \rangle$$

Each execution step $s_i$ corresponds to a grounded clinical criterion, demographic criterion, group criterion, temporal criterion, exclusion criterion, or aggregation step. These objects contain the execution-relevant fields required for SQL lowering: domain, concept set, temporal relation, anchor reference, polarity, value constraint, unit constraint, grouping logic, and grounding status.

The compiler constructs a sequence of common table expressions:

$$cohort_0, cohort_1, \dots, cohort_k$$

where $cohort_0$ represents the initial observed population, and each subsequent $cohort_i$ represents the patient set obtained after applying step $s_i$ to $cohort_{i-1}$.

The initial cohort is:

$$C_0 = P_{obs}(D)$$

In SQL, this is represented as:

```
WITH cohort_0 AS (
    SELECT DISTINCT person_id
    FROM observation_period
)
```

This choice reflects the event-time assumptions defined in Section 3.3. NS-CoT does not begin from all rows in the PERSON table by default. It begins from persons with valid observation periods, because eligibility-criteria-derived cohort definitions are evaluated over observed clinical history.

Each subsequent CTE applies one symbolic operation to the previous patient set:

$$C_i = Lower_\theta(s_i, C_{i-1})$$

where $Lower_\theta$ denotes the SQL lowering of a grounded symbolic step under the fixed compilation environment θ. The final query returns either the cohort count:

```
SELECT COUNT(*) AS cohort_count
FROM cohort_k;
```

or the final patient identifiers:

```
SELECT person_id
FROM cohort_k;
```

In feasibility assessment, the count form is the primary output. The patient-list form is used for inspection, sampling, and validation.

**3.7.2. Domain-to-Table Lowering.**

A grounded clinical criterion is lowered according to its OMOP domain. The domain field determines which OMOP table, concept column, and event date column are used. This is not decided by the LLM during SQL generation; it is determined by the domain-to-table mapping established in the symbolic layer.

| Domain | OMOP table | Concept column | Event date column |
|---|---|---|---|
| Condition | condition_occurrence | condition_concept_id | condition_start_date |
| Drug | drug_exposure | drug_concept_id | drug_exposure_start_date |
| Measurement | measurement | measurement_concept_id | measurement_date |
| Procedure | procedure_occurrence | procedure_concept_id | procedure_date |
| Observation | observation | observation_concept_id | observation_date |
| Visit | visit_occurrence | visit_concept_id | visit_start_date |
| Device | device_exposure | device_concept_id | device_exposure_start_date |

| Domain | OMOP table | Concept column | Event date column |
|---|---|---|---|
| Death | death | cause_concept_id, where applicable | death_date |

**Table 8. Domain-to-table mappings used for SQL compilation.**

*Note. For Death, cause_concept_id is relevant only when a cause-of-death concept is used; death occurrence itself is primarily represented by person_id and death_date.*

For a required grounded clinical criterion $s_i$ in domain $d_i$, with grounded concept set $\Gamma_i$, the compiler emits a CTE of the following form:

```
  , cohort_i AS (
SELECT DISTINCT t.person_id
FROM <domain_table> t
JOIN cohort_{i-1} prev
  ON t.person_id = prev.person_id
JOIN observation_period op
  ON t.person_id = op.person_id
 AND t.<event_date> >= op.observation_period_start_date
 AND t.<event_date> <= op.observation_period_end_date
WHERE t.<concept_column> IN (<concept_ids>)
)
```

This CTE implements three semantic constraints. First, the criterion is evaluated only among persons who remain in the previous cohort. Second, the grounded concept set is evaluated in the appropriate OMOP domain table. Third, the event is required to fall within the person's observation period.

When the criterion includes an absolute date window, the compiler adds predicates to the same event date column. For example:

```
    AND t.condition_start_date >= DATE '2023-01-01'
AND t.condition_start_date <    DATE '2024-01-01'
```

Absolute calendar windows are compiled as half-open intervals when appropriate. This convention makes adjacent windows composable and avoids ambiguity at the upper boundary.

### 3.7.3. Sequential Criteria as Patient-Set Intersection.

Top-level criteria are compiled sequentially. If a typed plan contains two top-level required criteria $s_1$ and $s_2$, the compiler first constructs $cohort_1$ from $cohort_0$, and then constructs $cohort_2$ from $cohort_1$. Thus, the second criterion is not evaluated independently over the entire database. It is evaluated only among persons who have already satisfied the first criterion.

Formally:

$$C_1 = REQUIRE(C_0, P_1)$$
$$C_2 = REQUIRE(C_1, P_2)$$

which is equivalent to:

$$C_2 = C_0 \cap \{p \mid P_1(p)\} \cap \{p \mid P_2(p)\}$$

The SQL realization of this semantics is the join against the previous CTE:

```
JOIN cohort_{i-1} prev
ON t.person_id = prev.person_id
```

This stepwise structure is not merely an implementation convenience. It corresponds to the patient-set narrowing semantics defined in Section 3.6. It also makes it possible to report intermediate cohort sizes after each criterion, which is useful for debugging, feasibility review, and explaining why a final cohort count is large or small.

**3.7.4. Group Criteria and OR Semantics.**

Disjunctive criteria require a different lowering strategy. Top-level sequential criteria are interpreted as conjunction by construction. Therefore, if a natural-language criterion contains OR logic, the alternatives must be represented as a group criterion before compilation. Otherwise, the alternatives would be incorrectly compiled as intersections.

For an OR group containing predicates $P_1, \dots, P_m$, the symbolic meaning is:

$$GROUP_OR(C; P_1, \dots, P_m) = \{p \in C \mid P_1(p) \lor \cdots \lor P_m(p)\}$$

The SQL compiler lowers this group into a single CTE whose branches are combined by UNION:

```
, cohort_i AS (
  SELECT DISTINCT t1.person_id
  FROM <domain_table_1> t1
  JOIN cohort_{i-1} prev
    ON t1.person_id = prev.person_id
  JOIN observation_period op
    ON t1.person_id = op.person_id
   AND t1.<event_date_column_1> >= op.observation_period_start_date
   AND t1.<event_date_column_1> <= op.observation_period_end_date
```

```
    WHERE <predicate_1>

    UNION

    SELECT DISTINCT t2.person_id
    FROM <domain_table_2> t2
    JOIN cohort_{i-1} prev
      ON t2.person_id = prev.person_id
    JOIN observation_period op
      ON t2.person_id = op.person_id
     AND t2.<event_date_column_2> >= op.observation_period_start_date
     AND t2.<event_date_column_2> <= op.observation_period_end_date
    WHERE <predicate_2>

    UNION

    SELECT DISTINCT t3.person_id
    FROM <domain_table_3> t3
    JOIN cohort_{i-1} prev
      ON t3.person_id = prev.person_id
    JOIN observation_period op
      ON t3.person_id = op.person_id
     AND t3.<event_date_column_3> >= op.observation_period_start_date
     AND t3.<event_date_column_3> <= op.observation_period_end_date
    WHERE <predicate_3>
  )
```

The use of UNION corresponds to patient-set union. Each branch evaluates one alternative against the same previous cohort, and the resulting patient identifiers are combined into a single next cohort. If an explicit AND group is represented within a group object, it may be lowered through intersection. However, ordinary top-level criteria do not require group-level AND because sequential CTE construction already implements conjunction.

This distinction is important for semantic fidelity. For example, “condition A or condition B” must not be compiled as requiring both conditions. By preserving OR as a group-level structure, the compiler maintains the logical meaning extracted in the typed IR.

#### 3.7.5. Value and Unit Predicate Lowering.

Although the central empirical focus of this research is on Condition and Drug criteria, the compiler can represent value-bearing predicates when the typed IR contains the necessary structure. Measurement criteria illustrate why typed decomposition is needed before SQL generation. A criterion such as “systolic blood pressure ≥ 140 mmHg” is not treated as a single concept-search phrase. It is decomposed into a measurement concept, comparator, threshold, and unit.

A measurement criterion may be represented by fields such as:

$$d_i = \text{Measurement}$$
$$\Gamma_i = \Gamma_{\text{Measurement}}(m; D)$$
$$\nu_i = (\text{operator} \geq, \text{threshold} = 140, \text{unit} = \text{mmHg})$$

The corresponding SQL branch has the form:

```
SELECT DISTINCT t.person_id
FROM measurement t
JOIN cohort_{i-1} prev
  ON t.person_id = prev.person_id
JOIN observation_period op
  ON t.person_id = op.person_id
 AND t.measurement_date >= op.observation_period_start_date
 AND t.measurement_date <= op.observation_period_end_date
WHERE t.measurement_concept_id IN (<measurement_concept_ids>)
  AND t.value_as_number >= <threshold>
  AND (
        t.unit_concept_id = <unit_concept_id>
        OR t.unit_concept_id IS NULL
      )
```

The exact unit policy depends on the configured handling of missing or heterogeneous units; the example above illustrates a permissive policy rather than a universal rule. Because this policy also admits measurements whose unit is unrecorded, it can over-include rows whose true unit differs from the threshold unit; it therefore favors sensitivity over specificity, which is acceptable for feasibility-oriented pre-screening but should be tightened when value semantics are safety-critical. The methodological point is that the measurement concept and numeric constraint are compiled as distinct execution fields. This prevents thresholds from being absorbed into concept retrieval and preserves value semantics as explicit SQL predicates.

#### 3.7.6. Temporal Relation Lowering.

Temporal criteria are compiled only after the temporal relation has already been represented in the typed and grounded plan. The compiler does not infer temporal logic from raw text. It receives a temporal specification such as anchor reference, temporal relation type, lower offset, upper offset, boundary policy, and anchor policy.

For anchor-relative criteria, the compiler first constructs an anchor CTE. Under the first-anchor policy, the anchor CTE selects the earliest qualifying anchor event date for each person:

```
    , anchor_i AS (
  SELECT
      anc.person_id,
      MIN(anc.<anchor_date_column>) AS anchor_date
  FROM <anchor_table> anc
  JOIN cohort_{i-1} prev
    ON anc.person_id = prev.person_id
  JOIN observation_period op
    ON anc.person_id = op.person_id
   AND anc.<anchor_date_column> >= op.observation_period_start_date
   AND anc.<anchor_date_column> <= op.observation_period_end_date
  WHERE anc.<anchor_concept_column> IN (<anchor_concept_ids>)
  GROUP BY anc.person_id
)
```

The target event CTE is then joined to this anchor CTE and receives an anchor-relative temporal predicate. For an inclusive bounded-after window, the compiler emits:

```
AND t.<event_date_column> >= a.anchor_date + INTERVAL '<lower> day'
AND t.<event_date_column> <= a.anchor_date + INTERVAL '<upper> day'
```

For a strict lower-only-after relation, such as “more than 30 days after diagnosis,” the compiler emits:

```
AND t.<event_date_column> > a.anchor_date + INTERVAL '<lower> day'
```

For an on-or-after relation, the predicate is:

```
AND t.<event_date_column> >= a.anchor_date
```

Boundary behavior follows the temporal operator defined in the grounded representation. Thus, same-day inclusion, strict-after behavior, bounded windows, and lower-only windows are not accidental properties of SQL comparison operators. They are compilation outcomes of explicit temporal policies defined before SQL generation.

Temporal lowering also preserves the distinction between anchor-relative temporal relations and co-occurrence relations. A criterion such as “drug exposure after diagnosis” is compiled as an outcome event evaluated relative to a diagnosis anchor. By contrast, expressions such as “drug A and drug B on the same day” or “drug A and drug B within N days” are represented as temporal joins between two event sets. For a same-day join, the compiler emits a date equality predicate:

```
t1.<left_date_column> = t2.<right_date_column>
```

For a within-days co-occurrence relation, the compiler emits a bounded date-difference predicate. The interval belongs to the join relation itself, not independently to each event. This prevents a common semantic error in which each event is incorrectly assigned its own absolute time window.

**3.7.7. Exclusion and Anti-Existence Lowering.**

A grounded fact with polarity $\pi_i$ = FORBID is lowered as an anti-existence predicate over the previous cohort. It is not compiled as a positive event CTE followed by an informal subtraction. The compiler emits a CTE that retains only persons for whom the forbidden event does not exist.

The general SQL form is:

```
    , cohort_i AS (
  SELECT prev.person_id
  FROM cohort_{i-1} prev
  WHERE NOT EXISTS (
    SELECT 1
    FROM <domain_table> t
    JOIN observation_period op
      ON t.person_id = op.person_id
     AND t.<event_date_column> >= op.observation_period_start_date
     AND t.<event_date_column> <= op.observation_period_end_date
    WHERE t.person_id = prev.person_id
      AND t.<concept_column> IN (<forbidden_concept_ids>)
  )
)
```

This lowering implements:

$$FORBID(C, P) = \{p \in C \mid \neg P(p)\}$$

When the forbidden criterion is temporal, the anchor CTE and temporal predicate are incorporated into the NOT EXISTS subquery. For example, “no exposure to drug $m_o$ more than N days after diagnosis” is compiled as the absence of a target drug event satisfying the lower-only temporal predicate relative to the anchor date:

```
    , cohort_i AS (
  SELECT prev.person_id
  FROM cohort_{i-1} prev
  WHERE NOT EXISTS (
    SELECT 1
```

```
        FROM drug_exposure t
        JOIN anchor_i a
          ON t.person_id = a.person_id
        JOIN observation_period op
          ON t.person_id = op.person_id
         AND t.drug_exposure_start_date >= op.observation_period_start_date
         AND t.drug_exposure_start_date <= op.observation_period_end_date
        WHERE t.person_id = prev.person_id
          AND t.drug_concept_id IN (<drug_concept_ids>)
          AND t.drug_exposure_start_date > a.anchor_date + INTERVAL '<N> day'
    )
)
```

This is the SQL realization of the temporal exclusion semantics defined in Section 3.6:

$$TEMPORAL_FORBID(C, a, o, \Theta) = \{p \in C \mid \neg\exists t_e \in T_o(p): \Theta(t_e, a(p))\}$$

The use of correlated NOT EXISTS is therefore not merely a SQL idiom. It is the concrete lowering of anchor-bound absence semantics. A patient is excluded only if the forbidden event exists in the specified temporal relation to the anchor. The patient is not excluded simply because the event appears somewhere in the observable record outside the relevant temporal scope. This anchor-bound exclusion presupposes that the anchor has been established as a required criterion earlier in the CTE chain $C_a = REQUIRE(C_0, P_a)$); consequently, a person with no qualifying anchor has no row in anchor_i, the correlated NOT EXISTS evaluates to true, and that person is retained by design rather than excluded on the basis of a missing anchor.

**3.7.8. Empty Executable Concept Sets.**

The compiler preserves the semantics of empty executable concept sets defined in Section 3.6. If grounding produces an empty concept set for a required clinical criterion, the event predicate is unsatisfiable. The SQL compiler represents this using an explicitly false predicate:

```
FALSE /* concept_ids empty */
```

For a required criterion, this produces an empty next cohort:

$$REQUIRE(C, P_\emptyset) = \emptyset$$

For a forbidden criterion, the interpretation is different. If there are no locally executable events that could satisfy the forbidden predicate, then the exclusion removes no patients:

$$FORBID(C, P_{\emptyset}) = C$$

In SQL, this is realized naturally through NOT EXISTS. Since the subquery predicate is unsatisfiable, the existential condition is false for every patient in the current cohort, and NOT EXISTS evaluates to true:

```
WHERE NOT EXISTS (
  SELECT 1
  FROM <domain_table> t
  WHERE t.person_id = prev.person_id
    AND FALSE /* concept_ids empty */
)
```

The same behavior applies to temporal exclusion:

$$TEMPORAL_FORBID(C, a, o_{\emptyset}, \Theta) = C$$

assuming the current cohort has already satisfied the required anchor. This distinction is essential in local OMOP CDM settings, where a vocabulary-valid drug class may have no locally observed descendant drug concepts. A required exposure to such a class should return no qualifying patients, whereas a forbidden exposure to such a class should not remove patients on the basis of an unobservable event.

### 3.7.9. Count and Patient-List Outputs.

The primary output for a feasibility-oriented cohort criterion $q$ is the final cohort cardinality:

$$|\ C_q(D)\ |$$

In count-output mode, the compiler emits::

```
SELECT COUNT(*) AS cohort_count
FROM cohort_k;
```

The same CTE chain can also be emitted in patient-list mode:

```
SELECT person_id
FROM cohort_k;
```

Patient-list output is useful for sampling, inspection, and event-level validation.

### 3.7.10. Output Artifacts.

SQL compilation produces more than a scalar count. Because the objective of NS-CoT is to preserve eligibility semantics across the transformation chain, the output includes artifacts that expose how the final cohort was constructed. These artifacts are not merely logs. They are part of the cohort-definition object that makes the result inspectable.

| Artifact | Description |
|---|---|
| Original query | Feasibility-oriented natural-language cohort criterion or feasibility-oriented query |
| Typed execution steps | Structured criteria after decomposition and normalization |
| Grounded concept sets | Domain-specific concept identifiers used in SQL predicates |
| Grounding metadata | Candidate concepts, selected seeds, descendant expansion, local filtering, and grounding status |
| Temporal specification | Anchor reference, temporal relation, offset, boundary policy, and anchor policy |
| Value constraints | Measurement comparator, threshold, unit, and emitted value predicate when applicable |
| Generated SQL | Deterministic CTE-chain SQL query |
| Cohort definition JSON | Structured representation of query, steps, concept sets, time windows, status, and counts |
| Execution output | Final cohort count and optional sample patient identifiers |

**Table 9. Output artifacts generated by the NS-CoT compilation pipeline.**

These artifacts allow the transformation to be reviewed at multiple levels. If a final count is unexpected, the analyst can examine whether the issue came from typed decomposition, concept grounding, descendant expansion, local concept filtering, temporal boundary interpretation, exclusion lowering, or SQL execution. This is the practical value of compiling from explicitly constructed cohort logic $h_q$ rather than generating SQL directly.

The role of SQL in NS-CoT is therefore deliberately constrained. SQL is the executable target language, not the semantic workspace. By the time SQL is emitted, the clinically relevant decisions have already been represented as typed facts, grounded concept sets, temporal predicates, polarity, and patient-set operations. The compiler's task is to serialize those decisions into OMOP CDM SQL in a reproducible CTE-chain form. This design makes the final feasibility count traceable to a cohort definition rather than to an opaque generated query.

## 3.8. Observability, Traceability, and Provenance Artifacts

The preceding sections defined the transformation from site feasibility patient-count questions with protocol-derived inclusion and exclusion constraints to grounded OMOP representations, patient-set algebraic cohort logic $h_q$, and deterministic SQL. For this transformation to be useful in site-level feasibility assessment, the final count must not stand alone. A scalar count is methodologically insufficient, and clinically difficult to interpret, unless it can be linked back to the semantic decisions that produced it: which facts were extracted, which concepts were selected, which temporal anchor was used, how exclusion was interpreted, and which SQL was executed.

NS-CoT therefore separates execution from observability. Execution refers to the deterministic lowering of cohort logic $h_q$ into OMOP CDM SQL and the subsequent execution of that SQL over database instance $D$. Observability refers to the ability to inspect the intermediate states of this transformation. In the reference implementation, observability is supported through structured local artifacts and, when enabled, Langfuse-based hierarchical traces[114]. These mechanisms are complementary. Local JSON and SQL artifacts provide programmatic records for scoring and error classification, whereas Langfuse traces provide a human-readable execution tree for post hoc review.

The core artifact of each saved run is a structured cohort-definition record. This record preserves the original cohort criterion, typed execution steps, grounded concept identifiers, grounding metadata, temporal specifications, exclusion polarity, generated SQL, execution status, final cohort count, elapsed time, and optional sample person identifiers. When tracing is enabled, the corresponding trace identifier is also stored so that the saved artifact can be cross-referenced with the run-level trace.

The pipeline can be inspected at several levels. At the top level, the run record identifies the input query, model configuration, execution status, final count, and elapsed time. At the semantic level, the typed decomposition records which clinical facts were extracted from the criterion and how they were assigned domains, roles, polarities, and temporal dependencies. At the grounding level, the artifact records concept candidates, selected seeds, descendant expansion, drug-class routing, local executable concept filtering, and grounding status. At the symbolic level, it records the cohort logic $h_q$, including temporal predicates, exclusion logic, and patient-set operation structure. At the compilation level, it records the generated CTE-chain SQL. At the execution level, it records whether the SQL returned a final scalar count and, when configured, example person identifiers for inspection.

This design is especially important for failure analysis. A mismatch between a system-derived count and a reference-defined site-level patient count can arise from multiple layers of the pipeline. The source may be natural-language decomposition, domain assignment, concept grounding, descendant expansion, local CDM filtering, temporal boundary interpretation, exclusion polarity, SQL compilation, or database execution. Without intermediate artifacts, all of these failures would appear only as an incorrect final count. With structured trace artifacts, the failure can be localized to a specific layer of the transformation.

For example, an error in interpreting "more than 30 days after diagnosis" as a different lower-bound predicate is conceptually distinct from an error in mapping a drug class to the wrong vocabulary concept set. Both can produce a wrong count, but they require different corrections. Traceability allows the system output to be evaluated not only by whether the count matches, but also by where and why a mismatch occurred.

Langfuse instrumentation, when enabled, provides an additional observability layer over the same pipeline. The reference implementation can record the run as a hierarchical trace, with phase-level observations corresponding to structured strategy planning, query decomposition, step construction, SQL compilation, and execution. Substeps may additionally capture domain-specific build operations, such as clinical criteria, demographic criteria, temporal joins, and group criteria. This trace structure makes the execution path inspectable from the original natural-language query to the final cohort count.

It is important to define the role of these traces carefully. Langfuse traces and saved JSON artifacts are experimental provenance artifacts. They support reproducibility review, debugging, and post hoc error decomposition. They do not, by themselves, constitute formal regulatory audit trails, nor do they guarantee clinical correctness of the generated cohort. Their methodological role in this study is narrower and more concrete: they make the transformation from feasibility-oriented criterion text to executable OMOP cohort logic inspectable. In this work, reproducibility refers to deterministic re-execution under the same CDM snapshot, vocabulary version, concept set, and compiler version; reproducibility after data updates or across institutions is a stronger property outside the present scope.

The output of NS-CoT is therefore not just:

$$\mid C_q(D) \mid$$

but a traceable tuple:

$$Output(q, D) = (IR_q, IR_q^g, h_q, SQL_q, C_q(D), \mid C_q(D) \mid, M_q)$$

Here, the tuple contains the typed intermediate representation, the grounded representation, the OMOP-executable cohort logic represented through patient-set algebra, the compiled SQL query, the resulting patient set, the final count, and the metadata and trace artifact associated with the run.

This traceable output structure is aligned with the central methodological claim of this study. A feasibility count becomes a reviewable research artifact not merely when it is returned by a syntactically valid SQL query, but when the cohort semantics behind that count can be inspected, reproduced, and revised.

## 3.9. Reference Implementation and Runtime Configuration

The formalization described in this chapter was implemented as a reference NS-CoT pipeline over a local OMOP CDM PostgreSQL database. The implementation instantiates the architectural separation described in Section 3.1: the language model is used for natural-language interpretation and structured semantic decomposition, while vocabulary grounding, database access, symbolic normalization, patient-set operations, SQL compilation, and execution are handled by deterministic software components.

The implementation is configuration-driven. Database connection parameters, ChromaDB retrieval parameters, concept-resolution settings, LLM backend settings, logging behavior, drug-resolution options, output settings, and cohort-persistence options are specified through a YAML configuration file and loaded by a configuration loader. Environment-variable placeholders are expanded at load time, allowing sensitive fields such as API keys to be supplied through the execution environment rather than hard-coded into the source code. The configuration loader exposes separate sections for database, ChromaDB, concept resolution, LLM, logging, drug resolution, output, and temporal defaults.

### 3.9.1. Database Environment.

The database backend used in the reference implementation is PostgreSQL. The implementation connects to the target OMOP CDM database through a database connector that sets the SQL search path to the configured OMOP schema and public. In the evaluation configuration used for this work, the target schema was synthetic_snuh_cdm.

The database connector provides a small set of execution primitives, including scalar query execution, row retrieval, dictionary-style retrieval, and general SQL execution. It also provides OMOP-specific helper functions for concept validation, descendant expansion through concept_ancestor, source-to-standard mapping through concept_relationship, unit lookup through UCUM concepts, and local concept-existence checks in domain-specific event tables. The connector includes a domain-to-table map for structured OMOP event domains such as Condition, Drug, Procedure, Measurement, Observation, Device, and Visit.

This structure is consistent with the event-time formalization in Section 3.3. The compiler does not require the LLM to choose table names or date columns. Instead, a grounded criterion's domain determines which OMOP table and concept column are used. The database connector and compiler then execute the resulting SQL against the configured OMOP schema.

### 3.9.2. LLM Backend for Interpretation.

The LLM backend is used for the neural interpretation portion of the pipeline. This includes structured strategy planning, natural-language criterion decomposition, typed fact generation, temporal expression extraction, and limited semantic validation of concept candidates. The LLM is not used to directly generate the final SQL query.

In the experimental configuration described here, the LLM endpoint was configured through an OpenAI-compatible API interface, with the model set to gpt-4.1-2025-04-14[115] and temperature set to 0.0. The low-temperature setting was used to reduce variation in structured outputs. However, as discussed earlier, this does not make the entire natural-language interpretation stage formally deterministic. The determinism claim of NS-CoT applies only after typed and grounded representations have been produced and are consumed by the symbolic layer.

The use of the OpenAI API for the LLM backend should be distinguished from the embedding backend used for ChromaDB retrieval. In this implementation, the language model responsible for decomposition and semantic reasoning and the embedding model used for vector retrieval are separate components.

### 3.9.3. ChromaDB Retrieval and Local Embedding Backend.

Concept candidate retrieval is supported by ChromaDB vector indexes over OMOP concept metadata. The ChromaDB configuration specifies the database path, collection name, embedding model, top-k retrieval size, and score threshold. The retrieval code uses an OpenAI-compatible client interface to request query embeddings, but the embedding endpoint can be served locally. In the configuration used here, ChromaDB query embeddings were generated using the nomic-embed-text-v1.5 embedding model served through LM Studio, rather than by the OpenAI LLM used for natural-language decomposition[116].

This distinction matters methodologically. The vector index is used to retrieve candidate OMOP concepts for a clinical mention. It is not the component that determines final cohort semantics. Retrieved candidates are later filtered, validated, expanded, and, for selected drug criteria, checked against local CDM evidence. Thus, ChromaDB provides candidate retrieval, not final concept-set authority.

For clarity, this retrieval stage should not be interpreted as a separate lexical-search or synonym-expansion module. The evaluated implementation retrieves candidate concepts by embedding the clinical mention and searching the configured ChromaDB collection over OMOP concept metadata. The returned candidates include metadata such as concept_id, concept_name, domain_id, vocabulary_id, and distance score. These candidates are then processed by downstream validation and grounding components. Although the OMOP vocabulary environment includes concept_synonym, concept_relationship, and concept_ancestor tables, the evaluated retrieval pathway should be described as ChromaDB-based candidate retrieval followed by OMOP vocabulary-backed validation, hierarchy or relationship expansion, and local CDM executability filtering.

The retrieval implementation supports domain-specific ChromaDB paths and optional vocabulary filters, allowing candidate retrieval to be constrained by the expected OMOP domain or vocabulary when configured.

#### 3.9.4. Concept Resolution Configuration.

The concept-resolution configuration controls the number of seed concepts, whether descendant expansion is enabled, how measurement candidates are retrieved, whether LOINC-specific measurement retrieval is merged, and how vocabulary priorities are applied. In the configuration used for this study, descendant expansion through concept_ancestor was enabled. This descendant expansion is applied after candidate retrieval and seed selection, not as the initial retrieval mechanism for raw natural-language terms. In other words, a clinical mention is first interpreted and matched to candidate OMOP concepts; selected seed concepts may then be expanded through concept_ancestor or other vocabulary relationships when the intended cohort definition requires descendant or class-level coverage. Measurement retrieval was configured with broader candidate collection and LOINC-prioritized merging because measurement concepts may not always appear in the default top-k results when only a general ChromaDB search is used.

The configuration also supports source-code expansion for vocabularies such as ICD-10-CM and ICD-9-CM. This allows source vocabulary concepts to be mapped to standard concepts when appropriate. However, the study's core empirical claims remain centered on the condition and drug criteria described in the previous sections, rather than unrestricted resolution of all possible source-code criteria.

For unit-bearing measurement criteria, the database connector resolves common unit strings against UCUM Unit concepts through exact or canonicalized lookup. This is intentionally separate from vector similarity retrieval because unit interpretation is better treated as terminology canonicalization than as approximate semantic search.

#### 3.9.5. Drug Resolution Configuration.

Drug grounding is controlled by a separate drug-resolution configuration. The implementation also supports dose-form filtering for route-specific drug criteria, with route-to-dose-form mappings for intravenous, oral, topical, and inhaled forms. This allows the system to restrict expanded drug concepts according to route or formulation when such information is represented in the typed criterion.

The database connector includes drug-specific utilities for ingredient lookup, expansion from ingredient concepts to clinical drug concepts, validation against ATC ancestry where applicable, and inspection of drug exposure concept classes in the local database. These utilities instantiate the drug grounding principles described in Section 3.5: a drug mention is not simply searched as a string, but resolved through vocabulary structure, descendant expansion, and local database evidence.

#### 3.9.6. Logging and Output.

The implementation supports local logging, SQL saving, and sample person-identifier output. In the configuration used here, generated SQL was saved, and a small number of sample person identifiers could be exported for inspection.

The output layer is important because the study evaluates not only whether a count is returned, but also whether the transformation path can be inspected. The saved SQL, cohort-definition JSON, grounding metadata, execution status, and optional sample person identifiers make it possible to connect a final count to the typed facts, grounded concepts, temporal predicates, and patient-set operations that produced it.

The reference implementation should therefore be understood as an instantiation of the formal architecture rather than as a separate methodological contribution. Its purpose is to demonstrate that the proposed formalization can be executed end-to-end over an OMOP CDM instance: from natural-language criterion to typed IR, to grounded concepts, to patient-set algebra, to deterministic SQL, and finally to a cohort count.

## 3.10. Scope, Assumptions, and Methodological Boundaries

The formalization and implementation described in this chapter are intentionally scoped. NS-CoT is designed to preserve and execute a selected class of eligibility semantics over structured OMOP CDM event tables. It is not presented as a general-purpose clinical trial recruitment system, nor as a complete interpreter of all possible ClinicalTrials.gov eligibility criteria.

The first assumption is that the target data environment is an OMOP CDM instance with structured event tables, observation periods, person identifiers, standard vocabulary tables, and concept hierarchy relationships. The framework depends on the ability to express clinical events as domain-specific event-time relations $E_d(p, c, t)$. Criteria whose meaning depends primarily on unstructured clinical notes, pathology narratives, imaging reports, subjective physician judgment, or protocol-specific interpretation that cannot be operationalized using structured CDM fields are outside the primary scope of the current system.

The second assumption is that cohort construction is performed over observed persons: $C_0 = P_{obs}(D)$ rather than over an abstract population of all possible patients. This assumption reflects the RWD setting in which feasibility counts are computed from observable longitudinal data. Absence of an event is therefore interpreted as absence of a recorded event within the relevant observation context, not as proof that the clinical event never occurred in reality.

The third assumption concerns vocabulary grounding. NS-CoT assumes that clinical mentions can be grounded to OMOP-compatible concept sets with sufficient functional fidelity for the selected criteria. It also assumes that standard vocabulary structure, descendant expansion, and relationship tables provide an appropriate basis for constructing executable concept sets. This assumption is strongest for criteria that map cleanly to structured OMOP concepts, such as disease diagnoses and drug exposures. It is weaker for vague, composite, subjective, or institution-specific criteria.

The fourth assumption is that the central temporal patterns can be expressed as single-anchor predicates. The current framework supports criteria in which an outcome event is evaluated relative to one anchor event, such as drug exposure after diagnosis, drug exposure within a bounded interval after diagnosis, or absence of drug exposure after diagnosis. It does not generally support unrestricted multi-anchor sequential temporal chains in which an intermediate outcome event becomes a new anchor for a later predicate. Supporting such criteria would require additional machinery for event-level state propagation, chained temporal joins, and anchor inheritance.

The fifth assumption concerns exclusion semantics. NS-CoT treats exclusion as an executable patient-set operation, often implemented as anchor-bound anti-existence. This is appropriate for criteria such as “no target drug exposure after diagnosis” or “no exposure more than N days after diagnosis.” However, the meaning of “no,” “without,” “prior,” or “history of” may vary across trial protocols. The framework can only preserve the exclusion semantics that have been explicitly represented in the typed IR and grounded representation. It does not resolve all ambiguities of clinical language automatically.

The sixth assumption concerns measurement and value predicates. The framework can represent measurement criteria when the typed IR separates the measurement concept, comparator, threshold, and unit, and applies them to recorded values. However, the main empirical evaluation of this study focuses on condition and drug criteria. Measurement support should therefore be understood as part of the extensible implementation rather than the central empirical claim.

The seventh boundary concerns clinical validity. The evaluation target of this framework is functional fidelity to predefined OMOP-executable cohort semantics. It is not chart-reviewed phenotype validity. A generated cohort may reproduce a reference OMOP cohort definition exactly while still requiring clinical validation before being used for recruitment, regulatory submission, or patient contact. The system operationally formalizes and executes cohort logic; it does not certify that the resulting patient set is a clinically definitive phenotype.

The eighth boundary concerns regulatory use. The observability and traceability artifacts described in Section 3.8 support inspection, debugging, reproducibility review, and error localization. They should not be interpreted as formal regulatory audit trails. Additional governance, validation, version control, access control, and institutional review processes would be required before such a system could be used in regulated clinical operations.

These boundaries do not weaken the central contribution of the study. They define it more precisely. The contribution is a neuro-symbolic framework for preserving executable cohort semantics when selected feasibility-oriented patient-count criteria are transformed into OMOP CDM cohort logic. Within that scope, NS-CoT provides an explicit and executable path from language to typed representation, from typed representation to grounded concepts, from grounded concepts to patient-set algebra, and from patient-set algebra to deterministic SQL and traceable cohort artifacts.

# Chapter 4. Evaluation Design and Study Setup

## 4.1. Evaluation Overview and Scope

Chapter 4 describes the evaluation design used to assess whether NS-CoT preserves executable patient-set semantics over OMOP CDM. The evaluation consists of four complementary components. First, CQ-Bench evaluates functional fidelity to internally constructed OMOP-executable reference cohort definitions that deliberately stress semantic components central to this study, including concept grounding, temporal anchoring, drug exposure, anchor-bound exclusion, and patient-set operations. Second, FOEM evaluates count and patient-set reproduction in an external template-derived OMOP query setting instantiated over the same SYN-SNUH database. Third, COVIC serves as an external multi-site feasibility simulation: the same COVIC-derived executable eligibility criterion component is applied across multiple OMOP CDM schemas treated as candidate sites, and the analysis asks whether system-generated site-level patient count preserves benchmark reference-defined counts, patient sets, site-feasibility labels, and site rankings. Fourth, three protocol-derived clinical trial case studies examine traceability, computability boundaries, and clinical interpretability when NS-CoT is applied to selected protocol-derived eligibility material.

These four components address related but distinct evaluation questions. CQ-Bench and FOEM are executable-reference benchmarks for scalar count and patient-set fidelity. COVIC extends the evaluation to criterion-site pairs and site-feasibility decision preservation across OMOP schemas. The NCT case studies are not statistical benchmarks; they are exploratory clinical case studies used to inspect whether generated cohort-definition artifacts expose the assumptions, computable components, and non-computable boundaries that matter for clinical review. Across all components, the target of evaluation is functional fidelity to executable cohort semantics and feasibility-oriented decision preservation, not chart-reviewed phenotype validity, regulatory-grade eligibility adjudication, or prospective recruitment performance.

The objective of this chapter is therefore not to evaluate one homogeneous benchmark task. Rather, the chapter defines a layered evaluation design. At the scalar-count level, the question is whether natural-language cohort feasibility queries can be transformed into executable OMOP cohort logic that reproduces predefined reference counts. At the patient-set level, the question is whether count reproduction corresponds to the same underlying person-id set. At the multi-site feasibility level, the question is whether system-generated count preserves reference-defined site-feasibility labels and site rankings. At the real-trial case-study level, the question is whether the generated artifacts expose computability boundaries and clinical assumptions in a form that can be reviewed.

## 4.2. Evaluation Data Environments

The evaluation used different data environments for different evaluation components. CQ-Bench and FOEM were instantiated over the SYN-SNUH OMOP CDM, which served as the controlled executable-reference environment for scalar count and patient-set reproduction. COVIC used three OMOP CDM schemas treated as candidate sites in an external multi-site feasibility simulation. The protocol-derived clinical trial case studies used selected ClinicalTrials.gov eligibility material as protocol-derived inputs for traceability analysis. Thus, the data environment should not be understood as a single benchmark database only, but as a set of evaluation-specific environments aligned with the purpose of each component.

For the CQ-Bench and FOEM benchmark components, the database was loaded into PostgreSQL and accessed through the synthetic_snuh_cdm schema. The OMOP vocabulary tables were available through the vocabulary schema, including concept, concept_synonym, concept_relationship, and concept_ancestor. The evaluation therefore used both local clinical event tables and standard OMOP vocabulary infrastructure.

The data environment contained structured event tables required for cohort execution, including condition_occurrence, drug_exposure, measurement, procedure_occurrence, visit_occurrence, and observation_period. The main empirical analyses focused on condition and drug criteria, with measurement support used where explicitly represented by a benchmark or system output. Table 10 summarizes the SYN-SNUH benchmark environment. The COVIC multi-site schemas and criterion-site construction are summarized separately in Section 4.6, and the protocol-derived clinical trial case-study inputs are summarized in Section 4.7.

| Component | Specification |
|---|---|
| CDM instance | Synthetic SNUH OMOP CDM |
| Database system | PostgreSQL |
| Primary schema | synthetic_snuh_cdm |
| CDM model | OMOP CDM v5.4 |
| Core event tables | condition_occurrence, drug_exposure, measurement, procedure_occurrence, visit_occurrence |
| Observation-period table | observation_period |

| Component | Specification |
|---|---|
| Initial cohort universe | Persons with valid observation periods |
| Vocabulary infrastructure | OMOP concept, concept_ancestor, concept_relationship, concept_synonym |
| Main evaluated domains | Condition and Drug |

**Table 10. SYN-SNUH benchmark data environment for CQ-Bench and FOEM.**

The use of SYN-SNUH has two implications. First, it provides a controlled OMOP CDM environment in which natural-language cohort criteria can be evaluated against actual execution results from the synthetic OMOP CDM instance[117]. Second, because the data are synthetic, the evaluation should not be interpreted as chart-reviewed clinical validation. These SYN-SNUH benchmark components assess whether a system can reproduce executable cohort definitions in the target CDM, not whether the selected persons are clinically adjudicated true cases.

## 4.3. Runtime Configuration and Evaluated Systems

NS-CoT was evaluated as a neuro-symbolic pipeline with separate components for language interpretation, concept retrieval, vocabulary grounding, symbolic compilation, and SQL execution. The runtime configuration specified the database connection, ChromaDB vector index, concept-resolution parameters, LLM backend, logging options, drug-resolution settings, output options, and cohort-persistence settings.

The natural-language interpretation component was connected to an OpenAI-compatible API endpoint. In the evaluated configuration, the model used for the NS-CoT interpretation pipeline was gpt-4.1-2025-04-14[115], with temperature set to 0.0. The low temperature was used to reduce variability in structured decomposition outputs. However, this setting does not make the natural-language interpretation stage formally deterministic. As defined in Chapter 3, determinism in NS-CoT applies after the typed and grounded representation has been produced and is consumed by the symbolic compiler.

Concept candidate retrieval used ChromaDB indexes over OMOP concept metadata. The Chroma configuration specified the collection name, top-k retrieval size, score threshold, and embedding model. The embedding model used for ChromaDB query embeddings was nomic-embed-text-v1.5, served through a local LM Studio OpenAI-compatible embedding endpoint[116]. This embedding backend should be distinguished from the OpenAI LLM backend used for natural-language interpretation. In other words, the system used the OpenAI API for the language-model reasoning and decomposition layer, while ChromaDB concept retrieval used a locally served embedding model.

This separation is methodologically important. ChromaDB retrieval provides candidate OMOP concepts for clinical mentions; it does not define the final cohort semantics. Candidate concepts are subsequently validated, expanded through OMOP vocabulary relations, and, where applicable, filtered according to local CDM evidence. The final concept sets used for SQL execution are therefore not simply top-k vector-search results. The database backend for the evaluation was PostgreSQL. The configured schema was synthetic_snuh_cdm, and the database connector set the search path accordingly. Generated SQL was saved for inspection, and a small number of sample person identifiers could be returned for event-level review.

The evaluation compared NS-CoT against a contemporary OMOP-oriented NL2SQL baseline. The baseline was FastOMOP, a multi-agent workflow for natural-language query execution over OMOP CDM[118]. FastOMOP separates semantic and database-oriented functions through an agentic architecture: a semantic component identifies OMOP-relevant concepts, and a database component generates and executes SQL. This makes FastOMOP a strong comparison point because it is designed specifically for OMOP-based RWE automation rather than for generic relational text-to-SQL.

The comparison is not framed as "prompting versus no prompting." Both systems rely on language-model-mediated interpretation at some stage. The key difference is where executable cohort semantics reside. In FastOMOP, many cohort-defining decisions, including concept selection, temporal interpretation, exclusion handling, and SQL composition, are mediated by the agentic LLM-driven query-generation process. In NS-CoT, these decisions are first represented as typed and grounded cohort semantics, then compiled through a deterministic symbolic layer.

The evaluated systems were compared under component-specific reference targets. In CQ-Bench and FOEM, each benchmark query was evaluated against a deterministic scalar reference count. In COVIC, each criterion-site pair was evaluated against a deterministic reference patient set and reference count. In the protocol-derived clinical trial case studies, NS-CoT artifacts were examined for traceability, computability boundaries, and clinical interpretability rather than scored as aggregate benchmark outcomes.

SQL exact match was not used as the primary metric because the evaluation target is functional cohort semantics, not string-level SQL identity. Two different

SQL queries may be semantically equivalent with respect to the final patient count, and two syntactically valid SQL queries may represent different cohort definitions. The evaluation therefore used count fidelity, patient-set fidelity, and, where applicable, site-feasibility decision preservation as the primary correctness criteria.

## 4.4. CQ-Bench

CQ-Bench was constructed as the internally designed executable-reference benchmark for evaluating whether a natural-language cohort definition system can reproduce predefined OMOP-executable cohort semantics over SYN-SNUH. It is a CDM-resident executable-reference benchmark whose released artifacts and construction scripts are available in the public repository cited in the bibliography. The repository does not include patient-level SYN-SNUH records; reproduction requires authorized access to the corresponding OMOP CDM environment and vocabulary resources. Its reference labels are deterministic cohort counts derived from OMOP concept slots and SQL templates. They are not chart-reviewed clinical gold standards.

The benchmark was designed from the target database outward. Rather than first writing arbitrary natural-language questions and then checking whether they happened to be executable, the construction process began by inspecting the SYN-SNUH resident inventory. The benchmark generator queried the OMOP CDM for resident standard condition concepts, RxNorm drug concepts, demographic strata, first-diagnosis year strata, condition-drug temporal pairs, and candidate drug-class exclusion patterns. These observations formed a capability matrix of concept and event patterns actually present in the target CDM.

Natural-language queries were then generated only from concepts and patterns represented in the capability matrix, except for the deliberately constructed negative-control tier. This design ensures that positive-control queries are executable in the target OMOP instance and that reference counts can be derived through deterministic SQL.

CQ-Bench consists of 100 natural-language cohort feasibility queries. Eighty queries are Tier A positive-control queries, and twenty queries are Tier B negative-control queries. Tier A queries are grounded in condition, drug, demographic, and temporal event patterns observed in SYN-SNUH. Tier B queries are constructed from clinically meaningful categories that were programmatically confirmed absent from the resident condition, procedure, or other related resident inventories at benchmark construction time. Tier B reference counts are therefore defined as zero by construction. This does not mean that the corresponding clinical entities are biologically impossible; it means that the benchmark's target OMOP instance does not contain resident evidence for those categories under the construction rules. The Tier A queries are stratified into five levels of semantic complexity.

| Stratum | Count | Reference semantics |
|---|---|---|
| L1 | 12 | Condition cohort |
| L2 | 18 | Condition + age/gender |
| L3 | 14 | Condition + first-diagnosis year |
| L4 | 22 | Condition diagnosis followed by drug exposure within an anchored window |
| L5 | 14 | Condition diagnosis followed by anchor-bound absence of a drug or drug class |
| Tier B | 20 | Negative controls, zero by construction |
| **Total** | **100** | |

**Table 11. CQ-Bench query strata.**

L1 evaluates simple condition grounding and condition-cohort retrieval. L2 adds demographic restrictions such as age or gender. L3 evaluates first-diagnosis year constraints, requiring the system to identify the earliest diagnosis date and apply a calendar-year predicate. L4 evaluates temporal inclusion: a drug exposure must occur within a post-diagnosis window. L5 evaluates anchor-bound temporal exclusion: a drug or drug-class exposure must be absent after the diagnosis anchor. Tier B evaluates whether the system avoids generating false-positive cohorts for concepts absent from the target CDM.

Each query is stored with metadata required for reference-count derivation and audit. A query record contains a stable identifier, natural-language query, tier, level, pattern label, condition name, drug name or drug class when applicable, OMOP seed concept identifiers, temporal flags, gold concept slots, reference-source labels, and selection-count metadata. These metadata are not model inputs. They are used to construct and audit the executable reference cohort.

The output of CQ-Bench construction was therefore not only a list of natural-language questions. It was a benchmark artifact bundle consisting of: (i) 100 natural-language cohort feasibility queries, (ii) query-level metadata including tier, stratum, pattern, clinical terms, OMOP seed concepts, temporal flags, and gold slots, (iii) deterministic executable reference counts, (iv) stratum-specific SQL derivation templates, (v) the SYN-SNUH capability matrix used for resident concept and event-pattern selection, and (vi) validation reports confirming output shape and tier-specific invariants. This bundle structure is important because it allows each natural-language query to be traced back to the executable OMOP semantics used to derive its reference count.

Reference counts were derived through a deterministic process. Each gold slot was interpreted as an OMOP standard seed concept or concept set. Seed concepts were expanded through 'concept_ancestor', restricted to valid standard concepts, intersected with resident concepts in the target clinical table, and then executed through tier-specific SQL templates. This process produced the reference count $g(q)$ for each query $q$.

The reference SQL templates were designed to match the intended semantics of each stratum. L1-L3 use a diagnosis CTE based on condition occurrences and earliest diagnosis date. L2 adds demographic predicates, and L3 adds a first-diagnosis year predicate. L4 uses an anchored 'EXISTS' predicate to determine whether drug exposure occurs within a bounded post-diagnosis window. In CQ-Bench, an L4 query is satisfied when a qualifying drug exposure occurs on or after the first diagnosis date and no later than N days after that diagnosis. This convention includes same-day diagnosis and drug exposure events.

L5 uses an anchor-bound 'NOT EXISTS' predicate to operationalize post-diagnosis drug absence. In this stratum, a patient is retained only when no qualifying drug exposure is observed on or after the first diagnosis date. Operationally, this is implemented as a correlated 'NOT EXISTS' condition over 'drug_exposure' using the first diagnosis date as the temporal anchor.

This convention intentionally evaluates inclusive anchored post-diagnosis windows and anchor-bound temporal non-existence semantics. It does not evaluate strict_after semantics unless explicitly encoded by the reference template.

Benchmark validation was performed after generation. The validation script checked that the query file and reference-count file contained 100 records, that query identifiers were complete and stable, that tier sizes matched the design, and that stratum counts corresponded to the intended distribution. It also checked that Tier A queries had non-empty executable concept slots and that Tier B queries had empty gold slots and zero reference counts. These checks verify that the benchmark artifact matches its design assumptions.

CQ-Bench is therefore not a generic clinical phenotype benchmark. It is a benchmark for functional fidelity to predefined OMOP cohort semantics in a specific CDM instance. Its design makes it possible to evaluate the semantic components that motivate NS-CoT: concept grounding, demographic filtering, first-anchor construction, temporal inclusion, temporal exclusion, drug-class handling, and resistance to false-positive generation in negative-control settings.

## 4.5. FOEM

FOEM was used as the external template-derived executable-reference benchmark component. Whereas CQ-Bench was constructed in this study to stress semantic components targeted by the proposed formalization, FOEM adapts the FOEM-style benchmark methodology associated with FastOMOP to the same SYN-SNUH OMOP CDM environment. This allows NS-CoT to be evaluated against a benchmark family not designed around its own formalization. In this study, FOEM is external in benchmark methodology and query-template family, while its concrete instantiation was made executable over the same SYN-SNUH CDM instance.

FOEM contained 201 externally derived query records. However, not all FOEM queries corresponded to a single scalar cohort count. Some queries asked for grouped or distributional outputs, such as counts by gender, race, year of birth, year of diagnosis, or year of prescription. Because the primary $R_0$ metric in this study is defined as exact reproduction of a final scalar cohort count, the paired exact-count analysis used the 166 FOEM queries that were scalar-count evaluable. The remaining 35 grouped or distributional queries were not included in scalar $R_0$ scoring. This exclusion was not a post-hoc performance filter; it followed from the predefined metric definition, which requires a single scalar count. Grouped or distributional queries would require group-wise output comparison rather than single-cohort exact-count comparison.

The construction process followed four steps. First, candidate query components were mined from the local OMOP CDM. These included drug concepts, condition concepts, demographic values, and combinations of concepts that produced executable SQL results in SYN-SNUH. Second, FOEM-style SQL templates were instantiated with these local concepts and demographic values. Third, the instantiated reference SQL was executed against SYN-SNUH to produce deterministic reference outputs. Fourth, each query was classified according to whether its output was a scalar cohort count or a grouped/distributional result. The scalar-count evaluable subset was then used for paired exact-count scoring.

## 4.6. COVIC Multi-Site Feasibility Simulation

COVIC was designed as an external multi-site feasibility simulation using criterion-site pair, patient-set, and site-feasibility metrics defined in Section 4.8. In this experiment, the same COVIC-derived executable eligibility criterion component was applied across multiple OMOP CDM schemas, each treated as a candidate site. The purpose was not only to test scalar count reproduction for each criterion-site pair, but also to evaluate whether system-generated count preserved the site-feasibility information implied by deterministic reference execution. Specifically, the analysis examined three related targets: count fidelity, patient-set fidelity, and site-level feasibility decision preservation. This distinguishes COVIC from CQ-Bench and FOEM: CQ-Bench evaluates targeted semantic components, FOEM evaluates external template-derived OMOP query reproduction, and COVIC evaluates how generated cohort logic behaves when used for multi-site-level patient counts comparison.

The paired unit in this experiment was a criterion-site pair, defined by a COVIC-derived executable criterion component and an evaluated OMOP schema. Sixty-five COVIC-derived criterion components were evaluated across three OMOP schemas, yielding 195 paired observations. For each criterion-site pair, deterministic reference SQL produced a reference patient set and reference count, whereas each system-generated SQL artifact produced a system patient set and system count. These outputs were then compared at three levels. First, count-level metrics evaluated whether the system reproduced the reference count for each criterion-site pair. Second, patient-set metrics evaluated whether the system identified the same patients as the reference execution. Third, site-feasibility metrics evaluated whether system-generated count preserved reference-defined feasible versus infeasible site labels and relative site rankings.

This experiment should be interpreted with an evaluation-unit caveat. During pilot audit, some COVIC rows used a full criterion line as the system input, whereas the deterministic reference SQL could be aligned to a selected entity-level COVIC concept. Therefore, patient-set fidelity in COVIC should be interpreted under a strict criterion-unit alignment assumption. This caveat does not make the experiment merely ancillary; rather, it clarifies what the experiment measures. COVIC evaluates the robustness of applying externally derived eligibility material to multiple OMOP schemas and asks whether system-generated count preserves reference-defined count, patient-set, and site-feasibility information under that strict alignment setting. It is therefore most appropriately interpreted as an external multi-site feasibility simulation with a unit-alignment caveat, rather than as a single-database benchmark of isolated temporal or exclusion semantics.

A second-pass NS-CoT repair analysis was also conducted after deterministic correction of local-resident drug filtering, partial OR execution, and typed-plan replay or fresh-fallback bookkeeping. FastOMOP was not rerun; its previously captured artifact-based outputs were reused as the baseline because no FastOMOP logic was changed. The purpose of this analysis was to test whether explicit NS-CoT artifacts could support error localization, implementation-level correction, and re-evaluation in the same multi-site feasibility setting. Therefore, the second-pass analysis should be interpreted as a repairability and decision-preservation analysis within COVIC, rather than as an independent new benchmark.

| Item | Value |
|---|---|
| Source corpus | COVIC, an OMOP CDM-based annotated corpus of eligibility criteria from real COVID-19 clinical trials |
| System input | Original criterion lines reconstructed from annotated trial text |
| Retained domains | Condition, Drug |
| Excluded components | Measurement predicates, numeric predicates, temporal predicates, and within-criterion linguistic negation patterns |
| Vocabulary scope | SNOMED for Condition; ATC, RxNorm, RxNorm Extension for Drug |
| Strictly aligned criteria | 65 |
| Unique NCT IDs | 46 |
| Schemas/sites | 3 |
| Criterion-site pairs | 195 |

**Table 12. COVIC dataset construction and evaluation unit.**

| Category | n (%) |
|---|---|
| Condition | 8 (12.31%) |
| Drug | 57 (87.69%) |
| Inclusion | 3 (4.62%) |
| Exclusion-trigger | 62 (95.38%) |
| Multi-annotation criteria | 34 (52.31%) |
| Multi-concept criteria | 33 (50.77%) |

**Table 13. COVIC criterion composition.**

| Schema | Population | Nonzero / 65 | Nonzero % | Median reference count | Mean reference count | Max reference count |
|---|---|---|---|---|---|---|
| public | 100,000 | 61 | 93.85% | 4,742 | 11,508.26 | 75,530 |
| synthea_cdm | 5,692 | 34 | 52.31% | 1 | 170.95 | 5,563 |
| synthetic_snuh_cdm | 10,000 | 10 | 15.38% | 0 | 296.83 | 5,589 |

**Table 14. Reference-count distribution across OMOP schemas.**

## 4.7. Protocol-Derived Clinical Trial Case Studies for Traceability Analysis

As the fourth evaluation component, three protocol-derived clinical trial case studies were conducted to examine how NS-CoT behaves when applied to selected ClinicalTrials.gov eligibility criteria. These case studies were run as implementation trace demonstrations rather than scored statistical benchmarks. Their purpose was to illustrate how the framework decomposes real eligibility text into execution-relevant semantic components, how computable and non-computable components are distinguished, and how stepwise trace artifacts identify where a criterion succeeds, collapses, or requires proxy interpretation.

The purpose of these case studies was not to establish a single correct cohort for each trial criterion. In real feasibility assessment, a free-text eligibility criterion does not always correspond to one self-evident gold-standard patient set. Different reviewers may reasonably choose different concept sets, temporal anchors, proxy variables, exclusion scopes, or observation windows, even when working from the same protocol text and the same clinical database. The case studies were therefore used to examine whether the proposed framework can externalize these interpretive choices as explicit executable semantic commitments. In this setting, $h_q$ should be understood not as a hidden clinical truth, but as an inspectable patient-set semantics object that can be reviewed, revised, and discussed before a site-level count is treated as feasibility evidence.

Three cases were selected to cover distinct semantic challenges. The first case tested value-bearing measurement semantics and exclusion handling using a blood-pressure feasibility criterion. The second case tested a procedure-linked intraoperative temporal anchor that is not directly available as a structured anchor in the evaluated OMOP event fields. The third case tested condition OR logic, drug-class exclusion, local drug-exposure availability, and empty executable concept-set behavior.

These probes should be interpreted as traceability analyses rather than clinical validation studies. They do not establish chart-reviewed eligibility, recruitment performance, or regulatory readiness. Instead, they demonstrate how the proposed architecture exposes the semantic and data-availability assumptions behind an site-level count estimate or a zero-count result. Box 1 summarizes the generic trace structure used in these probes. Its purpose is to make explicit which semantic commitments are made before SQL execution and how each commitment is carried forward into the final patient-set count.

| NCT | Trial criterion type | Main purpose |
|---|---|---|
| NCT04936022 | Blood pressure measurement + exclusions | Target computable semantics decomposition |
| NCT03152890 | Liver transplant procedure + glucose measurement + temporal anchor | Temporal-anchor ambiguity, proxy interpretation, and semantic-loss trace |
| NCT02958943 | AF/flutter OR logic + anticoagulant drug-class exclusion | Drug-class grounding, local executability, and empty-set risk |

**Box 1. Generic trace schema for exploratory real-trial criterion probes.**

| **Input**: a ClinicalTrials.gov eligibility criterion $q$ and an OMOP CDM instance $D$. |
|---|

**Step 1. Feasibility abstraction**
Identify the computable subset of $q$ relevant to site-level patient-count estimation. Distinguish directly computable components from non-computable, subjective, unstructured, or proxy-requiring components.

**Step 2. Interpretive commitment**
Externalize the chosen executable interpretation as a declared patient-set semantics object, $h_q$. Record whether each component is implemented directly, approximated by a proxy, deferred for review, or treated as not computable in the available structured OMOP CDM data.

**Step 3. Typed decomposition**
Decompose $q$ into typed semantic components: demographic criteria, clinical mentions, OMOP-compatible domains, inclusion or exclusion polarity, value constraints, temporal expressions, logical structure, and site-feasibility-relevant counting units.

**Step 4. Grounding and local executability**
For each clinical mention, retrieve candidate OMOP concepts, assess semantic and domain fit, expand through appropriate vocabulary relationships, and restrict to locally executable concepts where applicable. If the executable concept set is empty, record an executable-empty state rather than silently dropping the criterion or treating it as universally satisfied.

**Step 5. Patient-set logic construction**
Convert grounded components into patient-set operations: REQUIRE for inclusion, FORBID for exclusion, GROUP_OR for alternatives, value predicates for measurement thresholds, and TEMPORAL_REQUIRE or TEMPORAL_FORBID when anchor-relative temporal logic is available. For proxy-based components, record the proxy decision and the expected semantic limitation.

**Step 6. SQL compilation and execution**
Compile the declared patient-set logic into deterministic CTE-chain SQL and execute the query over $D$. Record intermediate cohort sizes after each operation.

**Step 7. Trace interpretation**
Classify the result as successful execution, count collapse due to local data unavailability, proxy-anchor limitation, executable-empty concept set, or non-computable component requiring review. The final count is interpreted as the cardinality of the declared executable patient set, not as an unqualified clinical gold-standard cohort.

## 4.8. Outcome Measures and Scoring Rules

Because the evaluation components differ in unit and purpose, the scoring rules were organized by evaluation level. CQ-Bench and FOEM used query-level scalar-count metrics, including liveness, conditional $R_0$, Strict $R_0$, $R_0$@5%, and zero-form agreement. Deterministic person-set verification was used to evaluate whether count-level agreement corresponded to the same underlying person-id set. COVIC used criterion-site pair metrics, extending count and patient-set fidelity to feasible-site discrimination and site-ranking preservation. The protocol-derived clinical trial case studies were evaluated qualitatively through trace artifacts, computability boundaries, and clinical interpretability rather than through aggregate benchmark scores.

For the executable-reference benchmark components, a reference site-level patient count is the result of executing a predefined SQL cohort definition over the target OMOP CDM instance. It is not an independent clinical truth label. This choice follows directly from the study's problem definition: the task is to reproduce predefined OMOP-executable cohort semantics from natural-language criteria. Therefore, a correct result means that the system reproduced the declared executable cohort semantics in the evaluated OMOP CDM environment, not that the selected persons were chart-reviewed clinically eligible patients.

Let $Q$ denote the set of benchmark queries. For each query $q \in Q$, let $g(q)$ be the reference count and $\hat{g}(q)$ be the count returned by the evaluated system. Define $L(q)$ as an indicator of whether the system returns a final integer count:

$$L(q) = \begin{cases} 1, & \text{if the system returns a final integer count for } q \\ 0, & \text{otherwise} \end{cases}$$

The overall liveness rate is:

$$\text{Liveness} = \frac{1}{|Q|} \sum_{q \in Q} L(q)$$

Liveness measures whether the system can complete the end-to-end workflow from natural-language query to scalar cohort count. It does not require the count to be correct.

Conditional $R_0$ measures exact-count accuracy among live runs. For a live query, the query-level conditional score is:

$$R_0^{cond}(q) = \begin{cases} 1, & \text{if } L(q) = 1 \text{ and } \hat{g}(q) = g(q) \\ 0, & \text{if } L(q) = 1 \text{ and } \hat{g}(q) \neq g(q) \end{cases}$$

The aggregate conditional $R_0$ is:

$$R_0^{cond} = \frac{\sum_{q \in Q} \mathbf{1}[L(q) = 1 \wedge \hat{g}(q) = g(q)]}{\sum_{q \in Q} L(q)}$$

This measure answers the question: when the system returns a count, how often is it exactly correct?

Joint $R_0$ treats non-live runs as failures and evaluates exact-count success over all attempted queries:

$$R_0^{joint}(q) = \begin{cases} 1, & \text{if } L(q) = 1 \text{ and } \hat{g}(q) = g(q) \\ 0, & \text{otherwise} \end{cases}$$

$$R_0^{joint} = \frac{1}{|Q|} \sum_{q \in Q} \mathbf{1}[L(q) = 1 \wedge \hat{g}(q) = g(q)]$$

Joint $R_0$ is the strictest metric used in this chapter. It reflects the proportion of all attempted queries for which the system both completed execution and reproduced the reference count exactly. This is the most relevant measure for feasibility workflows, because a system that frequently fails to return a count cannot be considered reliable even if its live runs are often correct.

In this study, Strict $R_0$ denotes joint exact-count fidelity. A query is scored as successful only when the system returns a final integer count and that count exactly matches the executable reference count; therefore, both non-live runs and live count mismatches are treated as failures. Conditional $R_0$ was computed only as a diagnostic decomposition metric to separate count accuracy among live runs from failures of execution, whereas Strict $R_0$ was used as the main end-to-end exact-count fidelity measure.

As secondary descriptive metrics, the evaluation also reports $R_0$@5% and Zero-form joint agreement. $R_0$@5% measures whether the returned count is within 5% of the executable reference count while still treating non-live runs as failures:

$$R_{0@5\%}^{joint} = \frac{1}{|Q|} \sum_{q \in Q} \mathbf{1}\,[L(q) = 1 \wedge A_{5\%}(q)]$$

where

$$A_{5\%}(q) = \begin{cases} \hat{g}(q) = 0, & \text{if } g(q) = 0, \\ \dfrac{|\hat{g}(q) - g(q)|}{g(q)} \leq 0.05, & \text{if } g(q) > 0. \end{cases}$$

Zero-form joint agreement measures whether the system preserves the zero-versus-nonzero answer form of the reference count while treating non-live runs as failures:

$$Z^{joint} = \frac{1}{|Q|} \sum_{q \in Q} \mathbf{1}\left[L(q) = 1 \wedge \mathbf{1}[\hat{g}(q) > 0] = \mathbf{1}[g(q) > 0]\right]$$

This metric does not require exact count agreement. It only evaluates whether the returned count and reference count have the same zero/nonzero form. Non-live runs are scored as failures, and the denominator is all scalar-count evaluable benchmark queries.

| Metric | Numerator | Denominator | Interpretation |
|---|---|---|---|
| Liveness | Queries with returned final count | All queries | End-to-end execution success |
| Conditional $R_0$ | Live queries with exact count match | Live queries | Accuracy among returned counts |
| Strict $R_0$ / Joint $R_0$ | Queries with returned count and exact match | All queries | End-to-end functional fidelity |
| $R_0$@5% | Returned count within 5% of reference | All queries | Approximate count fidelity |
| Zero-form joint | Live returned counts whose zero/non-zero form matches the reference | All attempted scalar-count queries | Joint zero/non-zero answer-form fidelity |

**Table 15. Evaluation metrics.**

The primary correctness measure uses exact count match rather than tolerance-based agreement. A one-patient difference is scored as incorrect. This strict rule is appropriate because the benchmark target is deterministic executable semantics. Even a small count difference may indicate a specific semantic discrepancy, such as a temporal boundary convention, concept-set expansion difference, or exclusion-scope error. The $R_0$@5% and Zero-form joint metrics are therefore reported only as secondary descriptive measures; they do not replace exact-count agreement as the primary evaluation criterion.

**Criterion-site metrics for COVIC multi-site feasibility simulation.**

COVIC required additional scoring rules because its evaluation unit was a criterion-site pair rather than a single benchmark query. Let $q$ denote a COVIC-derived executable eligibility criterion component, $s$ denote an OMOP schema treated as a candidate site, and m denote an evaluated system. The deterministic reference execution produced a reference patient set and reference count:

$$S_{ref}(q,s) = [[q]]_{ref,D_s}, \qquad N_{ref}(q,s) = |S_{ref}(q,s)|.$$

System m produced a system-generated patient set and system-generated count:

$$S_m(q,s) = [[q]]_{m,D_s}, \qquad N_m(q,s) = |S_m(q,s)|.$$

The same all-attempted count-fidelity logic was extended to the criterion-site unit in COVIC. Let $L_m(q,s)$ indicate whether system m returned an executable result for criterion $q$ at site s:

$$L_m(q,s) = \mathbf{1}[\text{system } m \text{ returned a final executable result for } (q,s)].$$

Strict exact-count fidelity for each criterion-site pair was defined as:

$$R_{0,m}(q,s) = \mathbf{1}\left[L_m(q,s) = 1 \;\wedge\; N_m(q,s) = N_{ref}(q,s)\right].$$

The overall strict $R_0$ for system m was computed as:

$$\overline{R}_{0,m} = \frac{1}{|\mathcal{Q}_C||\mathcal{S}|}\sum_{q\in\mathcal{Q}_C}\sum_{s\in\mathcal{S}} R_{0,m}(q,s).$$

The tolerance-based count metric was defined as:

$$R_{0@5\%,m}(q,s) = \mathbf{1}\left[L_m(q,s) = 1 \;\wedge\; A_{5\%,m}(q,s) = 1\right],$$

where

$$A_{5\%,m}(q,s) = \begin{cases} \mathbf{1}[N_m(q,s) = 0], & \text{if } N_{ref}(q,s) = 0, \\ \mathbf{1}\left[\dfrac{|N_m(q,s) - N_{ref}(q,s)|}{N_{ref}(q,s)} \le 0.05\right], & \text{if } N_{ref}(q,s) > 0. \end{cases}$$

This metric corresponds to $R_0$@5% at the criterion-site level.

Zero-form agreement at the criterion-site level was defined as:

$$Z_m(q,s) = \mathbf{1}\left[L_m(q,s) = 1 \;\wedge\; \mathbf{1}[N_m(q,s) > 0] = \mathbf{1}[N_{ref}(q,s) > 0]\right].$$

The overall zero-form agreement for system m was computed as the all-attempted mean of $Z_m(q,s)$ across all criterion-site pairs.

Patient-set fidelity was evaluated when reference and system patient sets were available. Exact patient-set agreement was defined as:

$$P_m(q,s) = \mathbf{1}\big[L_m(q,s) = 1 \ \wedge \ S_m(q,s) = S_{ref}(q,s)\big].$$

Jaccard similarity was defined as:

$$J_m(q,s) = \begin{cases} 1, & \text{if } S_m(q,s) = \varnothing \text{ and } S_{ref}(q,s) = \varnothing, \\ \dfrac{|S_m(q,s) \cap S_{ref}(q,s)|}{|S_m(q,s) \cup S_{ref}(q,s)|}, & \text{otherwise.} \end{cases}$$

Precision was defined as:

$$\pi_m(q,s) = \begin{cases} 1, & \text{if } S_m(q,s) = \varnothing \text{ and } S_{ref}(q,s) = \varnothing, \\ 0, & \text{if } S_m(q,s) = \varnothing \text{ and } S_{ref}(q,s) \neq \varnothing, \\ \dfrac{|S_m(q,s) \cap S_{ref}(q,s)|}{|S_m(q,s)|}, & \text{otherwise.} \end{cases}$$

Recall was defined as:

$$r_m(q,s) = \begin{cases} 1, & \text{if } S_m(q,s) = \varnothing \text{ and } S_{ref}(q,s) = \varnothing, \\ 0, & \text{if } S_m(q,s) \neq \varnothing \text{ and } S_{ref}(q,s) = \varnothing, \\ \dfrac{|S_m(q,s) \cap S_{ref}(q,s)|}{|S_{ref}(q,s)|}, & \text{otherwise.} \end{cases}$$

When system m did not return an executable patient set for a criterion-site pair, binary patient-set exactness was scored as 0 under all-attempted scoring. Continuous overlap metrics were reported under both all-attempted and both-evaluable summaries. In the all-attempted analysis, non-evaluable system outputs were assigned zero overlap. In the both-evaluable sensitivity analysis, only criterion-site pairs with evaluable patient sets for both systems were included.

For site-feasibility discrimination, the reference feasibility label for threshold tau was defined as:

$$Y_{ref}(q,s;\tau) = \mathbf{1}\big[N_{ref}(q,s) \geq \tau\big], \qquad \tau \in \{1,5,10,30,100\}.$$

The system-derived feasibility score was defined as:

$$u_m(q,s) = \log\big(1 + N_m(q,s)\big).$$

AUROC and AUPRC were computed across all criterion-site pairs for each threshold tau, using the reference feasibility label as the reference label and the system-derived feasibility score as the system score. These thresholds were not intended as universal clinical feasibility cutoffs. They were used as operational sensitivity thresholds spanning increasingly stringent definitions of an available patient pool, from any nonzero pool to a large candidate pool.

For site-ranking analysis, Spearman rank correlation was computed for each criterion $q$ when between-site variation was sufficient for both the reference and system-generated counts:

$$\rho_m(q) = \text{Spearman}\big(\{N_{ref}(q,s): s \in \mathcal{S}\}, \{N_m(q,s): s \in \mathcal{S}\}\big).$$

The mean Spearman correlation was computed over the subset of criteria for which rank correlation was defined.

For top-site agreement, criteria with all-zero reference counts across sites were excluded because no meaningful top feasible site exists. The applicable criterion set was defined as:

$$\mathcal{Q}_{\text{top}} = \left\{q \in \mathcal{Q}_{\text{C}}: \max_{s \in \mathcal{S}} N_{ref}(q,s) > 0\right\}.$$

For each criterion in the applicable top-site set, the reference top-site set was defined as the set of OMOP schemas with the largest reference count:

$$\text{Top}_{\text{ref}}(q) = \left\{s \in \mathcal{S}: N_{ref}(q,s) = \max_{s' \in \mathcal{S}} N_{ref}(q,s')\right\}.$$

The system-generated top-site set was defined using the system-derived feasibility score:

$$\text{Top}_m(q) = \left\{s \in \mathcal{S}: u_m(q,s) = \max_{s' \in \mathcal{S}} u_m(q,s')\right\}.$$

Because log(1+x) is monotonic for nonnegative counts, this is equivalent to ranking sites by the system-generated count while matching the implemented scoring procedure.

Top-site agreement was then scored as:

$$T_m(q) = \mathbf{1}[\text{Top}_{\text{ref}}(q) \cap \text{Top}_m(q) \neq \varnothing].$$

The overall top-site agreement for system $m$ was computed as:

$$\overline{T}_m = \frac{1}{|\mathcal{Q}_{\text{top}}|} \sum_{q \in \mathcal{Q}_{\text{top}}} T_m(q).$$

The use of set-valued top-site definitions allows ties in the largest site-level patient count. Therefore, a criterion was counted as top-site agreement when the reference and system-generated top-site sets shared at least one site.

## 4.9. Paired Comparison and Error Analysis

Paired comparison was applied at the unit appropriate to each evaluation component. For CQ-Bench and FOEM, the paired unit was a benchmark query. For COVIC, the paired unit was a criterion-site pair. In each paired comparison, NS-CoT and FastOMOP were evaluated against the same executable reference output for the same unit.

For CQ-Bench and FOEM, each query $q$ was evaluated as a paired unit. Each system received the same natural-language input and was scored against the same reference count $g(q)$. The primary paired comparison used query-level joint success indicators:

$$S_A(q) = \mathbf{1}[L_A(q) = 1 \wedge \hat{g}_A(q) = g(q)]$$
$$S_B(q) = \mathbf{1}[L_B(q) = 1 \wedge \hat{g}_B(q) = g(q)]$$

where $A$ denotes NS-CoT and $B$ denotes the baseline system. Thus, $S_A(q) = 1$ or $S_B(q) = 1$ only when the corresponding system both returned a final integer count and exactly reproduced the executable reference count. Secondary descriptive metrics such as $R_0$@5% and Zero-form joint agreement were not used for paired significance testing. This produces four paired outcome categories:

| Category | Condition |
|---|---|
| Both exact | $S_A(q) = 1, S_B(q) = 1$ |
| NS-CoT only exact | $S_A(q) = 1, S_B(q) = 0$ |
| Baseline only exact | $S_A(q) = 0, S_B(q) = 1$ |
| Neither exact | $S_A(q) = 0, S_B(q) = 0$ |

**Table 16. Paired outcome categories.**

The paired nature of the evaluation is important. Each system is tested on the same query set, so the relevant question is not only whether one system has a higher marginal accuracy, but whether the discordant query outcomes are systematically skewed toward one system.

McNemar's test was used to compare paired binary exact-count outcomes between NS-CoT and the baseline system on the same benchmark queries. Let $b$ denote the number of queries correct only for NS-CoT and $c$ denote the number of queries correct only for the baseline. McNemar's test evaluates whether the discordant counts $b$ and $c$ differ more than expected under the null hypothesis of equal paired performance. Because the number of discordant pairs can be small, the reported McNemar p-values were computed using the exact binomial version of McNemar's test. The continuity-corrected chi-square statistic, $\chi^2 = \frac{(|b-c|-1)^2}{b+c}$ is an asymptotic approximation when $b + c > 0$. For CQ-Bench and FOEM, McNemar's test was applied at the benchmark-query level. For COVIC binary outcomes, the same paired-testing logic was applied at the criterion-site pair level. Continuous or overlap-based COVIC outcomes, such as Jaccard similarity, precision, recall, AUROC, AUPRC, and site-ranking summaries, were analyzed using the paired procedures specified for those metrics rather than query-level McNemar testing.

Error analysis was performed at the relevant evaluation-unit level: query level for CQ-Bench and FOEM, and criterion-site level for COVIC where applicable. Errors were first divided into non-live failures and live count mismatches. Non-live failures indicate that the system did not return a final scalar count. Live mismatches indicate that the system returned a count, but the count differed from the reference count. For NS-CoT, live mismatches can be further inspected using its structured artifacts. The main error layers are:

- Decomposition error: incorrect extraction of clinical mention, role, polarity, or temporal relation.
- Grounding error: incorrect or incomplete OMOP concept resolution.
- Local executability error: a vocabulary-valid concept set fails to reflect the local CDM representation.
- Temporal boundary error: incorrect interpretation of inclusive, exclusive, bounded, or lower-only temporal relations.
- Exclusion-scope error: exclusion is applied globally rather than relative to an anchor, or vice versa.
- SQL compilation error: incorrect lowering of typed semantics into SQL.
- Database execution error: SQL failure, timeout, or unexpected execution output.

For the baseline system, error inspection was limited to returned SQL, execution status, logs, and observable output behavior, because it did not expose the same typed intermediate artifacts as NS-CoT. This layer-wise analysis supports the study's central methodological argument. A final count mismatch is not treated as an opaque failure. When sufficient artifacts are available, the mismatch can be traced to a specific transformation layer. For example, a temporal-boundary mismatch and a drug-class grounding mismatch may both produce incorrect counts, but they represent different methodological failures and require different corrections.

The paired comparison and error analysis together evaluate both performance and failure localization. Exact-count metrics quantify whether a system reproduces predefined executable semantics. Paired testing determines whether differences between systems are systematic on the same evaluation units. Error analysis then examines why mismatches occurred, particularly for temporal and exclusion-heavy criteria where semantic preservation is most important.

# Chapter 5. Results

## 5.1. Overview of Evaluation Runs

This chapter reports the empirical evaluation of NS-CoT across four complementary evaluation components: CQ-Bench, FOEM, COVIC, and three protocol-derived clinical trial case studies. CQ-Bench and FOEM were used as executable-reference benchmarks for scalar count and patient-set fidelity. COVIC was used as an external multi-site feasibility simulation to evaluate whether system-generated count preserves reference-defined counts, patient sets, site-feasibility labels, and site rankings across multiple OMOP schemas. The three NCT case studies were used to examine traceability, computability boundaries, and the clinical interpretability of the generated cohort-definition artifacts in realistic protocol-derived scenarios. As defined in Chapter 4, the target of evaluation is functional fidelity to executable cohort semantics and feasibility-oriented decision preservation, not chart-reviewed clinical phenotype validity, prospective recruitment performance, or final clinical eligibility adjudication.

Two systems were evaluated in the benchmark and multi-site simulation components: NS-CoT and FastOMOP. NS-CoT represents the proposed neuro-symbolic framework, in which feasibility-oriented natural-language criteria are decomposed into typed intermediate representations, grounded to OMOP-compatible concept sets, interpreted through patient-set algebra, and compiled into deterministic SQL. FastOMOP was used as the primary OMOP-oriented natural-language-to-SQL baseline. The comparison therefore examines whether explicitly representing cohort semantics before SQL generation improves executable count reproduction, patient-set preservation, and feasibility-decision preservation, particularly for criteria involving temporal anchoring, drug exposure, exclusion logic, and multi-site count comparison.

The evaluation components were designed to answer related but distinct questions. CQ-Bench consists of 100 natural-language cohort feasibility queries constructed from the SYN-SNUH OMOP CDM resident inventory. It was designed to test increasing levels of semantic complexity, from simple condition cohorts to demographic constraints, first-diagnosis-year constraints, post-diagnosis drug exposure, and post-diagnosis drug absence. FOEM was used as an external template-derived benchmark instantiated over the same SYN-SNUH database; the full FOEM benchmark contained 201 query records, but the exact-count results reported in this chapter are based on the 166 scalar-count evaluable queries used in the paired strict $R_0$ analysis. Whereas CQ-Bench evaluates the semantic components emphasized by this research, FOEM provides an external executable-reference setting for assessing whether the system generalizes beyond the internally designed query set.

COVIC extends the evaluation from single-database executable-reference count reproduction to external multi-site feasibility simulation. In this component, the same COVIC-derived executable eligibility criterion component was applied across multiple OMOP CDM schemas, each treated as a candidate site. For each criterion-site pair, deterministic reference SQL produced a reference patient set and reference count, whereas each system-generated SQL artifact produced a system-generated patient set and system-generated count. The analysis then evaluated count fidelity, patient-set fidelity, threshold-based feasible-site discrimination, and site-ranking preservation. This component asks whether a system can preserve the site-feasibility information implied by reference executable counts when the same criterion is evaluated across candidate sites.

The three protocol-derived clinical trial case studies were not intended as statistical benchmarks or superiority tests. Instead, they were used to examine how NS-CoT behaves when applied to real protocol-derived eligibility material. These case studies focus on whether the system produces interpretable intermediate artifacts, whether computable and non-computable components can be distinguished, whether concept grounding and temporal or exclusion assumptions can be reviewed, and whether the resulting cohort logic can support clinically meaningful feasibility interpretation. Their role is therefore complementary to the benchmark evaluations: they demonstrate clinical interpretability and traceability in realistic trial-facing scenarios.

The unit of evaluation differs across components. In CQ-Bench and FOEM, each natural-language query $q$ was evaluated against an executable reference count $g_{ref}(q)$. For system m, the system either returned a final integer cohort count $g_m(q)$ or failed to return a count. A query was scored as an exact-count success only when the system returned a final count and $g_m(q)$ exactly matched $g_{ref}(q)$. For the main Strict $R_0$ metric, no tolerance margin was applied; a one-patient difference was treated as a mismatch because the benchmark target is deterministic executable cohort semantics. In COVIC, the unit of evaluation was a criterion-site pair $(q, s)$, where q denotes a COVIC-derived executable criterion component and s denotes an OMOP schema treated as a candidate site. In the NCT case studies, the unit of analysis was the criterion component or case-level trace artifact rather than a benchmark query.

The results are reported in eight parts. First, overall exact-count performance is presented for CQ-Bench and FOEM using joint exact-count fidelity, reported in the result figures as Strict $R_0$. Second, liveness and conditional exact-count performance are reported as diagnostic decomposition metrics to distinguish execution failure from count mismatch. Third, secondary count-level metrics, including $R_0$@5% and zero-form agreement, are used to characterize approximate count fidelity and zero-versus-nonzero answer-form preservation. Fourth, CQ-Bench results are stratified by semantic complexity to examine whether NS-CoT's advantage is concentrated in temporal and exclusion-heavy criteria. Fifth, paired query-level comparisons are reported using McNemar's test, followed by a traceable failure case showing how an incorrect count can be localized to a specific temporal-boundary decision.

Sixth, post-hoc person-set verification is reported to examine whether exact count agreement corresponds to exact patient-set agreement and whether any count-matched queries conceal different underlying person sets. Seventh, COVIC is reported as an external multi-site feasibility simulation, evaluating count fidelity, patient-set fidelity, feasible-site discrimination, repairability, and site-ranking preservation across OMOP schemas. Eighth, the three protocol-derived clinical trial case studies are presented to illustrate traceability, computability boundaries, and clinical interpretability in realistic protocol-derived scenarios.

This organization reflects the methodological claim of the work. The question is not only whether a system can produce SQL or return a plausible count. The question is whether the system can preserve the cohort-defining semantics of natural-language trial criteria sufficiently to reproduce executable reference counts, recover the corresponding patient sets where available, preserve site-feasibility decisions across candidate OMOP schemas, and expose the assumptions and computability boundaries that matter for clinical review. In this framing, site-level patient count is not treated as a raw database number, but as the cardinality of a declared, executable, reviewable, and traceable patient-set definition.

## 5.2. Overall Count-Level Fidelity

The first analysis evaluated exact-count performance over all attempted benchmark queries. A query was scored as an exact-count success only when the system returned a final integer count and that count exactly matched the executable reference count. This measure corresponds to Strict $R_0$, or joint exact-count fidelity, because it treats both non-returned counts and incorrect returned counts as failures.

Figure 6 shows the benchmark-level exact-count performance of NS-CoT and FastOMOP. On CQ-Bench, NS-CoT reproduced the executable reference count for 83 of 100 queries, whereas FastOMOP reproduced the reference count for 60 of 100 queries. This corresponds to an absolute difference of 23 percentage points in favor of NS-CoT. The paired comparison for CQ-Bench was statistically significant by McNemar's test, as reported in Section 5.4.

On FOEM, NS-CoT reproduced the reference count for 151 of 166 queries, while FastOMOP reproduced the reference count for 158 of 166 queries. Although the numerical difference favored FastOMOP, the paired comparison was not statistically significant. Therefore, the FOEM result should not be interpreted as evidence of a systematic performance advantage for either system. Rather, it indicates that both systems achieved high exact-count performance on this external template-derived OMOP benchmark.

Secondary metrics showed a similar but more nuanced pattern. On CQ-Bench, NS-CoT achieved $R_0$@5% of 97.0% compared with 72.0% for FastOMOP, and Zero-form joint agreement of 98.0% compared with 97.0% for FastOMOP. Thus, NS-CoT's CQ-Bench advantage was not limited to exact count reproduction; it also appeared in approximate count fidelity, while both systems generally preserved the zero versus nonzero answer form. On FOEM, both systems performed strongly across secondary metrics. NS-CoT achieved $R_0$@5% of 96.4% and Zero-form joint agreement of 99.4%, whereas FastOMOP achieved $R_0$@5% of 97.6% and Zero-form joint agreement of 98.8%. These secondary results support the interpretation that FOEM did not show a systematic advantage for either system, whereas CQ-Bench exposed a larger gap in count-level functional fidelity.

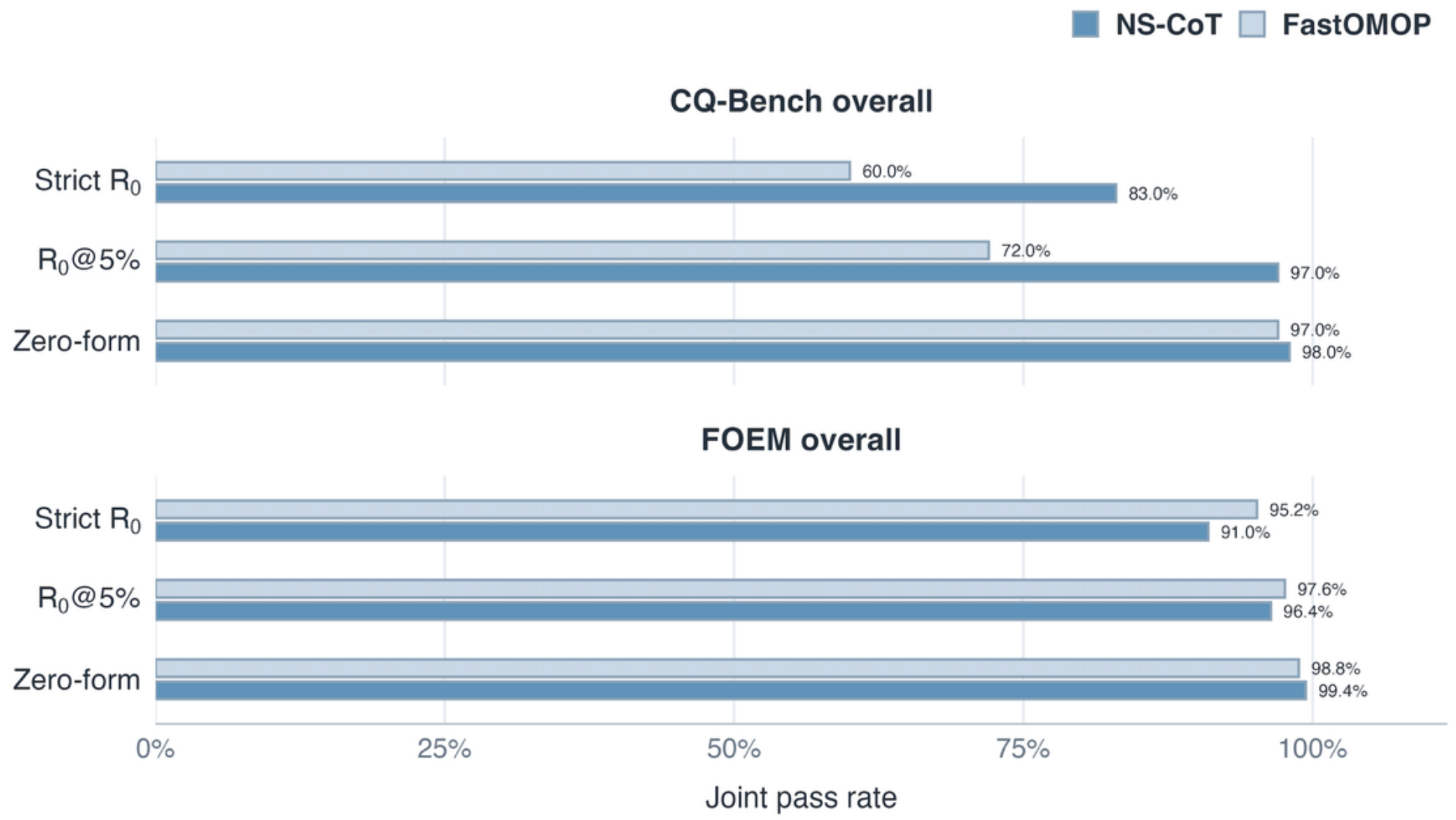

**Figure 6. Overall $R_0$ outcomes across CQ-Bench and FOEM.** Bars show joint pass rates for Strict $R_0$, $R_0$@5%, and Zero-form joint agreement under the evaluated runtime setting. CQ-Bench shows a larger NS-CoT advantage, whereas FOEM shows high performance for both systems.

The contrast between the two executable-reference benchmarks is important. CQ-Bench was designed to evaluate the semantic operations emphasized in this study, including first-diagnosis anchoring, post-anchor temporal windows, temporal exclusion, and negative-control handling. FOEM, by contrast, is an external template-derived benchmark instantiated over the same OMOP CDM environment. The overall results therefore suggest that the advantage of NS-CoT is not a uniform effect across all OMOP query settings. It is most visible in CQ-Bench, the benchmark designed to stress explicit operationalization of cohort semantics before SQL compilation.

However, overall exact-count performance alone does not distinguish between two different types of failure. A system may fail because it does not return a final count, or it may return a count that does not match the executable reference. To separate these cases, Figure 7 shows count-return liveness, exact-count fidelity among returned results, and overall exact-count fidelity across all attempted queries.

On CQ-Bench, NS-CoT returned a final count for 99 of 100 queries. Among those 99 live runs, 83 returned the exact reference count, yielding a conditional exact-count rate of 83/99. FastOMOP returned a final count for 97 of 100 queries, and 60 of those 97 returned counts matched the reference. Thus, both systems were generally able to execute on CQ-Bench, but their returned counts differed substantially in exact-count fidelity.

On FOEM, both systems returned a final count for all 166 queries. NS-CoT matched the reference count in 151 of 166 live runs, while FastOMOP matched the reference count in 158 of 166 live runs. Because liveness was 100% for both systems on FOEM, the overall exact-count difference in that benchmark reflects count mismatch rather than execution failure.

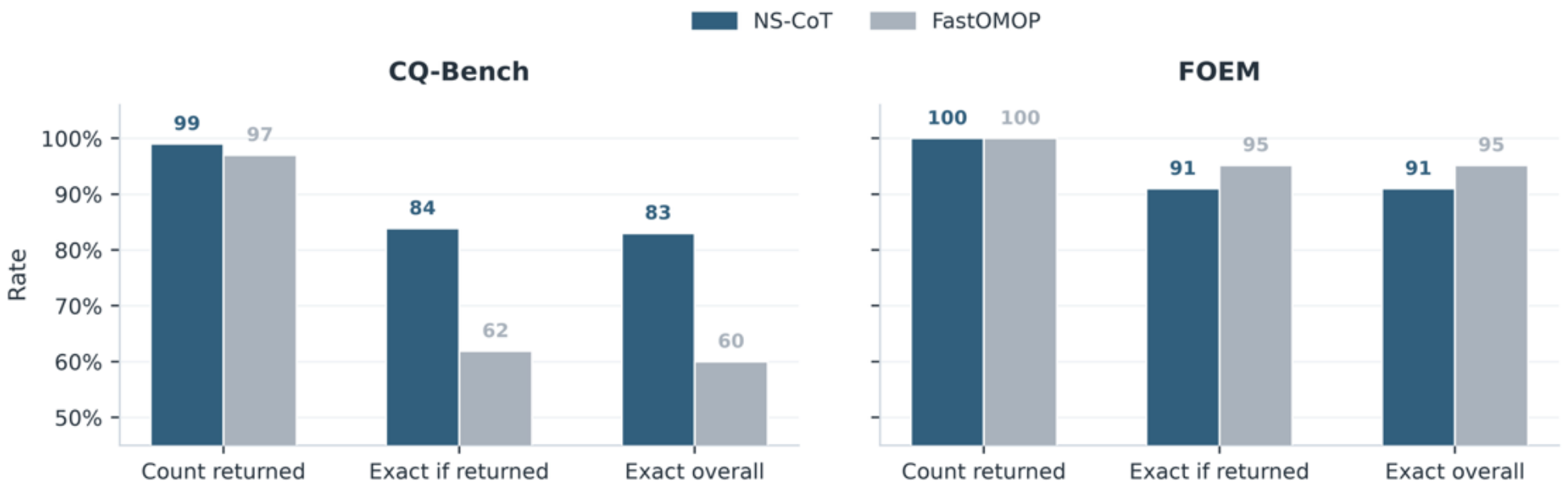

**Figure 7. Decomposing exact-count performance into execution and count fidelity.** Bars separate count-return liveness, exact-count fidelity among returned results, and overall exact-count fidelity across all attempted queries for CQ-Bench and FOEM. Values indicate audited rates for NS-CoT and FastOMOP under the GPT-4.1 evaluation setting.

This separation clarifies the nature of the observed performance differences. In CQ-Bench, the main difference between NS-CoT and FastOMOP was not simply that one system executed more often than the other. Liveness was high for both systems. The larger difference appeared after execution, in whether the returned count matched the executable reference. This indicates that the CQ-Bench performance gap primarily reflects semantic fidelity of the generated cohort logic, rather than basic runtime completion.

In FOEM, both systems reached complete liveness, and both achieved high exact-count performance. The absence of a statistically significant paired difference suggests that this external benchmark did not expose the same semantic gap observed in CQ-Bench. This pattern is consistent with the design of the two benchmarks: CQ-Bench deliberately stresses semantic decomposition and temporal/exclusion logic, whereas FOEM provides an external OMOP template-derived evaluation setting.

Taken together, the benchmark-level results show that NS-CoT's advantage is most apparent when the evaluation requires preservation of cohort-defining semantics beyond simple SQL executability. The next section examines this directly by stratifying CQ-Bench according to semantic complexity.

## 5.3. Performance by Semantic Complexity in CQ-Bench

To understand where the overall CQ-Bench performance difference came from, exact-count success was stratified by semantic complexity. CQ-Bench was designed with multiple query strata, ranging from simple condition cohorts to temporally anchored drug exposure and drug absence criteria. This stratification allows the evaluation to distinguish between failures of basic concept retrieval and failures of more complex cohort semantics.

Figure 8 shows CQ-Bench outcomes by semantic stratum for Strict $R_0$, $R_0$@5%, and zero-form joint agreement, with Strict $R_0$ used as the primary exact-count outcome. The simplest stratum, L1, consisted of single-condition cohort queries. Both NS-CoT and FastOMOP achieved exact-count success on all L1 queries. This indicates that both systems were able to handle simple condition grounding and basic condition-cohort retrieval when no additional demographic, temporal, or exclusion logic was required.

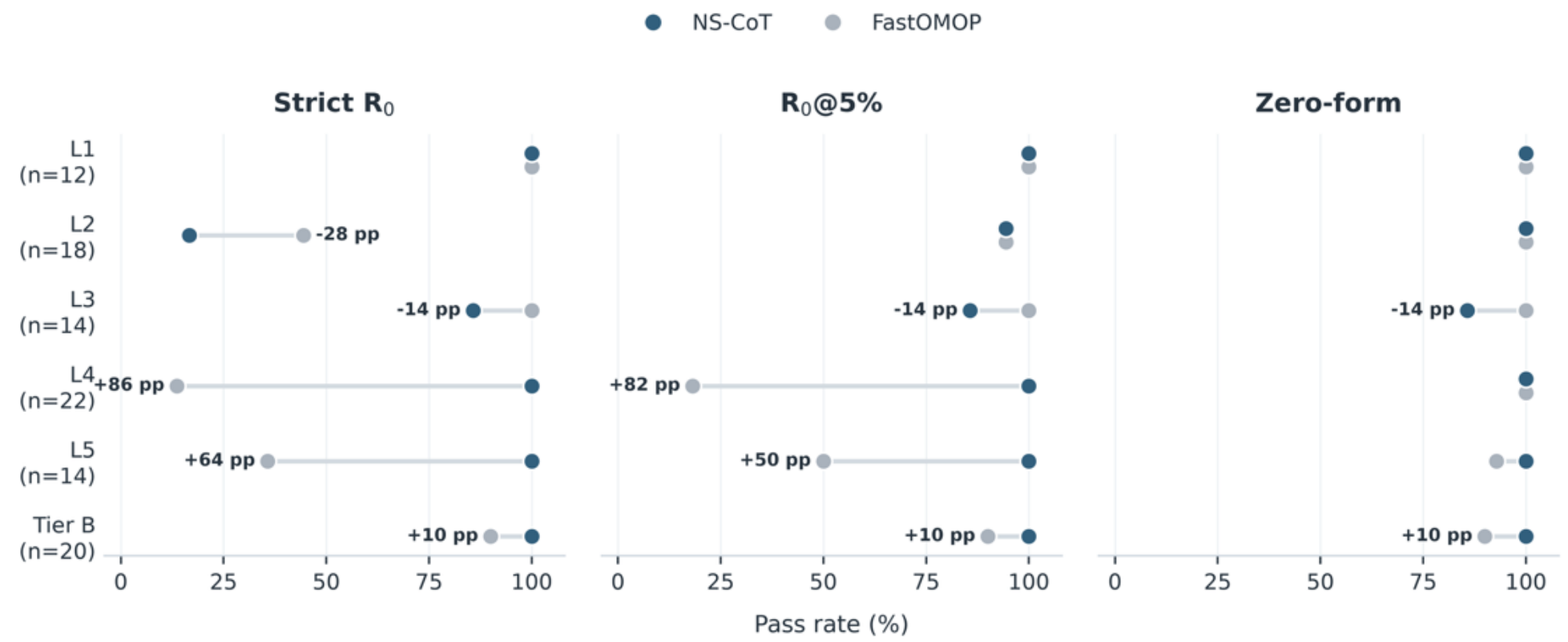

**Figure 8. CQ-Bench performance differences by semantic stratum.**
Dumbbell plots compare NS-CoT and FastOMOP pass rates within each CQ-Bench-SNUH stratum for strict $R_0$, $R_0$@5%, and zero-form joint agreement. Lines connect paired system results within the same stratum; labels indicate percentage-point differences when the absolute difference is at least 10 points.

The lower performance of NS-CoT in L2 shows that its advantage was not uniform across all query types. L2 queries required condition criteria combined with demographic filtering. In this stratum, NS-CoT achieved 3 of 18 exact matches, whereas FastOMOP achieved 8 of 18. However, the secondary metrics show that the L2 failures were not primarily zero-form failures. NS-CoT achieved $R_0$@5% on 17 of 18 L2 queries and Zero-form joint agreement on 18 of 18, while FastOMOP achieved the same $R_0$@5% and Zero-form joint results in this stratum. Thus, the L2 gap reflects strict exact-count disagreement in demographic filtering rather than failure to preserve the approximate count scale or zero versus nonzero answer form. This result indicates that the proposed architecture did not automatically improve every class of cohort query. In particular, demographic filtering errors are not the central target of the formalization described in Chapter 3, which focuses on event-time semantics, standard-vocabulary grounding, temporal anchoring, and exclusion. This result should therefore be interpreted as a limitation of the current implementation for strict demographic count alignment, rather than as a failure of liveness or zero/nonzero cohort-form preservation.

In L3, which involved condition criteria with first-diagnosis year constraints, both systems performed well. NS-CoT achieved 12 of 14 exact matches, while FastOMOP achieved 14 of 14. This suggests that first-diagnosis anchoring alone was not sufficient to create the largest difference between the systems. Although first-anchor construction is an important component of NS-CoT, the performance gap became more pronounced when the anchor had to be combined with downstream temporal drug logic.

The strongest contrast appeared in L4 and L5. L4 evaluated temporal inclusion: patients with a condition diagnosis followed by drug exposure within a specified post-diagnosis window. NS-CoT achieved 22 of 22 exact matches, whereas FastOMOP achieved 3 of 22. L5 evaluated temporal exclusion: patients with a condition diagnosis followed by anchor-bound absence of a drug or drug class after diagnosis. NS-CoT achieved 14 of 14 exact matches, whereas FastOMOP achieved 5 of 14.

These results are directly aligned with the methodological motivation of NS-CoT. L4 and L5 are the strata where natural-language criteria cannot be reduced to simple concept lookup. They require a condition concept set, a first-diagnosis anchor, a drug or drug-class concept set, an anchor-relative temporal predicate, and a patient-set operation. L4 requires a positive temporal existence predicate, whereas L5 requires anchor-bound temporal non-existence. In other words, these strata test precisely the part of the problem that Chapter 3 formalized as typed event-time predicates and patient-set algebra.

The L4 result indicates that NS-CoT preserved bounded post-diagnosis drug exposure semantics more reliably than the baseline. In CQ-Bench, the reference semantics for L4 required a qualifying drug exposure to occur on or after the first diagnosis date and no later than N days after that diagnosis. To reproduce this count, a system must not only identify the correct condition and drug concepts, but also construct the earliest diagnosis anchor and evaluate drug exposure within the correct temporal window. NS-CoT's exact success across all L4 queries suggests that the typed temporal representation and deterministic temporal lowering were effective for this supported pattern.

The L5 result is even more important for the study's central claim. Temporal exclusion is more difficult than simple negation because the absence condition is evaluated relative to an anchor. The benchmark reference semantics for L5 used an anchor-bound NOT EXISTS predicate: a patient is retained only when no qualifying drug exposure occurs on or after the first diagnosis date. NS-CoT's 14 of 14 exact matches indicate that the system consistently represented drug absence as anchor-bound temporal non-existence rather than as an unanchored global absence condition. FastOMOP's lower performance in this stratum suggests that prompt-guided SQL generation was less stable for this exclusion-heavy cohort logic.

Tier B negative controls provide a different kind of evidence. These queries were constructed to have zero reference counts because the target clinical categories were absent from the SYN-SNUH resident inventory under the benchmark construction rules. NS-CoT achieved 20 of 20 exact matches, while FastOMOP achieved 18 of 20. This result suggests that NS-CoT was able to preserve executable-empty or zero-count behavior without producing false-positive cohorts for most negative-control queries. In the context of cohort feasibility, this is important because a system should not create a nonzero cohort simply because a clinical term appears plausible in natural language. Because Tier B reference counts were zero by construction, exact-count success and Zero-form joint agreement coincide in this stratum.

The stratum-level results refine the interpretation of the overall CQ-Bench performance. NS-CoT was not uniformly better across all strata. Its strongest advantage appeared in L4 and L5, where the query semantics required explicit preservation of temporal anchoring and exclusion logic. This pattern supports the central claim of the study: the main value of NS-CoT lies not in replacing every OMOP query-generation method, but in making complex eligibility semantics explicit before SQL compilation.

Thus, the CQ-Bench results show that the performance gain was concentrated in the semantic region targeted by the proposed formalization. Simple condition queries were solved by both systems, and some demographic queries remained challenging for NS-CoT. Within the supported CQ-Bench L4 and L5 templates, however, temporal inclusion and temporal exclusion—the two strata most closely tied to the neuro-symbolic design—were reproduced with complete exact-count fidelity by NS-CoT.

## 5.4. Paired Query-Level Exact-Count Comparison for CQ-Bench and FOEM

The overall performance results in Sections 5.2 and 5.3 describe marginal exact-count rates. However, because NS-CoT and FastOMOP were evaluated on the same query IDs, the more appropriate comparison is paired at the query level. A paired analysis asks whether the two systems succeeded or failed on the same queries, and whether the discordant cases were systematically skewed toward one system. The paired comparison in this section focuses on the primary exact-count endpoint, strict $R_0$.

For each query $q$ and system $m$, a system-level exact-count success indicator was defined as:

$$S_m(q) = \mathbf{1}[L_m(q) = 1 \wedge g_m(q) = g_{\mathrm{ref}}(q)], \quad m \in \{\text{NS-CoT, FastOMOP}\}.$$

Here, $L_m(q) = 1$ indicates that system $m$ returned a final integer count, $g_m(q)$ is the returned count, and $g_{\mathrm{ref}}(q)$ is the executable reference count. Thus, $S_m(q) = 1$ only when the system both completed execution and returned the exact reference count.

For each benchmark query, the paired outcome between NS-CoT and FastOMOP falls into one of four categories: both systems exact, NS-CoT only exact, FastOMOP only exact, or neither system exact. McNemar's test was applied to the discordant pairs, that is, the queries for which only one of the two systems was exact. This test is appropriate because both systems were evaluated on the same query set, and the question is whether the one-system-only successes were balanced or significantly skewed. Figure 9 provides a visual summary of the paired query outcomes, and Table 18 reports the corresponding counts and McNemar p-values.

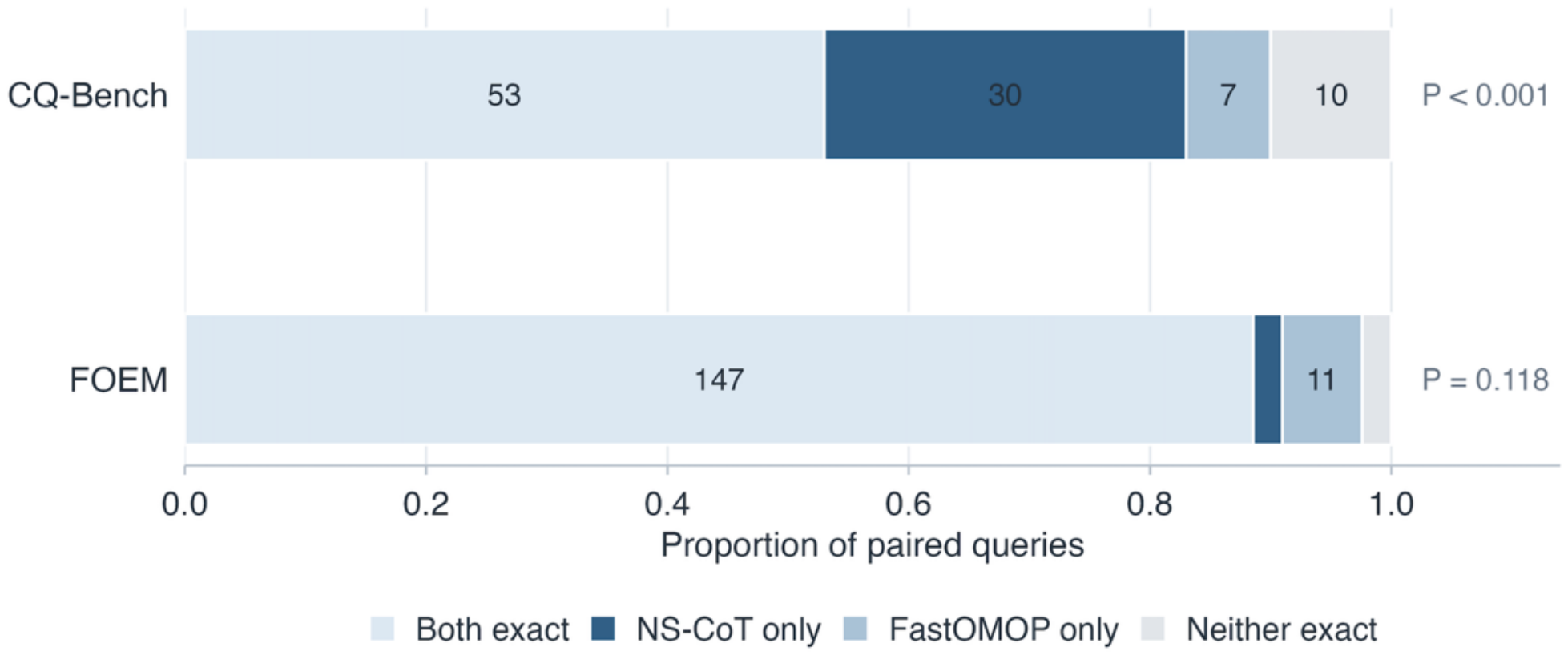

**Figure 9. Paired query outcomes behind McNemar's test.**

| Benchmark | Both exact | NS-CoT only exact | FastOMOP only exact | Neither exact | McNemar *p* |
|---|---|---|---|---|---|
| CQ-Bench | 53 | 30 | 7 | 10 | < 0.001 |
| FOEM | 147 | 4 | 11 | 4 | 0.118 |

**Table 17. Paired exact-count outcomes.**

*Note. McNemar p-values were computed using the two-sided exact binomial version of McNemar's test.*

In CQ-Bench, 53 queries were answered exactly by both systems, while 10 queries were not exact for either system. The decisive information for the paired comparison lies in the discordant cases: NS-CoT alone reproduced the reference count for 30 queries, whereas FastOMOP alone reproduced the reference count for 7 queries. This asymmetry yielded a significant McNemar result ($p < 0.001$), indicating that the CQ-Bench advantage of NS-CoT was not merely a marginal aggregate difference. It was also visible at the paired-query level.

This paired result is consistent with the stratum-level findings in Section 5.3. CQ-Bench contains query strata that explicitly evaluate temporal anchoring, post-diagnosis drug exposure, and anchor-bound drug absence. The 30-versus-7 discordance in favor of NS-CoT suggests that the typed IR, grounded concept representation, patient-set algebra, and deterministic SQL compilation were particularly helpful for queries in which cohort semantics had to be preserved across multiple transformation steps.

The FOEM paired comparison showed a different pattern. Both systems were exact on 147 of 166 queries, and neither system was exact on 4 queries. Among discordant cases, NS-CoT alone was exact on 4 queries, while FastOMOP alone was exact on 11 queries. Although the direction of the discordance favored FastOMOP numerically, the McNemar test was not statistically significant ($p = 0.118$). Therefore, the FOEM result does not support a conclusion that either system was systematically superior on this external template-derived benchmark.

The paired comparison strengthens the interpretation of the overall results. On CQ-Bench, NS-CoT's higher exact-count performance was statistically supported by the distribution of discordant query outcomes. On FOEM, the two systems were not statistically distinguishable despite a small numerical difference. This pattern reinforces the main empirical finding: NS-CoT's advantage is most evident in the benchmark setting designed to stress explicit cohort semantics, while its advantage is not universal across all OMOP query templates.

## 5.5. Traceability and Failure Localization

Exact-count evaluation treats each query as a binary exact-count success or failure. This binary scoring is necessary for benchmark comparison, but it does not explain why a mismatch occurred. One of the motivations of NS-CoT is that an incorrect count should not remain an opaque failure. Because the system preserves typed interpretation, grounding decisions, temporal predicates, generated SQL, and execution artifacts, a mismatch can be inspected to determine where the generated cohort semantics diverged from the executable reference. FOEM-129 provides a representative traceable failure case. The input query was: "How many people were treated by drug meloxicam more than 30 days after being diagnosed with condition Hyperlipidemia?"

The intended cohort semantics were to identify patients with a hyperlipidemia diagnosis anchor and then count those with meloxicam exposure more than 30 days after that diagnosis. NS-CoT completed the run successfully and returned a final count of 3,263. The corresponding executable reference count was 3,264, so the query was scored as an exact-count mismatch. The stepwise reconstruction and event-level audit are summarized in Tables 19 and 20. Figure 10 summarizes the traceable failure analysis for FOEM-129.

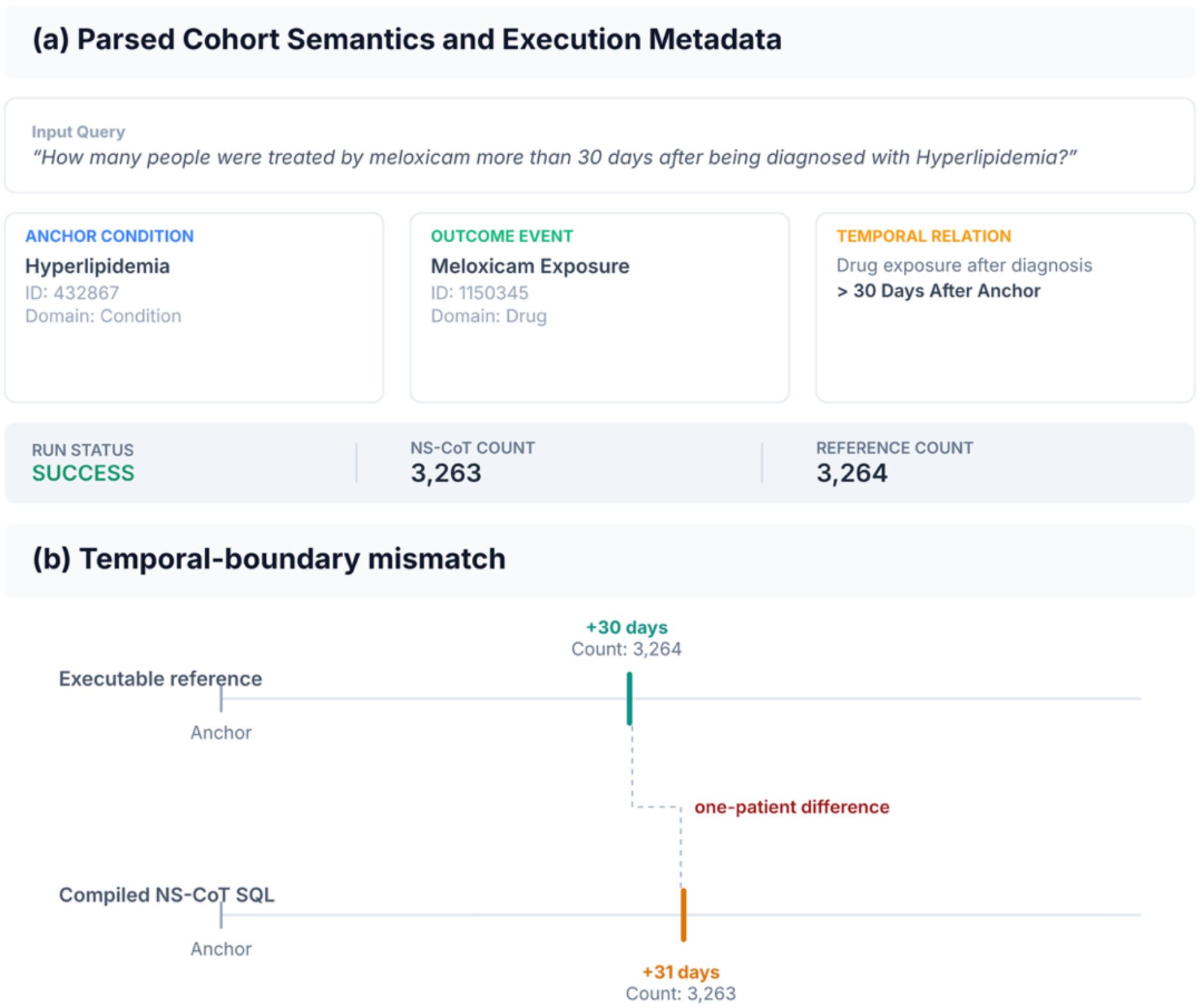

**Figure 10. Traceable failure analysis for FOEM-129.** (a) FOEM-129 was parsed as an anchor-relative drug exposure criterion, with hyperlipidemia diagnosis as the anchor and meloxicam exposure as the outcome event; (b) The exact-count mismatch was localized to temporal-boundary canonicalization: the executable reference evaluated exposure after anchor + 30 days, whereas the compiled NS-CoT SQL used a stricter anchor + 31 days predicate.

*Note. For (b), failure localized to temporal-boundary canonicalization. Concept grounding and SQL execution logic were verified as correct and are not the source of the mismatch.*

| Step | Meaning | N |
|---|---|---|
| Observed | Persons under observation | 10,000 |
| Condition set | Hyperlipidemia concept-set cohort | 6,675 |
| Anchor event | Earliest hyperlipidemia anchor | 6,675 |
| Exposure | Meloxicam after anchor | 3,385 |
| Generated Final cohort | Meloxicam >31 days after anchor | 3,263 |

**Table 18. Stepwise cohort reconstruction for FOEM-129.**

| Person ID | Anchor date | Drug exposure date | Days | Status |
|---|---|---|---|---|
| 22 | 2023-01-05 | 2023-01-21 | 16 | does not qualify |
| 22 | 2023-01-05 | 2023-06-15 | 161 | qualifies |
| 22 | 2023-01-05 | 2024-04-09 | 460 | qualifies |

**Table 19. Event-level audit example for FOEM-129.**

The event-level audit in Table 20 is not intended to identify the single boundary-flip patient responsible for the one-count difference. Rather, it illustrates how event-level records are aggregated into a distinct-person cohort count and why multiple qualifying drug exposure events for the same person contribute only one patient to the final count. The trace shows that this was not an architectural liveness failure. The run completed all major phases: strategy planning, query decomposition, clinical step construction, SQL compilation, and SQL execution. The root run was marked as successful, with a returned cohort count of 3,263 and two executed cohort steps. The trace also records the original natural-language query, the model configuration, the execution status, elapsed time, and the generated output. This confirms that the pipeline was live and that the mismatch arose within the semantics of the generated cohort logic rather than from failure to return a count.

The strategy and decomposition trace show that the system recognized the query as an anchor-relative temporal criterion. Hyperlipidemia was interpreted as a Condition-domain anchor, and meloxicam was interpreted as a Drug-domain outcome. The temporal expression “more than 30 days after diagnosis” was represented as a post-diagnosis drug exposure relation rather than as an unanchored drug criterion. This is the correct structural interpretation for the query: the drug exposure is not evaluated globally, but relative to a diagnosis anchor.

The concept-resolution trace also supports the conclusion that the mismatch was not primarily a grounding error. The event-level validation confirmed that the two key concepts were valid standard OMOP concepts: hyperlipidemia was represented by concept ID 432867 in the Condition domain, and meloxicam was represented by concept ID 1150345 in the Drug domain. The semantic gate selected meloxicam as the appropriate RxNorm ingredient concept and rejected narrower formulations or strength-specific candidates as too specific for the intended query. Thus, the clinical concept identity and OMOP domain routing were consistent with the intended criterion.

The generated SQL then constructed the cohort in the expected broad structure. It initialized the observed population from observation_period, selected patients with qualifying hyperlipidemia condition concepts, constructed a first-diagnosis anchor for concept ID 432867, and then selected patients with meloxicam exposure using drug concept ID 1150345. However, the temporal predicate in the compiled SQL used the following lower-bound condition: drug_exposure_start_date > anchor_date + INTERVAL '31 day'.

This predicate was stricter than the executable reference semantics and effectively shifted the lower bound by one additional day. Therefore, one patient who satisfied the reference temporal boundary was excluded from the NS-CoT-generated cohort, explaining the one-patient difference between the reference count of 3,264 and the NS-CoT count of 3,263.

The stepwise reconstruction confirmed the internal consistency of the generated SQL. The observed population contained 10,000 persons. The hyperlipidemia condition step yielded 6,675 persons, and the first-anchor reconstruction also yielded 6,675 persons. Among these patients, 3,385 had meloxicam exposure after the anchor, while 3,263 had meloxicam exposure after the stricter anchor_date + 31 days boundary. This stepwise result shows that the generated SQL executed coherently and that the final count followed from the compiled temporal predicate.

The event-level audit further supports the same interpretation. The validation query returned no invalid patients in the compiled cohort: every patient counted by the generated query had both a qualifying anchor and a qualifying meloxicam exposure satisfying the compiled temporal predicate. In other words, the 3,263 returned patients were not spurious. The issue was not that NS-CoT included patients who lacked the required condition or drug event. The issue was that the compiled temporal boundary was one day stricter than the reference boundary.

A single-patient audit illustrates the distinction between event-level validation and person-level cohort membership. For person ID 22, the hyperlipidemia anchor date was 2023-01-05. One meloxicam exposure occurred 16 days after the anchor and did not satisfy the compiled temporal predicate. Two later meloxicam exposures occurred 161 and 460 days after the anchor and satisfied the compiled predicate. The patient therefore contributes one distinct person to the final cohort count, not two drug-exposure events.

The FOEM-129 case therefore illustrates the practical value of traceability. Under ordinary exact-count scoring, this query is simply wrong. Under a traceable cohort-computation framework, the error can be localized. The decomposition was structurally appropriate, the concept grounding was appropriate, the SQL executed successfully, and the counted patients satisfied the generated SQL. The mismatch arose from temporal-boundary canonicalization: the interpretation of “more than 30 days after diagnosis” as a stricter lower-bound predicate than the executable reference.

This distinction matters for clinical research informatics. A one-patient mismatch could be caused by many different problems: wrong concept selection, missing descendant expansion, incorrect domain routing, wrong anchor definition, global rather than anchor-bound temporal logic, SQL execution failure, or boundary convention mismatch. FOEM-129 shows that NS-CoT can expose which of these layers caused the discrepancy. The error was not hidden inside an opaque SQL string; it was visible in the temporal predicate and could be validated against event-level data.

For this reason, FOEM-129 is best interpreted as a traceable failure rather than merely a failed query. It shows that the system's value is not limited to producing exact counts. It also lies in making count mismatches diagnostically useful. In feasibility workflows, this is important because cohort counts often inform protocol refinement and site feasibility decisions. A count that cannot be explained is difficult to trust; a count whose construction can be reconstructed can be corrected, reviewed, and reused.

## 5.6. Person-set Verification Across Executable-Reference Benchmarks

To examine whether count-level agreement corresponded to agreement at the patient-set level, a deterministic person-set verification was performed. For each evaluable query, the saved reference SQL and the saved NS-CoT SQL were re-executed as SELECT DISTINCT person_id, and the resulting patient sets were compared directly. No additional LLM interpretation or regeneration step was used in this verification. This verification does not constitute chart-reviewed phenotype validation or clinical eligibility adjudication. It evaluates whether the executable reference SQL and the saved NS-CoT SQL identify the same person_id set under the same OMOP CDM instance.

For each query, person-set similarity was summarized using exact set agreement and Jaccard similarity. Let $P_{\text{ref}}(q)$ denote the person_id set returned by the executable reference SQL and $P_{\text{NS}}(q)$ denote the person_id set returned by the saved NS-CoT SQL. Jaccard similarity was defined as $J(q) = \frac{|P_{\text{ref}}(q) \cap P_{\text{NS}}(q)|}{|P_{\text{ref}}(q) \cup P_{\text{NS}}(q)|}$, with exact set agreement defined as $P_{\text{ref}}(q) = P_{\text{NS}}(q)$. When both sets were empty, Jaccard similarity was treated as 1.0.

Person-set verification was performed for 99 CQ-Bench queries with saved executable NS-CoT SQL and for 166 scalar-count FOEM queries. Because one CQ-Bench query did not produce saved executable NS-CoT SQL, person-set verification used 99 executable CQ-Bench queries rather than the full set of 100 benchmark queries. The 35 FOEM group-by queries were also excluded because they require group-specific set comparison rather than single-cohort person-set comparison.

In CQ-Bench, 83 of 99 comparisons showed exact person_id set agreement, corresponding to an exact set-match rate of 83.8% with a 95% confidence interval of 75.3%-89.8%. The mean and median Jaccard similarities were 0.987 and 1.000, respectively. In FOEM scalar queries, 151 of 166 comparisons showed exact person_id set agreement, corresponding to an exact set-match rate of 91.0% with a 95% confidence interval of 85.6%-94.4%. The mean and median Jaccard similarities were 0.990 and 1.000, respectively. Equivalently, 16 of 99 CQ-Bench comparisons and 15 of 166 FOEM scalar comparisons showed person-set mismatches. No evaluated query in either benchmark showed exact count agreement with a different person_id set.

Here, false positives and false negatives refer to person_id differences relative to the executable reference set, not clinical false-positive or false-negative phenotype classifications. Aggregate false-positive and false-negative counts were 301 and 2,192 query-summed persons in CQ-Bench, and 12,826 and 7 query-summed persons in FOEM, respectively. These aggregate counts are query-summed error totals, not counts of unique persons across the full benchmark. The FOEM false-positive total was dominated by FOEM_165, where the reference cohort was empty but the system returned all 10,000 persons. This outlier explains why the FOEM aggregate false-positive total was large despite a high median Jaccard similarity. These results support the interpretation that exact-count agreement in this evaluation was not produced by compensating differences between distinct patient sets. They also show that saved SQL artifacts can be re-executed to localize patient-level differences when count mismatches occur.

| Benchmark | Scope | Exact set match | Exact rate, 95% CI | Jaccard mean/median | Count match but person-set mismatch |
|---|---|---|---|---|---|
| CQ-Bench | 99 executable queries | 83 / 99 | 83.8%, 75.3%-89.8% | 0.987 / 1.000 | 0 |
| FOEM | 166 scalar queries | 151 / 166 | 91.0%, 85.6%-94.4% | 0.990 / 1.000 | 0 |

**Table 20. Person-set verification across executable-reference benchmarks.**

*Note. Aggregate FP/FN were 301/2,192 query-summed persons for CQ-Bench and 12,826/7 query-summed persons for FOEM. One non-live CQ-Bench query was excluded from person-set comparison. FOEM group-by queries were excluded because they require group-specific set*

*comparison. FOEM aggregate false positives were dominated by FOEM_165, where the reference cohort was 0 and the system returned 10,000 persons.*

## 5.7. COVIC Multi-Site Feasibility Evaluation

The COVIC analysis evaluated the systems in an external multi-site feasibility simulation. Unlike CQ-Bench, which was designed to test targeted semantic components such as temporal anchoring and anchor-bound exclusion, and unlike FOEM, which evaluated external template-derived OMOP query reproduction, COVIC asked a site-feasibility question: when the same executable criterion component is applied across multiple OMOP schemas, does the system-generated count preserve the reference-defined count, patient set, site-feasibility label, and relative site ranking? The analysis was therefore organized around criterion-site pair fidelity, patient-set preservation, and site-level feasibility decision preservation.

The first-pass COVIC comparison showed a different pattern from CQ-Bench. FastOMOP achieved significantly higher live execution, strict exact-count fidelity, $R_0$@5%, and person-set exactness under paired McNemar tests. FastOMOP also achieved significantly higher Jaccard similarity, precision, and recall in the all-attempted patient-set analysis. Site-feasibility discrimination, measured by AUROC and AUPRC across reference-count thresholds, also favored FastOMOP. Top-site agreement was numerically higher for FastOMOP but was not statistically significant.

A second-pass NS-CoT repair analysis was conducted after deterministic correction of local-resident drug filtering, partial OR execution, and typed-plan replay or fresh-fallback bookkeeping. This repair analysis was designed to evaluate whether the explicit NS-CoT pipeline could turn first-pass COVIC failures into localizable and correctable implementation objects. The repair targeted general execution-policy issues identified during error inspection, including local-resident drug filtering, partial OR execution, and typed-plan replay or fresh-fallback bookkeeping. FastOMOP was not rerun; the previously captured FastOMOP artifacts were reused as the baseline because no FastOMOP logic was changed.

The COVIC analysis should be interpreted as a complementary multi-site feasibility evaluation rather than as a single-database scalar-count benchmark. Its operational target is different from CQ-Bench and FOEM: it asks whether system-generated count preserves reference-defined feasibility labels and site rankings when the same executable criterion component is applied across candidate OMOP schemas.

### 5.7.1. First-pass criterion-site fidelity and feasibility-decision preservation

The first-pass COVIC analysis evaluated the original NS-CoT implementation against FastOMOP in the external multi-site feasibility setting defined in Section 4.6. The paired unit was a criterion-site pair, and the analysis included 195 paired observations derived from 65 COVIC-derived executable criterion components evaluated across three OMOP schemas. This setting emphasized raw criterion-unit handling, broad OMOP vocabulary grounding, executable concept-set reproduction, multi-entity drug exclusions, and multi-schema site-level count comparison. Therefore, the first-pass analysis should be read as an evaluation of whether the initial system implementation preserved reference-defined count and site-feasibility information across candidate sites.

The first-pass results favored FastOMOP across all binary all-attempted outcomes. FastOMOP returned a live executable result for 182 of 195 criterion-site pairs, whereas NS-CoT returned a live result for 129 of 195 pairs. This corresponded to live execution rates of 93.3% and 66.2%, respectively. The paired discordance was also strongly asymmetric: FastOMOP succeeded where NS-CoT did not in 58 criterion-site pairs, whereas NS-CoT succeeded where FastOMOP did not in 5 pairs. The paired McNemar test was significant, indicating substantially greater execution robustness for FastOMOP in this external multi-site feasibility setting. Figure 11 summarizes the first-pass all-attempted count and patient-set fidelity pattern, and Table 22 reports the corresponding binary paired outcomes for live execution, count fidelity, and person-set exactness.

| Metric | NS-CoT | FastOMOP | Diff | 95% CI | p |
|---|---|---|---|---|---|
| Live | 129/195 (66.2%) | 182/195 (93.3%) | 27.2% | 20.0% to 34.4% | $<0.001$ |
| Strict $R_0$ | 61/195 (31.3%) | 109/195 (55.9%) | 24.6% | 17.9% to 31.3% | $<0.001$ |
| $R_0$@5% | 67/195 (34.4%) | 111/195 (56.9%) | 22.6% | 15.9% to 29.2% | $<0.001$ |
| Person-set exact | 61/195 (31.3%) | 108/195 (55.4%) | 24.1% | 17.4% to 30.8% | $<0.001$ |

**Table 21. First-pass binary paired outcomes for live execution, count fidelity, and person-set exactness in COVIC.**

The same pattern was observed for count-level fidelity. Strict $R_0$ was 55.9% for FastOMOP and 31.3% for NS-CoT. $R_0$@5% was 56.9% for FastOMOP and 34.4% for NS-CoT. Both paired count-fidelity comparisons favored FastOMOP with exact McNemar tests. Thus, the gap was not limited to execution failure. Even when approximate count agreement was considered, the first-pass NS-CoT implementation showed lower fidelity than FastOMOP in the COVIC multi-site feasibility simulation.

Patient-set fidelity showed the same direction. FastOMOP achieved exact person-id set agreement in 108 of 195 criterion-site pairs, whereas NS-CoT achieved exact person-id set agreement in 61 of 195 pairs. Mean Jaccard similarity in the all-attempted analysis was 0.646 for FastOMOP and 0.397 for NS-CoT. Precision and recall also favored FastOMOP in the all-attempted analysis. In the both-evaluable sensitivity analysis, FastOMOP remained higher in Jaccard similarity and precision, while recall was numerically higher for FastOMOP but was not statistically significant. Detailed Jaccard, precision, and recall results under all-attempted and both-evaluable analyses are reported in Table 23.

Figure 12 shows that strict-alignment fidelity varied across OMOP schemas, with the public schema remaining the most difficult setting for exact patient-set reproduction. This supports the interpretation that COVIC is a model-relative multi-schema feasibility evaluation rather than a single-database count benchmark.

| Metric | Analysis | n | NS-CoT | FastOMOP | Diff | 95% CI | p |
|---|---|---|---|---|---|---|---|
| Jaccard | All attempted | 195 | 0.397 | 0.646 | 0.249 | 0.180 to 0.319 | <0.001 |
| | Both evaluable | 123 | 0.612 | 0.712 | 0.100 | 0.029 to 0.175 | 0.019 |
| Precision | All attempted | 195 | 0.440 | 0.695 | 0.255 | 0.181 to 0.330 | <0.001 |
| | Both evaluable | 123 | 0.680 | 0.776 | 0.096 | 0.014 to 0.177 | 0.011 |
| Recall | All attempted | 195 | 0.465 | 0.718 | 0.254 | 0.180 to 0.327 | <0.001 |
| | Both evaluable | 123 | 0.715 | 0.791 | 0.076 | 0.000 to 0.154 | 0.311 |

**Table 22. First-pass patient-set overlap outcomes in COVIC.**

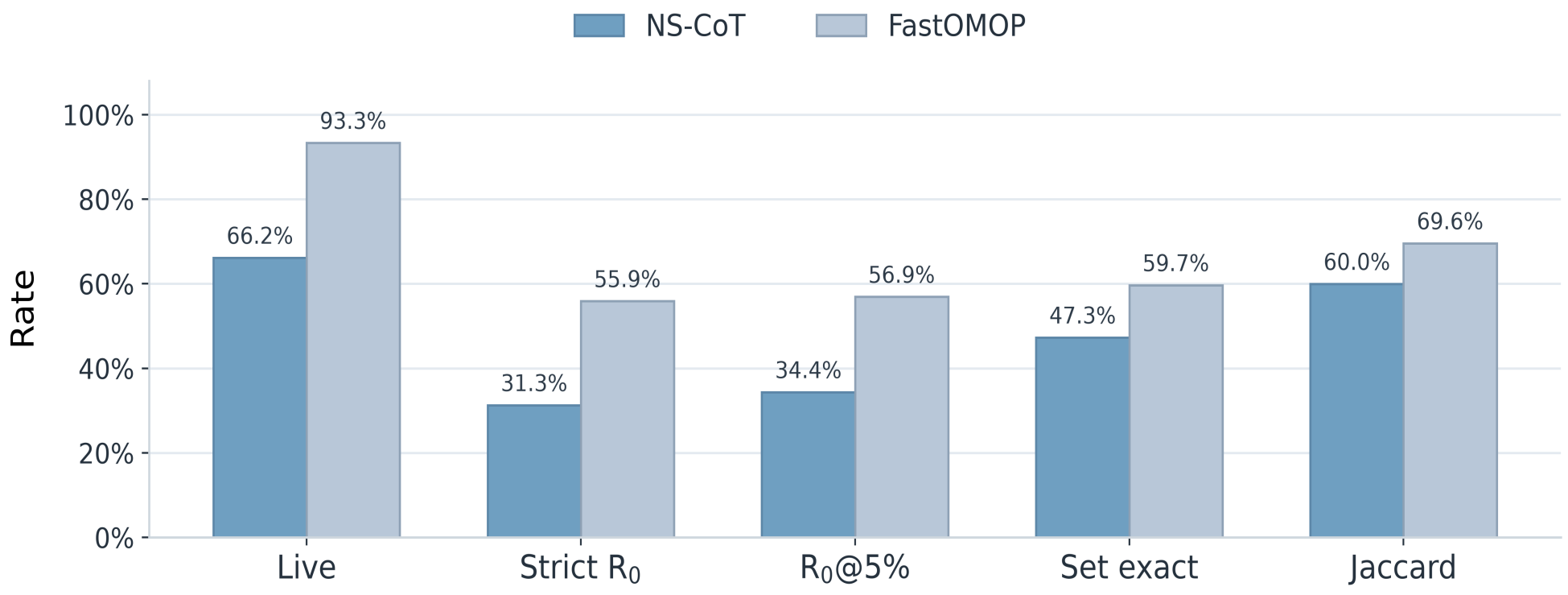

**Figure 11. First-pass COVIC count and patient-set fidelity.** Bars compare first-pass NS-CoT and FastOMOP on live execution, strict $R_0$, $R_0$@5%, person-set exactness, and mean Jaccard similarity under all-attempted scoring.

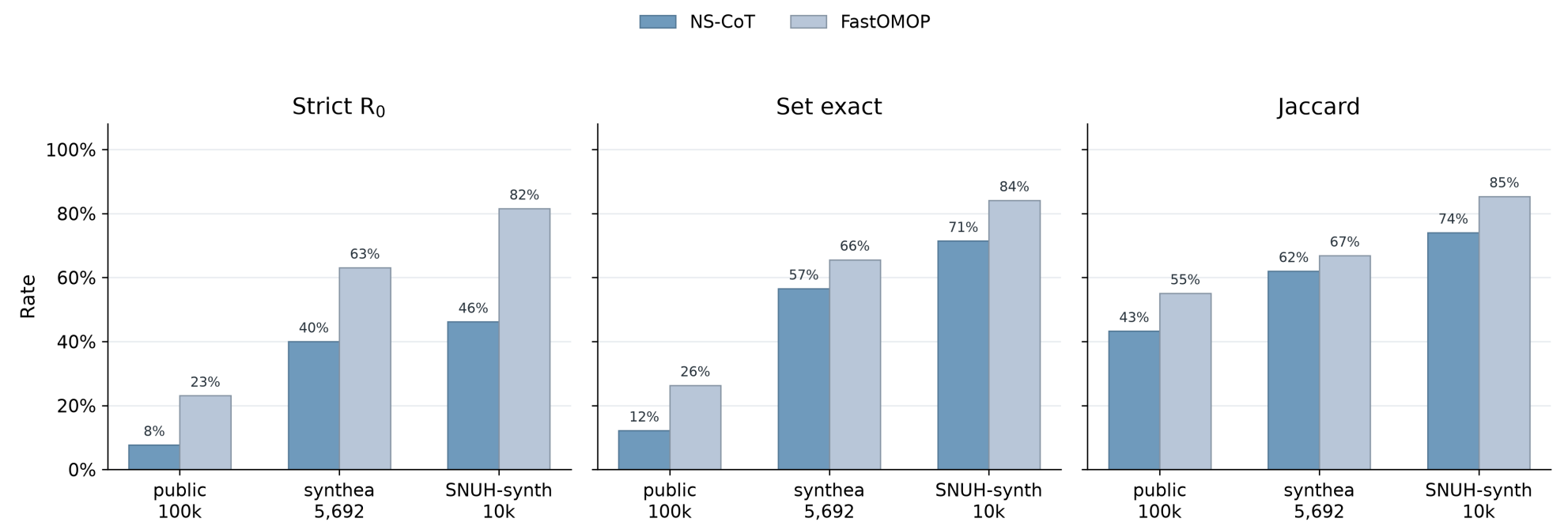

**Figure 12. First-pass COVIC fidelity by OMOP schema.** Bars compare first-pass NS-CoT and FastOMOP across evaluated OMOP schemas. The figure illustrates that strict-alignment fidelity varied by schema, reflecting the model-relative nature of site-level cohort execution.

Site-feasibility discrimination also favored FastOMOP. Across reference count thresholds $\tau$ = 1,5,10,30, and 100, FastOMOP achieved higher AUROC and AUPRC than NS-CoT. The paired cluster bootstrap confidence intervals for AUROC differences remained above zero across all tested thresholds. This indicates that, in this COVIC setting, FastOMOP better preserved the ability to distinguish criterion-site pairs with larger versus smaller reference-defined feasible patient pools. Top-site agreement showed a different pattern: FastOMOP was numerically higher than NS-CoT, but the paired difference was not statistically significant. Therefore, the strongest statistical evidence in the first-pass comparison was observed at the row-level fidelity and discrimination levels, rather than at the top-site ranking level alone. Table 24 reports the AUROC and AUPRC results for threshold-based site-feasibility discrimination. First-pass top-site agreement and Spearman site-ranking summaries are reported together with the repaired comparison in Table 26, Panel B.

| tau | NS AUROC | Fast AUROC | Diff | AUROC CI | p | NS AUPRC | Fast AUPRC | AUPRC diff |
|---|---|---|---|---|---|---|---|---|
| 1 | 0.703 | 0.848 | 0.145 | 0.073 to 0.217 | <0.001 | 0.727 | 0.884 | 0.156 |
| 5 | 0.680 | 0.843 | 0.163 | 0.081 to 0.244 | <0.001 | 0.688 | 0.872 | 0.184 |
| 10 | 0.684 | 0.841 | 0.157 | 0.079 to 0.240 | <0.001 | 0.690 | 0.870 | 0.180 |
| 30 | 0.677 | 0.837 | 0.160 | 0.077 to 0.243 | <0.001 | 0.681 | 0.865 | 0.184 |
| 100 | 0.676 | 0.852 | 0.176 | 0.093 to 0.259 | <0.001 | 0.661 | 0.871 | 0.210 |

**Table 23. First-pass preservation of reference-defined site-feasibility labels in COVIC.**

These results should be interpreted directly within the COVIC evaluation target. FastOMOP was more robust than the original NS-CoT implementation for first-pass execution, count reproduction, patient-set reproduction, and threshold-based site-feasibility discrimination in this external multi-site setting. This is an important negative finding for the initial NS-CoT implementation. At the same time, the result should not be interpreted as a rejection of the patient-set semantics framework. Rather, it shows that applying externally derived criterion components across multiple OMOP schemas introduces additional implementation challenges, including raw criterion-unit normalization, broad vocabulary grounding, executable concept-set reproduction, multi-entity drug exclusions, and site-specific local executability. These are precisely the types of challenges that a multi-site feasibility simulation is intended to expose.

### 5.7.2. Second-pass repair and recovery of count, patient-set, and site-decision fidelity

A second-pass NS-CoT repair analysis was conducted to determine whether the first-pass COVIC failures could be partially localized and corrected within the explicit pipeline. The repair targeted deterministic implementation issues identified during error inspection, including local-resident drug filtering, partial OR execution, and typed-plan replay or fresh-fallback bookkeeping. This repair analysis evaluated whether explicit NS-CoT artifacts could support error localization, implementation-level correction, and re-evaluation within the same COVIC multi-site feasibility setting. FastOMOP was not rerun; the previously captured FastOMOP artifacts were reused as the baseline because no FastOMOP logic was changed.

The repair substantially improved NS-CoT across all major COVIC outcomes. Live execution increased from 66.2% in the first pass to 78.5% in the second pass. Strict $R_0$ increased from 31.3% to 45.6%, $R_0$@5% from 34.4% to 48.7%, zero-form agreement from 23.1% to 33.8%, and all-attempted person-set exactness from 31.3% to 45.6%. Mean all-attempted Jaccard similarity increased from 0.397 to 0.558, and mean both-evaluable Jaccard similarity increased from 0.600 to 0.711. Figure 13 summarizes the first-pass, second-pass, and FastOMOP outcome pattern.

The paired first-pass versus second-pass comparisons confirmed that these improvements were not limited to a single metric. The second pass improved live execution, strict $R_0$, $R_0$@5%, zero-form agreement, person-set exactness, Jaccard similarity, precision, and recall. Exact McNemar tests favored the second pass for binary outcomes, and paired Wilcoxon signed-rank tests favored the second pass for continuous patient-set overlap metrics. Table 25 reports the first-pass versus second-pass repair effect.

Despite this improvement, the repaired NS-CoT implementation remained lower than FastOMOP on most all-attempted fidelity metrics. Compared with FastOMOP, second-pass NS-CoT still had lower live execution, strict $R_0$, $R_0$@5%,

zero-form agreement, and person-set exactness. FastOMOP also remained higher on all-attempted Jaccard similarity, precision, and recall. These differences remained statistically significant in the paired tests. Table 26, Panel A reports the paired comparison between repaired NS-CoT and FastOMOP.

Site-feasibility discrimination also improved after the second-pass repair. Across reference-count thresholds, NS-CoT second pass achieved higher AUROC and AUPRC than NS-CoT first pass, indicating better preservation of reference-defined feasibility ranking information after deterministic repair. However, FastOMOP remained higher than repaired NS-CoT across the evaluated discrimination thresholds. Figure 14 visualizes this threshold-dependent discrimination pattern.

The repair analysis also revealed a difference between row-level fidelity and site-ranking fidelity. Although repaired NS-CoT remained lower than FastOMOP on most row-level fidelity metrics, its site-ranking fidelity improved. Top-site agreement increased from 90.2% in the first pass to 95.1% in the second pass, and mean Spearman correlation increased from 0.699 to 0.871. In this run, repaired NS-CoT was numerically higher than FastOMOP for top-site agreement and mean Spearman correlation. Table 26, Panel B reports the site-ranking fidelity results. This finding should be interpreted cautiously because the strongest statistical evidence remained at the row-level fidelity metrics.

Taken together, the second-pass COVIC analysis has a repairability and implementation-diagnostic role. It shows that the current NS-CoT implementation has robustness gaps in COVIC-style raw criterion-line settings, but it also shows that explicit intermediate artifacts can make some of those gaps identifiable and partially repairable. The result therefore supports a narrower conclusion: the formal patient-set semantics framework remains methodologically useful, but the current implementation requires stronger vocabulary grounding, local executability handling, and criterion-unit alignment before it can be competitive in broad COVIC-style external multi-site feasibility settings.

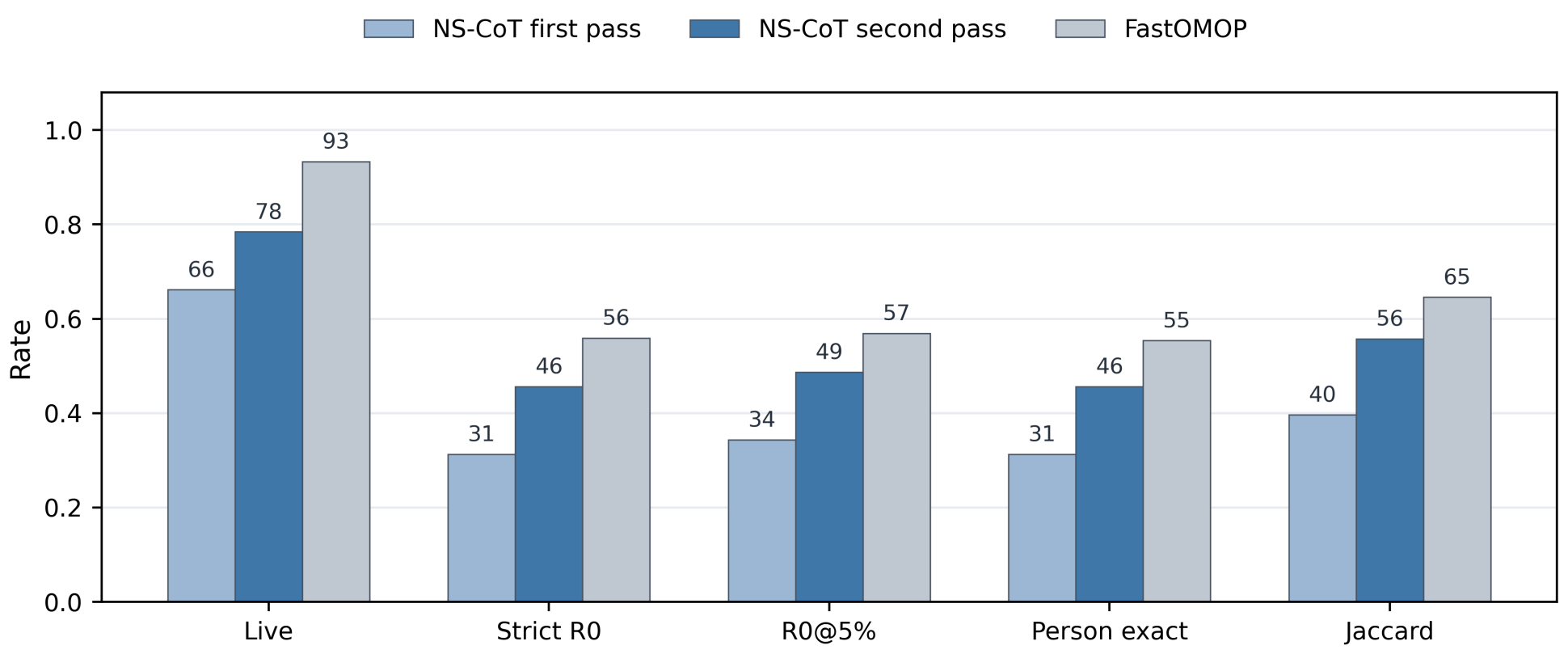

**Figure 13. Overall COVIC outcomes after NS-CoT second-pass repair.** Bars compare NS-CoT first pass, NS-CoT second pass, and FastOMOP on live execution, strict $R_0$, $R_0$@5%, zero-form agreement, person-set exactness, and mean Jaccard similarity.

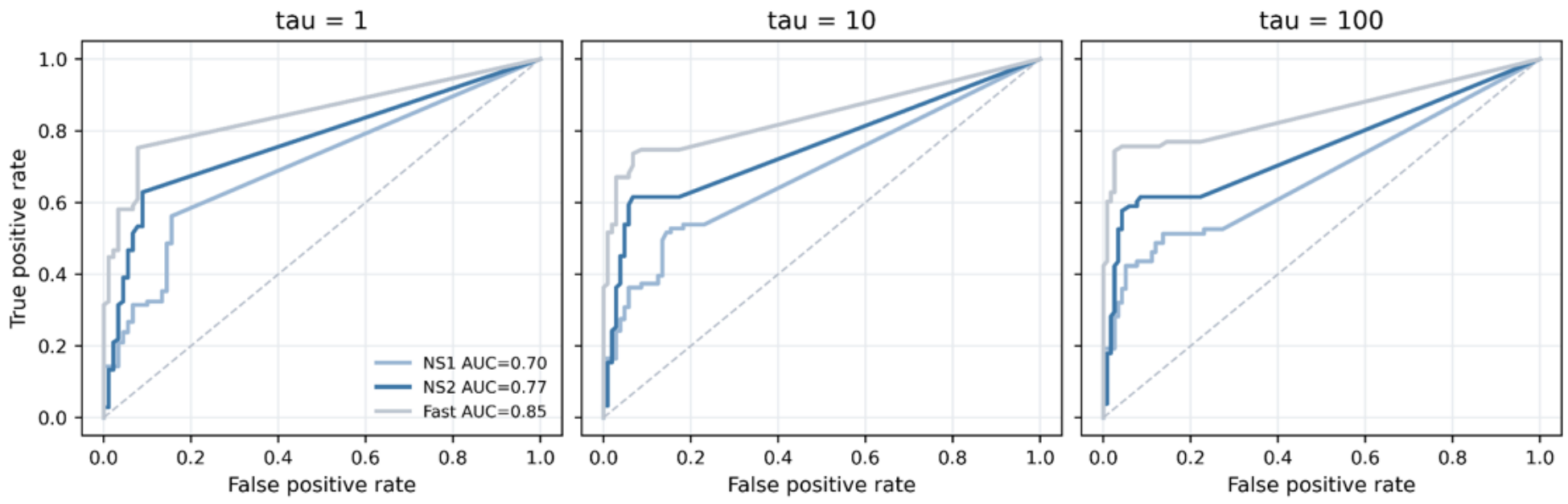

**Figure 14. COVIC site-feasibility discrimination across reference-count thresholds.** Curves compare AUROC and AUPRC for NS-CoT first pass, NS-CoT second pass, and FastOMOP across reference-count thresholds used to define feasible criterion-site pairs. These metrics evaluate whether system-generated count preserves reference-defined site-feasibility labels across increasingly stringent operational thresholds.

| Metric | Test | n | NS-CoT 1st | NS-CoT 2nd | 1st only | 2nd only | p |
|---|---|---|---|---|---|---|---|
| Liveness | McNemar exact | 195 | 129 | 153 | 15 | 39 | 0.001 |
| Strict $R_0$ | | 195 | 61 | 89 | 4 | 32 | <0.001 |
| $R_0$@5% | | 195 | 67 | 95 | 5 | 33 | <0.001 |
| Zero-form | | 195 | 45 | 66 | 3 | 24 | <0.001 |
| Person-set exact | | 195 | 61 | 89 | 4 | 32 | <0.001 |
| Jaccard | Wilcoxon signed-rank | 195 | 0.397 | 0.558 | | | <0.001 |
| Precision | | 195 | 0.440 | 0.609 | | | <0.001 |
| Recall | | 195 | 0.465 | 0.608 | | | <0.001 |

**Table 24. Repair effect: NS-CoT first pass vs second pass.** This table reports paired first-pass versus second-pass NS-CoT comparisons for live execution, strict $R_0$, $R_0$@5%, zero-form agreement, person-set exactness, and patient-set overlap metrics.

**Panel A. Repaired NS-CoT versus FastOMOP paired fidelity outcomes.**

| Metric | Test | n | NS-CoT 2nd | FastOMOP | NS only | Fast only | p |
|---|---|---|---|---|---|---|---|
| Liveness | McNemar exact | 195 | 153 | 182 | 6 | 35 | <0.001 |
| Strict $R_0$ | | 195 | 89 | 109 | 10 | 30 | 0.002 |
| $R_0$@5% | | 195 | 95 | 111 | 11 | 27 | 0.014 |
| Zero-form | | 195 | 66 | 77 | 5 | 16 | 0.027 |
| Person-set exact | | 195 | 89 | 108 | 11 | 30 | 0.004 |
| Jaccard | Wilcoxon signed-rank | 195 | 0.558 | 0.646 | | | 0.005 |
| Precision | | 195 | 0.609 | 0.695 | | | 0.006 |
| Recall | | 195 | 0.608 | 0.718 | | | 0.002 |

**Panel B. Site-ranking fidelity.**

| System | Query n | Applicable n | Top-site agreement | Mean Spearman | Spearman n |
|---|---|---|---|---|---|
| NS-CoT 1st | 65 | 61 | 90.2% | 0.699 | 40 |
| NS-CoT 2nd | 65 | 61 | 95.1% | 0.871 | 44 |
| FastOMOP | 65 | 61 | 93.4% | 0.855 | 56 |

**Table 25. Repaired NS-CoT versus FastOMOP: criterion-site fidelity and site-ranking preservation in COVIC.** Panel A reports paired all-attempted fidelity comparisons between second-pass NS-CoT and FastOMOP. Panel B reports site-ranking fidelity for NS-CoT first pass, NS-CoT second pass, and FastOMOP.

### 5.7.3. Row-level fidelity versus site-level decision preservation

The COVIC analysis adds a multi-site feasibility perspective to the study's evaluation. It should be interpreted differently from CQ-Bench and FOEM because its main operational question is different. CQ-Bench asks whether targeted semantic components are preserved. FOEM asks whether scalar OMOP query counts and patient sets can be reproduced in an external template-derived benchmark. COVIC asks whether the same externally derived executable criterion component, when applied across multiple OMOP schemas, yields system-generated count values that preserve the reference-defined site-feasibility information.

The first-pass results showed that FastOMOP was more robust than the original NS-CoT implementation at the criterion-site pair level. FastOMOP achieved stronger live execution, scalar count fidelity, patient-set exactness, patient-set overlap, and site-feasibility discrimination. This finding should be reported directly: the initial NS-CoT implementation was not sufficiently robust for broad COVIC-style multi-schema execution.

The second-pass repair analysis substantially improved NS-CoT across execution, count, zero-form, patient-set, and site-ranking outcomes. The repair did not reverse the main row-level performance conclusion, because FastOMOP remained higher on most all-attempted scalar and patient-set fidelity metrics. However, the repair showed that several COVIC failures could be localized to specific implementation layers and partially corrected through general pipeline changes rather than criterion-specific hardcoding.

The most important interpretation is that row-level fidelity and site-level decision preservation are related but not identical. Strict count fidelity, $R_0$@5%, person-set exactness, Jaccard similarity, precision, and recall evaluate whether the system reproduced the reference result for each criterion-site pair. Site-feasibility discrimination and site-ranking metrics ask a different operational question: whether the system preserves which sites appear feasible, or relatively more feasible, under reference-defined count. In clinical trial feasibility planning, both forms of fidelity matter. COVIC therefore strengthens the study by showing how cohort-generation systems can be evaluated not only as query generators, but also as tools for preserving multi-site feasibility decisions.

## 5.8. Traceability Analysis of Protocol-Derived Clinical Trial Criteria

The three protocol-derived clinical trial case studies were evaluated as traceability demonstrations rather than as scored statistical benchmarks or superiority tests. Case 1 examined a feasibility abstraction derived from a blood-pressure intervention trial. The case asked how many adults aged 18 years or older had a systolic blood pressure measurement between 140 and 159 mmHg, no history of diabetes mellitus, and no antihypertensive medication exposure. NS-CoT decomposed the query into four execution-relevant components: adult age, systolic blood pressure measurement, diabetes mellitus exclusion, and antihypertensive drug-class exclusion. The source criterion contained latest-measurement semantics, but this traceability operationalized the computable component as the presence of a qualifying SBP measurement; full latest-per-person selection was treated as an implementation extension.

The stepwise trace showed progressive cohort narrowing. The observed adult population contained 10,000 persons. Applying the systolic blood pressure value-and-unit criterion reduced the cohort to 3,792 persons. Excluding patients with diabetes mellitus reduced the cohort to 1,734 persons. Excluding patients with antihypertensive drug exposure reduced the final cohort to 114 persons. The measurement concept was grounded to systolic blood pressure, with value range 140–159 and unit handling through unit-source fallback. The antihypertensive drug-class criterion was resolved through the vocabulary graph to four locally executable drug concepts in the target drug_exposure table.

This case illustrates that the framework can represent value-bearing measurement criteria and exclusion semantics as traceable cohort operations. At the same time, it should be interpreted cautiously: the phrase “latest systolic blood pressure” was captured in the semantic interpretation, but full latest-per-person lowering was not part of the evaluated core benchmark functionality and should be treated as an implementation extension rather than a validated component of the core evaluation.

Case 2 examined a liver-transplant trial criterion involving intraoperative temporal meaning: liver recipients with blood glucose greater than 180 mg/dL after reperfusion of the liver graft, with pediatric patients excluded. This case was included not merely as an example of local data unavailability, but as an example of why a feasibility criterion must be externalized as an explicit executable semantic commitment before it can be counted. The phrase “after reperfusion of the liver graft” does not map automatically to a single OMOP CDM event. Different reviewers could reasonably choose different operational anchors, such as a directly recorded reperfusion timestamp if available, the liver-transplantation procedure date, or a visit-bounded perioperative window. These alternatives would produce different patient sets, and none should be treated as an implicit gold standard simply because it can be expressed in SQL.

NS-CoT decomposed the criterion into an adult-age component, a liver-transplantation context, a procedure-linked temporal anchor corresponding to the reperfusion context, and a blood-glucose measurement outcome with a value threshold and unit. The trace exposed the central computability issue: no OMOP event directly represented the intraoperative reperfusion timestamp in the evaluated structured event fields. The system therefore selected “Transplantation of liver” as the closest available standard procedure proxy for the reperfusion context, while flagging other candidate procedure concepts as less appropriate.

In the synthetic_snuh_cdm run, the cohort collapsed to zero at the procedure/proxy-anchor step, before the blood-glucose criterion could contribute additional narrowing. The subsequent liver-transplantation and glucose-measurement steps were skipped or remained at zero because the previous materialized cohort was already empty. This result should not be interpreted as evidence that no clinically eligible liver-transplant patients existed. Rather, it shows that the criterion reached a computability boundary where the clinically meaningful temporal anchor was not directly represented in the available structured OMOP event fields.

This case illustrates a central motivation of the proposed framework: the main risk in site-level feasibility counting is not only whether SQL can be generated or executed, but whether the clinical interpretation that precedes SQL has been made explicit enough to be reviewed. In this setting, the role of $h_q$ is to make the chosen interpretation explicit: what anchor was selected, what proxy was used, what semantic loss was accepted, and how that choice affected the resulting site-level patient count. By localizing the zero count to the proxy-anchor and local data-availability layer, the trace prevents the result from being misread as a clinically meaningful absence of eligible patients and instead turns it into a reviewable statement about the chosen executable interpretation.

Case 3 examined an atrial fibrillation or atrial flutter trial criterion with anticoagulant exclusion. The query was decomposed into adult age, an OR group over atrial fibrillation and atrial flutter condition concepts, and a forbidden anticoagulant drug-class exposure. The trace showed that the condition OR group was correctly represented as a group criterion rather than as a sequential conjunction, but the local synthetic CDM contained no qualifying AF/flutter cohort under the selected concept sets, reducing the cohort to zero at that stage.

The anticoagulant drug-class step was still informative. The system recognized “anticoagulant” as a drug-class mention and attempted OMOP vocabulary-graph resolution, but no locally observed drug_exposure descendants were available in the target database, resulting in an executable-empty drug concept set. This case therefore illustrates why local executability and empty-set semantics are required. A vocabulary-valid drug class should not be assumed to retrieve local exposure events unless corresponding standard drug concepts are present in the local CDM.

Box 2 provides a positive executable example to contrast with the proxy-anchor and executable-empty cases. It shows that even when a nonzero count is produced, the count should be interpreted as the cardinality of a declared patient set generated through explicit inclusion, exclusion, value, and drug-exposure operations, rather than as a direct database lookup.

Taken together, these NCT case studies show that NS-CoT’s trace artifacts can be useful even outside the scored benchmark setting. Case 1 illustrates successful construction of a computable patient-set trace for value-bearing measurement and exclusion criteria. Case 2 illustrates that some protocol criteria do not fail merely at the SQL generation stage, but at the earlier semantic-operationalization stage where a free-text temporal expression must be committed to an executable $h_q$. Case 3 illustrates how OR logic, drug-class recognition, local executability filtering, and executable-empty concept-set behavior can be exposed rather than hidden inside a final zero count. These case studies should therefore be interpreted as traceability demonstrations: they do not establish a single chart-reviewed gold-standard cohort but show how ambiguous protocol text can be transformed into explicit, reviewable, and executable patient-set commitments for site-level feasibility counting.

| Case | Free-text criterion feature | Semantic commitment exposed as $h_q$ | Stepwise result | Methodological interpretation |
|---|---|---|---|---|
| NCT04936022 (Case 1) | Adults with systolic blood pressure 140–159 mmHg, no diabetes mellitus history, and no antihypertensive exposure | A computable patient-set definition combining adult observation, value-bearing SBP measurement, diabetes-history exclusion, and antihypertensive drug-exposure exclusion | 10,000 observed adults → 3,792 with qualifying SBP value/unit criterion → 1,734 after diabetes exclusion → 114 after antihypertensive exposure exclusion | Demonstrates that a nonzero site-level count is not a direct database lookup, but the cardinality of an explicitly declared patient set produced through sequential inclusion, value, and exclusion operations. |
| NCT03152890 (Case 2) | Liver recipients with glucose >180 mg/dL after graft reperfusion; adults only | A procedure-linked temporal interpretation in which the clinically meaningful reperfusion anchor was not directly represented in structured OMOP event fields and required proxy interpretation | 10,000 adults → 0 at the procedure/proxy-anchor step in synthetic_snuh_cdm | Demonstrates that a zero count may arise from a computability boundary or proxy-anchor limitation rather than from true clinical absence. The case shows why the chosen temporal anchor and accepted semantic loss must be externalized before the count can be interpreted as feasibility evidence. |
| NCT02958943 (Case 3) | Adults with atrial fibrillation or atrial flutter and no active anticoagulant exposure | An OR-group condition definition combined with a forbidden drug-class exposure, with local executable concept-set filtering | 10,000 adults → 0 after atrial fibrillation/flutter OR condition group; anticoagulant class resolution produced no locally executable drug-exposure concepts | Demonstrates that vocabulary-valid criteria may still become locally non-executable. The trace exposes OR logic, drug-class grounding, local data availability, and executable-empty risk instead of hiding them inside a final zero count. |

**Table 26. Semantic trace demonstrations for protocol-derived clinical trial criteria.**

*Note. These case studies are traceability demonstrations rather than clinical validation studies. They do not assume that each free-text trial criterion has a single self-evident gold-standard patient set. Instead, they show how ambiguous protocol text is committed to explicit, reviewable, and executable patient-set semantics before the resulting count is interpreted as site-feasibility evidence. Trial eligibility criteria were accessed from ClinicalTrials.gov on 2026.06.10.*

**Box 2. Stepwise computable logic for Case 1.**

**Natural-language feasibility abstraction:**
Adults aged 18 years or older with systolic blood pressure between 140 and 159 mmHg, no history of diabetes mellitus, and no antihypertensive medication exposure.

**Declared executable interpretation:**
$h_q$ defines a patient set over OMOP CDM by combining an adult observation-period requirement, a value-bearing systolic blood-pressure measurement requirement, a diabetes-history exclusion, and an antihypertensive drug-exposure exclusion.

**Computable logic:**
1. Start from observed adults aged ≥18 years.
2. REQUIRE systolic blood pressure measurement with value_as_number between 140 and 159 and unit compatible with mmHg.
3. FORBID diabetes mellitus history using the grounded condition concept set.
4. FORBID antihypertensive drug exposure using the grounded drug-class concept set.
5. Count the remaining distinct persons.

**Stepwise trace:**
$C_0$ = observed adults: 10,000
$C_1 = C_0 \cap$ qualifying SBP measurement: 3,792
$C_2 = C_1 \setminus$ patients with diabetes mellitus history: 1,734
$C_3 = C_2 \setminus$ patients with antihypertensive exposure: 114

**Interpretation:**
The final count is not a direct database lookup. It is the cardinality of a declared patient set produced by sequential measurement, condition-exclusion, and drug-exclusion operations. This trace demonstrates how a site-level patient count can be inspected through its component semantic commitments and attrition steps.

# Chapter 6. Discussion

## 6.1. Principal Findings

The results of this study support a more specific and methodologically important conclusion than simple model superiority. NS-CoT did not outperform the baseline uniformly across all OMOP query settings. Rather, its advantage appeared most clearly in the setting for which it was designed: site feasibility patient-count questions with protocol-derived inclusion and exclusion constraints, especially those requiring temporal anchoring, drug exposure windows, and anchor-bound exclusion. This distinction is central to interpreting the findings. The study does not show that a neuro-symbolic framework is always better than LLM-centric OMOP NL2SQL workflows. It shows that when the clinical meaning of a feasibility-oriented cohort criterion must be decomposed before a site-level patient count can be produced, explicit semantic formalization becomes consequential.

In CQ-Bench, the difference between NS-CoT and FastOMOP was substantial. The overall exact-count performance favored NS-CoT, and the paired query-level comparison showed that the discordant cases were significantly skewed toward NS-CoT. However, the more informative result was not the aggregate percentage itself. The key pattern emerged across semantic strata. Both systems handled simple condition queries well, which suggests that the difference was not primarily about recognizing common disease concepts or producing executable SQL for basic cohort counts. The separation appeared in strata that required more than concept lookup: post-diagnosis drug exposure and post-diagnosis drug absence, which are central patterns in feasibility-oriented cohort-count criteria involving inclusion and exclusion constraints. These criteria require a condition anchor, a grounded drug or drug-class concept set, an anchor-relative temporal predicate, and a patient-set operation. In these strata, within the supported CQ-Bench L4 and L5 templates, NS-CoT reproduced the reference counts with complete exact-count fidelity, whereas the baseline showed lower exact-count fidelity.

This result is aligned with the methodological premise of the study. A criterion such as “patients with hyperlipidemia and no statin after diagnosis” is not a single database query in any simple sense. It must first be interpreted as a condition anchor, a drug-class outcome, an exclusion role, and an after-diagnosis temporal scope. The phrase “no statin” is not a global claim that the patient has never received statins; it is an absence condition relative to a diagnosis anchor and an observation context. Once represented in this way, the criterion becomes a patient-set operation: retain patients with the anchor condition and remove those with a qualifying forbidden event after that anchor. NS-CoT performed well in these cases because this is precisely the structure it makes explicit before SQL compilation.

The FOEM results provide an important counterweight to this interpretation. On the external template-derived benchmark, both systems achieved high exact-count performance, and the difference between them was not statistically significant. This finding prevents the results from being overstated. When query patterns are closer to OMOP SQL templates that can be instantiated through an agentic NL2SQL workflow, the additional structure of typed IR, patient-set algebra, and deterministic compilation may not produce a measurable advantage. This does not weaken the main claim. Instead, it clarifies its scope. NS-CoT is most valuable when the query requires preserving clinical semantics across several transformation layers; it is not presented as a universal replacement for every OMOP query-generation workflow.

Taken together, these findings clarify the object of computation in this study. In the site-feasibility setting, site-level patient count is not merely a number retrieved from a database. It is the cardinality of a declared executable patient set produced by a specified cohort definition under a fixed OMOP CDM instance, vocabulary snapshot, and compilation environment. The principal finding is therefore not only that NS-CoT improved exact-count reproduction in selected query strata, but that explicit cohort semantics made the meaning of those counts inspectable before they are interpreted as feasibility evidence.

The liveness results further sharpen this interpretation. In CQ-Bench, both systems generally returned counts. The main performance gap therefore did not arise simply because one system failed to run more often. It arose because returned counts did not always represent the same executable cohort semantics as the reference. This distinction matters for site feasibility workflows, where patient-count estimates may inform protocol refinement, site selection, and recruitment planning. A runnable SQL query can still diverge from the declared executable semantics if it chooses a different concept set, uses a different anchor, changes the temporal boundary, or treats an exclusion criterion as an unanchored negation. The fact that a count is produced does not make it a reviewable or reproducible feasibility estimate.

For this reason, the primary methodological object in this work is not the SQL string and not even the scalar count. It is the cohort definition behind the count. In NS-CoT, the count is the cardinality of a patient set constructed through typed interpretation, vocabulary grounding, temporal predicates, exclusion semantics, and patient-set operations. This reframes evaluation as a question of whether a declared operationalization of the criterion can be reproduced as the same executable patient set and count. The system does not claim to recover the unique true meaning of a natural-language eligibility criterion or to determine final enrollment eligibility. Natural language often permits multiple reasonable operationalizations. Instead, the system makes a specified operationalization explicit and then asks whether the resulting patient set and feasibility count can be reproduced under that declared interpretation.

This epistemic stance is important because natural-language clinical criteria and OMOP CDM inhabit different representational spaces. Site feasibility patient-count questions are written in language that carries ambiguity, implicit clinical knowledge, and context-dependent temporal meaning. OMOP CDM, by contrast, represents clinical data as discrete events indexed by person identifiers, concept identifiers, domains, and dates. There is no natural one-to-one mapping between these two spaces. A transformation from site feasibility patient-count questions to OMOP-executable cohort logic is therefore a process of semantic discretization. Clinical meaning must be committed to explicit concept sets, anchor dates, observation windows, exclusion scopes, and patient-set operations before it can be counted.

The traceability findings reinforce this point. In the failure case examined in the results, the final count did not exactly match the reference count, but the reason for the mismatch could be localized to a temporal-boundary decision. This is a different kind of evidence from simple accuracy. It shows that the system's output was decomposable: the typed interpretation, grounded concepts, generated SQL, and event-level evidence could be inspected to determine where the semantic divergence occurred. For clinical research informatics, this matters because cohort counts are often used as inputs to trial feasibility, protocol refinement, and downstream study design. A count that cannot be explained is difficult to trust; a count whose semantic construction can be reconstructed can be reviewed, corrected, and reused. The deterministic person-set verification further supports this interpretation: in the evaluated queries, exact count agreement did not conceal different underlying person_id sets, and saved SQL artifacts could be re-executed to compare reference and system-derived person_id sets directly.

The principal finding, then, is not that NS-CoT is a better SQL generator. It is that feasibility-oriented cohort computation benefits from moving the semantic work before SQL generation. The study shows that selected feasibility-oriented cohort criteria can be transformed into executable OMOP CDM patient-set logic through a neuro-symbolic architecture that separates language interpretation from execution semantics. Within the evaluated scope, this separation improved fidelity for temporal and exclusion-heavy criteria, exposed boundary states such as executable-empty concept sets, and produced cohort artifacts that could be inspected after execution. These findings motivate the broader discussion of methodological contribution and clinical utility in the following sections.

## 6.2. Interpretation of the Results

The benchmark results should be interpreted through the central representational problem addressed by the present study. Natural-language feasibility cohort criteria and OMOP-executable cohort logic do not occupy the same representational space. Site feasibility patient-count questions with protocol-derived inclusion and exclusion constraints are written in clinical natural language. They are compact, context-dependent, and often underspecified. A phrase such as "on insulin therapy," "no prior systemic treatment," or "diagnosed within the past six months" carries implicit assumptions about clinical intent, temporal scope, observation period, and operational convention. OMOP CDM, by contrast, represents clinical history through person identifiers, standard concept identifiers, domain-specific event tables, and event dates.

This difference is not merely an implementation inconvenience. It is a representational asymmetry. This interpretation is consistent with conceptual-space accounts of meaning and reasoning. Gärdenfors and Osta-Vélez argue that reasoning is not reducible to syntactic manipulation alone, but depends on conceptual structure, similarity, typicality, and diagnosticity. Gärdenfors's later account of the geometry and dynamics of meaning similarly frames meaning as structured and context-sensitive rather than reducible to isolated symbolic tokens. Such accounts help clarify why site feasibility patient-count questions cannot be treated as strings waiting for SQL translation. NS-CoT does not implement a continuous conceptual-space model; rather, conceptual-space theory is used here as an explanatory lens for the representational gap between flexible clinical meaning and discrete executable cohort logic.

Natural language permits a range of plausible meanings, whereas OMOP CDM requires a finite and explicit operationalization: a concept set, a table, a date column, an anchor event, a temporal window, an inclusion or exclusion rule, and a patient-set operation. There is no natural isomorphism between these two spaces. A natural-language criterion does not automatically determine a unique SQL query, and a syntactically valid SQL query does not automatically preserve the clinical meaning of the criterion. The transformation must therefore pass through an intermediate process of semantic discretization.

In this study, semantic discretization means converting a clinically meaningful but underspecified feasibility-oriented cohort criterion into a finite set of executable commitments over OMOP CDM. These commitments include typed role assignment, standard-vocabulary grounding, descendant expansion, domain routing, anchor selection, temporal boundary policy, exclusion scope, and patient-set composition. Once these decisions are declared, the resulting cohort count can be reproduced under the same database snapshot, vocabulary version, and compiler rules. The purpose is not to discover the single true meaning of the natural-language criterion or to adjudicate final patient eligibility. The purpose is to make the adopted interpretation explicit enough that the patient set and count are conditionally reproducible. The final SQL query is the executable image of this projection, not the projection itself.

This perspective explains why NS-CoT performed differently across CQ-Bench and FOEM. CQ-Bench was designed to stress the semantic discretization problem. It included feasibility-oriented cohort queries that required the system to move from natural-language expressions into explicit OMOP cohort semantics: first-diagnosis anchors, post-diagnosis drug exposure windows, post-diagnosis drug absence, and zero-count negative controls. In these cases, the challenge was not simply to retrieve a concept and write a query. The system had to decide what counted as the anchor, how the target event related to that anchor, and how the resulting patient set should be narrowed or preserved. NS-CoT's advantage in L4 and L5 is therefore not incidental. Those strata correspond closely to the part of the problem where semantic discretization is required.

L4 queries evaluated temporal inclusion. A natural-language phrase such as "drug exposure within 90 days after diagnosis" must be mapped to a diagnosis event, a first-anchor policy, a drug concept set, and a bounded post-anchor interval. If any of these commitments differ, the final patient count can change. NS-CoT's performance in this stratum is consistent with the value of making the temporal relation explicit: it was represented as a typed temporal predicate and then lowered deterministically into SQL, rather than being left inside an opaque generated SQL string.

L5 queries evaluated temporal exclusion, which is even more semantically delicate. A phrase such as "no statin after diagnosis" is not equivalent to "no statin ever." It is also not equivalent to the absence of a statin concept in a vocabulary search result. It means that, among patients who satisfy the diagnosis anchor, there should be no qualifying statin exposure event in the specified temporal scope. This is an anchor-bound non-existence condition. In patient-set terms, the system first constructs the anchor cohort and then removes patients for whom a positive forbidden event exists after the anchor. NS-CoT's performance in L5 shows that treating exclusion as patient-set subtraction, rather than as surface-level negation, is important for reproducing executable cohort semantics.

The FOEM result should be read differently. FOEM is an external template-derived benchmark instantiated over the same OMOP CDM environment. The FOEM results are consistent with the interpretation that many FOEM queries were closer to template-instantiated OMOP SQL patterns than to the temporally anchored exclusion-heavy criteria emphasized by CQ-Bench. In that setting, the additional structure of NS-CoT does not necessarily create a large performance advantage, because the representational distance between the natural-language question and the executable SQL template is smaller. The absence of a statistically significant difference on FOEM therefore does not contradict the main claim. It clarifies it. NS-CoT is not valuable because every query requires a full neuro-symbolic formalization. It is valuable because some feasibility-relevant cohort criteria derived from eligibility rules do require such formalization, and those are precisely the cases where direct text-to-SQL can hide meaning-changing decisions.

The distinction between liveness and exact-count fidelity further supports this interpretation. A system that returns a count has solved only part of the problem. It has produced an executable output, but not necessarily the cohort semantics encoded by the executable reference. In CQ-Bench, both systems generally returned counts, yet their exact-count fidelity differed substantially. This indicates that the main error mode was not simply runtime failure. It was semantic divergence: a generated query could run successfully while operationalizing a different concept set, temporal window, exclusion scope, or patient-set operation from the reference. From the standpoint of clinical trial feasibility, this matters more than syntactic validity. A plausible count can be misleading if the cohort definition behind it diverges from the declared or reference semantics.

The results also show that NS-CoT is not merely a wrapper around SQL generation. Its contribution is the placement of semantic commitments before SQL. In a direct generation architecture, concept selection, temporal interpretation, exclusion handling, and SQL composition are entangled in one generated artifact. When the result diverges from the reference, it can be difficult to determine whether the error came from clinical interpretation, vocabulary mapping, temporal boundary selection, or SQL composition. In NS-CoT, these decisions are separated. The typed IR records the semantic decomposition. The grounding layer records the concept-set decision. The symbolic algebra records inclusion, exclusion, and temporal operations. The SQL compiler lowers these decisions into executable SQL. This separation makes the transformation inspectable.

One of the most important implications is that exact-count correctness becomes conditional on declared semantics. In ordinary text-to-SQL evaluation, it is often assumed that the natural-language query has a single correct SQL target. For site feasibility patient-count questions with protocol-derived inclusion and exclusion constraints, this assumption is too strong. Many criteria admit several clinically reasonable and feasibility-relevant operationalizations. “Patients with diabetes on insulin therapy” can mean current insulin therapy, any prior insulin exposure, insulin initiation after diabetes diagnosis, or insulin exposure within a specified time window after diagnosis. Each interpretation is defensible in a different study context, and each produces a different cohort. NS-CoT does not eliminate this ambiguity. Instead, it changes the task: the system must declare which interpretation is being used and then compute the corresponding patient set reproducibly.

This is why the study evaluates functional fidelity rather than clinical truth. The benchmark asks whether the system reproduces predefined executable OMOP semantics, not whether the resulting cohort is a chart-reviewed clinical phenotype. This distinction is particularly important in site feasibility, where the immediate task is reproducible site-level patient-count estimation rather than final enrollment adjudication. Before a cohort definition can be clinically validated, it must first be executable, inspectable, and reproducible. The results show that explicit typed representation and symbolic compilation can improve this intermediate but essential step.

The deterministic person-set verification makes this distinction concrete. Exact count agreement was not treated as sufficient by itself; saved SQL artifacts were re-executed to compare the reference and system-derived person_id sets. This reinforces the interpretation that the scalar count is a summary of an underlying executable patient set, not an independent prediction detached from cohort semantics. This interpretation also clarifies why traceability is part of the result, not merely an engineering convenience. If the transformation is a semantic discretization, then each discretization decision should be visible. Which concept set represented the disease? Which vocabulary path represented the drug class? Which event date became the anchor? Was the lower bound inclusive or exclusive? Was the exclusion global or anchor-bound? These are not implementation details. They are the defining coordinates of the cohort definition in the OMOP event space. A feasibility count without these coordinates cannot be fully interpreted.

The traceable failure case in the results illustrates this point. The count mismatch was not treated as an opaque failure. It was localized to a temporal-boundary decision. This kind of evidence is important because it shows that the system produces artifacts that can be inspected at the level of semantic transformation. Even when the final count does not match the reference, the system can expose where the adopted operationalization diverged from the reference. That is the practical meaning of making cohort computation inspectable rather than opaque.

The implication is therefore conceptual rather than merely comparative. NS-CoT should not be interpreted as a universal replacement for all OMOP query-generation systems. Its contribution is clearest when the distance between natural-language criteria and executable cohort logic is semantically meaningful: when the criterion requires an explicit anchor, a temporal window, an exclusion scope, drug-class grounding, local executable concept filtering, or executable-empty concept-set behavior. In such cases, generating executable SQL is not sufficient unless the cohort-defining commitments behind that SQL are also declared and inspectable. In feasibility assessment, the question is therefore not only "How many patients are there?" but also "Under which concept set, time window, anchor definition, and exclusion rule are there this many patients?" This is why feasibility-oriented cohort computation should be treated as computable cohort-definition construction, not as text-to-SQL generation alone.

## 6.3. External Multi-Site Feasibility Simulation, Repairability, and Scope of Interpretation

The COVIC experiment extends the evaluation from single-database count reproduction to external multi-site feasibility simulation. Its purpose is to examine how system-generated cohort logic behaves when the same COVIC-derived executable criterion component is applied across multiple OMOP CDM schemas representing candidate sites. In this setting, the key question is not only whether each criterion-site count exactly matches the deterministic reference count, but also whether the generated count values preserve reference-defined site-feasibility labels and relative site rankings.

This interpretation requires an evaluation-unit caveat. During pilot audit, some COVIC rows used a full natural-language criterion line as the system input, whereas the deterministic reference SQL could correspond to a selected entity-level concept from that line. In such cases, a system that faithfully interprets the full line could appear to over-include patients relative to a narrower reference. Therefore, patient-set fidelity in COVIC should be interpreted under a strict unit-alignment assumption. This caveat limits the interpretation of exact patient-set mismatch, but it does not reduce the value of the experiment as a multi-site feasibility simulation. Rather, it clarifies that COVIC evaluates external criterion handling, multi-schema execution, site-level count preservation, repairability, and site-feasibility decision preservation under a strict reference-alignment setting.

Within this scope, the first-pass COVIC results showed a clear limitation of the original NS-CoT implementation. FastOMOP achieved higher live execution, strict count fidelity, tolerance-based count fidelity, zero-form agreement, exact person-set agreement, and patient-set overlap. Thus, the first-pass conclusion should be reported directly: NS-CoT was less robust than FastOMOP in this external COVIC-style strict-alignment setting. This finding limits any claim that NS-CoT is universally superior across all eligibility-query distributions. In particular, COVIC stressed raw criterion-line normalization, broad vocabulary grounding, executable concept-set reproduction, multi-entity drug exclusions, and multi-schema execution robustness more strongly than the anchor-bound temporal and exclusion-heavy semantic patterns emphasized in CQ-Bench.

However, the first-pass failure pattern was not a black-box SQL failure. Layer-wise audit showed that many failures occurred before SQL execution, particularly in vocabulary grounding, semantic validation, typed IR generation, OR-group handling, and local-resident drug filtering. This distinction is central to the interpretation of the experiment. If the system had failed only as an opaque SQL generator, the result would mainly indicate that a different generator performed better. Instead, the NS-CoT artifacts made it possible to localize failures to specific stages in the transformation from natural-language criterion to executable patient-set logic. The COVIC multi-site simulation therefore exposed implementation weaknesses in first-pass execution, but it also demonstrated why explicit intermediate artifacts matter in feasibility-oriented cohort computation: they make failures inspectable at the level of criterion decomposition, grounding, local executability, temporal or exclusion logic, and SQL lowering, rather than leaving them as opaque count mismatches.

The broad-expansion audit further refined this interpretation. NS-CoT did not simply fail because its vocabulary expansion was too broad in all cases. In some criterion-site pairs, broader expansion moved the result closer to the reference by recovering patients missed by the comparator. In other cases, especially in drug-domain exclusions and abstract drug-class expressions, broader expansion moved the result farther from the reference. This pattern suggests that expansion is not intrinsically beneficial or harmful. Its value depends on whether the expansion boundary preserves the intended executable criterion unit. The issue is therefore not whether concept expansion should be used, but whether expansion is constrained by domain, qualifier, local-resident evidence, exclusion role, and criterion-unit alignment.

The second-pass repair analysis was performed to test whether these first-pass failures were diagnosable and partially repairable through general pipeline corrections rather than benchmark-specific hardcoding. The repair introduced two main changes: partial execution for OR groups and stricter local-resident drug filtering. Under the OR partial execution policy, a failed branch in an OR group no longer caused the entire OR group to fail when other branches remained executable. Instead, executable branches were retained and skipped branches were recorded in structured artifacts. Under the drug local-resident filtering correction, drug concepts that existed in the OMOP vocabulary but had no resident evidence in the target site's drug exposure table were treated as empty event sets rather than being preserved through a broad fallback. These changes reflected general execution semantics rather than COVIC-specific term tuning.

The second-pass results showed substantial recovery. Live execution, strict count fidelity, tolerance-based count fidelity, zero-form agreement, exact person-set agreement, and patient-set overlap all improved compared with the first pass. This improvement supports the diagnostic value of the NS-CoT design: failures could be localized, translated into general execution-policy changes, and re-evaluated. Importantly, the repair did not merely increase execution coverage by returning more outputs. It also improved person-set fidelity, indicating that the additional coverage did not simply come from uncontrolled over-inclusion.

Nevertheless, the repaired NS-CoT implementation still remained below FastOMOP on most scalar count and row-level patient-set fidelity outcomes. FastOMOP retained higher live execution, strict count fidelity, tolerance-based count fidelity, exact person-set agreement, Jaccard similarity, precision, and recall in the all-attempted comparison. Therefore, the second-pass analysis should not be interpreted as reversing the main COVIC performance conclusion. The most accurate interpretation is that NS-CoT repair significantly improved functional fidelity, but FastOMOP remained stronger for raw COVIC-style criterion-line execution and exact patient-set reproduction in this benchmark.

A different pattern appeared for site-level decision preservation. Although repaired NS-CoT remained lower than FastOMOP in row-level scalar and patient-set fidelity, its site-ranking fidelity improved substantially after repair. Top-site agreement and mean Spearman correlation reached a level comparable to, and numerically slightly higher than, FastOMOP in this run. This finding should be interpreted cautiously, because row-level fidelity metrics still favored FastOMOP. However, it suggests that exact patient-count reproduction and site-selection decision preservation are related but not identical evaluation targets. Strict $R_0$, $R_0$@5%, exact person-set agreement, and Jaccard similarity measure how precisely the system reproduced the reference patient set for each criterion-site pair. Site-ranking fidelity asks a different operational question: whether the system preserved which site appeared to have the relatively larger available patient pool for a given criterion.

This distinction is important for clinical trial feasibility assessment. In some planning contexts, the exact count may be the central output. In other contexts, the more immediate operational question may be whether a system preserves the relative feasibility ordering among candidate sites. The COVIC results suggest that these two forms of fidelity can diverge. A system may remain imperfect at exact patient-set reproduction while still preserving useful site-level ranking information. Conversely, a system that is stronger at row-level count fidelity may not necessarily be uniformly superior for all feasibility decision artifacts. This does not reduce the importance of exact patient-set fidelity, but it clarifies that feasibility evaluation should distinguish scalar count accuracy, patient-set identity, and site-level decision preservation.

Taken together, the COVIC analysis identifies both an implementation limitation and a methodological value of NS-CoT in a multi-site feasibility setting. The limitation is clear: the current implementation requires stronger robustness for raw criterion-unit normalization, broad vocabulary grounding, drug-class boundary control, local executability handling, and criterion-unit alignment before it can be competitive with FastOMOP in broad external annotation-derived settings. The methodological value is also clear: because NS-CoT represents the transformation from criterion text to executable cohort logic through typed intermediate representations, grounding artifacts, local executability filters, patient-set operations, generated SQL, and structured trace logs, its failures can be inspected and converted into targeted repair hypotheses.

Therefore, the COVIC result should not be framed as evidence that NS-CoT is a generally stronger SQL generator. It is not. Nor should it be framed as evidence that formal patient-set semantics failed. Rather, it shows that multi-site count estimation introduces evaluation targets that are not captured by single-database scalar count benchmarks alone. External annotation-derived eligibility material exposes implementation-level robustness gaps, while the multi-schema design allows the evaluation to ask whether system-generated count preserves reference-defined site-feasibility labels and site rankings. In this sense, COVIC strengthens the study's central medical-informatics claim: reliable site-feasibility computation requires not only executable SQL, but also explicit, inspectable, and revisable patient-set semantics capable of supporting count fidelity, patient-set fidelity, and site-level decision preservation.

## 6.4. Methodological Contributions

The methodological contribution of the present study follows directly from the interpretation above. If feasibility-oriented cohort computation is a process of semantic discretization, then the central task is to define the intermediate objects through which that discretization occurs. NS-CoT contributes such an intermediate layer. It does not treat a feasibility-oriented natural-language cohort criterion as a sentence to be translated directly into SQL. It treats the criterion as a clinical-semantic object that must be decomposed, grounded, temporally anchored, converted into patient-set operations, and only then compiled into SQL. In this sense, NS-CoT operationalizes a neuro-symbolic form of knowledge engineering: neural interpretation is used to identify the candidate semantic structure of the criterion, while symbolic modules commit that structure to executable OMOP vocabulary, temporal, and patient-set representations[19,20,28,36,82,119].

This intermediate layer should be understood as a specification layer, not merely as a safeguard against imperfect SQL generation. Even if an LLM-generated SQL query is syntactically valid and executable, its reference consistency can only be judged relative to a declared cohort definition. NS-CoT contributes a way to make that cohort definition explicit before SQL is emitted. The methodological question is therefore not only whether the generated SQL runs, but what executable patient-set semantics the SQL is intended to implement.

The practical motivation is straightforward. A clinical researcher or trial designer may ask, "How many patients in our institution satisfy this protocol-derived cohort criterion?" On the surface, this appears to be a database query. But in practice, the question cannot be answered responsibly until the criterion has been operationalized. A count such as 235 patients is meaningful only if one can say what disease concept set was used, which drug concepts were included, whether descendants were expanded, which diagnosis date served as the anchor, what observation window was assumed, whether a drug exposure was required or forbidden, and how the final patient set was assembled. The methodological contribution of NS-CoT is to make these hidden cohort-defining decisions explicit[91,120].

This reframes the task from text-to-SQL generation to cohort logic construction. SQL is still necessary because OMOP CDM is ultimately queried through SQL or SQL-like cohort logic. However, SQL is not the primary methodological object. The primary object is the cohort definition behind the SQL: typed clinical facts, grounded concept sets, temporal predicates, exclusion semantics, patient-set operations, compiled query, and final count. In this framing, the central question changes from "Can the system generate executable SQL?" to "Can the system construct an explicit and reproducible OMOP cohort definition from a protocol-derived natural-language cohort criterion?"[62,63,87–90].

A key contribution is the minimally sufficient decomposition needed to preserve cohort semantics before execution. NS-CoT decomposes each criterion into typed representation, vocabulary grounding, temporal predicate, exclusion semantics, and patient-set algebra. This decomposition is not simply a software pipeline. Each layer preserves a different kind of clinical meaning that would otherwise be buried inside a generated SQL string. Prior work has represented or annotated eligibility criteria, mapped them to computable criteria, or implemented cohort definitions in structured tools, but the specific decomposition used here is directed toward OMOP-executable patient-set computation[29,33,36,73–75,92,93,119]. Table 28 summarizes the minimal semantic components required for executable OMOP cohort computation.

| Component | Role in cohort interpretation | What is lost if omitted |
|---|---|---|
| Typed representation | Identifies clinical mention, domain, role, polarity, value, and temporal dependency | Anchor/outcome roles, inclusion/exclusion direction, and logical structure become implicit |
| Vocabulary grounding | Links mentions to OMOP standard concepts, synonyms, descendants, and local executable concepts | Concept hierarchy, synonymy, and local data availability may be hidden, omitted, or incorrectly assumed |
| Temporal predicate | Represents anchor, event ordering, window bounds, and boundary policy | "after," "within," "prior," and "more than" collapse into ambiguous date logic |
| Exclusion semantics | Defines the scope of negation as patient-level or anchor-bound | "No exposure" may be misread as global absence or unscoped negation |
| Patient-set algebra | Specifies how cohorts are narrowed, subtracted, intersected, or unioned | AND/OR/NOT semantics become buried inside SQL structure |

**Table 27. Minimal semantic components required for executable OMOP cohort computation.**

This decomposition is the core informatics claim of the study. It states that feasibility-oriented natural-language cohort criteria become computable only after their clinical meaning has been committed to a finite set of execution-relevant fields.

The typed event-time representation is the first part of this contribution. OMOP CDM is often described as a standardized relational data model, but feasibility-oriented cohort definitions are not evaluated over tables in the abstract. They are evaluated over observed clinical events: a condition occurrence on a date, a drug exposure on a date, a measurement on a date with a value and unit, or a procedure on a date. This study abstracts such records as $E_d(p, c, t)$, where person $p$has an event in domain $d$, represented by concept $c$, at time $t$. This abstraction makes temporal reasoning explicit. A phrase such as "drug after diagnosis" is no longer treated as a text fragment to be inserted into SQL; it is represented as a relationship between an anchor event and an outcome event[16,56,57].

The second part is patient-set algebra as an intermediate semantics between feasibility-oriented cohort criteria and SQL. This algebra is not proposed as a replacement for OHDSI ATLAS or the OHDSI cohort model. It is a restricted representation for the specific task of transforming site feasibility patient-count questions with clinical trial inclusion and exclusion constraints into executable OMOP cohort logic. The cohort is modeled as a set of observed persons, and each criterion is interpreted as an operation over that set. $REQUIRE$ narrows the cohort, $FORBID$ removes patients who trigger an exclusion event, $GROUP_AND$ intersects patient sets, $GROUP_OR$ unions alternatives, $TEMPORAL_REQUIRE$ evaluates anchor-relative event existence, and $TEMPORAL_FORBID$ evaluates anchor-bound event absence.

This is especially important for exclusion. In natural language, words such as “no,” “without,” “excluding,” and “not on” appear to express negation. But in cohort computation, exclusion is usually not the direct evaluation of a negative clinical state. It is a set-difference operation triggered by a positive event. Let $P(p)$ denote the predicate that patient $p$ has an exclusion-triggering event within the relevant scope. Then exclusion can be represented as:

$$FORBID(C, P) = C \setminus \{p \in C \mid P(p)\}$$

This formulation prevents the system from treating “no statin” as a vague negative label. It forces the cohort definition to specify the positive event, the anchor, the observation context, and the time window in which that event would trigger exclusion. This is the difference between surface negation and computable exclusion semantics.

Another contribution is the treatment of vocabulary grounding as a semantic-to-executable mapping problem. A clinical mention must first be connected to a standard vocabulary structure, then expanded through appropriate vocabulary relationships, and finally constrained to concepts that are executable in the target CDM instance. This is most explicit in the implemented drug-class pathway. A pharmacologic class mention such as “statin” is not treated as a generic drug string and is not expanded by asking the LLM to list ingredients. Instead, the system routes the mention through a class-oriented vocabulary path, expands the selected class through OMOP vocabulary relationships such as concept_ancestor, and then restricts the result to standard drug concepts actually observed in the local drug_exposure table[31,97,121].

This local executability step can be expressed more simply by separating the vocabulary-derived concept set from the locally observed concept set. Let $S$ denote a vocabulary-derived concept set and let $O_D(D)$ denote the set of concepts observed in the relevant OMOP event domain d of database $D$:

$$O_d(D) = \{c \mid \exists p, t: E_d(p, c, t)\}$$

The locally executable concept set is then:

$$\Gamma_d^{exec}(S; D) = S \cap O_d(D)$$

Thus, $\Gamma_d^{\mathrm{exec}}(S; D)$ retains only the vocabulary-derived concepts that are actually observed in the target CDM instance. In the current implementation, this operation is most explicitly developed for drug-class grounding, where a vocabulary-derived drug concept set is intersected with the set of drug concepts present in the local drug_exposure table.

This distinction between vocabulary-level possibility and local executability is methodologically important. Feasibility counts are not computed in the abstract vocabulary space. They are computed in a particular database snapshot. A drug class may be clinically valid and properly represented in a standard vocabulary, but if none of its descendant concepts appears in the local drug_exposure table, it cannot retrieve local exposure events. Conversely, a local CDM may record exposures under more specific descendant concepts rather than under the seed concept identified from the natural-language mention. The grounding layer must therefore handle both semantic coverage and local data availability[14,31,97].

This design also gives the framework a scalable structure. Scalability here does not mean only that the system can process more queries. More importantly, it means that new vocabulary paths, domain-specific grounding rules, temporal operators, and local executability checks can be added without changing the core semantic pipeline. The deterministic person-set verification in Chapter 5 illustrates this methodological point: because the system preserved saved SQL and cohort-definition artifacts, the scalar count could be traced back to a re-executable person_id set rather than treated as an isolated model output. The same transformation remains in place.

This form of scalability is possible because different kinds of knowledge enter the system at different layers. The LLM interprets the clinical phrase and assigns typed fields. The grounding layer connects the phrase to vocabulary structures. The symbolic layer determines temporal and exclusion semantics. The compiler compiles the resulting patient-set logic into SQL. Because these responsibilities are not collapsed into one generated SQL string, each layer can be extended independently. A new vocabulary mapping can be added to the grounding layer without changing temporal algebra. A new temporal operator can be added without changing concept retrieval. A new local data-presence check can be added without changing the role of the LLM.

The drug-class grounding pathway is the clearest implemented example of this scalability pattern. Future extensions could incorporate richer relationships among ATC, RxNorm, SNOMED CT, MeSH, ICD vocabularies, and local source vocabularies as additional evidence in the vocabulary grounding layer. The LLM would not need to memorize these mappings or generate the full list of possible drugs. Its role would remain to identify the typed clinical mention and, when possible, the semantic class. The symbolic grounding layer would determine how that class should be expanded and made executable in the target CDM instance[97,121].

The same pattern can be extended cautiously to other OMOP domains. For condition criteria, expanded condition concept sets can be intersected with locally observed condition_occurrence concepts. For procedure criteria, procedure concept sets can be checked against procedure_occurrence. For measurement criteria, analytes can be grounded through LOINC or SNOMED and then checked against measurement, while numeric thresholds and units are handled as separate value predicates. These extensions are not automatic; each domain has its own coding practices, event semantics, and data completeness issues. The contribution is not that every domain has already been solved, but that the architecture defines where such domain-specific knowledge should enter the pipeline.

Measurement criteria illustrate this point. A criterion such as "HbA1c ≥ 7%" cannot be handled by concept grounding alone. It requires separation of the analyte concept, comparator, numeric threshold, and unit. Because NS-CoT already separates typed representation from vocabulary grounding and value predicates, future extensions can add LOINC-prioritized retrieval, UCUM unit normalization, unit conversion, or reference-range policies without changing the overall architecture. This is a different kind of scalability from adding more prompt examples. It expands the symbolic representation while preserving inspectability.

The architectural contribution is therefore not only the use of a neuro-symbolic pipeline, but the particular boundary drawn between neural and symbolic work. The LLM is used where linguistic flexibility is needed: interpreting the natural-language criterion, decomposing clinical facts, assigning tentative roles, and identifying potential temporal or exclusion structure. The LLM is not used as the final authority for cohort execution semantics. After the typed and grounded representation is formed, the execution semantics are handled by symbolic modules: vocabulary traversal, temporal predicate construction, exclusion handling, patient-set algebra, and SQL compilation[42,43,110,111].

This boundary changes the failure mode of the system. In a direct text-to-SQL architecture, concept selection, temporal interpretation, exclusion handling, and SQL composition may all be entangled in a single output. When the count diverges from the reference, the error may be difficult to localize. In NS-CoT, the error can be decomposed. The typed IR can be inspected for role or polarity errors. The grounding artifact can be inspected for concept-set errors. The temporal predicate can be inspected for boundary errors. The compiled SQL can be inspected as the deterministic realization of those earlier decisions. This does not guarantee correctness, but it makes the system's reasoning path reviewable[120].

The empirical contribution is that this formalization was implemented and evaluated rather than left as an abstract proposal. CQ-Bench was constructed to test the semantic components targeted by the method: concept grounding, demographic filtering, first-diagnosis anchoring, temporal inclusion, temporal exclusion, and negative-control handling[122]. FOEM provided an external template-derived benchmark setting[117,118,123]. Together, these evaluations showed where the method had a measurable advantage and where its advantage was not universal. This is important because the study's claim is not that NS-CoT is the best approach for every OMOP query. The claim is that explicit semantic representation improves fidelity where cohort-defining meaning would otherwise be hidden in SQL generation.

The final methodological contribution is the shift from opaque output correctness to conditional reproducibility. The present study does not claim that an automatically generated cohort is clinically valid in the chart-review sense. Nor does it claim that every natural-language criterion has one correct SQL representation. The claim is more precise: once the interpretation is declared through concept sets, temporal anchors, boundary policies, exclusion scopes, and patient-set operations, the resulting patient set and count should be reproducible. Clinical validity can then be evaluated as a separate downstream question. Conditional reproducibility is the prerequisite that makes such validation meaningful.

For this reason, the output of NS-CoT is not merely a count. It is a cohort-definition artifact. It includes the typed interpretation, grounded concepts, temporal and exclusion logic, SQL, final count, and trace metadata. This changes the feasibility question from "How many patients do we have?" to "Under this concept set, this temporal window, this anchor definition, and this exclusion rule, how many patients do we have?" That shift is the core clinical research informatics contribution of the method.

Existing OHDSI tools provide powerful mechanisms for authoring and executing cohort definitions once the analyst already knows how the cohort should be specified[62,63]. NS-CoT addresses the preceding step: how to move from a natural-language eligibility criterion to an explicit, executable, and inspectable cohort logic. The contribution is not to replace ATLAS, SQL, or OMOP. It is to formalize and operationalize the transformation that must occur before those execution tools can be used reliably from natural language.

In this sense, the work is best understood as clinical research informatics rather than only engineering. An engineering framing would say that the system uses an LLM to produce SQL and then reports execution accuracy. The informatics framing is different. It asks what representational objects are needed for eligibility criteria to become computable over standardized clinical data. The answer proposed here is typed event-time representation, standard-vocabulary grounding, local executability filtering, temporal and exclusion semantics, patient-set algebra, deterministic compilation, and traceable artifacts. The implementation and benchmark results show that this representation is not merely conceptual, but executable.

A natural-language eligibility criterion does not, by itself, determine an executable cohort. It must be committed to a declared executable patient-set semantics, $h_q$, defined over concepts, temporal anchors, exclusion scope, value constraints, and patient-set algebra. SQL is one execution of $h_q$, not the place where the meaning of the criterion is first decided; the operative question is therefore not whether SQL was generated, but which $h_q$ a given SQL implements and whether that semantics is preserved through execution and counting. Any language model or agent, however capable, produces $h_q$ candidates whose validity is itself another model output. The object that must remain available for human reading, independent recomputation, comparison, and revision is therefore $h_q$, external to the model. As language models become more fluent at generating executable queries, this externally reviewable semantic layer becomes more, not less, necessary.

Therefore, the methodological contribution of NS-CoT is therefore to make feasibility-oriented cohort computation explicit at the level where clinical meaning is lost or preserved. It does not remove the need for clinical review, and it does not solve all temporal reasoning. What it provides is a formal and executable bridge from site feasibility patient-count questions with protocol-derived constraints to OMOP CDM cohort logic, with semantic commitments visible before the count is trusted. This bridge is the core contribution of the study.

## 6.5. Foundational Clinical Utility as Reviewable Feasibility Evidence

This interpretation is consistent with broader clinical-trial feasibility discussions. The Clinical Trials Transformation Initiative has emphasized that real-world data can support eligibility-criteria planning and recruitment strategy by allowing trial teams to quantify how proposed criteria behave in real-world populations. Similarly, the EHR4CR protocol feasibility evaluation showed that automated patient-count systems can reduce the time needed to obtain feasibility counts, while also demonstrating that count accuracy depends on the availability, structure, and semantic consistency of local clinical data. The present study addresses the next layer of this problem: not whether patient counts can be generated automatically, but whether the cohort-defining assumptions behind an site-level count estimate can be specified, inspected, and reproduced.

This study does not demonstrate downstream clinical utility in recruitment outcomes or workflow performance; rather, it establishes a foundational utility layer by making feasibility counts reviewable. The clinical utility of NS-CoT is best understood at the feasibility stage of trial design. Before downstream trial-planning analyses can simulate outcomes, optimize recruitment, select sites, or compare protocol variants, it must first define the patient population to which those downstream analyses apply. If the upstream cohort definition is unstable, then every subsequent analysis is performed on an unstable patient set. NS-CoT addresses this upstream problem: it converts site feasibility patient-count questions containing protocol-derived inclusion and exclusion constraints into computable, inspectable, and reproducible cohort logic over OMOP CDM[1,4,46,48].

This places the study within a specific part of the clinical trial life cycle. The work does not address trial conduct, recruitment outreach, consent, screening adjudication, enrollment, monitoring, or outcome analysis. It addresses the data-dependent patient-pool feasibility layer that precedes those steps: estimating the site-level patient count implied by protocol-derived computable criteria at a candidate site. In this sense, the study uses observational EHR/CDM data to support an interventional-trial planning decision. Its clinical utility is therefore not recruitment automation, but the production of reviewable feasibility evidence before recruitment begins.

In practice, site feasibility often begins with a simple operational question: how many patients at this institution satisfy this protocol-derived cohort criterion? However, the useful form of that question is more specific. It is not merely "how many patients?" but "how many patients under this specified operationalization of the criterion?" A count becomes actionable only when the study team can see which clinical concepts were used, whether descendant concepts were included, which event served as the anchor, which time window was applied, how exclusion was scoped, and how the final patient set was assembled. A scalar count without this context may be fast, but it is difficult to trust, revise, or defend[91,120].

This distinction matters because inclusion and exclusion criteria are often adjusted during protocol design. A study team may begin with a strict criterion and discover that the eligible population is too small. The next question is not only whether the number is small, but why it is small. Did many patients fail because the disease concept set was too narrow? Did the drug exposure window exclude patients who were treated slightly later? Did a "no prior therapy" exclusion remove too many patients because it was interpreted globally rather than relative to an index date? Did a drug class resolve to concepts that are valid in the vocabulary but absent in the local CDM? These are protocol-design questions, not merely database questions. NS-CoT makes them answerable because it exposes the semantic steps that produced the count.

For example, a protocol-derived cohort criterion such as “patients with diabetes on insulin therapy” may appear straightforward, but it has different feasibility implications depending on how “on insulin therapy” is operationalized. It could mean current insulin treatment, any prior insulin exposure, insulin exposure after the diabetes diagnosis, or insulin exposure within a specified post-diagnosis window. Each interpretation is clinically plausible in a different protocol context, but each produces a different cohort. A direct text-to-SQL system may return a count without making this distinction visible[19,20]. NS-CoT instead makes the adopted interpretation explicit by representing the diabetes condition, insulin exposure, temporal relation, and patient-set operation as separate components of the cohort definition.

The same issue is even more important for exclusion criteria. Expressions such as “no prior treatment,” “without statin exposure,” or “not on anticoagulants” are not simple database negations. In an OMOP CDM setting, absence means absence of a recorded event within a defined observation context and temporal scope. “No statin after diagnosis” is therefore different from “no statin ever” and different again from “no statin within the past six months.” NS-CoT’s anchor-bound exclusion semantics allow these differences to be represented explicitly. This helps clinicians and analysts determine whether the system is excluding the intended patients for the intended reason.

The immediate clinical value is therefore the reviewability of feasibility counts. NS-CoT changes the output from a standalone number into a cohort-definition artifact. The system can produce the count, but it can also preserve the typed facts, grounded concept sets, temporal predicates, exclusion rules, and generated SQL behind the count. This allows the study team to review the operational definition before using the number for protocol decisions. If the count is unexpected, the team can inspect whether the issue came from the concept set, the temporal window, the exclusion scope, or local data availability[120].

This reviewability is particularly relevant for site feasibility assessment. When an investigator asks whether a protocol is feasible at a hospital, the answer often depends on local data representation[1,4]. A drug class may be standard in the vocabulary but sparsely represented in the institution’s drug_exposure table. A disease concept may have several descendants, some of which are recorded locally and others not. A procedure may be captured in one institution but absent in another because of source-system differences. NS-CoT’s separation of vocabulary validity from local executability makes this local dependence visible. It can distinguish between “the concept is not understood,” “the concept is valid but has no local events," and "the concept is valid and executable but yields a small cohort after temporal or exclusion constraints."

This is also where the framework's scalability becomes clinically relevant. The clinical setting does not remain fixed to one drug class, one condition, or one institution. Feasibility workflows must eventually handle new diseases, new drug classes, new laboratory criteria, new local coding practices, and new OMOP CDM instances. Because NS-CoT separates language interpretation, vocabulary grounding, local executability, temporal logic, patient-set algebra, and SQL compilation, new domain-specific knowledge can be added without changing the entire architecture. A new vocabulary mapping, a new drug-class relationship, a new measurement-unit rule, or a new temporal operator can be introduced at the layer where it belongs, while the final output remains an inspectable cohort definition.

This scalability is not only an engineering convenience. It is clinically important because feasibility counts are often compared across sites or across protocol versions. In a multi-institutional setting, the same natural-language criterion may be executable differently depending on local CDM contents, vocabulary version, formulary, measurement coding, and observation-period coverage[10,15]. A framework that exposes the concept sets, temporal anchors, and local executable subsets can support more transparent comparison across sites. It does not guarantee that every institution will produce the same count, but it helps clarify whether differences arise from true population differences, vocabulary mapping differences, or local data availability. This type of artifact-level transparency is also relevant to health systems in which feasibility information is aggregated from hospital EMR, clinical data warehouse (CDW), claims, or CDM-based data sources, because the comparability of site-level counts depends on whether their cohort-defining assumptions are visible.

NS-CoT can also support protocol refinement. Because each criterion is represented as a component of patient-set logic, one can imagine generating sensitivity analyses over cohort definitions. For example, a feasibility analyst may compare counts under a 90-day versus 180-day exposure window, include or exclude descendant concepts, change an exclusion from "ever" to "after diagnosis," or inspect the effect of broadening a drug class. The current study does not implement a full interactive protocol-design interface, but the representation makes such analyses conceptually straightforward. The important point is that the count is linked to modifiable semantic components.

The framework is also useful for human-in-the-loop cohort review. In realistic clinical research workflows, full automation is rarely the safest target[76]. A system should not silently decide every ambiguity in a protocol-derived cohort criterion.

Instead, it should surface the choices that require review. NS-CoT can expose whether a drug mention was treated as an ingredient or a class, whether the class was grounded through a vocabulary graph, whether local executable descendants were found, whether a temporal lower bound was inclusive, and whether an exclusion was anchor-bound. A clinician, informatician, or data analyst can then accept, modify, or reject those decisions before using the cohort count.

This emphasis on reviewable cohort logic is important because human oversight should not be treated as nominal reassurance. Recent critiques of human-in-the-loop oversight in AI-assisted health care argue that simply placing a clinician in the loop may individualize responsibility without giving that clinician the time, information, or institutional support needed to interrogate algorithmic outputs meaningfully. In the context of cohort feasibility, NS-CoT addresses this problem at the level of the artifact: it does not merely ask a human to approve a final count, but exposes the concept sets, temporal anchors, exclusion scope, SQL, and trace metadata that make substantive review possible.

This has implications for regulatory and audit-facing environments, although the distinction must be stated carefully. NS-CoT does not itself provide regulatory compliance, and the artifacts generated by the system should not be interpreted as formal regulatory audit trails. However, the architecture is aligned with the kinds of documentation and traceability that are needed when cohort counts influence protocol design, feasibility assessment, or post hoc review[98–101,103–108,124]. A reviewer may ask not only what count was produced, but how the count was produced. NS-CoT is designed to preserve the intermediate information needed to reconstruct the answer.

The clinical utility of NS-CoT therefore lies in making feasibility computation explainable at the level of cohort semantics. It supports a shift from an opaque count question—"How many patients satisfy this protocol-derived cohort criterion?"—to a reviewable feasibility question: "Under this concept set, this anchor date, this time window, and this exclusion rule, how many patients satisfy the criterion, and how did each component change the patient set?"

This shift is important because feasibility estimates are not neutral numbers. They influence whether a trial design appears practical, whether eligibility criteria are revised, whether a site is considered viable, and whether downstream analytics are built on a well-specified patient population. A count that cannot be explained is fragile. A count linked to an explicit cohort definition can be reviewed, corrected, compared, and reused.

Within the evaluated scope, NS-CoT provides this capability for selected OMOP-expressible eligibility criteria, especially condition and drug exposure criteria involving single-anchor temporal inclusion and exclusion. Its current clinical utility should therefore be understood as foundational rather than comprehensive. It does not replace clinician judgment, chart review, protocol review, or regulatory validation. It provides the computable cohort-definition layer that those later steps require. For clinical trial feasibility, that layer is essential: before a patient population can be predicted, simulated, recruited, or justified, it must first be defined in a way that is explicit, executable, and inspectable.

## 6.6. Limitations

This study should be interpreted within a deliberately defined scope. The study does not claim to solve all clinical trial eligibility interpretation or final enrollment screening. The evaluation target was functional fidelity to predefined OMOP-executable cohort semantics. In other words, the study tested whether NS-CoT could reproduce declared cohort logic over an OMOP CDM instance, not whether that logic established final clinical eligibility or clinical truth.

The primary executable-reference benchmark evaluations, CQ-Bench and FOEM, were performed on SYN-SNUH, a synthetic OMOP CDM database. This setting allowed controlled benchmark construction and deterministic reference-count derivation, but it does not replace validation in real institutional data. COVIC extended the evaluation to three OMOP schemas treated as candidate sites, but this should still be interpreted as a multi-schema feasibility simulation rather than prospective validation across real independent institutions. The NCT case studies used real protocol-derived eligibility material, but they were exploratory traceability demonstrations rather than chart-reviewed clinical validation studies.

The deterministic person-set verification should be interpreted within the same boundary. It compared SQL-derived person_id sets obtained by re-executing saved reference and NS-CoT SQL artifacts; it did not perform chart review or clinical adjudication of those persons. The finding that exact count agreement did not conceal different person_id sets applies only to the evaluated executable queries and under the same synthetic OMOP CDM instance. It should therefore be understood as verification of executable patient-set agreement, not as validation of clinical eligibility.

The benchmark design also imposes limitations. CQ-Bench was internally constructed to evaluate the semantic components targeted by this study, including first-diagnosis anchoring, post-diagnosis drug exposure, temporal exclusion, and negative-control behavior. This makes CQ-Bench appropriate for testing the proposed formalization, but it also means that its results should not be interpreted as performance over the full distribution of possible OMOP cohort queries. FOEM provided an external template-derived benchmark, but the exact-count analysis used only the scalar-count evaluable subset. Grouped or distributional FOEM queries were not evaluated with group-wise output metrics. Future work should therefore evaluate broader query distributions, additional external benchmarks, and grouped or longitudinal output types.

The current implementation focuses on structured OMOP-expressible feasibility-oriented cohort criteria with inclusion and exclusion constraints. The strongest evidence concerns condition and drug exposure criteria, especially single-anchor temporal inclusion and exclusion. Criteria that depend on unstructured clinical notes, imaging, pathology reports, subjective physician judgment, or complex multi-anchor temporal chains remain outside the evaluated scope. Likewise, measurement, procedure, visit, observation, and device criteria can be represented by the architecture, but they require further domain-specific validation before being treated as equally mature components. In particular, although nearly ubiquitous in real eligibility criteria, demographic filtering was a weakness of the current implementation (L2 strict $R_0$ 3/18 versus 8/18); the errors were boundary errors rather than scale or zero/non-zero errors ($R_0$@5% 17/18, zero-form 18/18), so cohort scale and feasibility direction were preserved while exact demographic alignment was not, and strengthening exact demographic predicates is a concrete implementation priority.

A further limitation concerns the interpretation of the count itself. The cardinality produced by NS-CoT is an estimate of patients who *potentially* satisfy the computable subset of a criterion in the institutional OMOP CDM. It is not a count of patients who are operationally available for recruitment, contactable, consented, or enrolled. Data-recorded eligibility, actual accessibility, a physician's willingness to approach a patient, patient consent, screen-failure outcomes, and competing trials are distinct stages, and the present count addresses only the earliest, data-computable stage of this funnel[1,4,5,18].

Relatedly, the count is derived from retrospective or current OMOP CDM records and does not predict prospective enrollment. A patient who satisfied a criterion in the past is not necessarily a future trial participant, and feasibility-stage prevalence does not by itself determine accrual rate. Projecting enrollment would additionally require incidence, visit continuity, consent rates, screen-failure rates, competing-trial load, and site operational capacity, none of which are modeled here[1,5,46,48].

The validity of any individual count is also conditional on the fitness-for-purpose of the underlying OMOP CDM instance. Because the primary scalar-count benchmarks used SYN-SNUH, this dependence was controlled rather than characterized in those benchmark evaluations. In real institutional data, ETL quality, vocabulary-mapping coverage, measurement and unit completeness, observation-period coverage, and source-capture patterns can all change the resulting count, and missing records may bias estimates in either direction depending on whether the affected criterion plays an inclusion or an exclusion role. A site-level fitness-for-purpose assessment would therefore be required before operational use[14,16,94,125].

The representation is also designed to improve portability by grounding cohort logic in standard vocabularies, domain routing, temporal rules, and patient-set algebra rather than in site-specific SQL. However, the present study does not demonstrate unmodified multi-institutional execution or result transportability: the primary scalar-count benchmarks were conducted in a controlled synthetic CDM environment, and COVIC provided only a multi-schema simulation rather than prospective execution across independent institutional OMOP CDM sites[10,15,60,126].

The evaluation also used a single LLM backend and configuration for the main paired comparison. This design controlled the model setting across NS-CoT and the baseline, but it does not establish robustness across LLM vendors, model versions, decoding settings, or prompt variants. In addition, the empirical comparison evaluated complete system architectures rather than isolating each NS-CoT component through a formal ablation study. Therefore, the results support the value of the integrated neuro-symbolic framework, but they do not quantify the marginal contribution of each individual module, such as typed decomposition, local executable grounding, temporal lowering, or deterministic SQL compilation. Nonetheless, because both systems used the same language model and execution environment, the L4 and L5 advantages cannot be attributed to a stronger interpreter and are most consistent with the symbolic layer rather than language-model capability; this is an architectural argument by shared-LLM control rather than a formal ablation.

The protocol-derived clinical trial case studies should also be interpreted within a limited scope. They were not designed as a formal benchmark and were not included in the paired statistical comparison. They illustrate traceability, component-level computability, and failure localization in selected ClinicalTrials.gov criteria, but they do not establish clinical eligibility validity or general performance across trial domains. In addition, criteria involving measurement recency, intraoperative timestamps, or active medication status require domain-specific validation before being treated as mature executable components.

Finally, NS-CoT reduces but does not eliminate the uncertainty of natural-language interpretation. The symbolic layer makes cohort execution reproducible after typed and grounded semantics have been declared. However, if the initial interpretation assigns the wrong domain, anchor, polarity, or temporal relation, the deterministic compiler will faithfully execute the wrong interpretation. For this reason, NS-CoT should be understood as a framework for inspectable and reproducible feasibility-oriented cohort computation, not as a substitute for clinical judgment or site-level operational assessment.

## 6.7. Future Directions

The next step is not simply to increase the size or coverage of NS-CoT. The more important direction is to extend the idea of semantic fidelity across the clinical data pipeline. The present study focused on one part of that pipeline: preserving meaning when site-feasibility patient-count questions become executable OMOP cohort logic. The broader research question is what happens whenever clinical meaning is operationalized as a concept set, cohort definition, SQL query, model input, patient set, or evidence claim. Future work should therefore treat semantic fidelity not as a narrow property of query generation, but as an infrastructure problem for trustworthy clinical computation.

A first direction is to measure semantic loss in standardized clinical data. Such metrics would be particularly relevant when the same feasibility criterion is executed across multiple OMOP CDM sites. OMOP CDM gives institutions a shared structure, but a shared structure does not guarantee shared meaning. Two hospitals may both use OMOP CDM and run the same cohort query, yet produce patient sets that differ because concepts were mapped differently, domains were used differently, or local events were encoded differently. Future work should therefore develop semantic data quality metrics: not only whether data conform to the schema, but whether the same clinical construct is represented comparably across sites. This direction extends prior work on EHR data-quality dimensions for clinical-research reuse and OHDSI data-quality evaluation from structural fitness-for-purpose toward semantic comparability[125,127]. This would move data quality assessment from structural validity toward semantic comparability.

A second direction is to extend semantic preservation beyond the condition-drug patterns evaluated here. The same architecture can be expanded to measurements, procedures, observations, visits, devices, and outcomes, but this is not a matter of adding more prompts. It requires domain-specific grounding logic, value and unit handling, local executability checks, and temporal operators appropriate to each OMOP domain. Laboratory criteria, for example, require LOINC grounding, numeric thresholds, UCUM unit normalization, and event-value predicates. Procedure and device criteria require different assumptions about occurrence, timing, and local coding practice. The architecture can support these extensions because its layers are explicit, but each domain will require separate validation before it can be treated as mature.

A third direction is richer temporal reasoning. The current framework focuses on single-anchor temporal logic, such as drug exposure after diagnosis or absence of exposure after diagnosis. Many real site feasibility patient-count questions with protocol-derived inclusion and exclusion constraints are more sequential: diagnosis, then treatment initiation, then laboratory response, then sustained control or outcome follow-up. Supporting such criteria requires intermediate events to become new anchors. Future work should therefore extend NS-CoT from patient-level single-anchor predicates to event-level temporal chains with produced anchors and anchor propagation.

A fourth direction is multi-site scalability and transportability. Although the primary scalar-count benchmarks were evaluated on SYN-SNUH and COVIC provided a multi-schema simulation, the proposed approach has not yet been validated through prospective execution across real independent institutional OMOP CDM sites. It is grounded in OMOP CDM, standard vocabularies, event tables, observation periods, and concept relationships. The natural-language layer may differ by language and clinical setting, but the execution layer can remain vocabulary-grounded and CDM-compatible. In distributed OMOP-based research networks, the same protocol-derived cohort criterion should not merely return a count at each site; it should return a count with its semantic construction exposed. Sites should be able to see which concepts were vocabulary-valid, which were locally executable, which temporal anchors existed, and where local data availability changed the result[10,15,60,126].

This multi-site direction is important because real-world evidence and clinical AI increasingly depend on distributed data. The difficulty is not only that data stay local. It is that meaning may change locally. A drug class may be well defined in one vocabulary but encoded differently in another institution. A measurement may be recorded with different units. A diagnosis may be mapped to different standard concepts. If these differences remain invisible, multi-site feasibility counts and downstream models may appear comparable while relying on different clinical constructs. NS-CoT points toward a way to make those semantic differences visible.

A fifth direction is to quantify downstream impact. Semantic errors are not merely technical mistakes. A wrong cohort definition may change recruitment estimates, screening burden, time-to-enrollment, model calibration, cost-effectiveness results, and policy decisions. Future work should connect semantic fidelity to downstream consequences: how much does a temporal-window error change the number needed to screen? How does an overly broad drug-class definition change trial feasibility? How does cohort-definition drift affect model performance across sites? This would translate semantic fidelity from an informatics property into a measurable clinical, economic, and policy concern.

This future direction is also consistent with recent discussions on the relationship between evidence-based medicine and artificial intelligence. Cruz-Suarez et al. argue that AI is beginning to reshape not only clinical prediction, but also the production, synthesis, and application of medical evidence[128]. From this perspective, semantic fidelity is not a narrow data-processing concern. It is an upstream condition for trustworthy AI-enabled evidence generation: if the clinical construct is distorted when it becomes a concept set, cohort definition, SQL query, or model input, the resulting evidence may be computationally precise but clinically misaligned.

A sixth direction is to develop NS-CoT from a cohort-generation framework into a semantic specification and verification layer for AI-generated cohort artifacts. As language models and agents improve, SQL generation, concept grounding, and self-checking may become faster and more capable. This does not eliminate the need for declared executable semantics. Even if an agent can propose concepts, anchors, proxy variables, and SQL on its own, those proposals still need to be externalized as an inspectable semantic object before the resulting count can be trusted as feasibility evidence. Future systems should therefore not only generate cohort artifacts, but also expose the declared patient-set semantics they implement and allow those semantics to be reviewed, revised, and independently checked. In this role, NS-CoT can develop from a cohort-generation framework into a specification and verification layer for AI-generated feasibility counts.

Future versions of NS-CoT should also support explicit uncertainty and abstention behavior. When a criterion cannot be grounded reliably, when local executable concepts are absent, or when a temporal expression admits multiple plausible operationalizations, the system should be able to surface uncertainty rather than force a single executable interpretation. This direction is consistent with recent calls for clinical AI systems to recognize and communicate when they do not know[129].

The liver-transplant case study illustrates a related future direction for proxy-selection research. In the present study, this case study was used to localize a computability boundary: the clinically meaningful anchor, graft reperfusion, was not directly available in the structured CDM data. This does not imply that the criterion has no clinical meaning, nor that any single proxy is automatically correct. Rather, it shows that multiple CDM-executable interpretations may be defensible, including a procedure-date anchor, an encounter-bounded perioperative proxy, or an anesthesia-record-derived timestamp if available.

Future work could extend this observation by representing such alternatives as competing executable interpretations and comparing their patient-set and count-level consequences. Expert-reviewed chart or anesthesia-record evidence may serve as comparison evidence in a specific institutional context, but it should not be treated as a universally self-evident gold-standard cohort. The practical goal is to support context-specific selection of a canonical executable interpretation by making explicit the semantic loss, expected bias, computability status, and site-level count differences associated with each proxy. Such proxy-selection studies would extend the present framework from preserving declared semantics to evaluating how alternative declared semantics should be selected before they are used for site selection.

# Chapter 7. Conclusion

This study addressed a central medical informatics problem in clinical trial site feasibility assessment: how to determine, in a transparent and reproducible way, how many patients at a candidate institution may potentially satisfy protocol-derived eligibility constraints. The main argument of this work is that this number, often treated operationally as the site-level patient count, is not a raw database output. It becomes meaningful only after a patient set has been explicitly defined. In other words, the fundamental problem is not simply how to generate SQL from a natural-language criterion, but how to specify what patient set that SQL is expected to denote over an OMOP CDM database.

Natural-language eligibility criteria and OMOP CDM represent clinical information in fundamentally different forms. Protocol criteria are written for human interpretation and often contain ambiguity, implicit assumptions, temporal dependencies, exclusions, and context-specific clinical meaning. OMOP CDM, by contrast, represents clinical evidence as discrete standardized events with concepts, domains, dates, values, units, and person identifiers. Because there is no natural one-to-one mapping between these two representation spaces, the same criterion can produce different patient sets depending on concept selection, vocabulary hierarchy expansion, local executability, temporal anchoring, exclusion scope, value constraints, and patient-set operations. This is why site-feasibility patient counting should be framed as a problem of executable patient-set semantics rather than as a direct text-to-SQL problem.

To address this problem, this study proposed NS-CoT, a neuro-symbolic methodology for transforming feasibility-oriented natural-language criteria into OMOP-executable cohort logic. In this framework, the language model is used for semantic interpretation, criterion decomposition, and typed intermediate representation. The execution semantics are then handled by deterministic symbolic modules, including OMOP vocabulary grounding, hierarchy expansion, local executable concept filtering, temporal and exclusion reasoning, patient-set algebra, and SQL compilation. This division of labor is central to the contribution of the work. It uses the language model where flexible interpretation is useful, while making the executable semantics explicit, inspectable, and reproducible before a count is interpreted as feasibility evidence.

This framing does not assume that a free-text protocol criterion has a single self-evident gold-standard patient set prior to interpretation. In site feasibility assessment, different clinicians, coordinators, sponsors, or data scientists may reasonably operationalize the same criterion differently, especially when local concept mapping, structured data availability, proxy variables, temporal anchors, or observation windows differ. The intersection of such interpretations may be overly conservative, and their union may be overly permissive; neither is automatically the unique truth of the criterion. The role of NS-CoT is therefore not to discover a hidden true cohort. Its role is to externalize the chosen interpretation as a declared executable patient-set semantics object that can be inspected, revised, and discussed before the resulting count is treated as feasibility evidence.

The evaluation showed that this formalization is most valuable when cohort semantics cannot be reduced to simple concept lookup. In CQ-Bench, NS-CoT achieved stronger exact-count fidelity than the direct SQL-generation baseline, especially for queries involving temporal anchoring, post-diagnosis drug exposure, and anchor-bound exclusion. In FOEM, both systems performed well, and the difference was not statistically significant, showing that NS-CoT should not be interpreted as a universally superior query-generation method across all OMOP query distributions. Instead, its value is concentrated in settings where the meaning of the cohort depends on preserving structured semantic decisions before SQL execution. Person-set verification further showed that exact count agreement did not conceal hidden patient-set mismatch in the evaluated cases, supporting the use of executable reference semantics as a functional evaluation target.

COVIC extended the evaluation from single-database count reproduction to a multi-site feasibility simulation. By applying the same externally derived executable criterion components across multiple OMOP schemas, this evaluation asked whether system-generated count preserved reference-defined counts, patient sets, site-feasibility labels, and site rankings. The first-pass COVIC results exposed robustness gaps in the initial NS-CoT implementation, particularly under broad external criterion-unit handling, vocabulary grounding, local executability, and multi-schema execution. However, the repair analysis showed that explicit intermediate artifacts could localize failures to specific layers and support targeted correction. This result is important because site feasibility is not only a question of whether one count can be reproduced in one database. It is also a question of whether the same criterion can be applied across candidate sites in a way that preserves the feasibility information needed for site comparison.

The protocol-derived clinical trial case studies further illustrated the role of traceability in realistic protocol-facing scenarios. These cases were not designed as statistical benchmarks. Rather, they demonstrated how the framework exposes computable components, proxy-dependent temporal anchors, local executability limits, executable-empty concept sets, and other boundaries that matter for clinical review. In this sense, the output of NS-CoT is not merely a count. It is a feasibility evidence artifact that records how the count was produced: which concepts were selected, which descendants were included, which local concepts were executable, which temporal anchors and exclusion scopes were applied, which patient-set operations were used, which SQL was generated, and where failures or partial execution occurred.

The clinical utility of this work lies in reviewability rather than prediction. NS-CoT does not predict who will ultimately enroll in a trial, does not replace clinical eligibility adjudication, and does not automate patient recruitment. Its utility is upstream: it makes the basis of site-feasibility patient counting explicit. Instead of asking only how many patients are present, the framework enables a more clinically meaningful question: how many patients are counted under this concept set, this temporal anchor, this observation window, this exclusion rule, and this executable interpretation? This supports protocol refinement, human-in-the-loop review, debugging of cohort definitions, and comparison of candidate sites under declared assumptions. It can also help distinguish whether differences in site-level patient count arise from true population differences, vocabulary mapping choices, local data availability, or implementation decisions.

The main contribution of this study is therefore a reframing of OMOP CDM-based clinical trial feasibility estimation. Site-level patient count is defined not as an opaque SQL result, but as the cardinality of a declared executable patient set. Natural-language criteria are transformed into typed, grounded, temporal, exclusion-aware patient-set logic. Criteria are represented not as a flat list of SQL predicates, but as patient-set operations such as intersection, union, difference, and anchor-bound absence. The resulting artifact preserves the assumptions and execution path behind the count. The evaluation framework also extends beyond SQL exact match by incorporating liveness, exact and approximate count fidelity, zero-form agreement, person-set fidelity, semantic-stratum analysis, failure localization, site-feasibility discrimination, and site-ranking preservation.

This work also has implications for the future use of language models and agents in clinical research informatics. As these systems become more capable of concept grounding, SQL generation, and self-checking, the central question will not disappear; it will move upstream. Even if an agent can propose concepts, anchors, proxy variables, and SQL on its own, its output still needs to be externalized as a declared executable patient-set semantics object that can be inspected, revised, and independently checked. NS-CoT should therefore be understood not only as a cohort-generation framework, but also as a step toward a semantic specification and verification layer for AI-generated cohort artifacts.

The scope of the current work remains intentionally bounded. The strongest evidence concerns selected OMOP-expressible condition and drug exposure criteria, executable reference semantics, and structured data available in the evaluated CDM environments. The study does not establish chart-reviewed phenotype validity, prospective recruitment performance, regulatory-grade auditability, or full transportability across independent real-world institutions. Future work should extend the framework to broader clinical domains, richer temporal chains, laboratory and measurement logic, unstructured clinical text, multimodal evidence, prospective multi-site OMOP deployments, and downstream recruitment outcomes.

In conclusion, reliable OMOP CDM-based site feasibility assessment requires more than the generation of executable SQL. It requires making explicit the semantic commitments that determine which patients are counted: the concept sets, local executability assumptions, temporal anchors, proxy choices, exclusion scopes, observation windows, value constraints, and patient-set operations. The contribution of this study is to make those commitments executable, reviewable, and traceable as declared patient-set semantics. The resulting site-level count is not presented as an unqualified clinical truth or a hidden gold-standard cohort. It is a transparent feasibility evidence claim: a statement about which patient set was counted, under which assumptions, through which executable logic, and with what limitations.

**A Note from the Author**

This research began with a question about how medical evidence is constructed: how clinical questions become computable definitions, and how the choices involved shape the conclusions we draw.

Working with clinical data brought this question into focus. Decisions about concepts, mappings, time windows, and inclusion criteria influence who becomes visible in an analysis and what its results can support. I believe these decisions deserve to be treated as part of the scientific argument, made explicit and open to examination.

This work matters deeply to me. It brings together questions that have shaped my thinking and the practical work of building systems through which those questions can be investigated. I am sharing it because I believe its methods and observations are worth examining, testing, and developing further.

The findings have a scope, and questions remain unresolved. Making that scope clear is part of the responsibility of research. By making this work publicly available, I want others to be able to follow its reasoning, challenge its assumptions, and build on what they find useful. I welcome that engagement, including findings that lead me to reconsider my own conclusions.

My deepest gratitude goes to the person whose insight brought an important problem in clinical practice into focus, and whose guidance and support enabled me to investigate it and develop an approach to addressing it through medical informatics.